\documentclass{book}
\usepackage{bm,amssymb,graphicx,xcolor,srcltx,multicol}
\usepackage{makeidx,authorindex}
\newdimen\inboxwidth \inboxwidth = \textwidth \advance \inboxwidth by -3mm
\newcommand{\kB}{k_{\rm B}}
\newcommand{\dR}{{\rm d}}
\newcommand{\eR}{{\rm e}}
\newcommand{\iR}{{\rm i}}
\newcommand{\mica}{Mathematica\,\textregistered}
\newcommand{\ave}[1]{\left\langle#1\right\rangle}
\newcommand{\Dbra}[1]{\left\langle#1\right|}
\newcommand{\Dket}[1]{\left|#1\right\rangle}
\newcommand{\Dbraket}[2]{\langle#1|#2\rangle}
\newcommand{\Qcommu}[2]{[#1,#2]}
\newcommand{\Qcommux}[2]{\left[#1,#2\right]}
\newcommand{\Qantico}[2]{\{#1,#2\}}

\renewcommand{\cite}[1]{\aicite{#1}}

\title{Quantum Field Theory\\as a Faithful Image of Nature}
\author{Hans Christian \"Ottinger}

\begin{document}
\bibliographystyle{plain}

\maketitle
\tableofcontents

\chapter*{Preface} \markboth{PREFACE}{PREFACE}
\addcontentsline{toc}{section}{Preface}
``ALL men by nature desire to know,'' states Aristotle in the famous first sentence of his \emph{Metaphysics}.\footnote{W.\,D.~Ross' translation of this major work, which initiated an entire branch of philosophy, can be found on the internet (classics.mit.edu/Aristotle/metaphysics.html); nowadays Aristotle would clearly say ``ALL human beings by nature desire to know.''} Knowledge about fundamental particles and interactions, that is, knowledge about the deepest aspects of matter, is certainly high if not top on the priority list, not only for physicists and philosophers. The goal of the present book is to contribute to this knowledge by going beyond the usual presentations of quantum field theory in physics textbooks, both in mathematical approach and by critical reflections inspired by epistemology, that is, by the branch of philosophy also referred to as the theory of knowledge.

This book is particularly influenced by the epistemological ideas of Ludwig Boltzmann: ``\ldots it cannot be our task to find an absolutely correct theory but rather a picture that is as simple as possible and that represents phenomena as accurately as possible'' (see p.~91 of \cite{BoltzmannPP}). This book is an attempt to construct an intuitive and elegant image of the real world of fundamental particles and their interactions. To clarify the word \emph{picture} or \emph{image}, the goal could be rephrased as the construction of a genuine \emph{mathematical representation} of the real world.

Consciously or unconsciously, the construction of any image of the real world relies on personal beliefs. I hence try to identify and justify my own personal beliefs thoroughly and in various ways. Sometimes I rely on philosophical ideas, for example, about space, time, infinity, or irreversibility; as a theoretical physicist, I have a limited understanding of philosophy, but that should not keep me from trying my best to benefit from philosophical ideas. More often I rely on successful physical theories, principles or methods, such as special relativity, quantum theory, gauge invariance or renormalization. Typically I need to do some heuristic mathematical steps to consolidate the various inputs adopted as my personal beliefs. All these efforts ultimately lead to an image of nature, in the sense of a mathematical representation, but they are not part of this image. The final mathematical representation should convince by logical clarity, mathematical rigor, and natural beauty.

Emphasis on the importance of beliefs, even if they are justified by a variety of philosophical and physical ideas, may irritate the physicist. The philosopher, on the other hand, is used to the definition of knowledge as \emph{true justified belief}. How can one claim truth for one's justified beliefs? This happens by confronting an image of nature with the real world.

According to Pierre Duhem \cite{Duhem}, known to thermodynamicists from the Gibbs-Duhem relation, and the analytic philosopher Willard Van Orman Quine \cite{Quine51}, only the whole image rather than individual elements or hypotheses can be tested against the real world. The confrontation of a fully developed image with the real world depends on all its background assumptions or an even wider web-of-belief, including the assumed logics (confirmation holism). Following Boltzmann's approach of ``deductive representation'' (see p.~107 of \cite{BoltzmannPP}), the present book makes an attempt to show how such a testable whole image of fundamental particle physics can be constructed within the framework of quantum field theory.

The focus of this book is on conceptual issues, on the clarification of the foundations of quantum field theory, and ultimately even on ontological questions. For our intuitive approach, we choose to go back to the origins of quantum field theory. In view of the many severe problems that had to be overcome on the way to modern quantum field theory, that may seem to be naive to the experts. However, with the deep present-day knowledge and with philosophical guidance, the intuitive origins can nicely be developed into a perfectly consistent image of the real world. On the one hand, there is a price to pay for this: practical calculations, in particular perturbation methods, may be less elegant and more laborious than in other approaches. Symbolic computation is the modern response to this challenge. On the other hand, there is a promising reward: a new stochastic simulation methodology for quantum field theory emerges naturally from our approach.

Hopefully, the present book motivates physicists to appreciate philosophical ideas. Epistemology and the philosophy of the evolution of science often seem to lag behind science and to describe the developments \emph{a posteriori}. As philosophical arguments here have a profound influence on the actual shaping of an image of fundamental particles and their interactions, our development should stimulate the curiosity and imagination of physicists.

This book can be used as an introductory textbook on quantum field theory for students of physics, as a supplementary resource in conjunction with one of the more comprehensive mainstream textbooks, or as a monograph for philosophers and physicists interested in the epistemological foundations of particle physics. The benefits of an approach resting on philosophical foundations is twofold: the reader is stimulated to critical thinking and the entire story flows very naturally, thus removing the mysteries from quantum field theory.

\vskip14pt\noindent
{\em\footnotesize Z{\"u}rich and Rafz}
\hfill
{\sc\small Hans Christian {\"O}ttinger}\\[-2pt]
{\em\footnotesize February 13, 2015}\\[1.2cm]

\chapter*{Acknowledgements} \markboth{ACKNOWLEDGEMENTS}{ACKNOWLEDGEMENTS}
\addcontentsline{toc}{section}{Acknowledgements}
I am indebted to Martin Kr\"oger for countless stimulating discussions and constructive comments during all stages of writing this book. Comments of Antony Beris and Jay Schieber helped to clarify the philosophical part at an early stage. The physical part was improved with the help of remarks by Pep Espa\~nol, Bert Schroer and Marco Schweizer. Discussions with Vlasis Mavrantzas and Alberto Montefusco helped to clarify a number of specific problems.

I would not have embarked on this book project without the inspiration from Suzann-Viola Renninger's philosophy courses. For the first time in my life I got the impression that philosophical ideas can support me in doing more solid and more beautiful work in physics. Her comments and questions on the philosophical part of this book added depth and substance.

I am very grateful that several philosophers with an interest in quantum field theory looked critically at a first version of this manuscript and provided encouraging and constructive feedback. In particular, I would like to thank Simon Friederich, Michael Esfeld, Antonio Augusto Passos Videira and Bryan W.\ Roberts for all their helpful comments.

\chapter{Approach to quantum field theory}\label{chapQFTapproach}
In this introductory chapter, which actually is the core of the entire book, we make a serious effort to bring together a variety of ideas from philosophy and physics. We first lay the epistemological foundations for our approach to particle physics. These philosophical foundations, combined with tools borrowed from the amazingly sophisticated, but still not really satisfactory, mathematical apparatus of present-day quantum field theory and augmented by some new ideas, are then employed to develop a mathematical representation of fundamental particles and their interactions. In this process, also the relation between `fundamental particle physics' and `quantum field theory' is going to be clarified.

This first chapter consists of two voluminous sections: a number of \emph{philosophical contemplations} followed by a discussion of \emph{mathematical and physical elements}. The reader might wonder why these massive sections are not presented as two separate chapters. The reason for keeping them together is the desire to emphasize the intimate relation between these two sections: the selection and construction of the mathematical and physical elements is directly based on our philosophical contemplations. The beauty of the entire approach is a fruit of this intimate relationship.

The presentation of the material in this chapter is based on the assumption that the reader has a basic working knowledge of linear algebra and quantum mechanics. If that is not the case, the reader might want to consult the equally entertaining and serious introduction \emph{Quantum Mechanics: The Theoretical Minimum} by Susskind and Friedman \cite{SusskindFriedman}. An almost equation-free discussion of the history and foundations of quantum mechanics can be found in the book \emph{Einstein, Bohr and the Quantum Dilemma: From Quantum Theory to Quantum Information} by Whitaker \cite{WhitakerA}. Complex vector spaces, Hilbert space vectors and density matrices for describing states of quantum systems, bosons and fermions, canonical commutation relations, Heisenberg's uncertainty relation, the Schr\"odinger and Heisenberg pictures for the time evolution of quantum systems, as well as a basic idea of the measurement process are all referred or alluded to in Section~\ref{secphilcontemp}; these basics of quantum mechanics are recapped only very briefly in Section~\ref{secMPelements}. The crucial construct of Fock spaces is explained in a loose way in Section~\ref{sectioninfinity} and later elaborated in full detail in Section~\ref{sectionFock}.

According to Henry Margenau \cite{Margenau}, ``[the epistemologist] is constantly tempted to reject all because of the difficulty of establishing any part of \index{Reality}reality'' (p.\,287). But, again in the words of Margenau, ``it is quite proper for us to assume that we know what a dog is even if we may not be able to define him'' (p.\,58). In this spirit, we try to resist the temptation of raising significantly more and more difficult questions than we can possibly answer, no matter how fascinating these questions might be. Philosophy shall here serve as a practical tool for doing better physics. We try to use philosophy in a relevant and convincing way, but we are certainly not in a position to do frontier technical research in philosophy.

\section{Philosophical contemplations}\label{secphilcontemp}
We begin this chapter with some general remarks on the methodology of science, where we heavily rely on the epistemological ideas of Ludwig Boltzmann. The representation of space and time is then considered in the light of Immanuel Kant's famous ideas. We further consider the more specific issues of infinity and irreversibility, and we conclude with some contemporary philosophic considerations about quantum field theory in its present form(s).

As a guideline for developing the mathematical and physical elements in Section~\ref{secMPelements} we condense our philosophical contemplations of the present section into four metaphysical postulates. \index{Metaphysical postulates}Metaphysical principles may not be particularly popular among contemporary physicists but, consciously or unconsciously, they play an essential role in any science. We here prefer the conscious approach, which is eloquently recommended by Henry Margenau in his philosophy of modern physics on pp.\,12--13 of \cite{Margenau}:

\begin{quotation}
To deny the presence, indeed the necessary presence, of metaphysical elements in any successful science is to be blind to the obvious, although to foster such blindness has become a highly sophisticated endeavor in our time. Many reputable scientists have joined the ranks of the exterminator brigade, which goes noisily about chasing metaphysical bats out of scientific belfries. They are a useful crowd, for what they exterminate is rarely metaphysics---it is usually bad physics. Every scientist \emph{must} invoke assumptions or rules of procedure which are not dictated by sensory evidence as such, rules whose application endows a collection of facts with internal organization and coherence, makes them simple, makes a theory elegant and acceptable.
\end{quotation}

In his verbose philosophical exploration of science and nature, also Simon Altmann \cite{Altmann} stresses the importance of \index{Metaphysical postulates}metaphysical principles: ``\ldots\ science requires the use of certain \emph{normative principles} that have a much greater generality than \emph{physical laws} \ldots'' (see p.\,30 of \cite{Altmann}). He actually distinguishes between metaphysical and meta-physical normative principles, where the former are beyond experience and the latter are directly wedded to experience (without actually being \emph{derivable} from it). We here use the conventional spelling but nevertheless claim that our metaphysical postulates are grounded in experience. Metaphysical postulates are used as a guideline for theory development, but they are themselves based on reflections on he evolving knowledge of physics. In the words of Cao (see p.\,267 of \cite{Caosr}), ``It [metaphysics] can help us to make physics intelligible by providing well-entrenched categories distilled from everyday life and previous scientific experiences. But with the advancement of physics, it has to move forward and revise itself for new situations: old categories have to be discarded or transformed, new categories have to be introduced to accommodate new facts and new situations.''\index{Metaphysics}

It is important to realize that the metaphysical postulates to be developed in this section are not meant as rigorous fundamental principles, but rather as helpful intuitive guidelines. The reader should interpret them with benevolence and should consider them as an invitation to personal reflections, with the goal of increasing the awareness of how we are doing modern science. I will, however, try to elaborate how these metaphysical postulates affect the present approach to quantum field theory in a deep and decisive way. The style of the presentation is a compromise between the philosopher's cherished culture of multifaceted discourse and the physicist's impatient desire to get to the core of the story.

\subsection{Images of nature}\label{secimagesBoltz}
Around 1900, the University of Vienna was a vivid center for agitated discussions about physics and philosophy, where the existence or nonexistence of atoms was one of the big topics. From 1895 to 1901, Ernst Mach held the newly created ``chair for philosophy, especially for the history and theory of the inductive sciences.'' From 1893 to 1900 and from 1902 to 1906, Ludwig Boltzmann was the professor of theoretical physics at the University of Vienna. The fact that Boltzmann left Vienna and returned only after the retirement of Mach was not just a matter of coincidence but a consequence of enervating quarrels with Mach and other colleagues. In 1897, after a lecture by Boltzmann, who was a leading proponent of atomic theory, at the Imperial Academy of Sciences in Vienna, Mach laconically declared: ``I don't believe that atoms exist!'' In 1903, while waiting for the faculty to propose candidates for Mach's replacement, the ministry gave Boltzmann the gratifying assignment to lecture every semester for two hours per week on the ``philosophy of nature and methodology of the natural sciences'' to fill the gap that had existed since Mach's retirement (actually Mach hadn't been teaching after a stroke he suffered in 1898). Boltzmann's philosophical lectures attracted huge audiences (some 600 students) and so much public attention that the Emperor Franz Joseph~I (reigning Austria from 1848 to 1916) invited him for a reception at the Palace to express his delight about Boltzmann's return to Vienna. So, Boltzmann was not only a theoretical physicist of the first generation, but also an officially recognized part-time philosopher. For the last years of his life he focused on philosophical ideas to defend his pioneering work on the foundations of statistical mechanics and the kinetic theory of gases, which heavily relied on the existence of atoms.

\begin{figure}
\centerline{\includegraphics[width=6.6cm]{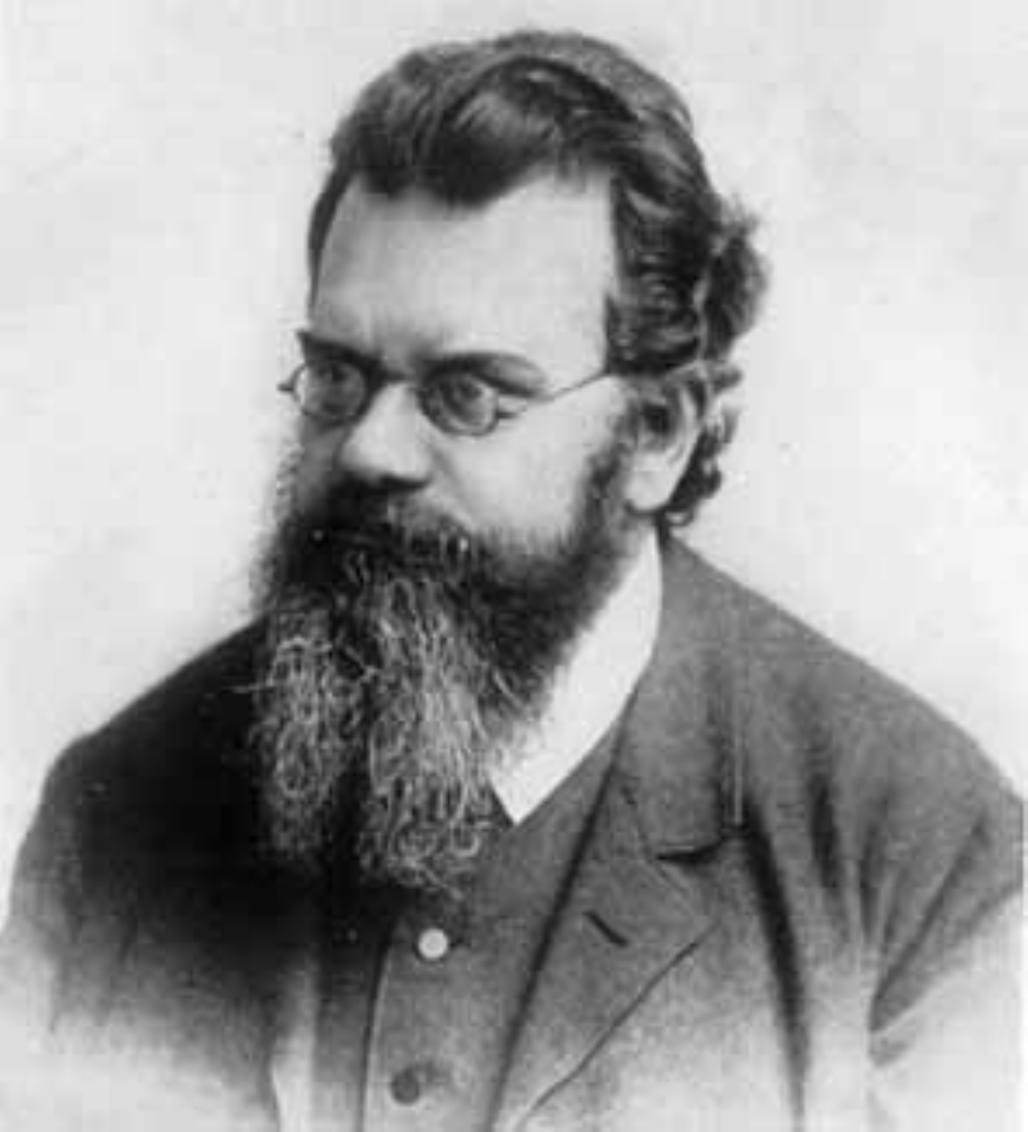}}
\caption[ ]{Ludwig Boltzmann, 1844--1906.} \label{figBoltzmann}
\end{figure}

In the very beginning of his very first philosophical lecture on October 26, 1903, Boltzmann stated that he had written only a single treatise with philosophical content in his entire life. He was referring to the article ``On the Question of the Objective Existence of Processes in Inanimate Nature,'' which had been published in 1897 (see essay 12 in \cite{BoltzmannPS}; an English translation is given on pp.~57--76 of \cite{BoltzmannPP}). However, Boltzmann had already made a number of contributions to the methodology of science that are clearly of epistemological content and would nowadays be classified as philosophical. For the subsequent discussion we actually rely on two such contributions dating from 1899. One of these contributions was an address to the meeting of natural scientists at Munich (``On the Development of the Methods of Theoretical Physics in Recent Times''), the other one a series of lectures given at Clark University in  Worcester (``On the Fundamental Principles and Equations of Mechanics''); both contributions were published in his writings addressed to the public in 1905 (as items 14 and 16 in \cite{BoltzmannPS}, translated in \cite{BoltzmannPP}; all the page numbers in the remainder of this section refer to the English translation \cite{BoltzmannPP} of his writings addressed to the public).

\paragraph*{``On the Development of the Methods of Theoretical Physics in Recent Times''} After describing the evolution of the theory of \index{Electrodynamics}electromagnetism, Boltzmann states, ``Whereas it was perhaps less the creators of the old classical physics than its later representatives that pretended by means of it to have recognised the true nature of things, Maxwell wished his theory to be regarded as a mere picture of nature, a mechanical analogy as he puts it, which at the present moment allows one to give the most uniform and comprehensive account of the totality of phenomena'' (p.\,83). Regarding physical theories as pictures of nature is a very fundamental idea. I prefer to call them \emph{images of nature} because imagination is exactly what theoretical physics should be about, with moral support from Einstein (quote from an interview given in 1929): ``Imagination is more important than knowledge. Knowledge is limited. Imagination encircles the world.'' Or, in my own simple words, imagination creates knowledge. Boltzmann elaborates the standing of images of nature in the following two paragraphs (pp.\,90--91), emanating from the example of the theory of \index{Electrodynamics}electromagnetism:

\begin{quotation}
Maxwell had called Weber's hypothesis a real physical theory, by which he meant that its author claimed objective truth for it, whereas his own account he called mere pictures of phenomena. Following on from there, Hertz makes physicists properly aware of something philosophers had no doubt long since stated, namely that no theory can be objective, actually coinciding with nature, but rather that each theory is only a mental picture of phenomena, related to them as sign is to designatum.

From this it follows that it cannot be our task to find an absolutely correct theory but rather a picture that is as simple as possible and that represents phenomena as accurately as possible. One might even conceive of two quite different theories both equally simple and equally congruent with phenomena, which therefore in spite of their difference are equally correct. The assertion that a given theory is the only correct one can only express our subjective conviction that there could not be another equally simple and fitting image. [Author: Note that here the German word `Bild' is actually translated as `image' rather than `picture.']
\end{quotation}

Images of nature are never meant to be absolutely correct and they should only be expected to cover a certain range of phenomena with a certain degree of accuracy. More complete images can always arise so that we can ask with Margenau (see p.\,171 of \cite{Margenau}): ``But why, after all, should scientific truth be a static concept?'' Or, in a beautiful formulation of William James (see p.\,x of \cite{James}),\footnote{There exist several online versions of this classical collection of writings first published in 1909.} ``The truth of an idea is not a stagnant property inherent in it. Truth \emph{happens} to an idea. It \emph{becomes} true, is \emph{made} true by events.''

Different images can do equally well on a certain range of phenomena, but one of the images may lead to the discovery of new phenomena and hence turn out to be more successful than the other ones, without making them useless. Boltzmann illustrates this point with the theories of \index{Electrodynamics}electromagnetic phenomena developed by Weber and by Maxwell (p.\,83), ``The phenomena known till then were equally well explained by both theories, but Maxwell's went much beyond the old theory [of Weber].'' The idea of electromagnetic waves emerged only from Maxwell's theory replacing long-range interactions by close-range effects, thus leading to a deeper understanding of light and to new technological applications, such as ``an ordinary optical telegraph.'' Also according to the philosopher Paul Feyerabend, ``it must be asserted that the discussion of possibilities and of alternatives to a current theory plays a most important role in the development of our physical knowledge'' (see p.\,233 of \cite{Feyerabend62ip}) and ``\emph{There is no way of singling out one and only one theory on the basis of observation}'' (see p.\,234 of \cite{Feyerabend62ip}). Such a tolerant view about the fruitful coexistence of old and new theories is at variance with Thomas Kuhn's more radical ideas about scientific revolutions (see p.\,98 of \cite{Kuhn}): ``Einstein's theory [of gravity] can be accepted only with the recognition that Newton's was wrong.'' Note that Altmann has criticized Kuhn's restrictive ideas in profound ways (see Chapter~20 of \cite{Altmann}).\footnote{We scientists seem to like Kuhn's ideas because, whenever one of our papers gets rejected, we can feel as the misunderstood heros of a scientific revolution hindered by conservative referees who are not yet ready for a paradigm shift.}

\index{Pluralism, scientific}Boltzmann's theoretical pluralism is the central topic in Videira's analysis \cite{Videira,RibeiroVideira98} of Boltzmann's philosophical works. Videira suggests that, by emphasizing the fundamental distinction between nature and its various representations, this theoretical pluralism is capable of counteracting dogmatic tendencies returning in modern science, for example, in 20th century cosmology. Actually, pluralism should be recognized as an enabling condition for progress in physics.\footnote{A.\,A.\,P.~Videira, private communication (October 2015).} The various images of nature should compete in a \index{Darwinism}Darwinistic sense. The idea of `evolutionary epistemology' has been expressed in a beautifully worded metaphor by van Fraassen (see p.\,40 of \cite{vanFraassen}):
\begin{quotation}
I claim that the success of current scientific theories is no miracle. It is not even surprising to the scientific \index{Darwinism}(Darwinist) mind. For any scientific theory is born into a life of fierce competition, a jungle red in tooth and claw. Only the successful theories survive---the ones which \emph{in fact} latched on to actual regularities in nature.
\end{quotation}

Boltzmann would presumably have questioned the concept of \index{Reality}reality because ultimately we don't even know how to distinguish reality from our various mental representations. As an extreme example, he once made the following oral statement (cited from p.\,213 of \cite{CercignaniB}):
\begin{quotation}
You see, it doesn't make any difference to me if I say that the atomic model is only a picture. I don't mind this. I don't require that they have absolute real existence. I don't say this. `An economic description' Mach said. Maybe the atoms are an economic description. This doesn't hurt me very much. From the viewpoint of the physicists this doesn't make a difference.
\end{quotation}
With a heavy heart, Boltzmann refrains from claiming \index{Reality}reality even for atoms, which are the most important concept in his favorite image of nature. This statement shows how indispensable scientific \index{Pluralism, scientific}pluralism is to Boltzmann. Instead of pluralism one can alternatively speak of an \index{Underdetermination}underdetermination of theories by empirical evidences (see p.\,3 of \cite{Caosr}).

The following paragraphs clarify that Boltzmann's images of nature are meant to be of mathematical character (pp.\,95--96):
\begin{quotation}
Mathematical phenomenology at first fulfils a practical need. The hypotheses through which the equations had been obtained proved to be uncertain and prone to change, but the equations themselves, if tested in sufficiently many cases, were fixed at least within certain limits of accuracy; beyond these limits they did of course need further elaboration and refinement. \ldots

Besides we must admit that the purpose of all science and thus of physics too, would be attained most perfectly if one had found formulae by means of which the phenomena to be expected could be unambiguously, reliably and completely calculated beforehand in every special instance; however this is just as much an unrealisable ideal as the knowledge of the law of action and the initial states of all atoms.

Phenomenology believed that it could represent nature without in any way going beyond experience, but I think this is an illusion. No equation represents any processes with absolute accuracy, but always idealizes them, emphasizing common features and neglecting what is different and thus going beyond experience.
\end{quotation}

How can one justify the fundamental mathematical equations adopted as an image of nature, how can one establish a theory as correct or true? Boltzmann answers these questions by essentially anticipating the ideas nowadays associated with the names of Duhem \cite{Duhem} and Quine \cite{Quine51} (see preface), ``He [Hertz] rightly points out that what convinces us of the correctness of all these equations is not, in mechanics, the few experiments from which its fundamental equations are usually derived, nor, in \index{Electrodynamics}electrodynamics, the five or six basic experiments of Amp\`ere, but rather their subsequent agreement with almost all hitherto known facts. He therefore passes a judgment of Solomon that since we have these equations we had best write them down without derivation, compare them with phenomena and regard constant agreement between the two as the best proof that the equations are correct'' (pp.\,94--95). As truth is considered to be a property of a mathematical image, not of an existent object, we here adopt what can be classified as the \index{Pragmatist's perspective on truth}pragmatist's perspective on truth (see p.\,xv of \cite{James}).

I would like to conclude the discussion of Boltzmann's essay ``On the Development of the Methods of Theoretical Physics in Recent Times'' with a beautifully inspiring quote (p.\,86): ``Given this enormous variety of [electromagnetic] radiations we are almost tempted to argue with the creator for making our eyes sensitive for only so minute a range of them. This, as always, would be unjust, for in all areas only a small range of a great whole of natural phenomena is directly revealed to man, his intelligence being made acute enough to gain knowledge of the rest through his own efforts.''\footnote{This remark nicely points to the biological origin of our cognitive faculties, adapted in response to our environment.} Margenau's entire philosophy of modern physics \cite{Margenau} is based exactly on the idea expressed in that quote: Starting from the plane of direct perception (sense data, immediate experience, nature), the field of valid rational constructs is obtained through so-called `rules of correspondence' or `epistemic correlations'; the field of constructs is subject to \index{Metaphysical postulates}metaphysical requirements and empirical verification. Or in the words of Altmann, ``Naked facts hardly exist at all: they are all processed by us through a network of theoretical constructs'' (see p.\,28 of \cite{Altmann}).

\paragraph*{``On the Fundamental Principles and Equations of Mechanics''}
Whereas the idea of regarding physical theories as images of nature should be sufficiently elaborated by now, Boltzmann's 1899 lectures at Clark University further clarify the process of creating images and the idea that only the fully developed image with all its possible consequences, rather than the basic hypotheses from which it was derived, should be tested against the facts of experience (pp.\,107--108):

\begin{quotation}
Some pictures were built up only gradually over centuries through the joint efforts of many enquirers, for example the mechanical theory of heat. Some were found by a single scientific genius, though often by very intricate detours, only then could other scientists illuminate them from various angles. Maxwell's theory of electricity and magnetism discussed above is one such. Now there is no doubt a particular mode of representation that has quite peculiar advantages, though it has its defects too. This mode consists in starting to operate only with mental abstractions, in tune with our task of constructing only internal mental pictures. In this we do not yet take account of facts of experience. We merely endeavour to develop our mental pictures as clearly as possible and to draw from them all possible consequences. Only later, after complete exposition of the picture, do we test its agreement with the facts of experience; it is, then, only after the event that we give reasons why the picture had to be chosen thus and not otherwise, a matter on which we give not the slightest prior hint. Let us call this deductive representation. Its advantages are obvious. For a start, it forestalls any doubt that it aims at furnishing not things in themselves but only an internal mental picture, its endeavours being confined to fashioning this picture into an apt designation of phenomena. Since the deductive method does not constantly mix external experience forced on us with internal pictures arbitrarily chosen by us, this is much the easiest way of developing these pictures clearly and consistently. For it is one of the most important requirements that the pictures be perfectly clear, that we should never be at a loss how to fashion them in any given case and that the results should always be derivable in an unambiguous and indubitable manner. It is precisely this clarity that suffers if we bring in experience too early, and it is best preserved if we use the deductive mode of representation. On the other hand, this method highlights the arbitrary nature of the pictures, since we start with quite arbitrary mental constructions whose necessity is not given in advance but justified only afterwards. There is not the slightest proof that one might not excogitate other pictures equally congruent with experience. This seems to be a mistake but is perhaps an advantage at least for those who hold the above-mentioned view as to the essence of any theory. However, it is a genuine mistake of the deductive method that it leaves invisible the path on which the picture in question was reached. Still, in the theory of science especially it is the rule that the structure of the arguments becomes most obvious if as far as possible they are given in their natural order irrespective of the often tortuous path by which they were found.
\end{quotation}

In the above quote, the word `clear' occurs four times and, in addition, the words `clarity,' `consistent,' `unambiguous,' and `indubitable' appear. Obviously the clarity and consistency of a mathematical image of nature is of greatest importance to Boltzmann. The role of experience in theorizing has been described by Feyerabend in a way that nicely reflects Boltzmann's deductive mode of representation (see p.\,226--227 of \cite{Feyerabend62ip}): ``Indeed the whole tradition of science from Galileo (or even from Thales) up to Einstein and Bohm is incompatible with the principle that `facts' should be regarded as the unalterable basis of any theorizing. In this tradition the results of experiment are not regarded as the unalterable and unanalyzable building stones of knowledge. They are regarded as capable of analysis, of improvement (after all, no observer, and no theoretician collecting observations is ever perfect), and it is assumed that such analysis and improvement is absolutely necessary.''

In the context of quantum field theory, mathematical consistency is a particularly serious concern raised even by its most famous proponents. In his Nobel lecture (1965), Feynman, in a catchy metaphorical statement, expressed the possible concern that renormalization ``is simply a way to sweep the difficulties of the divergences of [quantum] electrodynamics under the rug.'' Modern renormalization-group theory \cite{WilsonKogut74} has certainly provided a better understanding. But, in the words of the insistent critic Dirac \cite{Diracrem}, ``the quantum mechanics that most physicists are using nowadays [in quantum field theory] is just a set of working rules, and not a complete dynamical theory at all.'' Dirac felt that ``some really drastic changes'' in the equations were needed (see pp.~36--37 of \cite{Diraced}). In the end of the day, the mathematics of quantum field theory must be clear and consistent by the standards of Boltzmann for a theory to become acceptable as an image of nature. We hence adopt the following, even more far-reaching postulate.\\

\index{Metaphysical postulates!first metaphysical postulate}
\noindent\framebox[\textwidth]{\parbox{\inboxwidth}{\emph{First Metaphysical Postulate:}\label{metaphys1} A mathematical image of nature must be rigorously consistent; mathematical elegance is an integral part of any attractive image of nature.}}\\[1.1ex]

I would like to remind the reader that the metaphysical postulates should be read with benevolence, in particular, if they involve subjective judgements. If someone really doesn't know what `attractive' means, the word may be replaced by `acceptable'. If the word `acceptable' is unacceptable, it may be omitted. But it would be disappointing to give up the idea that we all recognize mathematical elegance when we encounter it. Let's try to approach this with the same attitude that makes us visit art museums.

The belief in the universal harmony of nature reflected in mathematical elegance, or even reflecting mathematical elegance, is in the tradition of Plato and Pythagoras. ``The latter took mathematics as the foundation of \index{Reality}reality and the universe as fundamentally mathematical in its structure. It was assumed that observable phenomena must conform to the mathematical structures, and that the mathematical structures should have implications for further observations and for counterfactual inferences which went beyond what were given'' (see p.\,xvii of \cite{Cao}). Note that the reliability and truth of mathematical images depends on the idea of `uniformity of nature', that is, the idea that the succession of natural events is determined by immutable laws.

According to Dworkin \cite{Dworkin}, the intrinsic beauty and sublimity of the universe belong to the characteristics of a religion without god. With or without (a personalized) god, these properties of the universe should be reflected in the elegance of the mathematical image.

Mathematical theories and concepts are most reliably introduced within the axiomatic approach. All objects are characterized by properties. The emphasis on mathematical images hence suggests to build \index{Ontology}ontology on properties. Some advantages of the mathematical formulation of physical theories for philosophical considerations have been emphasized by Auyang (see p.\,7 of \cite{Auyang}): ``Since physical theories are mathematical, their conceptual structures are more clearly exhibited. This greatly helps the philosophical task of uncovering presuppositions.''

\index{Metaphysics}Our \index{Metaphysical postulates!first metaphysical postulate}first metaphysical postulate covers several of the six metaphysical requirements formulated by Margenau in Chapter~5 of \cite{Margenau}: (A) logical fertility (``natural science is joined with logic through mathematics''); (B) multiple connections between constructs (otherwise a construct ``leads to no other significant knowledge''); (C) permanence and stability (where ``permanence \ldots\ extends over the lifetime of a given theory''); (D) extensibility of constructs (no ``special laws for special physical domains''); (E) causality (``constructs shall be chosen as to \emph{generate causal laws}''); (F) simplicity and elegance (``we bow to history and include simplicity \ldots\ there is also an aesthetic element''). Causality is the element that seems to be missing in our first metaphysical postulate. According to Bertrand Russell \cite{Russell12}, ``The law of causality, I believe, like much that passes muster among philosophers, is a relic of a bygone age, surviving, like the monarchy, only because it is erroneously supposed to do no harm.'' If introduced carefully, the principle of causality can, however, still be useful. Causality is also meticulously introduced and passionately motivated as a metaphysical principle by Altmann (see Chapter~4 of \cite{Altmann}). We here incorporate it in a simple-minded way by implicitly assuming that mathematical images for dynamic systems provide autonomous time-evolution equations, thus emphasizing that causality cannot be judged from a partial view of a system. However, we do not formulate autonomous time-evolution as a metaphysical postulate because there may be important physical theories that do not describe any time-evolution at all, such as equilibrium thermodynamics.\index{Metaphysics}

\subsection{Space and time}
As a student I was incredibly irritated when cosmologists wrote about a number of cosmogonic epochs in the first $10^{-32}$ seconds of the universe. Could units of time introduced by human beings and measured by sophisticated mechanical or electronic devices make any sense under extreme conditions where none of them could possibly exist? If so many dramatic events occurred within an incredibly short period of time, shouldn't one then consider a nonlinear function of time, a true time `felt by the universe,' in which cosmological events happen in a more uniform manner? Even human beings feel that time passes faster with increasing age. Moreover, we are used to all appearances happening in space and time. How shall we imagine the appearances of space and time themselves?
Of course, similar questions about space and time have been asked by philosophers and physicists long before the advent of Big Bang theory.

I would like to reflect on these questions with inspiration from a great philosopher who had deep things to say about space and time: Immanuel Kant (1724--1804). At the age of 46, Kant became the professor of logic and \index{Metaphysics}metaphysics at the university of his native city K\"onigsberg in East Prussia (nowadays Kaliningrad in an enclave of Russia). After his first application for this chair had failed in 1758, he later rejected a chair for the art of poetry, which indicates his admirable determination and persistence. At the age of 22, he self-consciously wrote in his first philosophical publication (entitled ``Thoughts on the True Estimation of Living Forces''): ``Ich habe mir die Bahn schon vorgezeichnet, die ich halten will. Ich werde meinen Lauf antreten, und nichts soll mich hindern ihn fortzusetzen.'' [I have already scribed the path that I want to follow. I will line up for this race, and nothing shall stop me from continuing it.] As a bachelor, he fully dedicated his life to work. Apparently Kant perceived two terms as the rector of his university and several calls to more prestigious (and better paid) professorships as annoying distractions from his scribed path. The anecdote that Kant never left K\"onigsberg, however, is not literally true---it must be wrong by roughly a hundred kilometers. Kant felt that the coronation city of the Prussian monarchy and the Hanseatic market town with a bustling harbor as a lively hub between the East and the West with a mixed population of 50,000--60,000 inhabitants was a decent place to gather knowledge of human existence and the world without any need for traveling. With increasing age, he followed an extremely regular daily schedule; as he walked the same route through K\"onigsberg every afternoon at the same time, people set their clocks according to his appearance. On the other hand, Kant must have been an inspiring and witty speaker, with a natural sense of humor, leaving a lasting impression by his deep and sublime thoughts expressed with great clarity and eloquence. In 1794, Kant was charged with misusing his philosophy to the distortion and depreciation of many leading and fundamental doctrines of sacred Scripture and Christianity [``Unsere h\"ochste Person hat schon seit geraumer Zeit mit gro\ss em Mi\ss fallen ersehen, wie Ihr Eure Philosophie zu Entstellung und Herabw\"urdigung mancher Haupt- und Grundlehren der Heiligen Schrift und des Christentums mi\ss braucht''] and was required by the government of the Kingdom of Prussia, following a special order of King Friedrich Wilhelm II, not to lecture or write anything further on religious subjects. Kant, who generally avoided annoying the guardians of a theologically and ecclesiastically interpreted Bible, followed this requirement.

\begin{figure}
\centerline{\includegraphics[width=6.6cm]{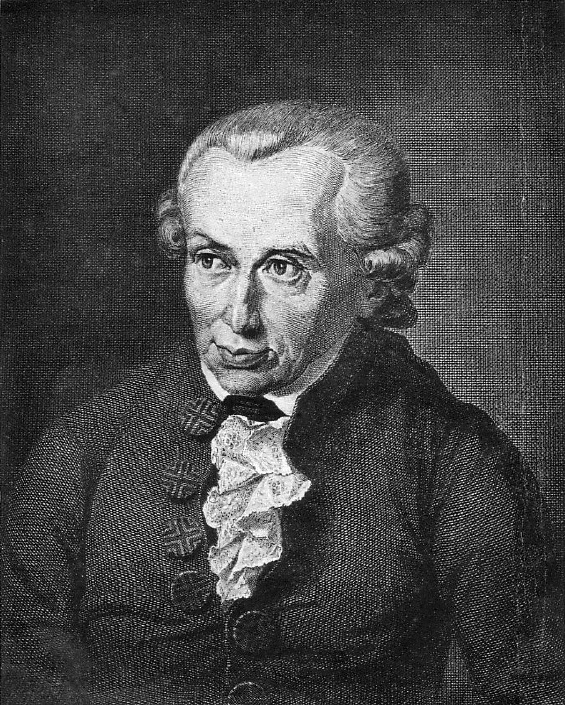}}
\caption[ ]{Immanuel Kant, 1724--1804.} \label{figKant}
\end{figure}

\begin{figure}[t]
\centerline{\includegraphics[width=\textwidth]{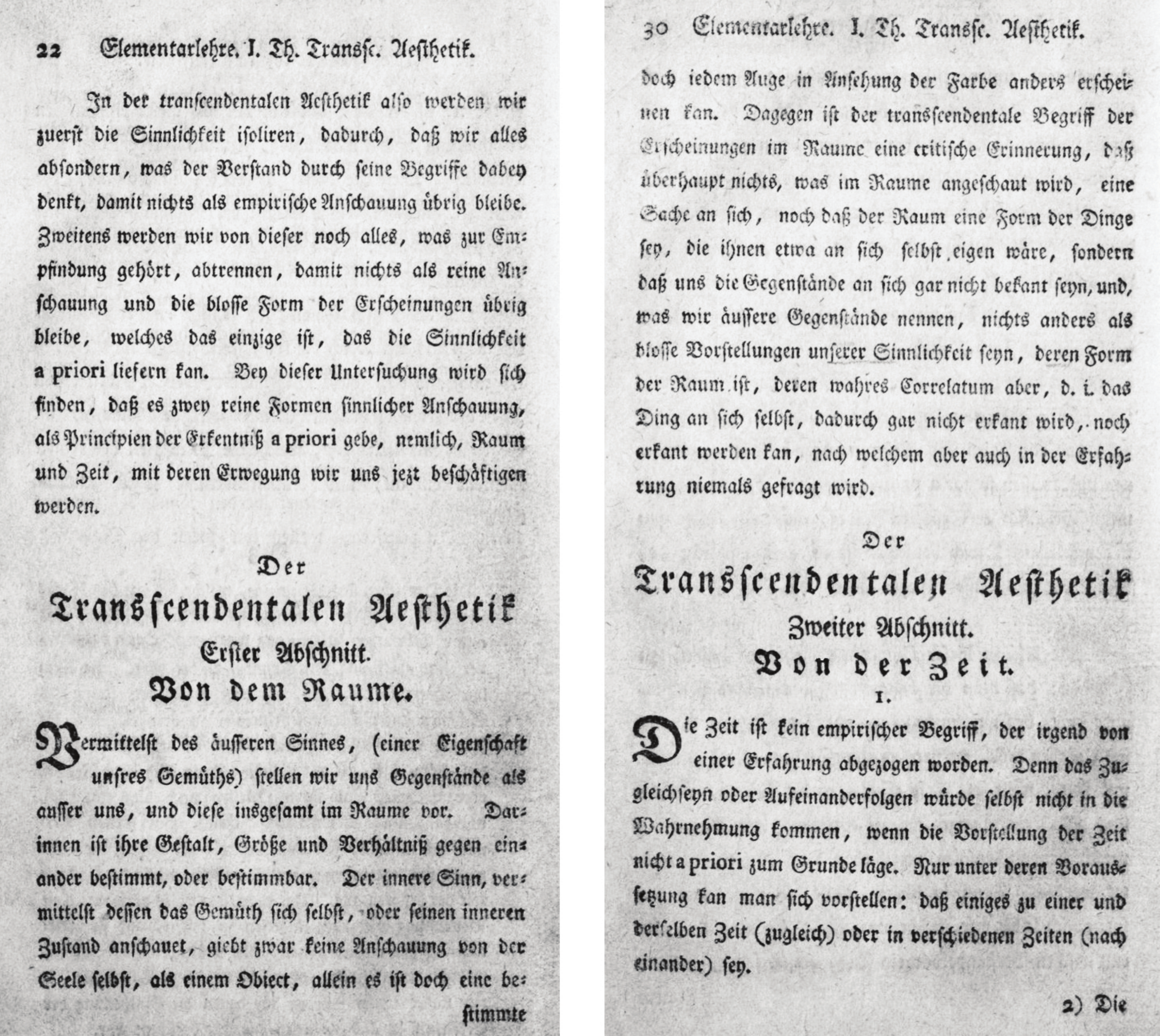}} \caption[ ]
{The two pages of Immanuel Kant's \emph{Critique of Pure Reason} (1781) on which the sections of his Transcendental Aesthetic on space and time begin.} \label{figkantvernunft}
\end{figure}

Immanuel Kant's \emph{Critik der reinen Vernunft} \cite{Kant}\footnote{For an online version, including a facsimile of the complete original 1781 edition, see www.deutschestextarchiv.de/book/view/kant{\_}rvernunft{\_}1781.} is generally considered as one of the most influential milestones in philosophy. The title of Kant's \emph{opus magnum} is usually translated into \emph{Critique of Pure Reason}, but it might better be called `a critical analysis of the capacity of mere reasoning, that is, independent of all practical experience.' The first part of this epistemological work, known as \emph{Transcendental Aesthetic} (pp.~19--49), is entirely dedicated to a discussion of space and time. After establishing some basic terminology, Kant first discusses \emph{space} (pp.~22--30) and then, in a highly parallel formulation followed by some explanatory remarks, \emph{time} (pp.~30--41). Some further remarks in the last few pages of the Transcendental Aesthetic (pp.~41--49) are meant to avoid misunderstandings of his challenging work, mainly by explicitly or implicitly comparing to previous philosophical works on the topic. The first page of each of the sections on space and time are shown in Figure~\ref{figkantvernunft}.

By devoting 30 pages of his widely recognized work to a thorough discussion of space and time, Kant has gained enormous attention from physicists. However, the interpretation of his work seems to be not at all straightforward, neither for physicists nor for philosophers. I hence try to rephrase the story in simple words, based on my own reading of the original edition \cite{Kant}. In order to support my interpretations and explanations, I offer quotes from the original text in modern German spelling in square brackets, always including page numbers referring to the original edition.

Of course, Kant was faced with the same problem as every philosopher or scientist. He had to distinguish himself from previous thinkers. The simplicity and clarity of his exposition may occasionally have suffered from the fact that he needed to emphasize the highly innovative character of his ideas about space and time compared to those of the greatest thinkers of the preceding century, Gottfried Wilhelm Leibniz (1646--1716), Isaac Newton (1643--1727), John Locke (1632--1704), Samuel Clarke (1675--1729), George Berkeley (1685--1753), and others. For our purposes, we do not need to go into a careful comparison of the various ideas and the subtle or serious differences between them. We rather wish to benefit from Kant's ideas by merely recognizing some essential issues about space and time in creating images of nature. The work of other philosophers could serve the same purpose.

Kant's fundamental postulate is that space and time should not be considered as empirical concepts abstracted from any kind of external experience [``Der Raum ist kein empirischer Begriff, der von \"au{\ss}eren Erfahrungen abgezogen worden.'' (p.\,23); ``Die Zeit ist kein empirischer Begriff, der irgend von einer Erfahrung abgezogen worden.'' (p.\,30)]. The perception of space and time rather resides in us [``\"Au{\ss}erlich kann die Zeit nicht angeschaut werden, so wenig wie der Raum, als etwas in uns.'' (p.\,23)]. In other words, we have an immediate intuitive view of space and time, an \emph{a priori} intuition, independent of all experience. Space and time, which are not properties of things-in-themselves [``keine Eigenschaft irgend einiger Dinge an sich'' (p.\,26)], are necessary prerequisites to all human experience unavoidably taking place in space and time. Kant refers to his philosophical approach as \emph{transcendental idealism}, where idealism asserts that \index{Reality}reality is a mental construct and transcendental indicates that we are not dealing with things-in-themselves.

Kant justifies his fundamental postulate by an indirect argument: If the perception of space referred to something outside us, that something would be separated from us in space and the sought-after representation of space would actually be a prerequisite [``Denn damit gewisse Empfindungen auf etwas au{\ss}er mich bezogen werden, ... dazu mu{\ss} die Vorstellung des Raumes schon zum Grunde liegen.'' (p.\,23)] And if we had no \emph{a priori} representation of time, we would not be able to perceive things happening simultaneously or sequentially in time [``Denn das Zugleichsein oder Aufeinanderfolgen w\"urde selbst nicht in die Wahrnehmung kommen, wenn die Vorstellung der Zeit nicht \emph{a priori} zum Grunde l\"age.'' (p.\,30)]. A further argument for the \emph{a priori} and transcendental nature of time goes as follows: One can remove all things from space until no thing is left, so that one can perceive an empty space, which is not a thing, but one cannot perceive no space [``Man kann sich niemals eine Vorstellung davon machen, da{\ss} kein Raum sei, ob man sich gleich ganz wohl denken kann, da{\ss} keine Gegenst\"ande darin angetroffen werden.'' (p.\,24)]. And similarly, one can remove all appearances, let's say events, from time, but one cannot annihilate time itself [``Man kann ... die Zeit selbst nicht aufheben, ob man zwar ganz wohl die Erscheinungen aus der Zeit wegnehmen kann.'' (p.\,31)]. Nothing is left, except the pure \emph{a priori} intuitions of space and time.

For a better understanding, one should realize that, in Kant's times, space and time were widely viewed to be `causally inert.' It is hence natural to consider them as imperceptible, which makes them inaccessible to direct experience.

Space and time are singular, in the sense of one-of-a-kind [``Denn erstlich kann man sich nur einen einigen Raum vorstellen \dots'' (p.\,25); ``verschiedene Zeiten sind nicht zugleich, sondern nach einander \ldots'' (p.\,31)]. Space and time are infinite in a sense that Kant explicates in some detail [``Der Raum wird als eine unendliche Gr\"o{\ss}e gegeben vorgestellt \dots Grenzenlosigkeit im Fortgange der Anschauung \ldots'' (p.\,25); `` Die Unendlichkeit der Zeit bedeutet nichts weiter, als da{\ss} alle bestimmte Gr\"o{\ss}e der Zeit nur durch Einschr\"ankungen einer einigen zum Grunde liegenden Zeit m\"oglich sei.'' (p.\,32)].

So far, not much about the properties of space and time has been mentioned, except their infinity. Kant points out that the \emph{a priori} nature of space implies an apodictic certainty of geometry [``Auf diese Notwendigkeit \emph{a priori} gr\"undet sich die apodiktische Gewi{\ss}heit aller geometrischen Grunds\"atze, und die M\"oglichkeit ihrer Konstruktionen \emph{a priori}.'' (p.\,24)]. If Kant speaks of geometry, of course, he can only think of Euclidian geometry. Whereas he does not explicitly refer to Euclid, a number of characteristic features of Euclidian geometry are repeatedly given throughout the entire book: two points uniquely determine a straight line, three points must lie in a plane, and the sum of the three angles in a triangle equals two right angles [``da{\ss} zwischen zweien Punkten nur eine gerade Linie sei'' (p.\,24); ``da{\ss} drei Punkte jederzeit in einer Ebene liegen'' (p.\,732); ``Da{\ss} in einer Figur, die durch drei gerade Linien begrenzt ist, drei Winkel sind, wird unmittelbar erkannt; da{\ss} diese Winkel aber zusammen zwei rechten gleich sind, ist nur geschlossen.'' (p.\,303)]. According to Kant, Euclidian geometry clearly inherits the \emph{a priori} one-of-a-kind status of space and hence frequently serves as the prototypical example for theorems of apodictic certainty.

It should be noted that Kant treats space and time in a highly parallel manner. The sentence ``Da{\ss} schlie{\ss}lich die transzendentale \"Asthetik nicht mehr, als diese zwei Elemente, n\"amlich Raum und Zeit, enthalten k\"onne, ist daraus klar, weil alle anderen zur Sinnlichkeit geh\"origen Begriffe, selbst der der Bewegung, welcher beide St\"ucke vereinigt, etwas Empirisches voraussetzen'' (p.\,41) is quite remarkable. Space and time are jointly required to describe motion and are brought together even more closely by recognizing them as the only two elements that transcendental aesthetic deals with; according to Kant, there is nothing else like space and time.

\begin{figure}[t]
\centerline{\includegraphics[width=\textwidth]{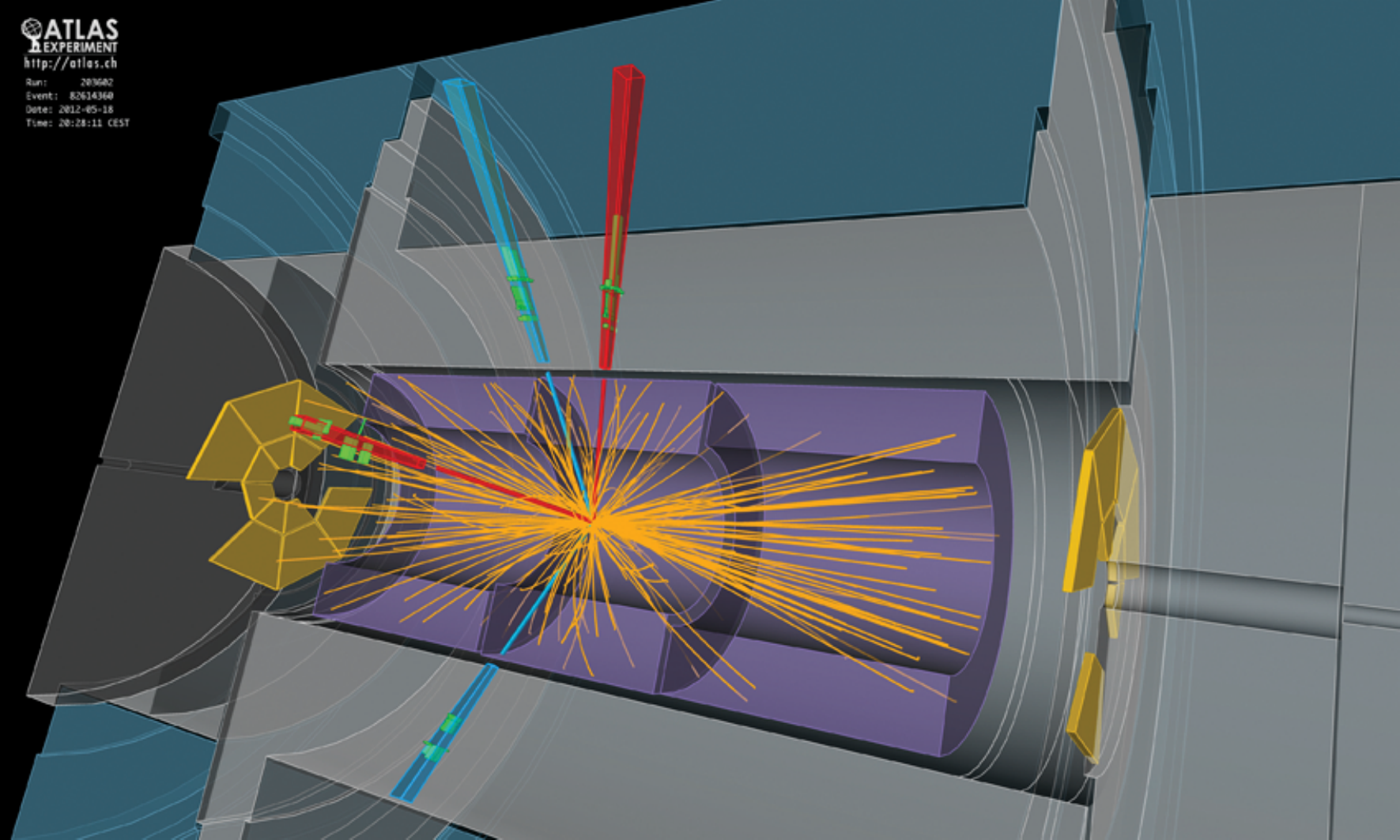}} \caption[ ]
{Particle tracks emerging from a proton-proton collision in the Large Hadron Collider at CERN. The tracks and clusters of two electron pairs, which can arise according to one of the decay channels of the Higgs boson, are colored red and blue, respectively. [cds.cern.ch/record/1459487; Copyright (2012) by CERN.]} \label{figHiggsCERN}
\end{figure}

The description of particle motion taking place in space and time is a core issue in fundamental particle physics. Particle tracks in space are the most basic output from detectors (see Figure~\ref{figHiggsCERN}), where a number of techniques are available to narrow down the identity of the particles. More about the temporal aspects of the particle motion along the trajectories can be learned from the curvature of the tracks in a magnetic field or from energy measurements by calorimeters.

Auyang's thorough modern analysis of space and time leads to interesting insight about their fundamental properties, which I would like to add to Kant's ideas (see p.\,170 of \cite{Auyang}): ``\emph{The primitive spatio-temporal structure is permanent; it is independent of temporal concepts.} It contains the time dimension as one aspect and makes possible the introduction of the time parameter, but is itself beyond time and change.''

According to Kant, all human experience takes place in space and time. Accordingly, physical theories should be formulated \emph{in} space and time, but a physical theory \emph{of} space and time would not make any sense. This is at variance with Einstein's theory of gravity, or general relativity, which is usually regarded as a theory \emph{of} space and time by introducing the geometry of space-time as an evolving variable. Therefore, Kant's philosophical ideas about space and time are nowadays generally considered as obsolete, wiped out by general relativity. However, in a wider sense, the situation is not so clear. Even in the presence of gravity, an \emph{a priori} space-time could exist in a topological sense; gravity as a physical theory would then merely introduce geometric structure into this \emph{a priori} topological space-time. Only measurability of space and time would no longer be \emph{a priori}. This line of thought was not only developed by several philosophers to the rescue of Kant, but Max von Laue, who was a most distinguished expert in special and general relativity, cherished it as a key step to a satisfactory understanding of relativity in the opening sentence of a presentation given in 1959 \cite{vonLaue61ip}.

Alternatively, one can assign an \emph{a priori} character to the \index{Minkowski space}Minkowski space-time of special relativity. Gravity can then be introduced as a gauge theory expressing the physical irrelevance of the particular choice of local coordinate systems in Minkowski space. Such a construction has been elaborated by Lasenby, Doran and Gull \cite{Lasenbyetal98}.

If we assume an underlying \index{Minkowski space}Minkowski space, this has far-reaching mathematical implications. We can then look at the group of inhomogeneous Lorentz transformations, that is, including translations. The irreducible representations of this group have been classified in a landmark paper by Wigner \cite{Wigner39}, which offers a more rigorous and complete version of earlier results by Dirac and Majorana. Wigner's representation theory basically implies that particles can be classified according to mass and spin, thus assigning a special role to these properties, where mass takes nonnegative real values and spin nonnegative integer or half-integer values. In our development of particle physics, we always need to assign masses and spins to the fundamental particles.

I personally find it very attractive to assume an underlying \index{Minkowski space}Minkowski space, although Kant's arguments seem to work naturally even in a topological space. Minkowski space comes with basic measures of length and time. However, these measures will always be distorted because, according to Einstein's theory of gravitation, as massive observers we unavoidably perturb the geometry around us. Still, the underlying space tells us by what properties we should label our fundamental particles. Minkowski space plays a similar role as the free particles to be considered below, which are unobservable but provide a good starting point for understanding the more complicated entities involved in observable quantities.

Although modern philosophers of science find Kant's classification of space and time as \emph{a priori} intuitions obsolete, they usually tend to comment in respectful words and  acknowledge some value in the great master's ideas. According to a summary offered by Auyang (see p.\,135 of \cite{Auyang}), ``the concepts of space and time that are required for the plurality and distinction of objects do not imply any specific geometric property such as those in Euclidean geometry. The approach agrees with Kant that space is not a substance but the precondition for particulars and individuals. It disagrees with Kant that space is a form of intuition.''

Also Margenau disagrees with the transcendental nature of space and time. He rather declares them to be constructs playing the same role as all the others, but ``more abstract than many other scientific constructs since they possess no immediate counterparts in direct perception'' (see p.\,165 of \cite{Margenau}). He accordingly characterizes Kant's philosophical approach with the sentences ``Kant's dichotomy of the a priori and the a posteriori has lost its basis in actual science,'' but ``if Kant's final conclusion were replaced by a milder one, stating that conceptual space is not compounded from immediate experiences, it would be wholly acceptable'' (see p.\,148 of \cite{Margenau}).

Stimulated by Kant's ideas about space and time, but taking into account the caveats resulting from relativity, we formulate the following postulate:\\

\index{Metaphysical postulates!second metaphysical postulate}
\noindent\framebox[\textwidth]{\parbox{\inboxwidth}{\emph{Second Metaphysical Postulate:}\label{metaphys2} Physical phenomena can be represented by theories \emph{in} space and time; they do not require theories \emph{of} space and time, so that space and time possess the status of prerequisites for physical theories.}}\\[1.1ex]

In the absence of gravity, we can choose a particular reference frame to recover Cartesian space and a well-defined time parameter, just as it is usually done for \index{Maxwell equations}Maxwell's equations governing \index{Electrodynamics}electromagnetic fields. We consider space and time as preconditions for physical theories but leave it open whether they are particularly fundamental constructs ultimately based on experience or \emph{a priori} intuitions independent of all experience. To maintain the above metaphysical postulate even when general relativity enters the scene, also gravity should be considered as a theory \emph{in} space and time, introducing a physically relevant geometry into a topological or \index{Minkowski space}Minkowski space.

One may ask whether, according to our present knowledge, there might be any further elements of transcendental aesthetic in addition to space and time (or constructs with a similarly fundamental status). In the same spirit that Kant's space is necessary to introduce motion and hence momentum, according to Wigner's representation theory, one should postulate an additional space as an arena for spin to manifest itself in (`spin-space'). Further potential elements of transcendental aesthetic are given by the `color space' for strong interactions and the `weak isospin space' for weak interactions. Are all these additional spaces on the same footing as Kant's space and time? Do we have an \emph{a priori} representation of these spaces or do they rather possess the status of constructs? Are certain interactions providing geometric structure to these additional spaces? Actually, the gauge theories for weak and strong interactions introduce geometry into the weak isospin and color spaces. The geometric interpretation of the gauge theories for all fundamental interactions has been elaborated in Section~11.3 of \cite{Cao}.

In the context of quantum field theory, a serious problem actually arises from Kant's assertion that space and time are infinite. Infinity is an intrinsic philosophical problem, which we need to address in the subsequent section to reveal the distinction between actual and potential infinity and the resulting fundamental importance assigned to limiting procedures.

\subsection{Infinity}\label{sectioninfinity}
Philosophical or logical reservations about the concept of infinity can be illustrated with an example from Euclid's mathematical and geometric treatise \emph{Elements}, written around 300\,BC. Instead of stating that the number of primes is infinite, Euclid prefers to say that ``the (set of all) prime numbers is more numerous than any assigned multitude of prime numbers.''\footnote{See Book~9, Proposition~20 on p.\,271 of R.~Fitzpatrick's translation of Euclid's \emph{Elements} (farside.ph.utexas.edu/books/Euclid/Euclid.html).} Euclid avoids the term infinite which, for ancient Greeks, had negative connotations since Zeno had shown that logical paradoxes arise from a naive use of the notion of infinity around 450\,BC. Around 350\,BC, Aristotle had tried to clarify the situation by distinguishing between actual infinity, which is ``a permanent actuality,'' and potential infinity, which ``consists in a process of coming to be, like time,'' where only the latter was felt to be a meaningful concept.\footnote{See Book~III of Aristotle's Physics, where the quotes are from Part~7, in the English translation of R.\,P.~Hardie and R.\,K.~Gaye (classics.mit.edu/Aristotle/physics.html).} Euclid clearly tries to describe a situation of potential infinity, which corresponds to a never-ending process in time. As Aristotle was a man of great authority, it took more than two thousand years, until pioneering work by Bernard Bolzano (1851), Richard Dedekind (1888), and Georg Cantor (1891) paved the way to a sound notion of actual infinity.

In field theories, one or several degrees of freedom are associated with each point of space which, according to Kant, is infinite. The collection of all degrees of freedom hence has the cardinality of the continuum. Both from a philosophical and from a mathematical point of view, one can ask whether such a large number of degrees of freedom can actually be handled. As an infinite number of degrees of freedom seems to be an intrinsic characteristic of field theories, one should at least try to restrict oneself to the smaller cardinality of the set of natural numbers. In other words, the number of degrees of freedom should, at most, be \emph{countably infinite}, that is, of the cardinality of the integers rather than the cardinality of the continuum.

In probability theory, the importance of keeping infinities countable has been recognized in the pioneering book by Kolmogorov (1933).\footnote{The original version of the book \cite{Kolmogoroff} was published in German.} More generally, in measure theory, following the ideas of Kolmogorov, the measures of a countably infinite number of disjoint measurable subsets can be added up to the measure of the full set obtained as the countable union of the subsets, but not for an uncountable union of subsets \cite{Bauer}. If one wishes to treat stochastic processes with continuous time evolution, the number of degrees of freedom must hence be limited to countably infinite by introducing the concept of \emph{separability}. This concept has also been introduced by von Neumann for limiting the dimensionality of the arena for quantum mechanics, called Hilbert space, where the number of base vectors must be limited to finite or countably infinite.

In our development of quantum field theory, we wish to make sure that the number of degrees of freedom remains countable at any stage. We will do so by using Fock spaces, which are based on occupation numbers for the possible quantum states of a single particle, as the basic arena. The detailed construction of Fock spaces is described in Section~\ref{sectionFock}.\footnote{Whereas Fock spaces are frequently used in quantum field theory (see, for example, Sections~1 and 2 of \cite{FetterWalecka} or Sections~12.1 and 12.2 of \cite{BjorkenDrell}), our actual construction of quantum field theories on Fock spaces will be somewhat unconventional.} Concerns about infinities appear repeatedly in the subsequent discussion of physical ideas and we then refer to them as the philosophical \emph{horror infinitatis}.

The infinite size of Kant's space nourishes our deep-rooted \emph{horror infinitatis}. More carefully, Kant actually infers the ``limitlessness of space in the progression of intuition'' by an indirect argument, in accordance with the idea of a potential infinity. In developing quantum field theory, we will hence consider large finite volumes together with a limiting procedure. Within a mathematical image, we can pass from position space to its dual Fourier space. The finite volume of position space corresponds to the discreteness and hence countability of Fourier space. We will further impose the finite size of Fourier space, which corresponds to eliminating small-scale features in position space, and introduce an associated limiting procedure. In Section~\ref{sectionirrev} we propose irreversibility as a more natural possibility of wiping out small-scale features, and in Section~\ref{sectiondynamics} we elaborate on its appealing mathematical formulation.

In the context of quantum field theory, we encounter various aspects of infinity. In discussing the number of degrees of freedom, we focused on the particular aspect of infinity that can be characterized as \emph{infinitely numerous}. The volume of space brings in the aspect of \emph{infinitely large} and, by simultaneously considering position space and the associated Fourier space, the dual aspect of \emph{infinitely divisible} comes up. \emph{Infinitely small} could be associated with the size of fundamental particles, \emph{infinitely large} could also be associated with the values of divergent integrals. We will make serious efforts to avoid such divergent integrals which, for example, result from the assumption of point particles in quantum field theory. According to the \index{Metaphysical postulates!first metaphysical postulate}metaphysical postulate that a mathematical image of the real world should be consistent and elegant, any subtle or artificial handling of divergencies should clearly be avoided.

As the `handling of infinities' is a key issue in quantum field theory, it is important to have some philosophical guidance:\\

\index{Metaphysical postulates!third metaphysical postulate}
\noindent\framebox[\textwidth]{\parbox{\inboxwidth}{\emph{Third Metaphysical Postulate:}\label{metaphys3} All infinities are to be treated as potential infinities; the corresponding limitlessness is to be represented by mathematical limiting procedures; all numerous infinities are to be restricted to countable.}}\\[1.1ex]

\noindent This metaphysical postulate has significant implications for or approach to quantum field theory. We will actually introduce four different limiting procedures:\label{listlimits}
\begin{enumerate}
  \item \emph{Thermodynamic limit:}\label{limitthermo} We will consider a finite number of momentum states for a quantum particle. The thermodynamic limit of an infinite number of states needs to be analyzed in such a way that critical behavior and symmetry breaking can be recognized.
  \item \emph{Limit of infinite volume:}\label{limitinfvol} We will consider a finite volume. This assumption provides a large characteristic length scale and infrared regularization. In the limit of infinite volume, we typically need to replace sums by integrals, which is useful for practical calculations.
  \item \emph{Limit of vanishing dissipation:}\label{limitvandiss} We will introduce a dissipative smoothing mechanism which provides ultraviolet regularization, even in the thermodynamic limit. The corresponding friction parameter is related to a small characteristic length scale and needs to go to zero.
  \item \emph{Zero-temperature limit:}\label{limitzerotemp} We will consider the dissipative evolution equations at a finite temperature, so that rigorous statements about a long-term approach to equilibrium can be made. Often one is interested in the limit of zero temperature.
\end{enumerate}
In field theories, one might expect a continuum of degrees of freedom, derivatives in the evolution equations for the fields and integrals in their solutions. All these expectations will only be fulfilled through the final limiting procedures. One should keep in mind that also differentiations and integrations are defined in terms of limiting procedures, so that the essential feature of our approach is that all limits are postponed to the end of the calculation. Postponing the limits does not have any disadvantages, at least as long as one does not perform practical calculations (sums are usually harder to evaluate than integrals). Performing premature limits, however, can introduce all kinds of singular behavior or paradoxes and, according to our \index{Metaphysical postulates!third metaphysical postulate}third metaphysical postulate, must be avoided.

A few comments on the thermodynamic limit appear to be worthwhile because an infinite number of momentum states might be considered as a hallmark of field theories. The extrapolation from finite to infinite systems is a most common and successful practice in computer simulations for analyzing critical behavior in statistical mechanics. It can be supported by finite-size scaling theory \cite{FisherBarber72,Privman}, which is based on renormalization group ideas. Even Nature herself works with large but finite numbers of degrees of freedom (of the order of $10^{23}$) so that, at least in statistical mechanics, the thermodynamic limit clearly is an idealization (see p.\,290 of \cite{Ruetsche}). Therefore, working with large finite systems is perfectly natural.

In the above list of limits, we have mentioned the limit of vanishing dissipation. The reasons for introducing a dissipation mechanism are our next topic.

\subsection{Irreversibility}\label{sectionirrev}
Kant was able to perceive an empty space, which we may think of as the vacuum state. In quantum field theory, electron-positron pairs---and, of course, other particle-antiparticle pairs---can spontaneously appear for a short time and then disappear (for example, together with a photon, if this process happens through electromagnetic interactions). In the words of Auyang (see p.\,151 of \cite{Auyang}), ``The vacuum is bubbling with quantum energy fluctuation and does not answer to the notion of empty space \ldots''. We do not know where and when such events occur but, the faster the annihilation of the electron-positron pair takes place, the larger is the number of events, because faster processes allow for a larger range of energies of the electron and the positron according to Heisenberg's uncertainty relation.\footnote{As time is not an operator in quantum mechanics, the proper interpretation of Heisenberg's \label{timeenergyuncert}time-energy uncertainty relation is subtle; a careful derivation and discussion, according to which the uncertainty in the energy of a quantum system reflects a corresponding uncertainty in the energy of the environment, can be found in \cite{Briggs08}.} These frequent events on very short length and time scales are clearly beyond the control of any experimenter.

The lack of mechanistic control of vacuum fluctuations should be recognized as a natural origin of irreversible behavior. In nonequilibrium thermodynamics \cite{hcobet}, uncontrollable fast processes are regarded as fluctuations, which unavoidably come with dissipation and hence irreversibility and decoherence. Therefore, irreversibility should be accepted as an intrinsic feature of quantum field theories, as one cannot control processes on arbitrarily small length and time scales.\\

\index{Metaphysical postulates!fourth metaphysical postulate}
\noindent\framebox[\textwidth]{\parbox{\inboxwidth}{\emph{Fourth Metaphysical Postulate:}\label{metaphys4} In quantum field theory, irreversible contributions to the fundamental evolution equations arise naturally and unavoidably.}}\\[1.1ex]

\index{Metaphysical postulates!fourth metaphysical postulate}
This fourth metaphysical postulate implies a major deviation from standard approaches to quantum field theory. We give up the idea that a theory of fundamental particles and interactions must be of the reversible or Hamiltonian form.
We no longer deal with the Schr\"odinger equation for a Hilbert space vector, but rather with a quantum master equation for a density matrix. In the rest of this section, we elaborate on a number of philosophical and physical implications of postulating irreversibility for the fundamental equations of quantum field theory, thus shedding more light on the role of our \index{Metaphysical postulates!fourth metaphysical postulate}fourth metaphysical postulate in developing a mathematical image of nature.

\subsubsection{Autonomous time evolution}\label{secautotimeevol}
We wish to describe the dynamic behavior of systems of our interest by time-evolution equations. More precisely, we would like to identify a list of variables for which a closed set of evolution equations can be formulated, without any need to specify external functions of time. Such equations are referred to as \emph{autonomous}. They encompass the predictive power of a theory. The search for proper variables and autonomous evolution equations is the fundamental task in creating mathematical images of nature. We first consider examples of autonomous evolution equations and then extract some important issues.

\begin{figure}
\centerline{\includegraphics[width=7.5cm]{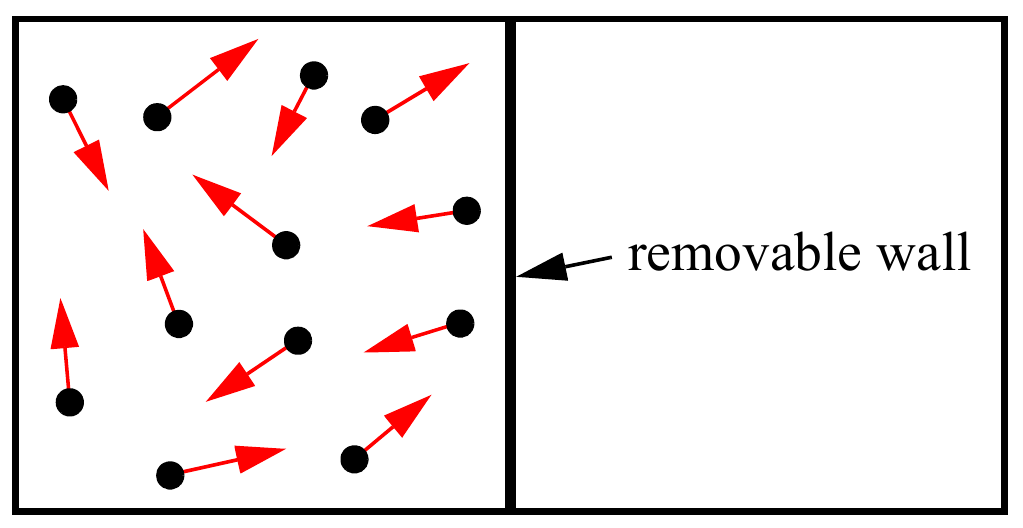}} \caption[ ]
{Gas initially confined to the left half of a container.} \label{figgasexpansion}
\end{figure}

Let us consider the system sketched in Figure~\ref{figgasexpansion}. A gas is confined to the left half of a container by means of a removable wall. Such an initial state can easily be prepared by means of a piston. We assume that the gas is in an equilibrium state. Equilibrium suggests that initially the gas particles are uniformly distributed over the accessible volume and the velocities are randomly chosen according to a Gaussian distribution with zero average and an isotropic covariance matrix corresponding to a certain temperature. Given particular, randomly chosen initial conditions, we remove the wall and solve Newton's equations of motion for the particles to obtain the evolution of the gas. We find that the particles move into the empty part of the container and eventually spread uniformly over the larger accessible volume. Newton's equations of motion provide an example of a set of autonomous evolution equations.

Now let's try to recognize some pattern in the evolution of the gas. An initially sharp density profile is smeared out by mass moving to the right. A density field together with a velocity field could provide an appropriate level of description. If viscous dissipation cannot be neglected, we need to consider a more complete hydrodynamic description including a temperature equation. For these hydrodynamic fields, a closed set of hydrodynamic evolution equations can be formulated. These equations describe how the initial constrained equilibrium state, after removing the constraining wall, develops into a new equilibrium state in the larger volume. In the transition phase, we encounter nonequilibrium density and velocity fields. The hydrodynamic equations governing them lead to entropy production, so that we have an example of a set of dissipative autonomous evolution equations. \emph{Pattern recognition} leads us to a less detailed or coarse-grained autonomous evolution that describes the transition to a new equilibrium state. From this example, we can draw the following conclusions:
\begin{enumerate}
  \item \emph{Levels of description:} Both on a detailed level of description (particle positions and velocities) and on a coarser level of description (hydrodynamic fields) we can formulate autonomous evolution equations that describe the same physical process of interest. Only on the coarser level these evolution equations are dissipative.
  \item \emph{Initial conditions:} The initial density and velocity profiles can easily be prepared by means of a piston. The initial particle positions and velocities need to be chosen randomly because we cannot measure or control all these positions and velocities. If we were able to control them, `abnormal' situations (for example, all particles moving with the same velocity to the right) could be produced. More generally, similar considerations apply to boundary conditions for a space-time domain.
  \item \emph{Role of large numbers:} In principle, `abnormal' situations could arise accidentally when choosing the initial particle positions and velocities randomly. However, whenever a large number of particles is involved, `normal' situations are vastly more likely than `abnormal' ones; for all practical purposes, the probability for `abnormal' situations is zero. (Our somewhat naive distinction between `normal' and `abnormal' situations follows Section~7 of \cite{Price13ip}.)
\end{enumerate}

Imagine that we would like to start our calculations not at the moment the wall is removed ($t=0$), but at a later time ($t_0>0$). We could obtain the hydrodynamic fields at $t_0$ by direct observation and use them as initial conditions for or calculations on the hydrodynamic level. However, we cannot find the particle positions and velocities required for solving Newton's equations of motion by direct observation. We would need to translate the hydrodynamic information into a stochastic model that is significantly more complicated than at equilibrium. Passing to the coarser level of description simplifies the problem of specifying the initial conditions, reduces the computational efforts, and enhances our understanding by focusing on the essence of a problem.

\subsubsection{Entropy}
If a normal person looks at the messy desk of a weird professor in terms of piles or kilograms of paper, that person would be inclined to think that the messy desk is in a state of large entropy. If the professor is able to pull out any desired sheet of paper in a second, this is a strong argument in favor of a perfectly organized desk without any entropy. Entropy is not a property of the desk, but of the level of description used for the desk. This is a fundamental feature of entropy, and this is very different for many other properties. For example, the normal person and the weird professor can easily agree on the total mass of the paper on the desk, but they will never agree on the entropy which is a matter of perspective. This point has been elaborated in the context of a different example which is illustrated in Figure~1.2 on p.\,12 of \cite{hcobet}.

For example, it makes no sense to speak about the entropy of the universe. A statement like ``It now seems clear that temporal asymmetry is cosmological in origin, a consequence of the fact that entropy was extremely low soon after the big bang'' (see p.\,78 of \cite{Price}) is as meaningless as a statement about the entropy of a messy desk. One first needs to specify the variables in terms of which the evolution of the universe can be described in a closed form.

Returning to the example of a freely expanding gas (see Figure~\ref{figgasexpansion}), the hydrodynamic level comes with a well-defined local-equilibrium entropy. This entropy grows during the expansion of the gas. On the other hand, like for the desk from the perspective of the weird professor, there is no need and no place for entropy on the detailed level of the particle positions and velocities.

Many if not most scientists feel that the concept of entropy is limited to equilibrium states. That is certainly not the tacitly assumed level of description for the early, the present, or any interesting state of the universe. Nonequilibrium thermodynamicists are willing to introduce a nonequilibrium entropy. It is associated with a set of autonomous evolution equations, where it actually plays the role of the generator of irreversible time evolution in the same spirit as energy is the generator of reversible time evolution in Hamiltonian dynamics \cite{hcobet}. Before speaking about the entropy of the universe one needs to solve the highly nontrivial problem of identifying a level of description that allows for an autonomous description of the evolution of the universe. Different levels of description might be appropriate at different stages of the evolution of the universe.

Evolution requires that we do not start in an equilibrium state, or that entropy is not at a maximum. It is not only interesting to know how low entropy is at an initial state, but also how large the entropy production rate is.

The very low entropy of the universe even after $14$ billion years of existence (assuming the availability of a proper level of description for the present universe, presumably of the hydrodynamic type) requires justification. According to Boltzmann, this (extremely) low entropy is related to an (extremely) improbable random fluctuation required for the existence of the world in its present state (including the existence of life). He assumes that such an improbable fluctuation of the size of the observable universe is rendered possible by the enormous size of the entire universe. Boltzmann presents this idea, which he actually attributes to his assistant Schuetz, in the following words (see pp.\,208--209 of his writings addressed to the public \cite{BoltzmannPP}):

\begin{quotation}
We assume that the whole universe is, and rests for ever, in thermal equilibrium. The probability that one (only one) part of the universe is in a certain state, is the smaller the further this state is from thermal equilibrium; but this probability is greater, the greater the universe itself is. If we assume the universe great enough we can make the probability of one relatively small part being in any given state (however far from the state of thermal equilibrium), as great as we please. We can also make the probability great that, though the whole universe is in thermal equilibrium, our world is in its present state. It may be said that the world is so far from thermal equilibrium that we cannot imagine the improbability of such a state. But can we imagine, on the other side, how small a part of the whole universe this world is? Assuming the universe great enough, the probability that such a small part of it as our world should be in its present state, is no longer small.

If this assumption were correct, our world would return more and more to thermal equilibrium; but because the whole universe is so great, it might be probable that at some future time some other world might deviate as far from thermal equilibrium as our world does at present. \ldots the worlds where visible motion and life exist.
\end{quotation}

In recent years, this argument has been taken much further to the so-called `Boltzmann brain paradox': The probability for the existence of our present world with many brains in an organized environment is vastly smaller than the probability for the existence of a single brain in an unorganized environment. We should then expect a huge number of lone Boltzmann brains floating in unorganized parts of the universe (or in many of the multiple copies of the universe). But if all our thinking and argumentation might simply take place in one of these numerous lone Boltzmann brains, if all our awareness resided in such a Boltzmann brain, any argument about existence or reality would be questionable. The very nature of this paradox suggests that it is not particularly meaningful to think too much about it.

\subsubsection{Effective field theory}\label{secefffieldtheo}
By postulating irreversibility we introduce a characteristic energy or length scale associated with dissipative phenomena. We assume that the fast short-scale (high-energy) degrees of freedom act as a heat bath on the slower degrees of freedom. By that assumption, the fast degrees are eliminated and autonomous evolution equations can be formulated for the slow degrees of freedoms. We thus arrive at an effective quantum field theory, which can more generally be characterized as follows: An effective field theory comes with a (small) characteristic length scale. It cannot resolve any phenomena on length scales shorter than this characteristic scale; any prediction of the theory is limited to length scales larger than its characteristic scale. As the large-scale properties should be independent of the particular choice of the characteristic length scale, the idea of renormalization appears naturally in effective field theories.

The characteristic length scale in an effective quantum field theory could lie anywhere between the smallest length scales resolvable in super-colliders ($10^{-20}$\,m) and the Planck\label{Plancklength} length ($10^{-35}$\,m), which can be defined in terms of three fundamental physical constants (the speed of light, the Planck constant, and the gravitational constant). The transition from effective quantum field theories to the most fundamental theory of nature unifying all interactions, including gravity, could result from the existence of a smallest length scale in nature (the Planck length). Dissipation could provide a plausible mechanism for putting a lower limit to physically resolvable length scales. Then dissipation would have to be implemented with all the proper symmetries because they cannot be restored in the limit of weak dissipation.

We have argued above that dissipation leads to an effective field theory. The reverse statement should also be true. If a theory is restricted to predictions on length scales larger than a characteristic length scale, small-scale features are eliminated. The elimination of small-scale features naturally leads to the occurrence of entropy, fluctuations and dissipation.

\subsubsection{Thermodynamics, decoupling and reductionism}\label{secdecoupreduc}
``How are the effects of the excluded high-energy processes upon the low-energy phenomena taken into account?'' This key question concerning effective field theories has been asked by Cao on p.\,341 of \cite{Cao} and further discussed as the `decoupling problem' on pp.\,345--350. Whenever we eliminate the high-energy processes from a field theory, we have to answer the above question in a convincing way. In our repeatedly mentioned thermodynamic approach, the fast high-energy processes are simply assumed to act as a heat bath on the slow low-energy degrees of freedom, whereas there is no feedback in the reverse direction. This thermodynamic coupling arises as a natural consequence of postulating the occurrence of irreversibility according to our \index{Metaphysical postulates!fourth metaphysical postulate}fourth metaphysical postulate. The thermodynamic formulation of an effective quantum field theory offers a specific, robust implementation of the decoupling of scales.

The distinction of reversible and irreversible contributions to dynamics in nonequilibrium thermodynamics, which is a core theme in the general framework for the thermodynamic description of nonequilibrium systems, is based on a separation of time scales. Given such a separation of time scales, one formulates autonomous equations for the slow variables on which the fast variables act as noise (fluctuations) and friction (dissipation). More generally, nonequilibrium thermodynamics deals with a hierarchy of levels of description for which there is a clear time scale separation between the variables eliminated (fast) and those kept (slow) in passing from one level to another one. The slow large-scale (low-energy) features are described by a thermodynamic quantum master equation for the evolution of the density matrix for our quantum system which, by construction, is driven to equilibrium (see Section~\ref{sectiondynamicsirr} for mathematical details about density matrices and quantum master equations).

At zero temperature, the equilibrium density matrix should be concentrated on the ground or vacuum state;\footnote{For the present discussion, we assume that there is a unique ground state to avoid the problem of symmetry breaking.} at higher temperatures, the equilibrium density matrix is characterized by Boltzmann factors. We assume that the dissipative coupling to the bath is very weak, except at short length scales. In other words, the dissipative coupling erases the short-scale features very rapidly, whereas it leaves large-scale features basically unaffected. In such an effective quantum field theory, we deal with a weakly non-unitary time evolution, whereas unitary evolution is the hallmark of the usual reversible formulation of quantum field theory (in the next section, a precise definition of unitary evolution is given in (\ref{Schroeunitary}) for Hilbert space vectors and in (\ref{HsuperopE}) for density matrices).

Note that, for quantum master equations, non-unitarity is not at variance with the conservation of the total probability obtained by summing over a complete set of eigenstates. This is different for the Schr\"odinger equation. By adding a small imaginary part to the Hamiltonian in order to describe irreversible decay, one simultaneously introduces non-unitary evolution and a loss of probability, no matter how small the imaginary contribution is. The rigorous conservation of probability is a significant advantage of the thermodynamic approach on the level of density matrices because philosophers like to link \index{Ontology}ontology to everlasting `substances'. A dissipative smearing mechanism moreover suggests that quantum field theory cannot be strictly local. According to Kuhlmann, a particle \index{Ontology!particle ontology}ontology would require strictly localized observables; even `almost localized observables' can be shown to be spread out in the whole universe and effectively require a field \index{Ontology!field ontology}ontology (see p.\,180 of Kuhlmann \cite{Kuhlmann}). However, the properties that can be expected of quantum particles still remain to be specified (see Section~\ref{secpartfield}).

In view of the hierarchical levels typically existing in nonequilibrium thermodynamics, the derivation of any less detailed from any more detailed level of description becomes an important issue. This observation should be seen in the context of methodological  reductionism. For a reductionist, all scientific theories should be reduced to a fundamental theory. According to Kuhlmann (see p.\,14 of \cite{Kuhlmann}), ``Whenever something is to be explained one has to start with the most basic theory and derive an explanation for the phenomenon in question by specifying a sufficient number of constraints and boundary conditions for the general fundamental laws.'' In thermodynamics, one has the well-defined wider task of deriving all less detailed from more detailed theories. Actually, in doing so, thermodynamicists distinguish between coarse-graining and reduction, depending on whether or not (additional) dissipation arises in decimating the number of degrees of freedom \cite{hcoredcg}. According to this distinction, coarse-graining comes with the emergence of irreversibility.

\subsubsection{Arrow of time}\label{arrowoftimesec}
\index{Arrow of time}According to our \index{Metaphysical postulates!fourth metaphysical postulate}fourth metaphysical postulate on p.\,\pageref{metaphys4}, irreversible dynamics must arise in quantum field theory. In other words, a dissipative mechanism leading to increasing entropy in the course of time and hence to an arrow of time occurs in the fundamental evolution equations of particle physics. We should emphasize that this arrow of time is \emph{not} the origin of the observable invariance of weak interactions under time reversal, also known as CP-violation. The dissipative mechanism leading to an effective field theory is felt only on length and time scales that are way too small to be observed experimentally.

If eventually it should turn out that there is some dissipation even on the smallest possible physical scale, that is, on the Planck scale, then the most fundamental equations of nature governing all interactions, including gravity, would come with entropy production and an arrow of time. Reversible equations would be an unrealizable idealization. This does not at all imply that all irreversibility of our macroscopic world is contained in such a fundamental irreversibility. Additional dissipative mechanisms can emerge on any scale, independently of all the irreversible phenomena on smaller scales. For example, diffusion is not a consequence of the fundamental irreversibility but an emerging phenomenon by itself arising on much larger length scales.

Irreversibility and \index{Arrow of time}time's arrow are fascinating topics for both physicists and philosophers (see, for example, \cite{Hoover} and \cite{Price}). In the remainder of this subsection we discuss some features of irreversible dynamics that are relevant to philosophical discussions.

According to our \index{Metaphysical postulates!second metaphysical postulate}second metaphysical postulate on p.\,\pageref{metaphys2}, space and time provide the arena in which mathematical images of nature can be formulated. Most fundamental is the topological space time in which there does not exist any preferred direction of time. Assuming a given space-time continuum corresponds to an external viewpoint, sometimes referred to as `block universe view' \cite{Price}. From such an external perspective, space time itself does not change in time so that it may be characterized as atemporal. In simpler words, one considers the entire universe from nowhere and nowhen \cite{Price}. If one introduces a flat \index{Minkowski metric}Minkowski metric, or a curved general relativistic metric in the presence of gravity, in the space-time continuum, there still is no arrow of time. Time is perfectly symmetric, exactly as argued so passionately by Huw Price \cite{Price}. It is important to realize that irreversibility is not a property of time \emph{per se}, but rather a property of autonomous time-evolution equations. The importance of autonomous time-evolution equations and their proper thermodynamic formulation for the discussion of irreversibility has not been considered in \cite{Price}. In the later publication \cite{Price13ip}, however, it is pointed out that ``the thermodynamic asymmetry is an asymmetry of physical processes \emph{in} time, not an asymmetry of time itself.'' According to the above arguments, asymmetry is not even a property of physical processes, but rather of the description of physical processes in terms of autonomous evolution equations. Price moreover admits in Section~5.4 of \cite{Price13ip} that ``we may need the notion of entropy to generalise properly,'' but he argues that such a generalized thermodynamic entropy is inessential to the essence of his arguments (``we can go on using the term entropy with a clear conscience, without worrying about how it's defined''). Let us briefly consider illustrative examples of autonomous evolution equations and how they can be used.

Let us reconsider the example illustrated in Figure~\ref{figgasexpansion}. What would happen if we looked at the evolution of the initially confined gas in the opposite direction of time? On the particle level, we would still start the calculation with the same initial positions of the particles but, in the usual interpretation of time reversal, we would reverse all velocities. As Newton's equation of motion are reversible and as the reversed velocities possess the same Gaussian distribution with zero mean, we would simply observe another realization of exactly the same evolution as in the original direction of time. Note that, for both directions of time, the initial conditions are equally special by being constrained to one half of the total volume (but not `abnormal'). As the process of interest is unchanged, we should use the same hydrodynamic equations for forward and backward time evolution. Entropy hence increases in both directions of time.

What would happen if we reversed time at $t_0>0$. On the particle level we must now use the truly `abnormal' initial conditions obtained by first evolving the randomly selected particle positions and velocities from $0$ to $t_0$ and then reversing all the velocities. In the time from $t_0$ to $2 t_0$, Newton's equations of motion would bring us back to the particle positions at $t=0$, but with reversed velocities. Further evolution from this `normal' state at $t=2 t_0$ would describe the original transition from constrained to unconstrained equilibrium. One might be tempted to say that the entropy decreases between $t_0$ and $2 t_0$, but increases after $2 t_0$. However, such a statement would be inappropriate because, on the level of particle positions and velocities, there is no entropy. Only on the hydrodynamics level we can speak about entropy. If we solve the hydrodynamic equations backward in time, we indeed find a decrease of entropy. At $2 t_0$ we recover the original constrained equilibrium with low entropy. However, a further backward calculation fails.

The above discussion shows how autonomous evolution equations cannot only be used for prediction, but also for retrodiction. Albert, who discusses many illustrative examples of retrodiction, underlines the difference between prediction and retrodiction very nicely (see p.\,116 of \cite{Albert}): ``The claim, then, is that whatever we take ourselves to know of the future, or (more generally) whatever we take to be \emph{knowable} of the future, is in principle ascertainable by means of \emph{prediction}. Some of what we take ourselves to know about the \emph{past}, (the past positions of the planets, for example) is no doubt similarly ascertainable by means of \emph{retrodiction}---but far from all of it; rather little of it, in fact. Most of it we know by means or \emph{records}.''

In this section, we have discussed that irreversible equations lead to an arrow of time. Whether irreversible evolution is also a necessary condition for an arrow of time is not a relevant question to us because irreversibility unavoidably occurs in quantum field theory.

\subsubsection{Causality}
Let us consider a reversible autonomous evolution equation. If we know the state of the system at any time $t$, we can calculate the state at any time before or after $t$. It is crucial to have complete knowledge of the state at $t$ so that the autonomous evolution can be used to obtain states in the future and in the past. We find perfect symmetry in time.

In the above situation, we can consider the state at time $t$ as the cause and the evolved state at an earlier or later time as an effect. Causality works perfectly in both time directions, in nice agreement with the ideas of \cite{Price}. However, the symmetry is a property of the autonomous evolution equation, not of time \emph{per se}. The level of description admitting autonomous equations determines what complete knowledge at a given time means. With incomplete knowledge, speaking about causality does not seem to make much sense. This innocuous observation seems to be at the heart of many philosophical discussions about causality. In the setting of autonomous evolution equations, the discussion of causality in reversible systems becomes almost trivial.

The story looks very different if we consider irreversible autonomous systems. They are ideal for forward predictions, where forward indicates that we have chosen a direction of evolution. Backward evolution quickly runs into practical or even fundamental problems because one would have to reconstruct details associated with fast degrees of freedom that, in the forward time direction, relax in an exponential manner. Eliminating details is much easier than reconstructing details, and this is the origin of the thermodynamic \index{Arrow of time}arrow of time. For irreversible systems, the aspect of complete knowledge is still essential for analyzing causality but, in view of the existence of an arrow of time, it is natural to say that the cause comes before the effect.

\subsubsection{Relation to previous work}\label{secpreviousirr}
The idea that irreversibility should be an intrinsic feature of quantum field theory occurs also in the work of Petrosky and Prigogine \cite{PetroPrigo96,PetroPrigo97,PetroPrigo88}, but the present approach is considerably less radical. Petrosky and Prigogine make the criticism that quantum mechanics is formulated by following the patterns of classical integrable systems too closely. In particular, for large nonintegrable systems with a continuously varying energy spectrum, they find it necessary to generalize the usual Hilbert-space formulation of quantum mechanics to a Liouville-space extension with a distinctly richer structure, thus allowing for a collapse of trajectories and singular density matrices. The extended spaces are similar to the rigged Hilbert spaces or Gelfand triples introduced for treating continuous spectra with non-normalizable eigenfunctions.\footnote{For an intuitive introduction to rigged Hilbert spaces see, for example, \cite{delaMadrid05}).} As a consequence of resonances in large nonintegrable systems, complex eigenvalues with imaginary parts signaling the occurrence of irreversibility (for example, caused by relaxation or diffusion) can arise in the spectrum of the evolution operator on the extended space. In other words, persistent rather than only transient interactions can occur in scattering processes. Technically speaking, the original group description splits into two semigroups, each one outside the range of the other one (one in which equilibrium is reached in the future, the other in which equilibrium is reached in the past). The approach of Petrosky and Prigogine can eliminate divergences from the usual unitary-evolution theory. In spite of the common goal of unifying dynamics and thermodynamics, thus introducing an \index{Arrow of time}arrow of time, and handling problems caused by the reversible approach the present approach seems to deviate fundamentally from the work of Petrosky and Prigogine, as the avoidance of environmental effects is essential to them \cite{PetroPrigo96}, whereas we deliberately represent the small-scale features of field theories by a heat bath. As an additional goal, Petrosky and Prigogine wish to describe the measurement process in dynamical terms.

As it is difficult to understand and appreciate the highly sophisticated and philosophically deep work of Prigogine and coworkers, I would like to quote from an obituary for Ilya Prigogine written by his close long-term friend Stuart Rice \cite{Rice04}:

\begin{quotation}
Perhaps the most daring---and most controversial---of Prigogine's work was his attempt to reconcile the microscopic-macroscopic irreversibility dichotomy by modifying the fundamental equations of motion. The conventional picture is that the equations of motion of quantum or classical mechanics are `exact' and that the second law of thermodynamics is to be interpreted as a macroscopic consequence of loss of correlations in the motions of the particles through averaging, or loss of information, or loss by some other means. Prigogine turned the question around and asked that if one accepted the second law of thermodynamics as `exact,' would it be possible to modify the equations of motion to preserve what is known about solutions to those equations and also have the second law emerge as an exact description of macroscopic behavior without use of further hypotheses. He and coworkers postulated such a modification and showed that, at least in solutions generated by perturbation theory, it had the desired features. It remains to be seen whether this development will fundamentally alter our worldview or will prove to be an interesting but fruitless theoretical byway.
\end{quotation}

There is another interesting variation of quantum mechanics that uses non-unitary time evolution: the Ghirardi-Rimini-Weber theory \cite{GhirardiRimWeb86}. The work of these authors focuses on the measurement problem rather than thermodynamic issues. The basic idea is that the collapse of a wave function should not be caused by a measurement; collapses are rather assumed to occur spontaneously. They focus on position measurements, so that the use of configuration space is essential, and assume that the probability for a spontaneous localization in a microscopic system is extremely small; more precisely, for a single particle, a spontaneous localization occurs only once every $10^8$--$10^9$\,yr. Therefore, a spontaneous localization is practically never observed in a microscopic system. In a macroscopic system containing many particles, for example, by including a measuring device, spontaneous localization becomes extremely probable. The Ghirardi-Rimini-Weber theory thus eliminates the special character of measurements in quantum mechanics at the expense of introducing a very small deviation from reversible, or unitary, time evolution. Interesting philosophical comments on the Ghirardi-Rimini-Weber theory, including \index{Ontology}ontological implications, can be found in \cite{Allorietal08}.

\subsection{On the measurement problem}\label{secmeasureprob}
Let us start with the question: What is the measurement problem? According to Maudlin (see p.\,7 of \cite{Maudlin95}), there is no straightforward answer to this question but rather considerable uncertainty: ``At least in the philosophical literature, there seems to be general agreement that there is a central interpretational problem in quantum theory, namely the measurement problem. But on closer examination, this seeming agreement dissolves into radical disagreement about just what the problem is, and what would constitute a satisfactory solution of it.'' Maudlin supports his statement by citing Richard Feynman (see p.\,471 of \cite{Feynman82}), ``\ldots\ we always have had a great deal of difficulty in understanding the world view that quantum mechanics represents. \ldots\ It has not yet become obvious to me that there's no real problem. I cannot define the real problem, therefore I suspect there's no real problem, but I'm not sure there's no real problem.''

Many physicists tend to ignore the measurement problem, most conveniently by pretending that the \emph{Copenhagen Interpretation} basically developed by Niels Bohr and Werner Heisenberg in the years 1925--1927 has solved all the problems. According to Whitaker (see Chapter~5 of \cite{WhitakerA}), the Copenhagen Interpretation is designed to avoid logical confusion and to provide a rigorous basis for the discussion of experimental results by modifying the usual forms of conceptual analysis and the usual modes of language. Bohr postulates that the interpretation of experimental results on microscopic systems rests essentially on classical concepts because measuring instruments cannot be included in the actual range of applicability of quantum mechanics. Experiments involve a microscopic, atomic region and a macroscopic classical region, where the division between the two comes with a certain ambiguity. In the words of Paul Feyerabend (see p.\,217 of \cite{Feyerabend62ip}), ``Bohr maintains that all state descriptions of quantum mechanical systems are \emph{relations} between the systems and measuring devices in action and are therefore dependent upon the existence of other systems suitable for carrying out the measurement.'' The insight that any given application of classical concepts precludes the simultaneous use of other classical concepts is at the heart of Bohr's new mode of description, called \emph{complementarity}. Position and momentum provide the fundamental example of this primary aspect of complementarity: if one of these quantities is examined experimentally or discussed theoretically, we are precluded from measuring or discussing the other (\emph{mutual exclusiveness}); together they provide a complete classical description of the motion of a particle (\emph{joint completion}). The wave-particle complementarity is a secondary aspect of Bohr's new mode of description, using two different limited classical concepts to describe quantum phenomena. The word `measurement' does not imply determination of a pre-existing property of the system. Instruments don't play a purely passive role---they rather allow observers to study the behavior of a system under different conditions, experimental conditions that exhibit complementary descriptions of the physics. In the Copenhagen Interpretation, quantum mechanics is restricted to systems set up and observed by experimenters.

In his monumental philosophical analysis of quantum theory and the Copenhagen Interpretation, which he admitted to be superior to a host of alternative interpretations, Paul Feyerabend stated (see p.\,192--193 of \cite{Feyerabend62ip}):
\begin{quotation}
   \ldots many physicists are very practical people and not very fond of philosophy. This being the case, they will take for granted and not further investigate those philosophical ideas which they have learned in their youth and which by now seem to them, and indeed to the whole community of practicing scientists, to be the expression of physical common sense. In most cases these ideas are part of the Copenhagen Interpretation.

   A second reason for the persistence of the creed of complementarity in the face of decisive objections is to be found in the \emph{vagueness} of the main principles of this creed. This vagueness allows the defendants to take care of objections by \emph{development} rather than a \emph{reformulation}, a procedure which will of course create the impression that the correct answer has been there all the time and that it was overlooked by the critic. Bohr's followers, and also Bohr himself, have made full use of this possibility even in cases where the necessity of a reformulation was clearly indicated. Their attitude has very often been one of people who have the task to clear up the misunderstandings of their opponents rather than to admit their own mistakes.
\end{quotation}

In other words, the interpretation of quantum mechanics provides a typical example for how a lack of \index{Pluralism, scientific}scientific pluralism through the identification of nature with a particular representation of nature leads to scientific dogmatism and stagnation \cite{RibeiroVideira98}. ``Dogmatism, however, should be alien to the spirit of scientific research, and this quite irrespective of whether it is now grounded upon `experience' or upon a different and more `aprioristic' kind of argument'' (see p.\,231 of \cite{Feyerabend62ip}). In his contribution to the 1976 Nobel Conference, Murray Gell-Mann describes the role of Bohr in a much more drastic way (see p.\,29 of \cite{GellMann79ip}): ``The philosophical interpretation of quantum mechanics is still probably not complete, but operationally quantum mechanics is in perfect shape. (The fact that an adequate philosophical presentation has been so long delayed is no doubt caused by the fact that Niels Bohr brainwashed a whole generation of theorists into thinking that the job was done 50 years ago.)'' Even Andrew Whitaker, who certainly appreciates the merits of Bohr's work and presents a thoughtfully balanced view of the development of quantum mechanics, closes with some critical remarks on Bohr and the Copenhagen interpretation (see pp.\,416--417 of \cite{WhitakerA}):
\begin{quotation}
   I have said that, at least at one level and in the short term, the general espousing of Bohr's position in the 1920s and 1930s was pragmatically useful, for it freed physicists to study the many applications of quantum theory. At a higher level, though, and in a rather longer term, it was anything but the most useful approach. It denied those who might have been interested the opportunity to think deeply, to progress further in understanding the theory, and just possibly to reach important new conclusions.

   \ldots

   It is when we move outside the strictly scientific area that the Bohrian perspective seems, at least in retrospect, badly flawed, and we may take from this an important general message. It can never be acceptable in science to stifle thought and inhibit discussion, as the advocates of Copenhagen certainly did.
\end{quotation}
Whitaker moreover asks whether Bohr's or Einstein's approach and suggestions stimulated and encouraged thoughtful considerations of the problems of quantum theory, the possibility of interesting and constructive ideas, and perhaps eventually important new physics, and he gives the clear answer (see p.\,416 of \cite{WhitakerA}): ``From this point of view, Einstein's approach must be judged far superior to that of Bohr.''

In conclusion, hiding behind Bohr and the Copenhagen Interpretation is not a viable option. There clearly is a need for understanding the world view represented by quantum mechanics and in particular, the measurement problem.

One way to describe the measurement problem is to recognize that there exists an inconsistency in the following three statements, so that not all three of them can be true \cite{Maudlin95}:\label{statementsmeasprob}
\begin{enumerate}
  \item The wave function contains complete information about all physical properties of a quantum system; no additional variables (often called `hidden variables') are required.
  \item The wave function always evolves according to the linear Schr\"odinger equation; no modifications (such as spontaneous collapses) are required.
  \item Measurements have determinate outcomes.
\end{enumerate}
Maudlin strongly suggests to give up the first or second statement. He recommends Bohm's theory with hidden variables (using particle trajectories in space in addition to wave functions) and the Ghirardi-Rimini-Weber theory (introducing spontaneous collapses or, more precisely, spontaneous localizations, as discussed in Section \ref{secpreviousirr}) as the most promising approaches (see also the more informal presentation in Chapter~7 of \cite{Albert}). Both approaches use position space in an essential way.\footnote{An elucidating mathematical formulation and a deep philosophical discussion of Bohmian mechanics and the Ghirardi-Rimini-Weber theory, emphasizing their common structure and revealing possible primitive \index{Ontology!primitive ontology}ontologies associated with these theories, can be found in \cite{Allorietal08}.} On the other hand, Maudlin opts vehemently against descriptions of collections of systems rather than individual systems because they would also deny the first statement. As a result of his fundamental philosophical analysis of quantum mechanics, Paul Feyerabend arrives at the contrary conclusion that it is ``very clear that the problem of measurement demands application of the methods of statistical mechanics \emph{in addition} to the laws of the elementary theory'' (see p.\,248--249 of \cite{Feyerabend62ip}).

Particle physics appears to be all about statistical results. The CERN press releases on the eve of the discovery of the Higgs boson in the end of 2011 nicely illustrate this point. Statements like
\begin{quotation}
   We do not exclude a Standard Model Higgs boson with a mass between $115$ GeV and $127$ GeV at $95$\% confidence level \ldots
\end{quotation}
or
\begin{quotation}
   The main conclusion is that the Standard Model Higgs boson, if it exists, is most likely to have a mass constrained to the range $116$--$130$ GeV by the ATLAS experiment, and $115$--$127$ GeV by CMS. Tantalising hints have been seen by both experiments in this mass region, but these are not yet strong enough to claim a discovery \ldots
\end{quotation}
clearly reveal the statistical character of the knowledge gain in fundamental particle physics. It is not our goal to describe the individual proton-proton collision process shown in Figure~\ref{figHiggsCERN}. It is the statistical features of many of such collisions that contain the exciting information about the existence and mass of the Higgs boson. As a further illustrative example, the celebrated, incredibly precise theoretical result for the electron magnetic moment provided by quantum electrodynamics is obtained by the perturbative evaluation of a suitable correlation function.

We finally point out that our discussion has not touched relativistic constraints which make the quantum measurement problem much harder in relativistic quantum mechanics \cite{Barrett14}. Barrett emphasizes that ``we have no idea whatsoever how to understand entanglement in a relativistic context'' (see p.\,173 of \cite{Barrett14}). The measurement problem then needs to be resolved jointly with an understanding of the entangled states of spacelike separated systems.\footnote{Maybe all these problems can be resolved in the wider context of algebraic quantum field theory by means of the promising idea of modular localization \cite{Schroer15}.} Barrett feels that the essential reliance of Bohmian mechanics and the Ghirardi-Rimini-Weber theory on configuration space renders them manifestly incompatible with relativistic constraints.

\subsection{Approaches to quantum field theory}\label{secQFTinterp}
Figure~\ref{figHiggsCERN} on p.\,\pageref{figHiggsCERN} suggests that a large number of particles can be created in a high-energy proton-proton collision. Moreover, as pointed out before, particle-antiparticle pairs can be created and annihilated spontaneously, for example, oppositely charged particles together with a photon through electromagnetic interactions. A theory of fundamental particles must hence be able to handle an arbitrarily large, variable number of particles. This requires going beyond quantum mechanics, which is only applicable to a fixed (usually small) number of particles. Such a generalization of quantum mechanics is one of the tasks of quantum field theory. In view of the high energy of the colliding particles and the possible occurrence of massless particles moving with the speed of light, a proper theory must obey the principles of special relativity.

The other task of quantum field theory is to provide rules for the quantization of classical field theories, such as the one given by Maxwell's equations for electromagnetic fields. Of course, for this task, valuable guidance is provided by the well-established quantization procedures of quantum mechanics. Before we discuss three different approaches to quantum field theory (referred to as pragmatic, rigorous, and dissipative), it is worthwhile to offer some remarks on the relation between particles and fields. In other words: Why should quantum field theory be the proper framework for fundamental particle physics? We also offer a brief summary of the main problems of pragmatic quantum field theory.

\subsubsection{Particles versus fields}\label{secpartfield}
Any basic textbook on classical physics contains an early chapter on the dynamics of point particles. The equations of motion for interacting point particles contain interaction forces, typically  between pairs of particles, which depend on the relative position of the particles. In a first step, one then presumes that all forces in nature correspond to instantaneous interactions between bodies a certain distance apart. In a second step, one argues that such interactions cannot really be instantaneous and that one needs to understand the mechanisms for the spatial transmission of interactions. In the context of \index{Electrodynamics}electromagnetic interactions, one eventually begins to understand what such talk really means. \index{Maxwell equations}Maxwell's equations imply (i) that a moving charged particle generates an electromagnetic field at the position of the particle, and (ii) that such a field propagates through space in a wavelike manner with the speed of light. The instantaneous field at the position of another charged particle then causes a Lorentz force on that particle. We thus arrive at a classical picture in terms of both particles and fields which play clearly distinct roles. Particles create fields and fields transmit interactions between particles.\footnote{In this simplified heuristic discussion of the roles of classical particles and fields for motivational purposes, we neglect the subtle problem of self-interactions for a charged particle.}

According to quantum mechanics, a particle can be described by the Schr\"odinger equation, the Klein-Gordon equation, or the Dirac equation for a single- or multi-component complex field known as the wave function. By quantization, we can thus go from particles to complex classical fields. Of course, the evolution of these fields is not given by the usual rules of classical field theories but by the above-mentioned equations.

On the other hand, in the quantization of electromagnetic fields, the concept of photons as field quanta arises. As demonstrated by Einstein's explanation of the photoelectric effect (1905), electromagnetic waves indeed have some particle-like properties associated with these photons, so that we can go from fields to particles. Of course, these field quanta are not particles in a classical sense, but rather are characterized by the properties imposed by quantization rules.

Moreover, we can formally quantize wave functions with the same ideas as electromagnetic fields to introduce the quanta associated with the original particles; this procedure is known as \index{Second quantization}second quantization. In the quantum world, interactions then occur in `collisions' between various types of field quanta.

We have seen that, by quantization, we can go back and forth between particles and fields (even if some of the steps admittedly are quite \emph{ad hoc}). The classical story of particles and fields becomes a quantum story in which particles develop some field-like properties and fields develop some particle-like properties. The resulting particle-field duality (or complementarity) should not necessarily be considered as bad news. The distinct classical concepts of particle and field do not become obfuscated, but they get nicely unified in the single concept of field quantum. The attractiveness of such an unified approach to resolve the particle-wave duality in an elegant mathematical formulation was first recognized by Pascual Jordan in 1926--1927 (for more details, see the remarks on the early history of quantum field theory in Section~\ref{sechistroots}). Whether we perceive macroscopic numbers of field quanta as classical particles or fields depends on whether they are bound or free. In terms of the two classes of particles in quantum mechanics, fermions are associated with what we tend to think of as particles or matter, bosons describe the fields mediating interactions. From the unified point of view of quantum field theory, both bosons and fermions constitute the material world.

The question ``particles or fields?'' seems to be a big issue in philosophical discussions of quantum field theory, with severe implications for \index{Ontology}ontology. If we wish to insist on our classical intuition for particles and fields, the answer is clear: ``neither particles, nor fields!'' It actually is one of the tasks of quantum physics to extend our macroscopically limited classical intuition to atomic and subatomic scales. We can only try to describe some of the aspects of the subatomic world in terms of the classical concepts of our direct experience, but we should ultimately rely more on a consistent mathematical image than on only partially adequate intuitive terms. The intuition develops in working with the image. Instead of lamenting the particle-field dualism, one should celebrate the unifying monism of all matter brought about by field quanta based on the idea of \index{Second quantization}second quantization.

A more drastic warning that we should not think or speak too classically about quantum systems has been issued by Auyang (see p.\,64 of \cite{Auyang}): ``Sure, quantum systems have no classical properties. But why can't they have quantum properties? Is it more reasonable to think that quantum mechanics is necessary because the world has properties that are not classical?'' She further emphasizes the point (see p.\,78 of \cite{Auyang}): ``Unlike classical properties, quantum properties are not visualizable, but visualizability is not a requirement for physical properties. The peculiarity of quantum systems lies in the specific features of their properties, not in the violation of the concept of properties.''

As implied by this warning, the above remarks on going back and forth between particles and fields clearly are a bit naive. After all, the properties of particles and fields are incompatible. Cao rightly criticizes the lax approach of physicists and, for several reasons, finds the idea of \index{Second quantization}second quantization inadequate (see Section~7.3 of \cite{Cao}). However, after a lengthy discussion of a variety of subtleties, on p.\,172 of \cite{Cao}, he eventually arrives at the surprisingly simple conclusion ``Here is the bridge connecting the continuous field and the discrete particle: radiation as the quantum physical reality is both waves and particles.'' A few pages later (p.\,178), Cao concludes that ``the classical distinction between the discrete particle and the continuous force field is dimmed in quantum theory'' and he summarizes: ``Thus the distinction between matter and force field vanishes from the scene, and is to be replaced by a universal particle-field duality affecting equally each of the constituent entities.'' Having said that and being fully aware of the fact that quantum fields are not fields in any usual sense (``as locally quantized fields, they to great extent lost their continuity''), Cao nevertheless states on p.\,211: ``In sum, considering the fact that in QED [quantum electrodynamics] both interacting particles and the agent for transmitting the interaction are the quanta of fermion fields and the electromagnetic field respectively, that is, they are the manifestation of a field \index{Ontology!field ontology}ontology, QED should be taken as a field theory.'' This postulate is understandable because, in interacting theories, typically three or four particles can be `created out of nothing'. Particles, or field quanta, can hence hardly be regarded as a fundamental substance. One might hence be inclined to assume that the fundamental substance is the field out of which field quanta can be generated. In any case, the choice of the lesser of two evils is not a healthy basis for \index{Ontology}ontology, and we should hence simply acknowledge our classical limitations and accept the particle-field duality as a scanty classical depiction of quantum \index{Reality}reality.

One should not too easily conclude that quantum field theory deserves to be classified as a theory of fields merely because it deals with operator-valued fields. In classical field theory, one is interested in the temporal evolution of the actual values of fields, such as the values of temperatures, velocities, or electromagnetic fields as functions of position and time. For operator-valued fields, the operators play the role of observables, not of measured values of observables. For example, whether or not these fields of operators evolve in time depends on our choice of the Heisenberg or Schr\"odinger picture and hence has nothing to do with the evolution of measured values. Instead of operator-valued fields one should hence consider the expectation values of the field operators evaluated with a wave function or a density matrix. These position- and time-dependent expectation values would then be the counterpart of the evolving classical field configurations. This point has been elaborated in great detail by Teller (see Chapter Five of \cite{Teller}).

As the \index{Ontology}ontology of quantum field theory cannot easily by resolved in classical terms, we can at least ask the more modest question about the most useful language. In view of arguments provided by Teller \cite{Teller}, we find the particle language quite appropriate for a theory developed on Fock space. The basic argument in favor of particles is that we deal with discrete, countable entities for which we assume the energy-momentum relationship for relativistic particles. As our Fock space will be constructed from momentum eigenstates, according to the principles of quantum mechanics, we cannot localize the particles within the total volume considered. In that sense, a particle picture is not really justified because, according to our classical intuition, it should be possible to localize particles. A famous no-go theorem for a relativistic quantum mechanical theory of (localizable) particles has been derived by Malament \cite{Malament96ip}. Hegerfeldt \cite{Hegerfeldt98,Hegerfeldt98ip} has shown that, under amazingly general assumptions, there occurs an instantaneous spreading of wave functions over all space. Our quantum particles must hence be non-localizable and are actually spread over the entire space.\footnote{As Hegerfeldt discusses on p.\,244 of \cite{Hegerfeldt98ip}, ``\ldots\ if all systems were spread out over all space to begin with, then no problems would arise. \ldots\ there would be no self-adjoint position [operators] satisfying causal requirements \ldots''} Also the indistinguishability of particles is a non-intuitive quantum feature, although it had been recognized by Gibbs (1902) by thermodynamic considerations well before the advent of quantum mechanics.\footnote{``If two phases differ only in that certain entirely similar particles have changed places with one another, are they to be regarded as identical or different phases? If the particles are regarded as indistinguishable, it seems in accordance with the spirit of the statistical method to regard the phases as identical. In fact, it might be urged that in such an ensemble of systems as we are considering no identity is possible between the particles of different systems except that of qualities, and if n particles of different systems are described as entirely similar to one another and to $n$ of another system, nothing remains on which to base the identification of any particular particle of the first system with any particular particle of the second.'' (See p.\,187 of \cite{Gibbs}).} As anticipated, the particle picture has its limitations for quantum particles, or better, field quanta; nevertheless, the particle language is useful.

The construction of the Fock space relies on single-particle states, which we take as momentum eigenstates of a particle. The particle language suggested by the Fock space is based on independent particles, often called free particles, but eventually we want to consider interactions between these particles. According to Fraser \cite{Fraser08}, the particle concept breaks down for interacting theories. Within our approach, it is natural to consider free particles together with their collision rules. It is important to realize that the free particles associated with a Fock space do not coincide with the phenomenological particles (see p.\,249 of \cite{Ruetsche}). As a consequence of the omnipresent collisions, we actually see clouds of free particles. The concept of a cloud of free particles is somewhat ambiguous because the distinction between a single or several clouds depends on the level of resolution admitted by the strength of the dissipative mechanism. The smaller the smearing effect, the better a cloud can be resolved. In accordance with the idea of effective field theories, the identification and counting of clouds depends on a cutoff or smearing parameter. To get consistent results for different choices of the smearing parameter, the other parameters of the theory have to be properly adjusted, which is the idea of renormalization (to be discussed in detail in Section \ref{secrenormalization}). The ambiguity of clouds is limited to unobservably small scales; on observable scales the clouds are unambiguous---for example, the three clouds associated with the quarks in a proton or neutron, which are surrounded by further clouds representing gluons and other particles.

As the cloud concept is ambiguous, one cannot really introduce well-defined, countable particles for an interacting theory. Unambiguous are only the unobservable free particles and the collision rules which, together with the ambiguous dissipation mechanism, lead to the observable clouds of free particles. We hence cannot speak about states with a well-defined number of interacting particles and, in particular, not about a vacuum state characterized by the condition that it contains zero interacting particles. Note, however, that the difference between free particles and clouds on the scale set by the dissipation mechanism is experimentally inaccessible. The distinction between the particles of the free and interacting theories is an important theoretical step whereas, from an experimental point of view, they cannot be distinguished and there is a direct correspondence between them.

Finally, two remarks concerning the collision process shown in Figure~\ref{figHiggsCERN} are in order. (i) Whereas we cannot localize particles with well-defined momenta in our volume of interest, the collision rules are always chosen such that a collision can only occur if all particles involved simultaneously are at the same position; this is why tracks emerge from centers. (ii) The tracks in Figure~\ref{figHiggsCERN} are not associated with free particles but with clouds; the scales associated with dissipative smearing and clouds are actually much smaller than the resolution of such a diagram, so that each track represents a number of interacting clouds rather than a free particle.

\subsubsection{Pragmatic quantum field theory}
Quantization procedures are traditionally based on a canonical Hamiltonian formulation of the evolution equations for the underlying classical systems. In the canonical approach to quantum mechanics, rooted in Dirac's pioneering work, the canonical Poisson brackets of classical mechanics are replaced by the commutators of quantum mechanics \cite{Diracorig}. This canonical procedure has been adapted to quantum field theory (see, for example, Sections 11.2 and 11.3 of \cite{BjorkenDrell}, Sections 2.4 and 3.5 of \cite{PeskinSchroeder}, or Section I.8 of \cite{Zee}). Even in Feynman's alternative path-integral formulation of quantum field theory \cite{Feynman48}, the justification of the proper action needs to be supported by the canonical approach (see, for example, the introduction to Section~9 of \cite{WeinbergQFT1}). Understanding its Hamiltonian structure is hence indispensable for the quantization of any classical system.

In classical mechanics, one has the choice between the Lagrangian and Hamiltonian formulations; both formulations can also be used for classical field theories (see, for example, Chapters~2, 8, and 12 of \cite{Goldstein}). The Lagrangian approach is based directly on the variational principle for the action, which is defined as the time integral of the Lagrangian. In contrast, the equivalent Hamiltonian approach needs two structural elements, a Hamiltonian and a Poisson bracket required to turn the gradient of the Hamiltonian into a vector describing time evolution. The canonical Poisson structure, which in the non-degenerate case is also know as symplectic structure, is the key to formulating the proper commutators in the quantization procedure. The existence of a non-degenerate Poisson structure is crucial for establishing an underlying variational principle and hence the equivalence of the Lagrangian and Hamiltonian formulations of classical mechanics and field theory \cite{Santilli}.

In view of the equivalence of the Lagrangian and Hamiltonian formulations, the canonical approach and the path-integral formulation, which is based on the time integral of the Lagrangian, may jointly be referred to as \emph{Lagrangian quantum field theory} instead of pragmatic or conventional quantum field theory. When faced with nature, Lagrangian quantum field theory turns out to be an extremely successful theory. In particular, incredibly accurate predictions have been made in the context of electrodynamics. Many of the most striking predictions rely on perturbation theory. With the recent discovery of the Higgs boson, the success story of the so-called standard model of electromagnetic, weak, and strong interactions has continued with another spectacular highlight of Lagrangian quantum field theory.

\subsubsection{Main problems of quantum field theory}\label{secmainproblemsQFT}
The key problem in going from quantum mechanics to quantum field theory is the step from a finite (usually small) to an infinite number of degrees of freedom. An exhaustive discussion of the fundamental problems of quantum mechanics with infinitely many degrees of freedom has been attempted by Laura Ruetsche \cite{Ruetsche}. In this section, we only sketch the most important defects.

\noindent (i) According to \emph{Haag's theorem}, the interaction picture cannot exist for a relativistic quantum field theory because, for a nontrivial translation-invariant local interaction, it is impossible that the free and the interacting fields act on the same Hilbert space and coincide at some initial time. This problem arises because there is only one translation invariant vacuum state so that the interacting theory must have the same vacuum as the free theory. The fact that a single Hilbert space cannot properly accommodate both free and interacting fields has further unpleasant consequences. For example, ``No single fundamental particle notion embraces the ingoing and outgoing particles encountered in the iconic phenomena of particle physics, as well as the interacting particles portrayed in Feynman diagrams'' (see p.\,250 of \cite{Ruetsche}) and ``The incommensurability of these particle notions precludes extending a single fundamental particle notion over the entire microhistory of a scattering experiment'' (see p.\,251 of \cite{Ruetsche}). A possible workaround to escape the consequences of Haag's theorem is to restrict the local interaction in space, to calculate vacuum correlation functions for the restricted interaction, and to eliminate the restrictions by a final limiting procedure.

\noindent (ii) A more general problem arises because, if one quantizes field theories by specifying canonical commutation relations, \emph{inequivalent representations} do exist (see Section~3.3 of \cite{Ruetsche}). This situation is very different from the case of mechanic systems with a finite number of degrees of freedom, for which all possible irreducible representations of canonical commutation relations are unitarily equivalent. This lack of uniqueness in the representation of quantum fields suggests a fundamental ambiguity that needs to be resolved by choosing the proper physical representation among uncountably many inequivalent representations.

\noindent (iii) Of course, the \emph{measurement problem} discussed in Section~\ref{secmeasureprob} does not get any simpler by going from a finite to an infinite number of degrees of freedom. As pointed out before, this problem is particularly hard for the relativistic systems we are dealing with in fundamental particle physics.

\noindent (iv) Early quantum field theory was plagued by many divergent expressions. \emph{Renormalization} was developed into a powerful mathematical tool to remove these divergencies. However, renormalization should not appear as a mysterious trick to fix disastrous problems, but as a perfectly natural ingredient to effective quantum field theories. According to Ruetsche, ``Renormalization Group techniques, validated by taking thermodynamics quite seriously, are instrumental in identifying plausible future directions for QFTs [quantum field theories]'' (see p.\,339 of \cite{Ruetsche}).

In spite of all these problems, Wallace \cite{Wallace06} finds Lagrangian quantum field theory, which he refers to as the naive quantum field theory used in mainstream physics, to be ``a perfectly respectable physical theory'' and ``a legitimate object of foundational study.'' However, the majority of philosophers working on the foundations of quantum field theory prefer to rely on the more rigorous mathematical formulations to be discussed next.\footnote{Philosophers have been seriously interested in quantum field theory as the most fundamental theory of matter since about 1990; the pioneering books by Auyang \cite{Auyang} and Teller \cite{Teller} appeared in 1995.}

\subsubsection{Rigorous quantum field theory}
The problem sketched in Section~\ref{secmainproblemsQFT} stimulated the search for a mathematically rigorous, axiomatic formulation of quantum field theory. There are two main lines of this development. One line is strongly influenced by von Neumann's algebraic formulation of quantum mechanics, proposed in the 1930s, and the further developments of Gelfand, Neumark, and Segal in the mid-1940s. In the context of quantum field theory, this approach was largely developed by Rudolf Haag (a decade of work culminated in the famous paper \cite{HaagKastler64}; for an updated and more complete exposition see \cite{Haag}). As abstract C$^*$-algebras (corresponding to the bounded operators on Hilbert spaces) are the starting-point for this approach, it is known as \emph{algebraic quantum field theory}. The other line, which is influenced by Dirac's $\delta$ function, Schwartz's more general theory of distributions, and the development of the concept of rigged Hilbert spaces, was initiated by the famous work of Arthur S.\ Wightman in 1956 \cite{Wightman56} and refined by Nikolay N.\ Bogoliubov in the late 1960s. This line became known as \emph{axiomatic quantum field theory} (although axiomatic is not really the distinguishing feature). The notion of a field, in the sense of an operator-valued distribution, plays a central role in axiomatic quantum field theory, whereas the algebraic approach focuses on local observables.

Whereas mathematical rigor certainly is an essential feature (see our \index{Metaphysical postulates!first metaphysical postulate}first metaphysical postulate on p.\,\pageref{metaphys1}), the rigorous approaches have a severe problem: a derivation of empirical consequences is very difficult. We have beautiful mathematical rigor on one hand and a spectacular success story on the other hand, where the relation between the rigorous and pragmatic approaches is often perceived as unclear. Whereas mainstream physics clearly goes with the successful pragmatic approach, the supposed necessity to choose between ``no (mathematical) rigor'' and ``no (practical) relevance'' is much harder for philosophers, as they should insist on both (so should physicists). The powerful concept of \emph{modular localization}, which is based on the Tomita-Takesaki modular theory of operator algebras, has a promising potential for achieving the big goal of bringing the algebraic and pragmatic approaches together, thus even revealing the reasons for the success of the pragmatic approach and at the same time leading to a deeper understanding of the quantization of gauge theories and to a demystification of the Higgs mechanism (for an overview, see \cite{Schroer15}). Moreover, the ensemble aspect and statistical mechanics type of probability follow naturally from modular localization, although in a more radical way compared to how we here implement these features by an irreversible contribution to time evolution on a finite space.

\subsubsection{Dissipative quantum field theory}
The present approach to quantum field theory is an alternative attempt to combine \emph{rigor} and \emph{relevance}. In essence, the irreversible contribution to the evolution of an effective quantum field theory according to our \index{Metaphysical postulates!fourth metaphysical postulate}fourth metaphysical postulate (see p.\,\pageref{metaphys4}) plays the role of a natural, intrinsic, dynamical cutoff in Lagrangian quantum field theory. Concerning relevance, our goal is to reproduce the celebrated results of Lagrangian quantum field theory and to offer some new tools (in particular, a new stochastic simulation methodology): as a first step, we consider the scalar field theory known as $\varphi^4$ theory (see Chapter~\ref{chapphi4}); a preliminary test of the simulation methodology for quantum electrodynamics has been performed in \cite{hco214}. Concerning rigor, the present development rests on the systematic use of the conventional Fock space (see Section~\ref{sectionFock}). We begin with a finite number of possible momentum states for a single particle; only after evaluating the quantities of interest (correlation functions) we go to the limit of infinitely many momentum states.

According to Ruetsche, our approach could be labeled as Hilbert space conservatism (see Section~6.2 of \cite{Ruetsche}). Using Fock spaces as the basic arena of quantum field theory may be considered as equivalent to a particle notion: ``A fundamental particle interpretation is available only when there is an irreducible Fock space representation comprehending all physically possible states'' (p.\,259 of \cite{Ruetsche}) and ``Empirical successes mediated by the particle notion, and explanations relating the energy of quantum field to that of their classical predecessors, require the structure of a privileged Fock space representation'' (p.\,348 of \cite{Ruetsche}).

In several crucial aspects, our approach deviates from standard Lagrangian quantum field theory:
\begin{itemize}
  \item The problem of inequivalent representations is avoided by using only a finite number of momentum states for a single particle and postponing the limit of infinitely many states to the end.
  \item Dissipation leads to a rapid spatial smearing of the small-scale features and hence provides a natural ultraviolet regularization mechanism (see the discussion of irreversibility in Section \ref{sectionirrev}). Our fundamental evolution equation is a quantum master equation for a density matrix in Fock space. By relying on a thermodynamically consistent quantum master equation, rigorous tools for the analysis of the qualitative solution behavior become available.
  \item The quantum master equation describes collisions of free particles (field quanta) and the interaction with a heat bath representing small-scale features. The construction of spatial fields is not a part of our mathematical image of particle physics. Spatial fields are only used for a heuristic motivation of the collision rules respecting fundamental principles, such as those imposed by special relativity.
  \item The concrete implementation of dissipation leads to a nonperturbative connection between the free and interacting theories. It is natural that such a connection arises through the dissipative mechanism which defines the clouds of free particles that cannot be resolved and may be interpreted as particles of the interacting theory.
  \item We do not make use of the interaction picture. In general, the thermodynamic quantum master equation is nonlinear so that there is no alternative to the Schr\"odinger picture, which we use throughout this book.
  \item The steady states of the quantum master equation are equilibrium density matrices characterized by the Boltzmann factors for a given temperature. An equilibrium density matrix plays the role of an underlying vacuum state, at least, in the low-temperature limit.
  \item Infinities are avoided by considering weak dissipation (providing an ultraviolet cutoff) and a large finite volume of space (providing an infrared cutoff). Limits are only performed on the final predictions for a countable set of quantities of interest. We focus entirely on correlation functions rather than algebras of observables. The existence of well-defined limits is deeply related to the usual renormalization program.
  \item To obtain a stochastic unraveling of the fundamental quantum master equation, we consider stochastic trajectories in Fock space. The stochastic representation substantiates the idea of quantum jumps and collisions and moreover leads to a new simulation methodology in quantum field theory.
  \item We do not feel the need to provide the rules for quantizing an arbitrary classical field theory, but rather focus on the intuitive formulation of the quantum theories of the fundamental particles and their interactions. Once all interactions are unified, only a single quantum field theory needs to be formulated.
\end{itemize}

Of course, these remarks provide only a first glimpse at the structure and standing of dissipative quantum field theory and all the details remain to be elaborated. The basic mathematical and physical elements required to implement dissipative quantum field theory are described in Section \ref{secMPelements}. How this approach works for scalar field theory ($\varphi^4$ theory) is then discussed in detail in Chapter~\ref{chapphi4}.

\section{Mathematical and physical elements}\label{secMPelements}
With guidance from philosophical ideas, we can now lay the mathematical and physical foundations of quantum field theory. We first develop the Fock space description of quantum fields and then introduce time evolution. A careful discussion of `quantities of interest' is crucial for the comparison with the real world. Symmetries are another important topic because they guide the choice of concrete models. Stochastic unravelings are developed as a starting point for a new simulation methodology. Whenever we feel the need to illustrate the basic ideas, we do that for the example of $\varphi^4$ theory. Generalizations required, for example, in the presence of polarization or spin are postponed to later developments.

Before going into any mathematical and physical details, a general question concerning the proper characterization of the approach to be outlined in the present section should be addressed: Does our approach deserve to be classified as \emph{intuitive} or might it be disqualified as \emph{naive}? It is natural that intuition played an important role in the pioneering steps towards quantum field theory, right after the advent of quantum mechanics in the mid-1920s. Therefore, we heavily rely on the early approach. As the early developments got stuck in severe mathematical problems for about two decades, we need to argue why, nevertheless, it would be wrong to consider our approach based on the early ideas and concepts as unduly naive.

The enormous obstacles to early quantum field theory have eventually been overcome by a rather formal renormalization procedure. Also the quantization of field theories with gauge degrees of freedom involves some very formal ideas. It so happened that, in present-day quantum field theory, \emph{formal} almost became synonymous with \emph{advanced} or even \emph{worldly-wise}. It is the main goal of the present book to show that the key developments in quantum field theory are nicely compatible with an intuitive approach. Renormalization occurs most naturally in the effective quantum field theories implied by the dissipative approach to fundamental particle physics. Gauge degrees of freedom can be handled within an elegant version of the modern ideas known as \index{BRST quantization}BRST quantization. The issue of inequivalent representations is avoided by employing limiting procedures. Fundamental developments rather than tricky rules should be at the heart of quantum field theory. We would like to (re)discover the simplicity of quantum field theory. In other words, our goal is to establish \emph{intuitive}, \emph{advanced}, and \emph{worldly-wise} as nicely compatible, mutually reinforcing aspects of quantum field theory.

The present book is significantly different from the enormously successful standard textbooks on quantum field theory of the last 50 years: Bjorken and Drell \cite{BjorkenDrellQM,BjorkenDrell}, Itzykson and Zuber \cite{ItzyksonZuber}, Peskin and Schroeder \cite{PeskinSchroeder}, or Weinberg \cite{WeinbergQFT1,WeinbergQFT2,WeinbergQFT3}. In the preface to his more recent textbook \emph{The Conceptual Framework of Quantum Field Theory} \cite{Duncan}, Duncan blames the standard books for following a purely pragmatic desire to `start with a Lagrangian and compute a process to two loops' rather than addressing the important conceptual issues. In his opinion, ``if the aim is to arrive at a truly deep and satisfying comprehension of the most powerful, beautiful, and effective theoretical edifice ever constructed in the physical sciences, the pedagogical approach taken by the instructor has to be quite a bit different from that adopted in the `classics' enumerated above'' (see p.\,iv of \cite{Duncan}; I added the book by Peskin and Schroeder to the list of `classics'). Whereas Duncan's conceptual framework covers a wide spectrum of standard material, the present work rather focuses on philosophical considerations for conceptual clarification and it might hence be advisable to use it together with one of the more comprehensive `classics'.

\subsection{Fock space}\label{sectionFock}
Every textbook on quantum mechanics tells us that the proper arena for describing quantum phenomena is provided by complete separable \emph{Hilbert spaces}. These are complex vector spaces with an inner product, which implies a norm and hence a metric. Separability requires that there exists a countable basis, completeness means that every Cauchy sequence in a Hilbert space has a limit in that space. These spaces allow us to do some powerful mathematical analysis. In particular, the philosophical \emph{horror infinitatis} (see Section~\ref{sectioninfinity}) is addressed by restricting the dimension of Hilbert spaces to countable and by allowing for limiting procedures.

We assume that the possible quantum states of a single particle are labeled by $\nu=1,2,\ldots$. For convenience, we speak about particles, but the construction works for any kind of quantum entity. We then introduce the following states representing $N = \sum_{\nu=1}^\infty n_\nu$ independent particles in two interchangeably used forms,
\begin{equation}\label{Npartstates}
    \Dket{n_\nu} = \Dket{n_1,n_2,\ldots} \,,
\end{equation}
where $\Dket{\ldots}$ indicates vectors in Hilbert space, $n_1$ is the number of particles in state $1$, $n_2$ is the number of particles in state $2$, $\ldots$. No labeling of the particles is required in introducing (\ref{Npartstates}), which relies on counting only; in the words of Teller \cite{Teller}, these quantum entities do not possess `primitive thisness,' whereas physicists naively refer to `indistinguishability.' For bosons, each occupation number $n_\nu$ is a nonnegative integer; for fermions, each $n_\nu$ must be $0$ or $1$ because the Pauli principle forbids double occupancies. If all occupation numbers $n_\nu$ vanish, the resulting state is denoted by $\Dket{0}$ and referred to as the vacuum state. As the states (\ref{Npartstates}) are characterized by a countable set of integer numbers, the set of all states (\ref{Npartstates}) is also countable. We take the states of the form (\ref{Npartstates}) as basis vectors of the \emph{Fock space} $\cal F$, which is a complex vector space.

Finally, we define a canonical inner product $s^{\rm can}$ by assuming that the states (\ref{Npartstates}) form an orthonormal basis of the Fock space $\cal F$,
\begin{equation}\label{caninnerproddef}
    s^{\rm can}(\Dket{n'_{\nu'}},\Dket{n_\nu}) = \prod_{\nu=1}^\infty \delta_{n_\nu n'_\nu} \,,
\end{equation}
where we have used Kronecker's $\delta$ symbol. The inner product of any two vectors is then implied by antilinearity in the first argument and by linearity in the second argument,
\begin{eqnarray}
    s^{\rm can}(c_1 \Dket{\phi_1}+c_2 \Dket{\phi_2},\Dket{\psi}) &=&
    c_1^* \, s^{\rm can}(\Dket{\phi_1},\Dket{\psi}) + c_2^* \, s^{\rm can}(\Dket{\phi_2},\Dket{\psi}) \,,\\
    s^{\rm can}(\Dket{\phi},c_1 \Dket{\psi_1}+c_2 \Dket{\psi_2}) &=&
    c_1 \, s^{\rm can}(\Dket{\phi},\Dket{\psi_1}) + c_2 \, s^{\rm can}(\Dket{\phi},\Dket{\psi_2}) \,, \qquad\qquad
\label{scanlinantilin}
\end{eqnarray}
for all $\Dket{\phi_1}, \Dket{\phi_2}, \Dket{\phi}, \Dket{\psi_1}, \Dket{\psi_2}, \Dket{\psi} \in \cal F$ and for all complex numbers $c_1, c_2$, where $c_j^*$ is the complex conjugate of $c_j$. These equations define $s^{\rm can}$ as a sesquilinear form. Note that the canonical inner product is positive definite, that is,
\begin{equation}\label{scanposdef}
    s^{\rm can}(\Dket{\phi},\Dket{\phi}) > 0 \,,
\end{equation}
for any $\Dket{\phi} \in \cal F$ different from the zero vector. This completes our construction of the Fock space.

For our further developments, it is convenient to introduce Dirac's bra-ket notation. We consider the dual of the Fock space, which is defined as the set of linear forms on $\cal F$. In particular, for any $\Dket{\phi} \in \cal F$, we can consider the linear form $s^{\rm can}(\Dket{\phi},\cdot)$, which we denote by $\Dbra{\phi}$. In Dirac's elegant notation, the result of evaluating the form $\Dbra{\phi}$ on the vector $\Dket{\psi}$ can then be written as
\begin{equation}\label{dualdef}
    \Dbra{\phi} \big( \Dket{\psi} \big) = \Dbraket{\phi}{\psi} = s^{\rm can}(\Dket{\phi},\Dket{\psi}) \,.
\end{equation}
The inner product $s^{\rm can}$ thus provides a natural mapping between the Fock space and its dual.

As a further convenient tool, we introduce \emph{creation and annihilation operators},\footnote{Some authors prefer to call them \emph{raising and lowering operators} but, in our interpretation, the names creation and annihilation operators are perfectly appropriate.} where we first consider bosons. As for all linear operators, it is sufficient to define them on the base vectors. The creation operator $a_\nu^\dag$ increases the number of particles in the state $\nu$ by one,
\begin{equation}\label{creationopdef}
    a_\nu^\dag \Dket{n_1,n_2,\ldots} =  \sqrt{n_\nu+1} \, \Dket{n_1,n_2,\ldots, n_\nu+1, \ldots} \,,
\end{equation}
whereas the annihilation operator $a_\nu$ decreases the number of particles in the state $\nu$ by one,
\begin{equation}\label{annihilationopdef}
    a_\nu \Dket{n_1,n_2,\ldots} = \left\{ \begin{array}{lll}
    \sqrt{n_\nu} \, \Dket{n_1,n_2,\ldots, n_\nu-1, \ldots} & \mbox{for} & n_\nu > 0 \,,\\
    0 & \mbox{for} & n_\nu = 0 \,.
    \end{array} \right.
\end{equation}
The outcome $0$ for $n_\nu = 0$ is the zero vector of the Fock space and should not be confused with the vacuum state $\Dket{0}$, which is a unit vector describing the absence of particles. The prefactors in the definitions (\ref{creationopdef}) and (\ref{annihilationopdef}) are chosen to be real and nonnegative, to make the operator $a_\nu^\dag$ the adjoint of $a_\nu$ for the inner product $s^{\rm can}$, and to obtain the particularly simple and convenient commutation relations
\begin{equation}\label{ancrecomrel}
    \Qcommu{a_\nu}{a_{\nu'}^\dag} = \delta_{\nu\nu'} \,,
\end{equation}
where $\Qcommu{A}{B}=AB-BA$ is the commutator of two linear operators $A$ and $B$ on $\cal F$. We further note the commutation relations
\begin{equation}\label{ancrecomrelx}
    \Qcommu{a_\nu}{a_{\nu'}} = \Qcommu{a_\nu^\dag}{a_{\nu'}^\dag} = 0 \,,
\end{equation}
which are an immediate consequence of the definitions (\ref{creationopdef}) and (\ref{annihilationopdef}). Moreover, the creation operators allow us to create all Fock base vectors from the vacuum state,
\begin{equation}\label{Fockstatecreate}
    (a_1^\dag)^{n_1} (a_2^\dag)^{n_2} \ldots \Dket{0} =
    \left( \prod_{\nu=1}^\infty \sqrt{n_\nu!} \right) \, \Dket{n_1,n_2,\ldots} \,.
\end{equation}
This result follows by repeated application of the definition (\ref{creationopdef}). Further note that all annihilation operators annihilate the vacuum state. This is a characteristic of the vacuum state for independent particles. The operator $a_\nu^\dag a_\nu$ counts the number of particles in the state $\nu$,
\begin{equation}\label{partnucountop}
    a_\nu^\dag a_\nu \Dket{n_1,n_2,\ldots} = n_\nu \Dket{n_1,n_2,\ldots} \,.
\end{equation}

The above discussion of creation and annihilation operators holds only for bosons. For fermions, creation operators $b_\nu^\dag$ and annihilation operators $b_\nu$ are defined by
\begin{equation}\label{creationopdeff}
    b_\nu^\dag \Dket{n_1,n_2,\ldots} =  \left\{ \begin{array}{lll}
    0 & \mbox{for} & n_\nu =1 \,,\\
    (-1)^{s_\nu} \Dket{n_1,n_2,\ldots, n_\nu+1, \ldots} & \mbox{for} & n_\nu = 0 \,,
    \end{array} \right.
\end{equation}
in accordance with the Pauli principle, and
\begin{equation}\label{annihilationopdeff}
    b_\nu \Dket{n_1,n_2,\ldots} = \left\{ \begin{array}{lll}
    (-1)^{s_\nu} \Dket{n_1,n_2,\ldots, n_\nu-1, \ldots} & \mbox{for} & n_\nu =1 \,,\\
    0 & \mbox{for} & n_\nu = 0 \,,
    \end{array} \right.
\end{equation}
where $s_\nu=n_1+n_2+\ldots+n_{\nu-1}$. The commutation relations (\ref{ancrecomrel}) and (\ref{ancrecomrelx}) are replaced by anticommutation relations,
\begin{equation}\label{ancreanticomrel}
    \Qantico{b_\nu}{b_{\nu'}^\dag} = \delta_{\nu\nu'} \,,
\end{equation}
and
\begin{equation}\label{ancreanticomrelx}
    \Qantico{b_\nu}{b_{\nu'}} = \Qantico{b_\nu^\dag}{b_{\nu'}^\dag} = 0 \,,
\end{equation}
where $\Qantico{A}{B}=AB+BA$ is the anticommutator of two linear operators $A$ and $B$ on $\cal F$. Instead of (\ref{Fockstatecreate}) for bosons, we use the following convention consistent with the definitions (\ref{creationopdeff}) and (\ref{annihilationopdeff}) for fermions,
\begin{equation}\label{Fockstatecreatef}
    (b_1^\dag)^{n_1} (b_2^\dag)^{n_2} \ldots \Dket{0} = \Dket{n_1,n_2,\ldots} \,.
\end{equation}

A more formal construction of the Fock space can be carried out in terms of Hilbert spaces. The $N$-particle Hilbert space
is given by the properly symmetrized or antisymmetrized tensor product of $N$ single-particle Hilbert spaces. The Fock space $\cal F$ is then obtained as the direct sum of all $N$-particle Hilbert spaces. The symmetrization and antisymmetrization appear as extra rules because, in the tensor products, particles get labeled, in contradiction to the absence of `primitive thisness' for quantum entities. In that sense, the construction based on the states (\ref{Npartstates}) is more direct and more appropriate for quantum entities. In a nice formulation of Auyang (see p.\,162 of \cite{Auyang}), the labels ``say too much'' so that we have to ``unsay'' something (``unsaying something is much harder than saying''). The present approach nicely eliminates the need to ``unsay'' anything because an artificial labeling of indistinguishable particles is avoided. These considerations suggest that we should consider the Fock space with base vectors (\ref{Npartstates}) as more fundamental than the creation and annihilation operators, which we have introduced only in a second step.

To conclude the discussion of Fock spaces, we should be more specific about the choice of the single-particle states, so far only labeled by a positive integer $\nu$. We would actually like to choose momentum eigenstates. However, the momentum operator possesses a continuous spectrum with a continuum of generalized eigenstates, which cannot be normalized and hence do not belong to the single-particle Hilbert space. This situation is at variance with the idea that we would like to see the countable dimension of the Fock space at any stage of the development. We hence do not follow the possible option of using the construction of a rigged Hilbert space \cite{delaMadrid05} to handle generalized eigenstates.

If we consider fields in a $d$-dimensional position space, the corresponding space of momentum states is also $d$-dimensional. Instead of choosing a continuum of momentum states, we here work with a finite discrete set of momentum states from a $d$-dimensional lattice,
\begin{equation}\label{Kdlatticedef}
    K^d = \left\{ \bm{k} = (z_1,\ldots,z_d) K_L \, | \, z_j \mbox{ integer with } |z_j| \le Z_L
    \mbox{ for all } j=1,\ldots,d\right\} ,
\end{equation}
where $K_L$ is a lattice constant in momentum space, which is assumed to be small, and the large integer $Z_L$ limits the magnitude of each component of $\bm{k}$ to $Z_L K_L$. For $d=3$, $K^d$ is a cube with center at the origin of a cubic lattice; for $d=2$, $K^d$ is a square centered at the origin of a square lattice. The physical connection of $K_L$ with a finite position space is discussed in the subsequent section. Symmetry with respect to the origin suggests that the fixed inertial system we are working in should be taken as the center-of-mass system. The finite number of elements in $K^d$ correspond to the label $\nu$ of the general construction of Fock spaces. With the truncation parameters $K_L$ and $Z_L$ we keep the set $K^d$ discrete and finite. In the end, we are interested in the limits $Z_L \rightarrow \infty$ and $K_L \rightarrow 0$, so that the entire momentum space is densely covered. In the limit $Z_L \rightarrow \infty$, we use the symbol \label{Kbardefpage} $\bar{K}^d$ for the infinitely large lattice of momentum vectors (there are still countably many). For massless particles, the momentum state $\bm{k}=\bm{0}$ must be excluded because massless particles cannot be at rest. If the origin is excluded from the lattice, we use the symbols $K^d_\times$ and $\bar{K}^d_\times$ instead of $K^d$ and $\bar{K}^d$, respectively.

In general, in addition to momentum, further properties may be needed to characterize the single-particle states. For example, for $d=3$, an electron needs to be characterized by an additional spin state and a photon possesses an additional polarization state. Like momentum, the choice of labels should be based on suitable elements of an extended version of Kant's transcendental aesthetic. If we deal with a number of different fundamental particles, such as electrons, photons, quarks, and gluons, each species comes with its own Fock space. The choice of these particles requires an \index{Ontological commitment}ontological commitment. Ideally, the entire spectrum of fundamental particles, no less than $61$ particles for the widely accepted and remarkably successful `standard model' (see Section \ref{secapproachsum} for details), would follow from a still unrevealed element of transcendental aesthetic. All the Fock spaces for different particles can be combined into a single product space with a common vacuum state corresponding to no particle of any kind. The corresponding Fock states represent an ensemble of independent particles of different kinds.

In our discussion of Fock spaces, we have refrained from speaking about noninteracting or free particles. We rather used the term `independent particles,' where `independent' refers to the product structure of the Fock space and the label `particle' still needs to be earned. So far, we only have a space designed for independent, discrete, and countable quantum entities (and a few operators that allow us to jump around between the base states). By construction, a Fock space allows us to go from the Hilbert space for a single entity to a Hilbert space for many independent entities, where the number of these entities can vary---no more, no less. We have not yet made any reference to any Hamiltonian, so that we cannot speak about interacting, noninteracting, or free particles; nor have we provided any information about time evolution in Fock spaces. Why the Fock space for independent particles plays such a fundamental role even for interacting theories will be explained in Section \ref{sectiondynamicsirr} (see p.\,\pageref{freeFockrelevance}).

\subsection{Fields}\label{secFourierfields}
As discussed at the end of Section \ref{sectionFock}, the Fock space can quite naturally be associated with the concept of independent particles. In the present section, we introduce a spatial field of operators in terms of creation and annihilation operators. This field should not be considered as part of the mathematical image we develop. The field is used only as an auxiliary quantity for the heuristic motivation of collision rules and quantities of interest with the proper symmetries---and to establish contact to the usual formulation of Lagrangian quantum field theory. As our approach is based on a representation in terms of momenta rather than positions, any discussion of length scales should be interpreted in terms of inverse momentum scales.

Every sufficiently regular function $f(x)$ defined on the finite interval $-L/2 \leq x \leq L/2$ can be represented by a Fourier series,
\begin{equation}\label{Fourierseries}
    f(x) = \sum_z f_z \, \eR^{-2 \pi \iR z x/L} \,,
\end{equation}
where the summation is over all integers $z$ and the coefficients $f_z$ are given by the integrals
\begin{equation}\label{Fouriercoeff}
    f_z = \frac{1}{L} \int_{-L/2}^{L/2} f(x) \, \eR^{2 \pi \iR z x/L} \dR x \,.
\end{equation}
Equation (\ref{Fourierseries}) tells us that a function of an argument varying continuously in the interval $-L/2 \leq x \leq L/2$ can be represented by a lattice of momentum states with lattice spacing,
 \begin{equation}\label{KLchoiceL}
    K_L=\frac{2\pi}{L} \,.
\end{equation}
The $d$-dimensional generalization is given by the lattice of momentum vectors $K^d$ defined in (\ref{Kdlatticedef}) in the limit $Z_L \rightarrow \infty$, in which it is possible to represent a function of a continuous position-vector argument in a $d$-dimensional hypercube of volume $V=L^d$. On a finite volume, the eigenstates of momentum are normalizable and actually form the basis of a Hilbert space. For finite $Z_L$, the spatial resolution has a lower limit of order $L/Z_L$.

As a generalization of (\ref{Fourierseries}), we use the following Fourier series representation of position-dependent field operators in a finite space continuum,
\begin{equation}\label{phiexpression}
    \varphi_{\bm{x}} = \frac{1}{\sqrt{V}} \sum_{\bm{k} \in K^d} \frac{1}{\sqrt{2\omega_k}}
    \left( a^\dag_{\bm{k}} + a_{-\bm{k}} \right) \eR^{- \iR \bm{k} \cdot \bm{x}} \,,
\end{equation}
where proper weight factors $\omega_k$ depending on $k=|\bm{k}|$ remain to be chosen. The combination of creation and annihilation operators in (\ref{phiexpression}) is introduced such that $\varphi_{\bm{x}}$ becomes self-adjoint. We actually choose
\begin{equation}\label{relenergmomrel}
    \omega_k = \sqrt{m^2+k^2} \,,
\end{equation}
which is the relativistic energy-momentum relation for a particle with mass $m$. Throughout this book, we use units with
\begin{equation}\label{hbarcconv}
     \hbar=c=1 \,,
\end{equation}
where $\hbar$ is the reduced Planck constant and $c$ is the speed of light. This convention implies an equivalence of units of time, length, and mass, and allows us to identify momenta with wave vectors, as we have anticipated in (\ref{KLchoiceL}).

The significance of including the factor $1/\sqrt{2\omega_k}$ into the definition of the Fourier components of the field will become clear only when we calculate correlation functions that we wish to be Lorentz invariant. These factors prevent us from interpreting the representation (\ref{phiexpression}) as the straightforward passage from momentum eigenstates to position eigenstates. This observation implies a fundamental difficulty for localizing particles in a relativistic theory (for a detailed discussion, see pp.\,85--91 of \cite{Teller}). In the nonrelativistic limit, that is, for velocities small compared to the speed of light, or $k \ll m$, $\omega_k$ can be replaced by the constant $m$, the fields (\ref{phiexpression}) correspond to position eigenstates, and the localization problem disappears. An independent argument in favor of the factor $1/\sqrt{2\omega_k}$ can be given once we have introduced the Hamiltonian (see end of Section~\ref{sectiondynamicsrev}). Proper relativistic behavior can only be recognized once time-dependencies are provided in addition to space-dependencies.

To obtain information about the normalization of the field (\ref{phiexpression}), we consider
\begin{equation}\label{phiexpressionnorm}
  \int_V \Dbra{N} : \varphi^\dag_{\bm{x}} \varphi_{\bm{x}} : \Dket{N} \dR^dx =
  \sum_{\bm{k} \in K^d} \frac{1}{\omega_k} \Dbra{N} a^\dag_{\bm{k}} a_{\bm{k}} \Dket{N} \,,
\end{equation}
where $\Dket{N}$ is a Fock space eigenvector with a total of $N$ particles and the colons around an operator indicate normal ordering, that is, all creation operators are moved to the left and all annihilation operators are moved to the right. As the operator $a^\dag_{\bm{k}} a_{\bm{k}}$ counts the number of particles with momentum $\bm{k}$, the operator $m : \varphi^\dag_{\bm{x}} \varphi_{\bm{x}} :$ can be interpreted as the particle number density, at least, if we ignore the relativistic subtleties discussed in the preceding paragraph.

\subsection{Dynamics}\label{sectiondynamics}
After setting up the proper Hilbert space, we would now like to introduce dynamics on this Hilbert space. For that purpose, we rely on the Schr\"odinger picture. We first need to introduce a Hamiltonian to implement the reversible contribution to time evolution and to recognize interactions. In view of our \index{Metaphysical postulates!fourth metaphysical postulate}fourth metaphysical postulate (see p.\,\pageref{metaphys4}), we subsequently need to introduce an irreversible contribution to time evolution.

\subsubsection{Reversible dynamics}\label{sectiondynamicsrev}
In the Schr\"odinger picture, the evolution of a time-dependent state $\Dket{\psi_t}$ in Hilbert space is governed by the Schr\"odinger equation,
\begin{equation}\label{Schroe}
    \frac{\dR}{\dR t} \Dket{\psi_t} = - \iR H \Dket{\psi_t} \,,
\end{equation}
where $H$ is the Hamiltonian. The formal solution of this linear equation for the initial condition $\Dket{\psi_0}$ at $t=0$ is given by
\begin{equation}\label{Schroeunitary}
    \Dket{\psi_t} = \eR^{- \iR H t} \Dket{\psi_0} \,.
\end{equation}
If the Hamiltonian $H$ is self-adjoint, the time-evolution operator $\eR^{- \iR H t}$ is unitary, and we hence refer to reversible evolution as \emph{unitary evolution}. We write the full Hamiltonian $H$ as the sum of a free term and a collision term, $H = H^{\rm free} + H^{\rm coll}$, which we now discuss separately.

\paragraph*{Free Hamiltonian.}
The single-particle states $a^\dag_{\bm{k}} \Dket{0}$ are supposed to be eigenstates of the free Hamiltonian, that is,
\begin{equation}\label{Schroe1}
    \frac{\dR}{\dR t} \, c(t) \, a^\dag_{\bm{k}} \Dket{0} =
    - \iR \omega_k \, c(t) \, a^\dag_{\bm{k}} \Dket{0} \,,
\end{equation}
with the relativistic energy-momentum expression (\ref{relenergmomrel}) and the explicit solution $c(t) = \eR^{-\iR \omega_k t} c(0)$ for the initial condition $\Dket{\psi_0} = c(0) \, a^\dag_{\bm{k}} \Dket{0}$. According to the commutation relations (\ref{ancrecomrel}) and (\ref{ancrecomrelx}) for bosons, this evolution equation can be written in the form (\ref{Schroe}) with the free Hamiltonian
\begin{equation}\label{Hfree}
    H^{\rm free} = \sum_{\bm{k} \in K^d} \omega_k \, a^\dag_{\bm{k}} a_{\bm{k}} \,.
\end{equation}
Note that the same result would also be obtained for fermions. For the moment, however, we restrict ourselves to the scalar field theory for bosons. With (\ref{ancrecomrel}) and (\ref{ancrecomrelx}), we obtain the useful commutators
\begin{equation}\label{Hfreecomakdag}
    \Qcommu{H^{\rm free}}{a^\dag_{\bm{k}}} = \omega_k \, a^\dag_{\bm{k}} \,,
\end{equation}
and
\begin{equation}\label{Hfreecomak}
    \Qcommu{H^{\rm free}}{a_{\bm{k}}} = - \omega_k \, a_{\bm{k}} \,.
\end{equation}
If $H^{\rm free}$ acts on a multi-particle state of the form (\ref{Fockstatecreate}), now with the discrete momentum label $\bm{k}$ instead of the integer label $\nu$, the energy eigenvalue is the sum of all the single-particle eigenvalues, which justifies the idea of free or noninteracting particles.

\paragraph*{Collisions.}
To describe interactions between four particles in $d$ space dimensions, we fall back on the $\varphi^4$ theory, most simply defined in terms of fields,
\begin{equation}\label{Hcolx}
    H^{\rm coll} =  \frac{\lambda}{24} \, \int_V \varphi_{\bm{x}}^4 \, \dR^dx \,,
\end{equation}
where the interaction parameter $\lambda$ characterizes the strength of the quartic interaction. By inserting the representation (\ref{phiexpression}) for $\varphi_{\bm{x}}$, we obtain
\begin{eqnarray}
    H^{\rm coll} &=& \frac{\lambda}{96} \, \frac{1}{V}
    \sum_{\bm{k}_1,\bm{k}_2,\bm{k}_3,\bm{k}_4 \in K^d}
    \frac{\delta_{\bm{k}_1+\bm{k}_2+\bm{k}_3+\bm{k}_4 , \bm{0}}}{
    \sqrt{\omega_{k_1}\omega_{k_2}\omega_{k_3}\omega_{k_4}}} \nonumber \\
    && \Big( a_{-\bm{k}_1} a_{-\bm{k}_2} a_{-\bm{k}_3} a_{-\bm{k}_4}
    + 4 a^\dag_{\bm{k}_1} a_{-\bm{k}_2} a_{-\bm{k}_3} a_{-\bm{k}_4}
    + 6 a^\dag_{\bm{k}_1} a^\dag_{\bm{k}_2} a_{-\bm{k}_3} a_{-\bm{k}_4} \nonumber \\
    &&
    + 4 a^\dag_{\bm{k}_1} a^\dag_{\bm{k}_2} a^\dag_{\bm{k}_3} a_{-\bm{k}_4}
    + a^\dag_{\bm{k}_1} a^\dag_{\bm{k}_2} a^\dag_{\bm{k}_3} a^\dag_{\bm{k}_4} \Big) \nonumber \\
    &+& \frac{\lambda'}{2} \sum_{\bm{k} \in K^d} \frac{1}{\omega_k}
    \left( a_{\bm{k}} a_{-\bm{k}} + 2 a^\dag_{\bm{k}} a_{\bm{k}} + a^\dag_{\bm{k}} a^\dag_{-\bm{k}} \right)
    + \lambda'' \, V \,,
\label{Hcolk}
\end{eqnarray}
where we have introduced a $d$-component version of Kronecker's $\delta$ symbol and the parameters
\begin{equation}\label{z12def}
    \lambda' = \lambda \, \frac{1}{V} \sum_{\bm{k} \in K^d} \frac{1}{8 \omega_k} \,, \qquad
    \lambda'' = 2 \lambda \left( \frac{1}{V} \sum_{\bm{k} \in K^d} \frac{1}{8 \omega_k} \right)^2 \,.
\end{equation}

In our reformulation of (\ref{Hcolx}), we have used the convention of normal ordering, that is, with the help of the commutation relations (\ref{ancrecomrel}) and (\ref{ancrecomrelx}), all creation operators are moved to the left and all annihilation operators are moved to the right. Closer inspection of the expressions for $\lambda'$ and $\lambda''$ reveals that these quantities become infinite for $Z_L \rightarrow \infty$. As we are determined to avoid infinities in our construction of quantum field theory, we simply treat $\lambda'$ and $\lambda''$ as further free interaction parameters in the Hamiltonian (\ref{Hcolk}) in addition to $\lambda$. We always do that as an integral part of the normal ordering procedure for a Hamiltonian. Of course, the passage from one to three independent interaction parameters might violate some symmetries of the original model defined by (\ref{Hcolx}) which we would then have to reintroduce in the further process. Clearly $\lambda$ should be regarded as the fundamental interaction parameter, whereas $\lambda'$ and $\lambda''$ should be regarded as correction parameters; $2 \lambda'$ has the interpretation of an additional contribution to the square of the mass, $\lambda''$ represents a constant background energy per unit volume or a vacuum energy density associated with collisions.

For the further development, it is important that the spectrum of $H = H^{\rm free} + H^{\rm coll}$ is bounded from below. For the validity of such an assumption it is crucial that we consider a finite volume $V$. We choose $\lambda''$ such that the lowest energy eigenvalue is equal to zero. If there is more than one zero-energy ground state, symmetry breaking can occur, provided that we handle the limiting procedures properly.

The \emph{ansatz} (\ref{Hcolk}) with three interaction parameters is part of our mathematical image of nature. On the other hand, the expression (\ref{Hcolx}) is only used for motivational purposes and is not part of the image.

Note that the interaction (\ref{Hcolk}) can change the number of free particles only by an even number ($0$, $\pm 2$, or $\pm 4$). Therefore, the subspaces with only even or odd numbers of particles are not mixed in the course of the Hamiltonian time evolution. We have thus found an example of a \index{Selection rule}`selection rule'\label{selectrule} for $\varphi^4$ theory.

\paragraph*{Ground state.}
We write any ground state of the interacting theory as $\Dket{\Omega} = \Dket{0} + \Dket{\omega}$, where $\Dket{\omega}$ has no component along $\Dket{0}$. Note that $\Dket{\Omega}$ is not a unit vector; its normalization is implied by assuming the component $1$ along the free vacuum state (as long as the component of a ground state along the free vacuum state is nonzero, this assumption is without loss of generality). The ground state condition is given by
\begin{equation}\label{groundstatecond1}
    H \Dket{\omega} = - H^{\rm coll} \Dket{0} \,.
\end{equation}
The condition of zero ground-state energy for fixing $\lambda''$ can be written as
\begin{equation}\label{groundstatecond0}
    \Dbra{0} H^{\rm coll} \Dket{\omega} = - \Dbra{0} H^{\rm coll} \Dket{0}
    \qquad \mbox{or} \qquad \Dbra{0} H^{\rm coll} \Dket{\Omega} = 0 \,.
\end{equation}
The form of $\Dket{\omega}$ is determined by the projected condition
\begin{equation}\label{groundstatecond2}
    P_0 H \Dket{\omega} = - P_0 H^{\rm coll} \Dket{0} \,,
\end{equation}
with the projector $P_0 = 1 - \Dket{0} \Dbra{0}$. A formal solution of (\ref{groundstatecond2}) is given by
\begin{equation}\label{groundstatecond2sol}
    \Dket{\omega} = \sum_{n=1}^\infty \big[ - (H^{\rm free})^{-1} P_0 H^{\rm coll} \big]^n \Dket{0} \,,
\end{equation}
where $(H^{\rm free})^{-1}$ is well-defined on the image of the projector $P_0$. Of course, there is no guarantee that the perturbation series (\ref{groundstatecond2sol}) converges. We assume that the inhomogeneous linear equation (\ref{groundstatecond2}) for $\Dket{\omega}$ has at least one solution. Note that (\ref{groundstatecond2sol}) can be rewritten as
\begin{equation}\label{groundstatecond2solO}
    \big[ 1 + (H^{\rm free})^{-1} P_0 H^{\rm coll} \big] \Dket{\Omega} = \Dket{0} \,,
\end{equation}
which leads to the following form of the condition (\ref{groundstatecond0}) for fixing $\lambda''$,
\begin{equation}\label{groundstatecond0O}
    \Dbra{0} H^{\rm coll} \big[ 1 + (H^{\rm free})^{-1} P_0 H^{\rm coll} \big]^{-1} \Dket{0} = 0 \,.
\end{equation}
More explicitly, we have
\begin{equation}\label{groundstatecondX}
    V \lambda'' = - \sum_{n=1}^\infty
    \Dbra{0} H^{\rm coll} \big[ - (H^{\rm free})^{-1} P_0 H^{\rm coll} \big]^n \Dket{0} \,,
\end{equation}
which begins with a term of second order in $H^{\rm coll}$.

\paragraph*{Conjugate momenta.}
The total Hamiltonian consisting of (\ref{Hfree}) and (\ref{Hcolk}) can be used to evaluate the generalized velocity
\begin{equation}\label{genmomentadef}
    \pi_{\bm{x}} = \raisebox{.7em}{``} \, \frac{\partial \varphi_{\bm{x}}}{\partial t}\raisebox{.6em}{''}
    = \iR \Qcommu{H}{\varphi_{\bm{x}}} \,,
\end{equation}
where $\varphi_{\bm{x}}$ is defined in (\ref{phiexpression}). We have put the time-derivative in quotation marks because we work exclusively within the Schr\"odinger picture so that the field operators do not depend on time. Nevertheless, for heuristic arguments, it can be useful to think of $\iR\Qcommu{H}{\cdot}$ as a formal `would-be time derivative'. We have further introduced the notation $\pi_{\bm{x}}$ because, for scalar field theory, the Lagrangian is quadratic in the generalized velocity and the conjugate momentum coincides with the generalized velocity.

A straightforward calculation based on the fundamental commutation relations (\ref{ancrecomrel}) and (\ref{ancrecomrelx}) and, in particular, on the resulting commutation relations (\ref{Hfreecomakdag}) and (\ref{Hfreecomak}), gives
\begin{equation}\label{genmomentaeval}
    \pi_{\bm{x}} = \iR \Qcommu{H}{\varphi_{\bm{x}}} = \iR \Qcommu{H^{\rm free}}{\varphi_{\bm{x}}} =
    \frac{\iR}{\sqrt{V}} \sum_{\bm{k} \in K^d} \sqrt{\frac{\omega_k}{2}}
    \left( a^\dag_{\bm{k}} - a_{-\bm{k}} \right) \eR^{- \iR \bm{k} \cdot \bm{x}} \,.
\end{equation}
With this expression for the conjugate momenta, we can obtain the further commutator
\begin{equation}\label{posmomcom}
    \Qcommu{\varphi_{\bm{x}}}{\pi_{\bm{x}'}} = \iR \delta(\bm{x}-\bm{x}') \,.
\end{equation}
This canonical commutation relation between fields and their conjugate momenta is the generalization of the commutator between positions and momenta in quantum mechanics. It is hence considered as a fundamental ingredient in the quantization of fields. From our perspective, it dictates the occurrence of the factor $1/\sqrt{2\omega_k}$ in (\ref{phiexpression}) (except for a possible factor of $-1$). In particular, it explains why the frequencies $\omega_k$ associated with the Hamiltonian (\ref{Hfree}) must also occur in the definition (\ref{phiexpression}) of the field.

\subsubsection{Irreversible dynamics}\label{sectiondynamicsirr}
Dissipative quantum systems are most often described by master equations for the density matrix \cite{BreuerPetru}. We here rely on a thermodynamically consistent quantum master equation \cite{hco199,hco221}. Thermodynamic consistency is an important criterion for well-behaved equations, which we want to build on in our image of particle physics. Actually one may say that it is the task of thermodynamics to formulate equations for which the existence and uniqueness of solutions can be proven.

\paragraph*{Thermodynamic quantum master equation.} According to \cite{hco221}, the general thermodynamic master equation for a quantum system in contact with a heat bath with temperature $T$ is
\begin{eqnarray}
    \frac{\dR \rho_t}{\dR t} &=& -\iR \Qcommu{H}{\rho_t} \nonumber\\
    &-& \sum_\alpha \int\limits_0^1 f_\alpha(u) \bigg(
    \Qcommux{Q_\alpha}{\rho_t^{1-u} \Qcommu{Q^\dag_\alpha}{\mu_t} \rho_t^u}
    + \Qcommux{Q^\dag_\alpha}{\rho_t^u \Qcommu{Q_\alpha}{\mu_t} \rho_t^{1-u}}
    \bigg) \dR u \,, \nonumber\\ &&
\label{QMEthermogen}
\end{eqnarray}
where the coupling or scattering operators $Q_\alpha$ are labeled by a discrete index $\alpha$, the rate factors $f_\alpha(u)$ are real and non-negative, and $\mu_t = H+ \kB T \ln \rho_t$ is a free energy operator driving the irreversible dynamics. The second term in the integral of (\ref{QMEthermogen}) is chosen to keep the density matrix self-adjoint without any need for further restrictions on the function $f_\alpha(u)$.

For a pure state $\Dket{\psi_t}$, the equivalent density matrix is given by $\rho_t = \Dket{\psi_t}\Dbra{\psi_t}$, so that the reversible term in the first line of (\ref{QMEthermogen}) corresponds to the Schr\"odinger equation (\ref{Schroe}). More generally, $\rho_t$ can represent a number of different pure states occurring with certain probabilities. The average of an observable $A$ is given by $\ave{A} = {\rm tr}(A \rho_t)$ where, within the Schr\"odinger picture, the operator $A$ is time-independent and all the time-dependence resides in $\rho_t$. For a consistent probabilistic interpretation, we need ${\rm tr}(\rho_t) = 1$; it is hence important to note that the quantum master equation (\ref{QMEthermogen}) leaves the trace of $\rho_t$ unchanged.

The irreversible contribution in the second line of (\ref{QMEthermogen}) is given in terms of double commutators involving the free energy operator $\mu_t$ as a generator. The contributions involving single and double commutators in (\ref{QMEthermogen}) play roles that are analogous to the first- and second-order-derivative contributions in classical diffusion or Fokker-Planck equations. The multiplicative splitting of $\rho_t$ into the powers $\rho_t^u$ and $\rho_t^{1-u}$, with an integration over $u$, is introduced to guarantee an appropriate interplay with entropy and hence a proper steady state or equilibrium solution. The structure of the irreversible term is determined by general arguments of nonequilibrium thermodynamics or, more formally, by a modular dynamical semigroup.\footnote{The modular dynamical semigroup can be employed to treat irreversible systems that do not possess a discrete spectrum; density matrices cannot be used for such systems and Gibbs equilibrium states need to be replaced by the more general KMS states (see Kubo \cite{Kubo57}, Martin and Schwinger \cite{MartinSchwinger59}).} Detailed arguments can be found in \cite{hco199,hco221}. The thermodynamic approach relies on the quantization of the equations of classical nonequilibrium thermodynamics rather than on the formal derivation of the emergence of irreversibility from reversible quantum mechanics, which I consider as extremely difficult and generally questionable. As in classical nonequilibrium thermodynamics, we do not consider memory effects in the irreversible term; whenever there seems to be a need for memory effects, one should rather look for a more detailed level of description for the system of interest (see Section~1.1.3 of \cite{hcobet}).

The entropy production rate $\sigma_t$ associated with the irreversible contribution to the quantum master equation (\ref{QMEthermogen}) is given by
\begin{eqnarray}
    \sigma_t &=& \frac{2}{T} \sum_\alpha \int\limits_0^1 f_\alpha(u) \,
    {\rm tr} \Big( \iR\Qcommu{Q^\dag_\alpha}{\mu_t} \, \rho_t^u \,
    \iR\Qcommu{Q_\alpha}{\mu_t} \, \rho_t^{1-u} \Big)
    \, \dR u \nonumber \\
    &=& - \frac{1}{T} \, {\rm tr} \left( \mu_t \frac{\dR \rho_t}{\dR t} \right)
    = \kB \frac{\dR}{\dR t} {\rm tr} ( \rho_t \ln \rho_{\rm eq} - \rho_t \ln \rho_t ) \,,
\label{QMEthermogenep}
\end{eqnarray}
where the equilibrium density matrix is given by
\begin{equation}\label{rhoeqdef}
    \rho_{\rm eq} = \frac{\eR^{-H/(\kB T)}}{{\rm tr} (\eR^{-H/(\kB T)})} \,.
\end{equation}
For the equilibrium density matrix to exist, the spectrum of the Hamiltonian needs to be bounded from below. Of course, in accordance with the laws of thermodynamics, $\sigma_t$ is non-negative and vanishes at equilibrium. The quantum master equation (\ref{QMEthermogen}) hence implies convergence to the equilibrium density matrix. We thus avoid the long-run universal warming occurring in the Ghirardi-Rimini-Weber theory (see p.\,481 of \cite{GhirardiRimWeb86}). The total entropy production associated with the relaxation of an initial density matrix $\rho_0$ to the equilibrium density matrix $\rho_{\rm eq}$ depends only on the initial and final density matrices; it is given by the entropy gain $-\kB ( \ln \rho_{\rm eq} - \ln \rho_0 )$, averaged with the initial density matrix $\rho_0$. This result is independent of the time and length scales on which dissipation takes place. I follows from the structure of the thermodynamic quantum master equation.

Note that, in general, thermodynamic quantum master equations are nonlinear in $\rho_t$. This feature clearly distinguishes them from the popular Lindblad master equations \cite{Lindblad76} (see also \cite{BreuerPetru}). One might actually call thermodynamic master equations semilinear because, although the additivity rule fails, scalar multiplication is a linear operation.

\paragraph*{Linearized quantum master equation.} To construct the linearization of the thermodynamic quantum master equation (\ref{QMEthermogen}) around equilibrium, we introduce the linear super-operators ${\cal K}$ and ${\cal K}^{-1}$ by
\begin{equation}\label{Ksupopdef}
    {\cal K} A = \int_0^1 \rho_{\rm eq}^u A \rho_{\rm eq}^{1-u} \dR u \,,
\end{equation}
and
\begin{equation}\label{Kinvsupopdef}
    {\cal K}^{-1} A = \int_0^\infty (\rho_{\rm eq}+s)^{-1} A \, (\rho_{\rm eq}+s)^{-1} \dR s \,.
\end{equation}
Acting on an arbitrary operator $A$, these super-operators insert or remove a factor of $\rho_{\rm eq}$ in a symmetrized form. The fact that these super-operators are inverses to each other can be verified by an explicit calculation in an eigenbase of the Hamiltonian, which is also an eigenbase of the equilibrium density matrix. Note that ${\cal K} 1 = \rho_{\rm eq}$ and hence ${\cal K}^{-1} \rho_{\rm eq} = 1$.

The dimensionless free energy operator can be written as $\beta \mu_t = \ln \rho_t - \ln \rho_{\rm eq}$, where $\beta=1/(\kB T)$ is the inverse temperature and an irrelevant constant has been neglected. Near equilibrium, $\beta \mu_t$ is small. We hence obtain the following expansion of $\rho_t$ around $\rho_{\rm eq}$,
\begin{equation}\label{rhoexpansioneq}
    \rho_t = \exp \{ \ln \rho_{\rm eq} + \beta \mu_t \} \approx
    \rho_{\rm eq} + \beta {\cal K} \mu_t \,,
\end{equation}
which implies $\beta \mu_t \approx {\cal K}^{-1} \rho_t - 1$. Moreover, as $\beta \mu_t$ is small, we can replace $\rho_t$ in the irreversible contribution to (\ref{QMEthermogen}) by $\rho_{\rm eq}$ to arrive at the linearized thermodynamic quantum master equation
\begin{eqnarray}
    \frac{\dR \rho_t}{\dR t} &=& -\iR \Qcommu{H}{\rho_t} \nonumber\\
    && \hspace{-4em} - \, \sum_\alpha \int\limits_0^1 \frac{f_\alpha(u)}{\beta} \bigg(
    \Qcommux{Q_\alpha}{\rho_{\rm eq}^{1-u} \Qcommu{Q^\dag_\alpha}{{\cal K}^{-1}\rho_t} \rho_{\rm eq}^u}
    + \Qcommux{Q^\dag_\alpha}{\rho_{\rm eq}^u \Qcommu{Q_\alpha}{{\cal K}^{-1}\rho_t} \rho_{\rm eq}^{1-u}}
    \bigg) \dR u \,. \nonumber\\ &&
\label{QMEthermogenlin}
\end{eqnarray}

\paragraph*{Concrete form of quantum master equation.} For $\varphi^4$ theory, we choose the coupling operators $Q^\dag_\alpha, Q_\alpha$ as the creation and annihilation operators $a^\dag_{\bm{k}}, a_{\bm{k}}$, so that a dissipative `smearing of the fields' can arise. These coupling operators are further motivated by the fact that dissipative events should take us around in Fock space (to guarantee ergodicity).\footnote{To respect the \index{Selection rule}selection rule associated with even and odd numbers of particles (see p.\,\pageref{selectrule}), we should actually create and annihilate pairs of particles; for reasons of simplicity, however, we here prefer to restore this \index{Selection rule}selection rule only in the limit of vanishing dissipation.} We hence rewrite the thermodynamically consistent quantum master equation (\ref{QMEthermogen}) for the density matrix $\rho_t$ as
\begin{eqnarray}
    \frac{\dR\rho_t}{\dR t} &=& -\iR \Qcommu{H}{\rho_t} \nonumber\\
    &-& \!\! \sum_{\bm{k} \in K^d} \beta \gamma_k
    \int\limits_0^1 \eR^{-u \beta \omega_k} \bigg(
    \Qcommux{a_{\bm{k}}}{\rho_t^{1-u} \Qcommu{a^\dag_{\bm{k}}}{\mu_t} \rho_t^u}
    + \Qcommux{a^\dag_{\bm{k}}}{\rho_t^u \Qcommu{a_{\bm{k}}}{\mu_t} \rho_t^{1-u}}
    \bigg) \dR u \,, \nonumber\\ &&
\label{QMEthermo}
\end{eqnarray}
where $\gamma_k$ is a decay rate that is negligible for small $k$ and increases rapidly for large $k$. The rapid decay of modes with large momenta implements what we have so far referred to as `smearing' in position space. The exponential factor chosen for $f_\alpha(u)$ produces the proper relative weights for transitions involving the creation or annihilation of free particles (detailed balance). For the concrete functional form of the decay rate $\gamma_k$ we propose
\begin{equation}\label{gammakconcr}
    \gamma_k = \gamma_0 + \gamma k^4 \,.
\end{equation}
In real space, $k^2$ corresponds to the Laplace operator causing diffusive smoothing; the occurrence of both coupling operators $a_{\bm{k}}$ and $a^\dag_{\bm{k}}$ in the double commutators in (\ref{QMEthermo}) suggests the power $k^4$. The parameter $\gamma_0$ has been added so that also the state with $k=0$ is subject to some dissipation. As we are eventually interested in the limit $\gamma_k \rightarrow 0$, the parameters $\gamma_0$ and $\gamma$ should be regarded as small; according to the convention (\ref{hbarcconv}), $\gamma_0$ actually has units of mass and $\gamma$ has units of ${\rm mass}^{-3}$. A small characteristic length scale associated with dissipation can hence be introduced as
\begin{equation}\label{lengthdissipdef}
    \ell = \gamma^{1/3} \,.
\end{equation}
Other choices of $\gamma_k$ would be possible but should lead to the same results in the limit $\gamma_k \rightarrow 0$.

With irreversible dynamics, temperature comes naturally into quantum field theory. It is associated with the fact that local degrees of freedom cannot fully be resolved in an effective field theory and are hence treated as a heat bath. They therefore need to be incorporated by thermodynamic arguments.

The factor $\eR^{-u \beta \omega_k}$ under the integral in (\ref{QMEthermo}) is new compared to the dissipation mechanism for quantum fields previously proposed in \cite{hco200}. Such a factor is allowed according to the general class of modular thermodynamic quantum master equations introduced in \cite{hco221}, but was not considered in the entirely phenomenological original formulation \cite{hco199} (see also \cite{hco201}). The exponential factor is chosen such that, in the absence of interactions, (\ref{QMEthermo}) becomes a linear quantum master equation of the so-called Davies type \cite{Davies74}. This can be seen by means of the identities
\begin{eqnarray}
    \frac{\dR}{\dR u} \left( \eR^{-u \beta \omega_k}
    \rho^u  \, a_{\bm{k}} \, \rho^{1-u} \right) &=& \nonumber\\
    && \hspace{-9em} - \eR^{-u \beta \omega_k}
    \rho^u  \big( \beta \omega_k a_{\bm{k}}
    + \Qcommu{a_{\bm{k}}}{\ln \rho} \big) \, \rho^{1-u} = \nonumber\\
    && \hspace{-9em} - \eR^{-u \beta \omega_k}
    \rho^u  \Qcommux{a_{\bm{k}}}{\beta H^{\rm free} + \ln \rho} \, \rho^{1-u} \,,
\label{QMEthermos1}
\end{eqnarray}
and similarly
\begin{equation}\label{QMEthermos2}
    \frac{\dR}{\dR u} \left( \eR^{-u \beta \omega_k}
    \rho^{1-u}  \, a^\dag_{\bm{k}} \, \rho^u \right) =
     \eR^{-u \beta \omega_k} \rho^{1-u}
     \Qcommux{a^\dag_{\bm{k}}}{\beta H^{\rm free} + \ln \rho} \, \rho^u \,,
\end{equation}
which allow us to rewrite the fundamental quantum master equation (\ref{QMEthermo}) in the equivalent form
\begin{eqnarray}
    \frac{\dR \rho_t}{\dR t}  &=& -\iR \Qcommu{H^{\rm free}}{\rho_t} - \iR \Qcommu{H^{\rm coll}}{\rho_t} \nonumber\\
    &+& \sum_{\bm{k} \in K^d} \gamma_k
    \left( 2 a_{\bm{k}} \rho_t a^\dag_{\bm{k}} - \Qantico{a^\dag_{\bm{k}} a_{\bm{k}}}{\rho_t}
    + \eR^{-\beta \omega_k}
    (2 a^\dag_{\bm{k}} \rho_t a_{\bm{k}} - \Qantico{a_{\bm{k}} a^\dag_{\bm{k}}}{\rho_t}) \right)
    \nonumber\\
    && \hspace{-3.5em} - \sum_{\bm{k} \in K^d} \beta \gamma_k
    \int\limits_0^1 \! \eR^{-u \beta \omega_k} \bigg(
    \Qcommux{a_{\bm{k}}}{\rho_t^{1-u} \Qcommu{a^\dag_{\bm{k}}}{H^{\rm coll}} \rho_t^u}
    + \Qcommux{a^\dag_{\bm{k}}}{\rho_t^u \Qcommu{a_{\bm{k}}}{H^{\rm coll}} \rho_t^{1-u}}
    \bigg) \dR u .
    \nonumber\\ &&
\label{QMEthermos}
\end{eqnarray}

The equivalent master equations (\ref{QMEthermo}) and (\ref{QMEthermos}) are significantly different in structure. For example, verifying the equilibrium solution is trivial for (\ref{QMEthermo}) because $\mu_t$ vanishes at equilibrium (another normalization wouldn't matter for the argument). To verify that $\rho_{\rm eq}$ is a solution of (\ref{QMEthermos}), we need to make use of the identities
\begin{equation}\label{QMEthermos1eq}
    \frac{\dR}{\dR u} \left( \eR^{-u \beta \omega_k}
    \rho_{\rm eq}^u  \, a_{\bm{k}} \, \rho_{\rm eq}^{1-u} \right) =
    \beta \eR^{-u \beta \omega_k} \rho_{\rm eq}^u
    \Qcommu{a_{\bm{k}}}{H^{\rm coll}} \, \rho_{\rm eq}^{1-u} \,,
\end{equation}
and
\begin{equation}\label{QMEthermos2eq}
    \frac{\dR}{\dR u} \left( \eR^{-u \beta \omega_k}
    \rho_{\rm eq}^{1-u}  \, a^\dag_{\bm{k}} \, \rho_{\rm eq}^u \right) =
    - \beta \eR^{-u \beta \omega_k} \rho_{\rm eq}^{1-u}
    \Qcommu{a^\dag_{\bm{k}}}{H^{\rm coll}} \, \rho_{\rm eq}^u \,,
\end{equation}
which follow from (\ref{QMEthermos1}) and (\ref{QMEthermos2}) by setting $\rho$ equal to $\rho_{\rm eq}$. On the other hand, (\ref{QMEthermos}) is ideal for perturbation theory because it nicely separates the linear part for the free theory from the collisional part, which is nonlinear. The choice of the factor $\eR^{-u \beta \omega_k}$ in (\ref{QMEthermo}) is motivated by the desire to keep the master equation for the free theory linear. Also the linearization around equilibrium works differently for (\ref{QMEthermo}) and (\ref{QMEthermos}). By the same arguments employed to obtain the linearization (\ref{QMEthermogenlin}) in the general setting, (\ref{QMEthermo}) leads to the linear master equation
\begin{eqnarray}
    \frac{\dR\rho_t}{\dR t} &=& -\iR \Qcommu{H}{\rho_t} \nonumber\\
    && \hspace{-4.4em} - \sum_{\bm{k} \in K^d} \gamma_k
    \int\limits_0^1 \eR^{-u \beta \omega_k} \bigg(
    \Qcommux{a_{\bm{k}}}{\rho_{\rm eq}^{1-u} \Qcommu{a^\dag_{\bm{k}}}{{\cal K}^{-1}\rho_t} \rho_{\rm eq}^u}
    + \Qcommux{a^\dag_{\bm{k}}}{\rho_{\rm eq}^u \Qcommu{a_{\bm{k}}}{{\cal K}^{-1}\rho_t}
    \rho_{\rm eq}^{1-u}} \bigg) \dR u . \nonumber\\ &&
\label{QMEthermolin}
\end{eqnarray}
The deviation from equilibrium, $\Delta\rho_t = \rho_t - \rho_{\rm eq}$, satisfies exactly the same equation. In (\ref{QMEthermos}), we need to linearize the collisional part by means of
\begin{equation}\label{rhoulinearize}
    \rho_t^u = \rho_{\rm eq}^u + {\cal K}_u \Delta \rho_t \qquad \mbox{with \ \ }
    {\cal K}_u \Delta \rho_t =
    u \int_0^1 \rho_{\rm eq}^{u u'} {\cal K}^{-1}(\Delta\rho_t) \rho_{\rm eq}^{u (1-u')} \dR u' \,,
\end{equation}
which is a generalization of (\ref{rhoexpansioneq}). In the energy eigenbase of the full Hamiltonian we find
\begin{equation}\label{Kuopeigenbaseev}
  \Dbra{m} {\cal K}_u \Delta \rho_t \Dket{n} =
  \frac{p_m^u-p_n^u}{p_m-p_n} \, \Dbra{m} \Delta \rho_t \Dket{n} \,,
\end{equation}
where $p_m$ is the eigenvalue associated with the eigenvector $\Dket{m}$ of $\rho_{\rm eq}$. The linearization of (\ref{QMEthermos}) can now be written in the form
\begin{eqnarray}
    \frac{\dR\rho_t}{\dR t}  &=& -\iR \Qcommu{H^{\rm free}}{\rho_t} - \iR \Qcommu{H^{\rm coll}}{\rho_t} \nonumber\\
    && \hspace{-3em} + \sum_{\bm{k} \in K^d} \gamma_k
    \left( 2 a_{\bm{k}} \Delta\rho_t a^\dag_{\bm{k}} - \Qantico{a^\dag_{\bm{k}} a_{\bm{k}}}{\Delta\rho_t}
    + \eR^{-\beta \omega_k}
    (2 a^\dag_{\bm{k}} \Delta\rho_t a_{\bm{k}} - \Qantico{a_{\bm{k}} a^\dag_{\bm{k}}}{\Delta\rho_t}) \right)
    \nonumber\\
    && \hspace{-4.5em} - \! \sum_{\bm{k} \in K^d} \beta \gamma_k
    \int\limits_0^1 \! \eR^{-u \beta \omega_k} \bigg(
    \Qcommux{a_{\bm{k}}}{\rho_{\rm eq}^{1-u} \Qcommu{a^\dag_{\bm{k}}}{H^{\rm coll}} {\cal K}_u \Delta \rho_t}
    + \Qcommux{a^\dag_{\bm{k}}}{{\cal K}_u \Delta \rho_t \Qcommu{a_{\bm{k}}}{H^{\rm coll}} \rho_{\rm eq}^{1-u}}
    \nonumber\\
    && + \Qcommux{a_{\bm{k}}}{{\cal K}_{1-u} \Delta \rho_t \Qcommu{a^\dag_{\bm{k}}}{H^{\rm coll}} \rho_{\rm eq}^u}
    + \Qcommux{a^\dag_{\bm{k}}}{\rho_{\rm eq}^u \Qcommu{a_{\bm{k}}}{H^{\rm coll}} {\cal K}_{1-u} \Delta \rho_t}
    \bigg) \dR u .
    \nonumber\\ &&
\label{QMEthermoslin}
\end{eqnarray}
Again, the equivalence of the linearized master equations (\ref{QMEthermolin}) and (\ref{QMEthermoslin}) is not obvious. The verification can be based on the identities obtained by linearization of (\ref{QMEthermos1}) and (\ref{QMEthermos2}) around equilibrium.

\paragraph*{Low-temperature limit.} In the low-temperature limit, $\beta \rightarrow \infty$, the free energy operator $\mu$ is dominated by energetic effects and $\beta \omega_k \eR^{-u \beta \omega_k}$ approaches a $\delta$ function at $u=0$. If we apply these ideas to the full nonlinear quantum master equation (\ref{QMEthermo}) or (\ref{QMEthermos}), we obtain the linear equation
\begin{eqnarray}
    \frac{\dR\rho_t}{\dR t} &=& -\iR \Qcommu{H}{\rho_t}
    + \sum_{\bm{k} \in K^d} \gamma_k
    \left( 2 a_{\bm{k}} \rho_t a^\dag_{\bm{k}} - a^\dag_{\bm{k}} a_{\bm{k}} \rho_t
    - \rho_t a^\dag_{\bm{k}} a_{\bm{k}} \right)
    \nonumber \\
    &-& \sum_{\bm{k} \in K^d} \frac{\gamma_k}{\omega_k} \bigg(
    \Qcommux{a_{\bm{k}}}{\rho_t \Qcommu{a^\dag_{\bm{k}}}{H^{\rm coll}}}
    + \Qcommux{a^\dag_{\bm{k}}}{\Qcommu{a_{\bm{k}}}{H^{\rm coll}} \rho_t} \bigg) \,.
\label{QMEthermolowT}
\end{eqnarray}
If, on the other hand, we apply the same procedure to the linearized quantum master equation (\ref{QMEthermoslin}) and use the results ${\cal K}_1 \Delta \rho_t = \Delta \rho_t$, ${\cal K}_0 \Delta \rho_t = 0$ following from (\ref{Kuopeigenbaseev}), we arrive at
\begin{eqnarray}
    \frac{\dR\Delta\rho_t}{\dR t} &=& -\iR \Qcommu{H}{\Delta\rho_t}
    + \sum_{\bm{k} \in K^d} \gamma_k
    \left( 2 a_{\bm{k}} \Delta\rho_t a^\dag_{\bm{k}} - a^\dag_{\bm{k}} a_{\bm{k}} \Delta\rho_t
    - \Delta\rho_t a^\dag_{\bm{k}} a_{\bm{k}} \right)
    \nonumber \\
    &-& \sum_{\bm{k} \in K^d} \frac{\gamma_k}{\omega_k} \bigg(
    \Qcommux{a_{\bm{k}}}{\Delta\rho_t \Qcommu{a^\dag_{\bm{k}}}{H^{\rm coll}}}
    + \Qcommux{a^\dag_{\bm{k}}}{\Qcommu{a_{\bm{k}}}{H^{\rm coll}} \Delta\rho_t} \bigg) \,. \qquad
\label{QMEthermolowTi}
\end{eqnarray}
The occurrence of two different master equations demonstrates that the low-temperature limit is subtle and should not be performed as naively as suggested. Equation (\ref{QMEthermolowTi}) is preferable because, contrary to (\ref{QMEthermolowT}), it has the proper equilibrium solution $\rho_t = \rho_{\rm eq}$.

In the absence of collisions, (\ref{QMEthermolowT}) and (\ref{QMEthermolowTi}) are identical. From both equations, we arrive at the simpler master equation of the free theory,
\begin{equation}\label{QMEthermolowTfree}
    \frac{\dR\rho_t}{\dR t} = -\iR \Qcommu{H^{\rm free}}{\rho_t} + \sum_{\bm{k} \in K^d} \gamma_k
    \left( 2 a_{\bm{k}} \rho_t a^\dag_{\bm{k}} - a^\dag_{\bm{k}} a_{\bm{k}} \rho_t
    - \rho_t a^\dag_{\bm{k}} a_{\bm{k}} \right) \,.
\end{equation}
This suggests a well-defined zero-temperature master equation for the free theory, as will be confirmed in Section~\ref{secunravelfreef}. For interacting particles, however, the low-temperature limit calls for caution and deeper investigation.

In summary, we have the thermodynamically founded full nonlinear quantum master equation [in the equivalent forms (\ref{QMEthermo}) and (\ref{QMEthermos})] and the corresponding linearized equation [in the equivalent forms (\ref{QMEthermolin}) and (\ref{QMEthermoslin})]. The nonlinear equation is more robust and makes no reference to the equilibrium density matrix, whereas the linear equation has the practical advantage of being more tractable. Even simpler to handle is the zero-temperature master equation (\ref{QMEthermolowTi}); it may be suited best for perturbation theory because an expansion for the equilibrium density matrix (ground state) is available.

\paragraph*{Discussion.} By introducing an irreversible contribution to the time evolution, we have also introduced a thermodynamic \index{Arrow of time}arrow of time. In particular, there is a preferred direction of time for causality. A crucial benefit of irreversibility is that time evolution in the long-time limit is perfectly controlled by the approach to equilibrium, whereas the long-time limit of the unitary time-evolution operator for purely reversible dynamics is notoriously subtle. Moreover, dissipative smoothing regularizes the theory.

Even in the presence of interactions, we choose our coupling operators as the creation and annihilation operators of the free theory satisfying the commutation relations (\ref{Hfreecomakdag}) and (\ref{Hfreecomak}). This \emph{ansatz}\label{freecollboth} goes nicely together with working in the Fock space of free particles. Both $H^{\rm free}$ and $H^{\rm coll}$ separately matter for the irreversible dynamics, not only the sum of both. The relevance of $H^{\rm free}$ to the interacting theory hence goes beyond perturbation theory. The dissipation term is constructed such that it leads to unresolvable clouds of free particles, which can then be interpreted as the particles of the interacting theory. The possibility of such an interpretation actually is the deeper reason for choosing the coupling operators $a^\dag_{\bm{k}}, a_{\bm{k}}$, thus giving importance to the free theory and ultimately justifying the fundamental role of the Fock space \label{freeFockrelevance} of the free theory even in the presence of interactions. Note that the equilibrium density matrix (\ref{rhoeqdef}) depends only on the total Hamiltonian $H^{\rm free} + H^{\rm coll}$.

The commutators $\Qcommu{a^\dag_{\bm{k}}}{H^{\rm coll}}$ and $\Qcommu{a_{\bm{k}}}{H^{\rm coll}}$ play a prominent role in the nonlinear and linear quantum master equations (\ref{QMEthermos}) and (\ref{QMEthermolowTi}). We hence provide these commutators for the example of the $\varphi^4$ Hamiltonian (\ref{Hcolk}) for future reference,
\begin{eqnarray}
    \Qcommu{a_{\bm{k}}}{H^{\rm coll}} &=& \frac{\lambda'}{\omega_k}
    \Big( a_{\bm{k}} + a^\dag_{-\bm{k}} \Big)
    + \frac{\lambda}{24 V} \sum_{\bm{k}_1,\bm{k}_2,\bm{k}_3 \in K^d}
    \frac{\delta_{\bm{k}_1+\bm{k}_2+\bm{k}_3,\bm{k}}}{\sqrt{\omega_k
    \omega_{k_1}\omega_{k_2}\omega_{k_3}}} \nonumber \\
    &\times& \! \Big( a_{\bm{k}_1} a_{\bm{k}_2} a_{\bm{k}_3}
    + 3 a^\dag_{-\bm{k}_1} a_{\bm{k}_2} a_{\bm{k}_3}
    + 3 a^\dag_{-\bm{k}_1} a^\dag_{-\bm{k}_2} a_{\bm{k}_3}
    + a^\dag_{-\bm{k}_1} a^\dag_{-\bm{k}_2} a^\dag_{-\bm{k}_3}
    \Big) \,, \nonumber \\ &&
\label{Hcollcommua}
\end{eqnarray}
and
\begin{equation}\label{Hcollcommuad}
    \Qcommu{H^{\rm coll}}{a^\dag_{\bm{k}}} = \Qcommu{a_{\bm{k}}}{H^{\rm coll}}^\dag \,.
\end{equation}

\subsection{Quantities of interest}\label{subsecobservables}
Having a fundamental quantum master equation, we now need to address the question which quantities we want to calculate. We refer to them as `quantities of interest'. The choice of the quantities of interest is clearly subjective. However, we propose a general class of correlation functions to choose from. The calculation of any of these correlation functions is unambiguous.

The quantities of interest shall eventually serve to confront theory with nature. One might hence ask how correlation functions can actually be measured. The discussion of the measurement problem in Section~\ref{secmeasureprob} suggests that measurements on quantum systems are a notoriously subtle problem. One should however note that, by all experience, the proper matching of correlation functions with experimentally measured results is not a serious problem.

The subsequent discussion of correlation functions is inspired by Sections 2.2 and 2.3 of \cite{GardinerZoller}. The quantities of interest that we are going to introduce do all possess a statistical character. This is natural as we work with density matrices, also known as statistical operators, in order to be able to describe irreversible dynamics. This observation is important for appreciating the discussion of the measurement problem in Section~\ref{secmeasureprob}.

\subsubsection{Definition of multi-time correlation functions}\label{secdefcorfcts}
Quantities of interest involving multiple times can be related to nested expressions of the form
\begin{equation}\label{observableform1}
    {\rm tr} \left\{ {\cal N}_n A_n {\cal E}_{t_n-t_{n-1}}\Big( \ldots \, {\cal N}_2 A_2 {\cal E}_{t_2-t_1} \big( {\cal N}_1 A_1 {\cal E}_{t_1-t_0}(\rho_0) A^\dag_1 \big)  A^\dag_2 \, \ldots \Big) A^\dag_n \right\} \,,
\end{equation}
which should be read from the inside to the outside, starting from the initial density matrix $\rho_0$ at time $t_0$. In this expression, ${\cal E}_t$ is the evolution super-operator obtained by solving a quantum master equation over a time period $t$, the $A_j$ are linear operators associated with the times $t_j$ with $t_0 < t_1 < \ldots < t_n$, and the normalization factors ${\cal N}_j$ are chosen such that, after every step, we continue the evolution with a density matrix. Note that, as the trace of a density matrix, the expression given in (\ref{observableform1}) actually equals unity. The important information about the outcomes of a time series of `measurements' is hence contained in the sequence of numbers ${\cal N}_j$.

The operators $A_j$ in (\ref{observableform1}) are not assumed to be projection operators associated with the eigenvalues found in certain measurements. We thus allow for a generalization to imperfect measurements that can only be achieved on the level of density matrices. The repeated occurrence of ${\cal N}_j A_j \rho_j A^\dag_j$ in (\ref{observableform1}) seems to lead to the most general construction of density matrices that can be further evolved by a master equation. Of course, one could also add several of such contributions and still keep a density matrix.

Even for the nonlinear thermodynamic master equation (\ref{QMEthermo}), multiplying a solution by a constant positive factor leads to another solution of the same equation so that the domain of the super-operator ${\cal E}_t$ can be extended from density matrices to unnormalized self-adjoint operators with nonnegative eigenvalues. The numbers ${\cal N}_j$ can be pulled out of the trace and could be reproduced at any stage so that we can focus on the correlation functions
\begin{equation}\label{observableform2}
    {\rm tr} \left\{ A_n {\cal E}_{t_n-t_{n-1}}\Big( \ldots \, A_2 {\cal E}_{t_2-t_1} \big( A_1 {\cal E}_{t_1-t_0}(\rho_0) A^\dag_1 \big)  A^\dag_2 \, \ldots \Big) A^\dag_n \right\} \,,
\end{equation}
which, by construction, are equal to $({\cal N}_1 {\cal N}_2 \ldots {\cal N}_n)^{-1}$.

In concrete calculations, we often consider evolution on a large but finite time interval, $-\tau/2 \leq t \leq \tau/2$, that is, we set the initial time to $t_0 = -\tau/2$ and assume $t_n \le \tau/2$. According to the theory of Fourier series given in (\ref{Fourierseries}) and (\ref{Fouriercoeff}), this leads to a countable set of frequencies required for the representation of any time-dependent function, and we can thus naturally keep the number of frequency-dependent quantities of interest countable.

If we further evolve the expression (\ref{observableform2}) from $t_n$ to $\tau/2$ by applying the super-operator ${\cal E}_{\tau/2-t_n}$ before taking the trace, then the value of that expression, which is ${\cal V} = ({\cal N}_1 {\cal N}_2 \ldots {\cal N}_n)^{-1}$, does not change change in time. In the limit of large $\tau$ (for this argument, it is convenient to keep $t_0$ fixed), the evolution super-operator ${\cal E}_{\tau/2-t_n}$ drives the system to the equilibrium solution or, more precisely, to the properly normalized limit ${\cal V} \rho_{\rm eq}$. Instead of taking the trace, we can hence project on the vacuum state and divide by $\Dbra{0} \rho_{\rm eq} \Dket{0}$ to obtain ${\cal V}$. Under the reasonable assumption that $\Dbra{0} \rho_{\rm eq} \Dket{0}$ does not vanish, these arguments prove rigorously that our correlation functions (\ref{observableform2}) are equivalently given by
\begin{equation}\label{observableform2l}
    \frac{\lim\limits_{\tau \rightarrow \infty} \Dbra{0} {\cal E}_{\tau/2-t_n}\left( A_n {\cal E}_{t_n-t_{n-1}} \Big(
    \ldots \, A_2 {\cal E}_{t_2-t_1} \big( A_1 {\cal E}_{t_1-t_0}(\rho_0) A^\dag_1 \big)  A^\dag_2 \, \ldots
    \Big) A^\dag_n \right) \Dket{0}}{\Dbra{0} \rho_{\rm eq} \Dket{0}} \,.
\end{equation}
For a nonlinear thermodynamic master equation, this is as far as we can go.

To develop the discussion of quantities of interest further, we assume that we deal with a linear quantum master equation, that is, that the super-operator ${\cal E}_t$ is linear. Under that assumption, we can naturally extend the domain of the linear super-operator ${\cal E}_t$ from unnormalized self-adjoint operators with nonnegative eigenvalues to general linear operators. In view of the polarization identity,
\begin{eqnarray}
    A X B^\dag &=& \frac{1}{4} \Big[ (A+B)X(A+B)^\dag - (A-B)X(A-B)^\dag \nonumber \\
    &+& \iR(A+\iR B)X(A+\iR B)^\dag - \iR(A-\iR B)X(A-\iR B)^\dag \Big] \,,
\label{polarident}
\end{eqnarray}
it is clear that the set of quantities (\ref{observableform2}) are then equivalent to the seemingly more general correlation functions
\begin{equation}\label{observableform3}
    {\rm tr} \left\{ A_n {\cal E}_{t_n-t_{n-1}}\Big( \ldots \, A_2 {\cal E}_{t_2-t_1} \big( A_1 {\cal E}_{t_1-t_0}(\rho_0) B^\dag_1 \big)  B^\dag_2 \, \ldots \Big) B^\dag_n \right\} \,,
\end{equation}
where the linear operators $B_j$ are allowed to differ from the linear operators $A_j$. The expression (\ref{observableform3}) defines the most general quantities of interest that we consider in this book. The fact that the operators $A_j$ and $B^\dag_j$ are introduced at the same times $t_j$ is without loss of generality because some of these operators may be chosen as identity operators for desynchronization of the descending and ascending times associated with $A_j$ and $B^\dag_j$, respectively. The equivalence of (\ref{observableform2}) and (\ref{observableform2l}) implies that our most general quantities of interest (\ref{observableform3}) can alternatively be written as
\begin{equation}\label{observableform3l}
    \frac{\lim\limits_{\tau \rightarrow \infty} \Dbra{0} {\cal E}_{\tau/2-t_n}\left( A_n {\cal E}_{t_n-t_{n-1}}\Big(
    \ldots \, A_2 {\cal E}_{t_2-t_1} \big( A_1 {\cal E}_{t_1-t_0}(\rho_0) B^\dag_1 \big)  B^\dag_2 \, \ldots
    \Big) B^\dag_n \right) \Dket{0}}{\Dbra{0} \rho_{\rm eq} \Dket{0}} \,.
\end{equation}
In short, for the special case of linear quantum master equations, we can achieve an `$A$-$B$ decoupling'.

Actually, we mostly consider correlation functions in which all the $B_j$ are chosen to be identity operators. In that case, our equivalent expressions for the correlation functions become
\begin{equation}\label{observableform3s}
    {\rm tr} \left\{ A_n {\cal E}_{t_n-t_{n-1}}\Big( \ldots \, A_2 {\cal E}_{t_2-t_1} \big( A_1 {\cal E}_{t_1-t_0}(\rho_0)
    \big) \ldots \Big) \right\} \,,
\end{equation}
and
\begin{equation}\label{observableform3sl}
    \frac{\lim\limits_{\tau \rightarrow \infty} \Dbra{0} {\cal E}_{\tau/2-t_n}\left( A_n {\cal E}_{t_n-t_{n-1}}\Big(
    \ldots \, A_2 {\cal E}_{t_2-t_1} \big( A_1 {\cal E}_{t_1-t_0}(\rho_0) \big)  \, \ldots
    \Big) \right) \Dket{0}}{\Dbra{0} \rho_{\rm eq} \Dket{0}} \,.
\end{equation}

For large differences $t_1-t_0$, which occur in the limit of large negative $t_0$, the expression ${\cal E}_{t_1-t_0}(\rho_0)$ occurring in all these correlation functions evolves into the equilibrium density matrix $\rho_{\rm eq}$. A convenient choice for theoretical considerations is $\rho_0 = \rho_{\rm eq}$ because we then have the rigorous identity ${\cal E}_{t_1-t_0}(\rho_{\rm eq}) = \rho_{\rm eq}$ even without going to any limit. A convenient choice for practical calculations is $\rho_0 = \Dket{0}\Dbra{0}$ because we then do not need to know $\rho_{\rm eq}$. For future reference, we define the multi-time correlation functions in two equivalent forms,
\begin{eqnarray}
    \hat{C}^{A_n \dots A_1}_{t_n \ldots t_1} &=&
    {\rm tr} \left\{ A_n {\cal E}_{t_n-t_{n-1}}\big( \ldots \, A_2 {\cal E}_{t_2-t_1} ( A_1 \rho_{\rm eq} ) \ldots \big) \right\}
    \nonumber \\
    &=& \frac{\lim\limits_{\tau \rightarrow \infty} \Dbra{0} {\cal E}_{\tau/2-t_n}\Big( A_n {\cal E}_{t_n-t_{n-1}}\big(
    \ldots \, A_2 {\cal E}_{t_2-t_1} ( A_1 \rho_{\rm eq} )  \, \ldots
    \big) \Big) \Dket{0}}{\Dbra{0} \rho_{\rm eq} \Dket{0}} \,.
    \nonumber \\ &&
\label{multitimecordef}
\end{eqnarray}

\subsubsection{Limit for Hamiltonian systems}
For purely Hamiltonian dynamics one deals with the evolution super-operators
\begin{equation}\label{HsuperopE}
    {\cal E}_t(\rho) = \eR^{-\iR Ht} \rho \, \eR^{\iR Ht} \,,
\end{equation}
which express the unitary time-evolution of density matrices, and our most general correlation functions introduced in (\ref{observableform3}) become
\begin{equation}\label{observableform4}
    {\rm tr} \left\{ \eR^{\iR Ht'_1} B^\dag_1  \eR^{-\iR Ht'_1} \ldots \eR^{\iR Ht'_n} B^\dag_n  \eR^{-\iR Ht'_n} \,\,
    \eR^{\iR Ht'_n} A_n  \eR^{-\iR Ht'_n} \ldots \eR^{\iR Ht'_1} A_1  \eR^{-\iR Ht'_1} \, \rho_0 \right\} \,,
\end{equation}
with $t'_j=t_j-t_0$. In (\ref{observableform4}), time-dependent Heisenberg operators can be recognized. As we usually consider the time-ordered correlation functions (\ref{observableform3s}) with all $B_j$ being identity operators and assume $\rho_0 = \Dket{0}\Dbra{0}$, we deal with the quantities
\begin{equation}\label{observableform5}
    \Dbra{0}  \eR^{-\iR H(t_0-t_n)} A_n  \eR^{-\iR H(t_n-t_{n-1})} A_{n-1} \ldots
    A_2  \eR^{-\iR H(t_2-t_1)} A_1 \eR^{-\iR H(t_1-t_0)} \Dket{0} \,.
\end{equation}
The latter equation is written such that we can easily interpret it as alternating Schr\"odinger evolution of a state and application of an operator $A_j$.

Instead of going back from $t_n$ to $t_0$ with the left-most exponential $\eR^{-iH(t_0-t_n)}$ in (\ref{observableform5}) one would prefer to progress to the final time $\tau/2$ and one hence likes to rewrite these correlation functions as
\begin{equation}\label{observableform6}
    \frac{\Dbra{0}  \eR^{-\iR H(\tau/2-t_n)} A_n  \eR^{-\iR H(t_n-t_{n-1})} A_{n-1} \ldots
    A_2  \eR^{-\iR H(t_2-t_1)} A_1 \eR^{-\iR H(t_1+\tau/2)} \Dket{0}}{\Dbra{0} \eR^{-\iR H\tau} \Dket{0}} \,.
\end{equation}
This actually is a crucial step for establishing the compatibility of quantum field theory with the principles of special relativity. The integrations over all space in the Hamiltonian $H$ need to be matched by an evolution through the entire time domain. In the spirit of potential infinity, an infinite space-time is achieved in the limit $V \rightarrow \infty$ and $\tau \rightarrow \infty$.

Of course, the usually claimed equivalence of (\ref{observableform5}) and (\ref{observableform6}) requires justification. Peskin and Schroeder replace $\eR^{-\iR Ht}$ by $\eR^{(-\iR H-\epsilon H)t}$ with a small parameter $\epsilon$, which corresponds to a weak damping mechanism isolating the ground state of the interacting theory from that of the free theory for large $t$, where the correct normalization leads to the denominator in (\ref{observableform6}) (see pp.\,86--87 of \cite{PeskinSchroeder}). Even more formally, Bjorken and Drell argue that $\Dket{0}$ is an eigenstate of $\eR^{\pm \iR Ht}$ for large $t$, where the eigenvalues are eliminated by the denominator in (\ref{observableform6}) (see pp.\,179--180 of \cite{BjorkenDrell}). In the words of Teller, ``Common practice assumes unproblematic limiting behavior of unitary time evolution operators, but it turns out that this assumption is inconsistent with other assumptions of the theory'' (see p.\,123 of \cite{Teller}). It is a major advantage of our dissipative approach to quantum field theory that the long-time evolution is controlled by the approach to equilibrium and that the above arguments, which are \emph{ad hoc} or dubious, can be made perfectly rigorous. Indeed, if the equality of (\ref{observableform3s}) and (\ref{observableform3sl}) is considered in the limit of the reversible super-operator (\ref{HsuperopE}), it is recognized as the rigorous version of the equivalence between (\ref{observableform5}) and (\ref{observableform6}).

\subsubsection{Laplace-transformed correlation functions}\label{secLapltranscorrfunc}
In Section \ref{secdefcorfcts}, we had discussed the most general quantities of interest and, in particular, we had introduced the multi-time correlation functions in (\ref{multitimecordef}). All these quantities have been given in terms of the evolution super-operator ${\cal E}_t$. For the purpose of constructing perturbation expansions and other theoretical developments, it turns out to be easier to consider Laplace-transformed correlation functions. We hence introduce the Laplace-transformed super-operators
\begin{equation}\label{Rsupopdef}
    {\cal R}_s(\rho) = \int_0^\infty {\cal E}_t(\rho) \, \eR^{- s t} \, \dR t \,,
\end{equation}
where the real part of $s$ must be larger than zero to ensure convergence. We actually assume that ${\rm Re}(s) \, \tau \gg 1$ so that the integrand decays within our basic time interval of length $\tau$ (see Section \ref{secdefcorfcts}).

If $\rho$ is a density matrix, as the result of time-evolution we obtain another density matrix, ${\cal E}_t(\rho)$. The integration in (\ref{Rsupopdef}) corresponds to a superposition of density matrices with positive weights. The proper normalization is found from ${\rm tr} [s {\cal R}_s(\rho)] =1$, so that $s {\cal R}_s(\rho)$ is recognized as another density matrix. The convergence of ${\cal E}_t(\rho)$ to $\rho_{\rm eq}$ implies the limits
\begin{equation}\label{Rsupoplim}
    \lim_{s \rightarrow 0} s \, {\cal R}_s(\rho) =
    \lim_{t \rightarrow \infty} {\cal E}_t(\rho) = \rho_{\rm eq} \,,
\end{equation}
which provides a very useful representation of $\rho_{\rm eq}$, in particular, for $\rho = \Dket{0} \Dbra{0}$. For finite $s$, the stationarity of the equilibrium density matrix $\rho_{\rm eq}$ implies the identity
\begin{equation}\label{Rsupopstationary}
    s \, {\cal R}_s(\rho_{\rm eq}) = \rho_{\rm eq} \,.
\end{equation}

We would like to introduce Laplace-transformed correlation functions in terms of the super-operator ${\cal R}_s$. By using the relationship (\ref{Rsupoplim}) in the definition (\ref{multitimecordef}), we obtain the multi-time correlation functions in the form\footnote{In a more rigorous approach, this argument should be applied on the level of the correlation functions (\ref{observableform2l}) and the polarization identity should be used only afterwards.}
\begin{eqnarray}
    \hat{C}^{A_n \dots A_1}_{t_n \ldots t_1} &=&
    {\rm tr} \left\{ A_n {\cal E}_{t_n-t_{n-1}}\big( \ldots \, A_2 {\cal E}_{t_2-t_1} ( A_1 \rho_{\rm eq} ) \ldots \big) \right\}
    \nonumber \\
    &=& \frac{\lim\limits_{s \rightarrow 0} \, \Dbra{0} s {\cal R}_s \Big( A_n {\cal E}_{t_n-t_{n-1}}\big(
    \ldots \, A_2 {\cal E}_{t_2-t_1} ( A_1 \rho_{\rm eq} )  \, \ldots
    \big) \Big) \Dket{0}}{\Dbra{0} \rho_{\rm eq} \Dket{0}} \,.
    \nonumber \\ &&
\label{multitimecorpert}
\end{eqnarray}
Both expressions for the correlation function depend on the $n-1$ nonnegative time differences $t_2-t_1, \ldots t_n-t_{n-1}$ only. It is natural to introduce Laplace-transformed correlation functions depending on the variables $s_1, \ldots s_{n-1}$ associated with the time differences $t_2-t_1, \ldots t_n-t_{n-1}$,
\begin{eqnarray}
    \tilde{C}^{A_n \dots A_1}_{s_{n-1} \ldots s_1} &=&
    {\rm tr} \left\{ A_n {\cal R}_{s_{n-1}}\big( \ldots \, A_2 {\cal R}_{s_1} ( A_1 \rho_{\rm eq} ) \ldots \big) \right\}
    \nonumber \\
    &=& \frac{\lim\limits_{s \rightarrow 0} \, \Dbra{0} s {\cal R}_s \Big( A_n {\cal R}_{s_{n-1}} \big(
    \ldots \, A_2 {\cal R}_{s_1} ( A_1 \rho_{\rm eq} )  \, \ldots
    \big) \Big) \Dket{0}}{\Dbra{0} \rho_{\rm eq} \Dket{0}} \,.
    \nonumber \\ &&
\label{multitimecorR1}
\end{eqnarray}

The first line of (\ref{multitimecorR1}) may look simpler than the second line, but the latter expression has the advantage that, in perturbation expansions, there can occur a systematic cancelation of terms between numerator and denominator. This observation becomes particularly useful if we write the denominator in an even more complicated form by inserting identity operators $s {\cal R}_s$ (see (\ref{Rsupopstationary})) to match all the super-operators of the numerator in the denominator and then pass from a ratio of limits to the limit of a ratio,
\begin{equation}\label{multitimecorR3}
    \tilde{C}^{A_n \dots A_1}_{s_{n-1} \ldots s_1} = \lim_{s \rightarrow 0}
    \frac{\Dbra{0} s{\cal R}_s
    \Big( A_n {\cal R}_{s_{n-1}}\big( \ldots \, A_2 {\cal R}_{s_1} ( A_1 \rho_{\rm eq} )  \ldots \big) \Big) \Dket{0}}{
    \Dbra{0} s{\cal R}_s \Big( s_{n-1}{\cal R}_{s_{n-1}}\big( \ldots \, s_1 {\cal R}_{s_1} (
    \rho_{\rm eq} )  \ldots \big) \Big) \Dket{0}} \,.
\end{equation}
For the equilibrium density matrix $\rho_{\rm eq}$ occurring both in the numerator and in the denominator of (\ref{multitimecorR3}) we can use the representation $s'{\cal R}_{s'} \big( \Dket{0} \Dbra{0} \big)$ in the additional limit $s' \rightarrow 0$.

\subsubsection{Fields in correlation functions}
Up to this point, the discussion of quantities of interest has been very general. In a field theory, of course, we would like to express the operators $A_j$ occurring in correlation functions in terms of field operators. In our Fock space approach it is natural to use the Fourier components of the field, which we obtain from (\ref{phiexpression}),
\begin{equation}\label{phiFourier}
    \varphi_{\bm{k}} = \frac{1}{\sqrt{2\omega_k}} \left( a^\dag_{\bm{k}} + a_{-\bm{k}} \right) \,.
\end{equation}
In other words, we choose well-defined, simple combinations of creation and annihilation operators as our basic building blocks. However, these operators create and annihilate free particles and we have learned in Section~\ref{secpartfield} that free particles are not observable. We should rather consider correlation functions of clouds. We hence base our correlation functions on the `cloud operators'
\begin{equation}\label{phiFourierZ}
    \Phi_{\bm{k}} = \sqrt{\frac{Z}{2\omega_k}} \left( a^\dag_{\bm{k}} + a_{-\bm{k}} \right) \,,
\end{equation}
where the factor $\sqrt{Z}$ translates between free particle and cloud. Only in the absence of interactions we should expect $Z=1$. According to the cloud idea, $Z$ may depend on the coupling constants and the dimensionless friction constant $\gamma m^3$ (see text around (\ref{lengthdissipdef})). As it results from collisions between free particles, the cloud factor $Z$ is part of the physics we are interested in and should be determined in the process of calculating physical properties. We expect $Z$ to be large compared to unity because a cloud consists of many free particles.

\subsection{Renormalization}\label{secrenormalization}
Renormalization is sometimes perceived as a tricky toolbox to remove annoying divergences from quantum field theory. The present section emphasizes that the renormalization group should rather be considered as a profound tool to refine perturbation expansions which would be entirely useless without renormalization. We first try to develop some intuition in the context of a much simpler example borrowed from polymer physics. Also historically, renormalization has been much better understood as a result of deep relations between quantum field theory and statistical physics (see the review \cite{WilsonKogut74} by Wilson and Kogut). We then present a few equations that provide practical recipes for refining perturbation expansions and allow us to calculate the `running coupling constant' of renormalization-group theory. Finally, we discuss some nonperturbative aspects of renormalization.

\subsubsection{Intuitive example}
Plain perturbation theory is clearly not appropriate for problems involving a large number of interactions. However, if a problem exhibits self-similarity on different length scales, perturbation theory can be refined to obtain useful results by successively accounting for more and more interactions. This refinement may be considered as a summation procedure guided by a renormalization-group analysis for self-similar systems.

Intuitive examples of refined perturbation expansions can be found in the theory of linear polymer molecules. The beauty of polymer physics actually stems from the self-similarity of polymers \cite{deGennes}. If we model polymer molecules in dilute solution as linear chains of beads connected by springs, hydrodynamic interactions between the beads arise because the motion of each bead perturbs the solvent flow around it and, after the fast propagation of the perturbation through the solvent, it almost instantaneously affects the motion of the other beads. The bead friction coefficient determines the strength of such hydrodynamic-interaction effects.

\begin{figure}
\centerline{\includegraphics[width=\textwidth]{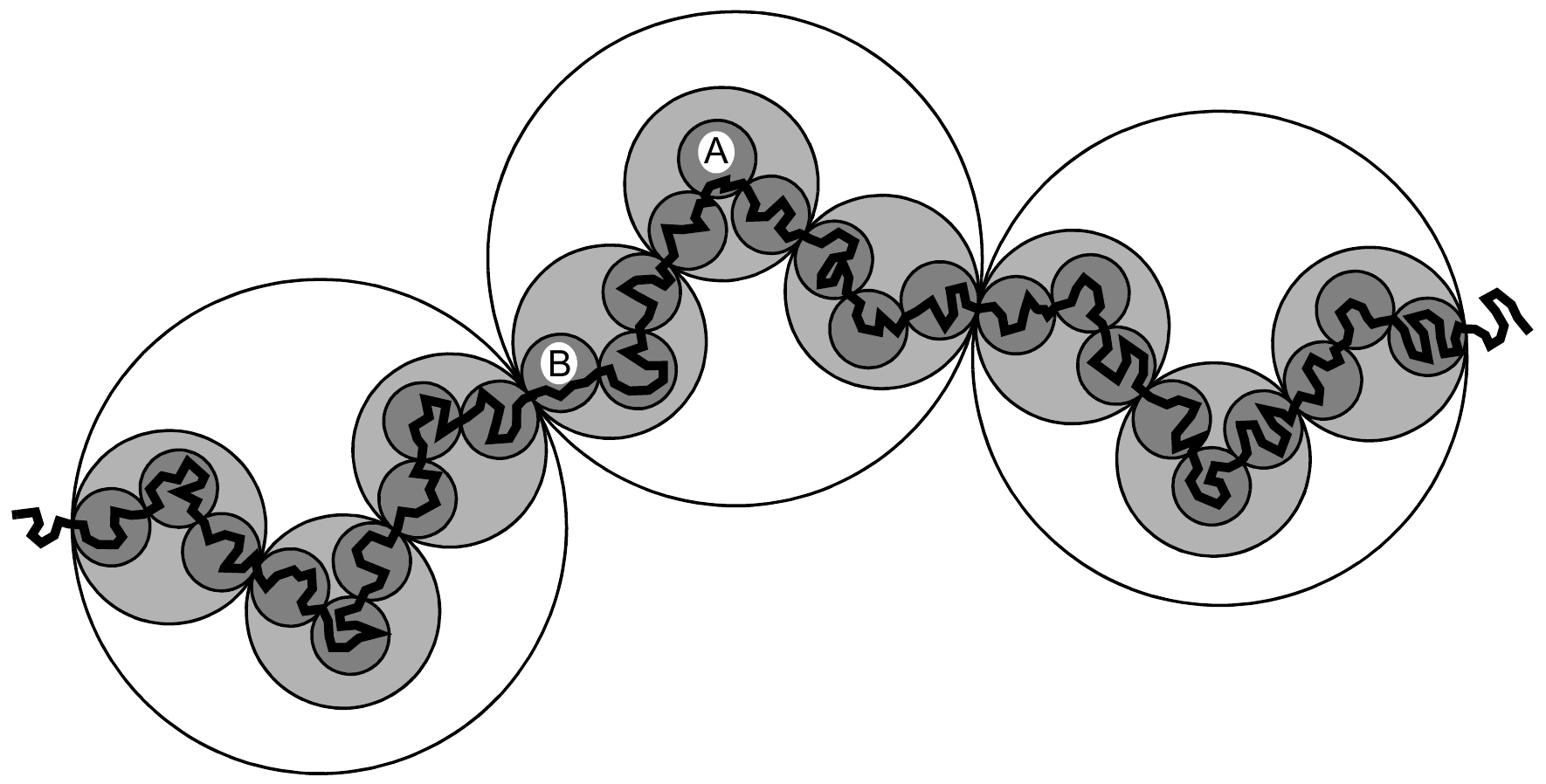}} \caption[ ]
{Modeling a polymer chain with three different bead sizes.} \label{figpolyreno}
\end{figure}

The beads of such mechanical polymer models, however, are fictitious objects consisting of many chemical monomer units. The modeling of the same polymer molecule with three different bead sizes is illustrated in Figure~\ref{figpolyreno}, showing the self-similarity of polymer molecules on different length scales. If one uses larger beads, hydrodynamic interactions between different smaller beads inside a single larger bead have to be incorporated into the effective friction coefficient of the larger bead. For example, the interaction between beads A and B in Figure~\ref{figpolyreno} can be resolved as an inter-bead interaction for the two smaller bead sizes, but becomes an intra-bead interaction changing the friction coefficient of the largest bead. This consideration offers the possibility of incorporating more and more interactions by passing to successively larger beads. Small increments in bead size implying only few interactions inside the larger beads can be handled by perturbation theory; renormalization-group theory allows us to accumulate a very large number of interactions via many small increments of the bead size. As a result of this process, a nontrivial rescaling associated with the self-similarity of polymer molecules is expected. The transformation behavior between effective interaction strengths on different length scales contains important information, in particular, about the critical exponents associated with self-similarity and the limiting value of the effective coupling strength on large scales. With the large scale model and the effective coupling strength, one can finally perform a perturbative calculation for any quantity of interest. A number of both static and dynamic properties of polymers in solution have been computed along these lines \cite{desCloizeauxJannink,Freed,Oono85,hco185,hco38}. Such refined perturbation expansions are remarkably successful. As we shall see below, renormalization is motivated by but not restricted to perturbation theory.

Note that divergences are not an issue in the above discussion of hydrodynamic interactions in dilute polymer solutions. They would only arise if, to establish contact with field theory, we considered the limit of an infinitely large number of infinitesimally small beads. In the next subsection, we show how such intuitive ideas can be translated into tractable equations.

The ambiguous nature of the beads corresponds to the ambiguity of the particle concept in the presence of interactions. The choice of a particular bead size corresponds to the choice of a particular friction parameter in the fundamental quantum master equation of particle physics.

\subsubsection{Basic equations}\label{secrenobaseq}
Let us introduce a small length scale $\ell$, which could be a bead size, a lattice spacing, an inverse momentum cutoff, or the characteristic length scale of dissipative smoothing defined in (\ref{lengthdissipdef}). If some model of interest contains a dimensional parameter $\lambda$, say the strength of some interaction, the proper choice of $\lambda$ for modeling the same large-scale physics with different $\ell$ typically depends on $\ell$. For the further discussion, it is convenient to introduce the dimensionless coupling constant
\begin{equation}\label{lambdadimless}
    \tilde{\lambda}(\ell) = \ell^\epsilon \lambda(\ell) ,
\end{equation}
with a suitable exponent $\epsilon$ obtained from dimensional analysis. The intuitive ideas explained in the preceding subsection are implemented by constructing a perturbation theory of the rate of change of $\tilde{\lambda}(\ell)$ with $\ell$ rather than for some measurable quantity or $\tilde{\lambda}(\ell)$ itself. We assume that there actually exists a perturbation expansion of the function describing the rate of change of the dimensionless coupling constant,
\begin{equation}\label{betadef}
    \beta(\tilde{\lambda}\big(\ell)\big) = - \ell \, \frac{\dR\tilde{\lambda}(\ell)}{\dR\ell}
    = - \epsilon \tilde{\lambda}(\ell) - \ell^{\epsilon+1} \frac{\dR\lambda(\ell)}{\dR\ell} ,
\end{equation}
that is, for the so-called $\beta$ function for the running coupling constant of polymer physics or quantum field theory. We always display the $\beta$ function with its argument to avoid confusion with the inverse temperature. If the free theory remains free on all length scales,that is, $\beta(0)=0$, the most general second-order expansion of the $\beta$ function for small $\tilde{\lambda}$ is given by
\begin{equation}\label{beta2nd}
    \beta(\tilde{\lambda}) = - \alpha \tilde{\lambda} \left( 1 - \frac{\tilde{\lambda}}{\lambda^*} \right) ,
\end{equation}
where the numerical parameters $\alpha$ and $\lambda^*$ remain to be determined. The second identity in (\ref{betadef}) suggests $\alpha = \epsilon$, which we assume from now on. This value of $\alpha$ implies that the changes proportional to $\tilde{\lambda}$ result from the factor $\ell^\epsilon$ in (\ref{lambdadimless}), whereas changes of $\lambda(\ell)$ are responsible for higher-order effects due to interactions. The linear contribution arises directly from straightforward dimensional analysis, whereas the higher-order terms imply a nontrivial renormalization and scaling behavior.

\begin{figure}
\centerline{\includegraphics[width=8cm]{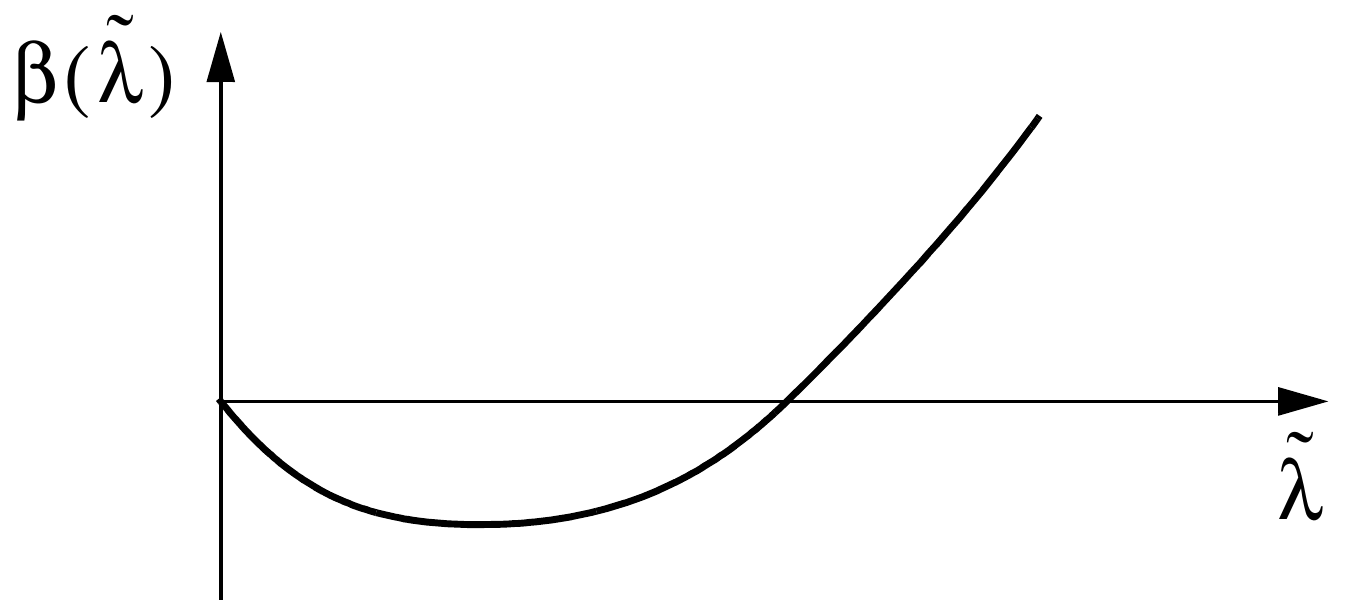}} \caption[ ]
{A typical $\beta$ function for the running coupling constant.} \label{figbetafunc}
\end{figure}

A typical $\beta$ function is shown in Figure~\ref{figbetafunc}. For increasing $\ell$, the dimensionless coupling constant $\tilde{\lambda}$ approaches the value $\lambda^*$ with $\beta(\lambda^*)=0$, which determines a critical point at which $\tilde{\lambda}(\ell)$ no longer changes with $\ell$. In other words, the coupling constant approaches a critical value that is independent of the choice of the coupling constant at much smaller scales.

According to Eq.~(\ref{betadef}), the second-order perturbation theory (\ref{beta2nd}) for $\beta$ implies the following nontrivial dependence of the running dimensional coupling constant on the length scale,
\begin{equation}\label{lambdaform}
    \lambda(\ell) = \frac{\lambda^* \, \lambda(\ell_0)}{\lambda(\ell_0)(\ell^\epsilon-\ell_0^\epsilon)+\lambda^*} ,
\end{equation}
for all $\ell \ge \ell_0$, where the coupling constant $\lambda(\ell_0)$ at a small length scale $\ell_0$ be given. For large $\ell/\ell_0$, we find the expected behavior $\lambda(\ell) = \ell^{-\epsilon} \lambda^*$, or $\tilde{\lambda}(\ell) = \lambda^*$. Note the useful identity
\begin{equation}\label{lambdaformc}
    \frac{\lambda(\ell)}{ 1 - \ell^\epsilon \lambda(\ell)/\lambda^* } =
    \frac{\lambda(\ell_0)}{ 1 - \ell_0^\epsilon \lambda(\ell_0)/\lambda^* } = \hat{\lambda} ,
\end{equation}
where $\hat{\lambda}$ is independent of $\ell$. In other words, we can enrich $\lambda(\ell)$ by higher-order terms to obtain the unambiguous coupling constant $\hat{\lambda}$. Equation (\ref{lambdaformc}) is the basis for a resummation of perturbation theory guided by renormalization.

We want any quantity of interest $P$ to be independent of the choice of $\ell$, at least for sufficiently small $\ell$. It is hence natural to assume that $P$ possesses a perturbation expansion of the form
\begin{equation}\label{genpertexp1}
    P = P_0 + P_1 \hat{\lambda} + P_2 \hat{\lambda}^2 + \ldots .
\end{equation}
In practice, however, we construct perturbation expansions for a particular $\ell$ and the corresponding $\lambda(\ell)$. We hence use (\ref{lambdaformc}) to rearrange (\ref{genpertexp1}) into the form
\begin{equation}\label{genpertexp3}
    P = P_0 + P_1 \lambda(\ell) + \left( P_1 \frac{\ell^\epsilon}{\lambda^*} + P_2 \right) \lambda(\ell)^2 + \ldots .
\end{equation}
This means that the second-order perturbation expansion in $\lambda(\ell)$ has to have a particular structure in which a correction term of order $\ell^\epsilon$ occurs, from which we can actually identify $\alpha=\epsilon$ and $\lambda^*$. In other words, the $\beta$ function (\ref{beta2nd}) can be reproduced from the $\ell$-dependent corrections to perturbation theory.

Of course, these arguments can be generalized to construct the general form of higher-order perturbation expansions of quantities of interest. Generalizing Eq.~(\ref{lambdaformc}), a polynomial expansion of the $\beta$ function in $\tilde{\lambda}$ leads to a nonpolynomial expression
\begin{equation}\label{lambdahatdef}
    \hat{\lambda} = \lambda(\ell) \exp \left\{ - \int_0^{\tilde{\lambda}(\ell)}
    \left( \frac{\epsilon}{\beta(\tilde{\lambda}')}
    + \frac{1}{\tilde{\lambda}'} \right) \dR\tilde{\lambda}' \right\} ,
\end{equation}
which can be checked by differentiating with respect to $\ell$. For polynomial $\beta(\tilde{\lambda}')$, the integration in (\ref{lambdahatdef}) can be performed in closed form. The power-law expansion of (\ref{lambdahatdef}) can be used to rewrite a higher-order expansion of the quantity $P$ in $\hat{\lambda}$ into an expansion in $\lambda(\ell)$, thus generalizing (\ref{genpertexp3}) to higher orders. Note that such an expansion contains various corrections from the small length scale $\ell$. The consistency of all the corrections expresses the renormalizability of the theory, their detailed form contains the coefficients of the $\beta$ function from which, in particular, we obtain the critical coupling constant $\lambda^*$. The explicit value for the dimensionless coupling constant is a major outcome of the procedure and can be used to evaluate perturbation expansions, where $\ell$ is to be taken as the physical length scale set by the parameters of the model.

\subsubsection{Self-similar versus hierarchical}
As we have learned in Section~\ref{secdecoupreduc}, the emergence of irreversibility in nonequilibrium thermodynamics relies on a separation of time scales leading to clearly distinct levels of description. In decoupling high-energy and low-energy processes in quantum field theory, however, we typically do not have such a well-defined separation of time scales. Rather than a hierarchical structure of clearly distinct levels of description, we have self-similarity. The degrees of freedom that we eliminate in the renormalization procedure are faster than those we keep, but they are not separated by a clear gap. Nevertheless, we do not need to keep the faster degrees of freedom because, as a consequence of self-similarity, we know that they behave in a similar way as the slower degrees of freedom. We simply assume that also in such a situation, where fast degrees of freedom can be eliminated, nonequilibrium thermodynamics is applicable. In other words, we assume that nonequilibrium thermodynamics works for self-similar systems as well as for hierarchical systems \cite{hco185}.

According to (\ref{lengthdissipdef}), increasing the characteristic length scale $\ell$ is equivalent to increasing the rate parameter $\gamma$, which implies a larger entropy production rate. An increasing entropy production rate with increasing $\ell$ indicates a coarse-graining process. Of course, the increasing rate parameter implies a more rapid approach to equilibrium. This is consistent with the result (\ref{QMEthermogenep}) implying a total entropy production in the approach to equilibrium that is independent of the rate parameter.

Natural scientists and philosophers seem to like hierarchical structures. Atoms were supposed to be at the bottom end of a hierarchical picture but, of course, they eventually turned out to be at the beginning of particle physics. The hierarchy continues with nuclei, nucleons, quarks. According to the idea of effective quantum field theories, at some point, this hierarchy ends and self-similarity takes over.

We note in passing that cosmology provides us with a system where the distinction between hierarchical and self-similar does not seem to be entirely resolved. One usually groups stars into galaxies, clusters of galaxies, galaxy clouds, and superclusters. However, whether such a hierarchical description of the universe is appropriate seems to be an open question. Some cosmologists believe that the distribution of galaxies rather follows a self-similar pattern \cite{Pietronero87}.

\subsubsection{Asymptotic safety}
\begin{figure}[t]
\centerline{\includegraphics[width=10cm]{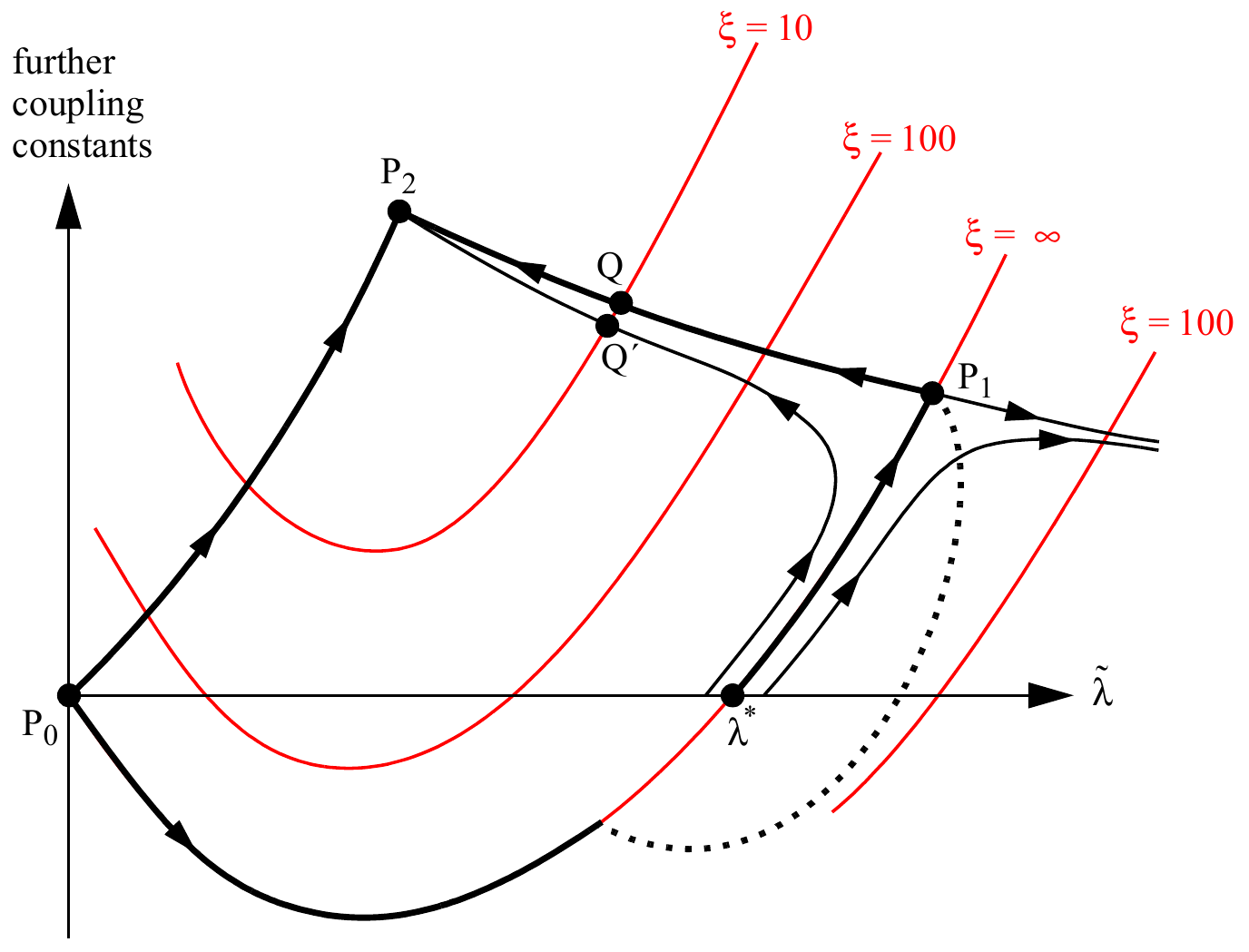}} \caption[ ]
{Renormalization-group flow in theory space, assuming a topology that is appropriate for $\varphi^4$ theory; the arrows indicate how the dimensionless coupling constants change with increasing reference length scale $\ell$.} \label{figRGflow}
\end{figure}

So far, our discussion of renormalization was focused on perturbation theory. At the end of the day, (\ref{lambdaformc}), or the higher-order generalization (\ref{lambdahatdef}), allow us to refine perturbation expansions by guided resummation. We now consider nonperturbative renormalization. Our discussion is based on the qualitative analysis of renormalization-group flow in theory space (see Figure~\ref{figRGflow}) and inspired by Section 12.2 of \cite{WilsonKogut74}.

Figure~\ref{figRGflow} illustrates how the dimensionless coupling constants change if we increase our small reference length scale $\ell$ on which the model is defined, but keep the same large-scale properties. For example, if we start with a $\varphi^4$ theory, that is, with a point on the $x$~axis in Figure~\ref{figRGflow}, with increasing $\ell$ the model develops further coupling constants and does not stay a pure $\varphi^4$ theory. For $\tilde{\lambda} < \lambda^*$, we move to the fixed point $P_2$ upon rescaling. If we fine-tune $\tilde{\lambda}$ to $\lambda^*$, the model moves to the fixed point $P_1$. The further fixed point $P_0$ at the origin corresponds to a massless noninteracting theory and is often referred to as the Gaussian fixed point. The fixed points $P_0$, $P_1$, $P_2$ have been labeled by the number of stable directions for which the renormalization-group flow moves into the fixed point (as counted in the simplified two-dimensional representation of the renormalization-group flow in theory space in Figure~\ref{figRGflow}).

For the further discussion it is useful to consider the hyper-surfaces of constant dimensionless correlation length $\xi$, which is defined as the physical correlation length divided by the reference length $\ell$. If we increase $\ell$ and keep the physical correlation length $\xi \ell$ constant, then $\xi$ decreases on moving in the flow direction. Any finite value of $\xi$ eventually flows to $\xi=0$. Fixed points must hence correspond to $\xi=\infty$ or $\xi=0$. The critical value $\lambda^*$ corresponds to $\xi=\infty$ and the entire trajectory from the $\varphi^4$ theory with $\tilde{\lambda} = \lambda^*$ to the fixed point $P_1$ hence lies in the critical hyperplane with $\xi=\infty$. If the value of $\tilde{\lambda}$ is only slightly below $\lambda^*$, the trajectory comes close to $P_1$ and then moves along the curve $P_1 \, Q \, P_2$. This curve defines the nonperturbatively renormalized theory, which is parameterized by the dimensionless correlation length $\xi$. For given reference length scale $\ell$, the parameter $\xi$ is associated with the physical correlation length and hence with the physical mass parameter; for that reason we did not need to pay explicit attention to the mass parameter in discussing the renormalization-group flow in the space of coupling constants.

Let us assume that we would like to write down a model on a length scale $\ell$ that is ten times smaller than the physical correlation length, that is, with $\xi=10$ (if more spatial resolution was required, we could also choose $\xi=100$ or an even larger value bringing us very close to the critical point $P_1$). The renormalized model with $\xi=10$ is indicated by $Q$ in Figure~\ref{figRGflow}. We obtain the approximate renormalized model $Q'$ by starting from a $\varphi^4$ theory with $\tilde{\lambda}'$ very close to $\lambda^*$ on a much smaller reference scale, $\ell' \ll \ell$, after suitable rescaling (see Figure~\ref{figRGflow}); the length scale $\ell'$ can easily be obtained from the value $\xi'$ of the hyper-surface to which $\tilde{\lambda}'$ belongs according to $\xi'\ell'=10\ell$, which is the physical correlation length. The closer $\tilde{\lambda}'$ is to $\lambda^*$, the larger is $\xi'$, the smaller is $\ell'$, and the closer is $Q'$ to the renormalized model $Q$. In the limit $\ell' \rightarrow 0$, $Q'$ converges to the renormalized model $Q$ with the desired physical correlation length. Reproducing the same large-scale phenomena now has the more precise meaning of approximating $Q$. Keeping the large scale physics invariant while letting the length scale $\ell'$ of the underlying model go to zero implies that the model has to approach a critical point with diverging correlation length in units of $\ell'$. This well-known relationship between field theory and critical phenomena has been discussed extensively, for example, in the two classical review articles \cite{BrezinetalDombGreen,WilsonKogut74}.

If we start with small values of $\tilde{\lambda}'$ on the line of $\varphi^4$ theories, the model moves away from the free theory $P_0$ and crosses over to some fixed point. The simplest assumption is a crossover to $P_1$, rather than having two different fixed points reached from $\varphi^4$ theory for $\tilde{\lambda}'$ close to zero or $\tilde{\lambda}'$ close to $\lambda^*$. Contrary to what Figure~\ref{figRGflow} would suggest, we do not have to pass through the $\varphi^4$ theory labeled by $\lambda^*$ in crossing over from $P_0$ to $P_1$ because, in that figure, we have not resolved the extra dimensions associated with further coupling constants. The dotted line indicates that we can cross over from $P_0$ to $P_1$ without getting close to a $\varphi^4$ theory by leaving the plane of the figure in the direction of other coupling constants, but staying in the hyper-plane with $\xi=\infty$.

We have now constructed the quantum field theory associated with the nontrivial fixed point $P_1$ in Figure~\ref{figRGflow} by qualitative arguments. We started from $\varphi^4$ theory which, however, is merely a particular model within the universality class associated with $P_1$. Many other models on the critical surface with $\xi=\infty$ could have been used to define the same model. As $\varphi^4$ theory is particularly simple and convenient, we can think of it as a minimal model in the universality class of the fixed point $P_1$. However, renormalized $\varphi^4$ theory is a much more universal model and does not really deserve to be referred to as $\varphi^4$ theory.

The graphical (qualitative) construction of the quantum field theory associated with the fixed point $P_1$ is nonperturbative. Weinberg \cite{Weinberg78ip} calls such a quantum field theory associated with a nontrivial fixed point that controls the behavior of the coupling constants \emph{asymptotically safe}. Perturbation theory is limited to the $x$ axis near the fixed point $P_0$ in Figure~\ref{figRGflow}. We have seen how to refine the perturbation theory near the origin by means of renormalization-group theory. A perturbation analysis can be useful if the dimensionless coupling constant is actually very small, as for quantum electrodynamics, or if $\lambda^*$ is small. The latter situation can typically be achieved by changing the space dimension $d$; for example, for $\varphi^4$ theory with $d \lesssim 3$, the critical parameter $\lambda^*$ is close to zero. Figure~\ref{figRGflow} suggests that the critical regions around $0$ and $\lambda^*$ should be considered separately because, with increasing $\tilde{\lambda}$, the correlation length $\xi$ first decreases and then increases, no matter how small $\lambda^*$ is. As $\xi$ always decreases along the renormalization-group flow, there cannot be a natural flow from $0$ to $\lambda^*$ along the $\tilde{\lambda}$-axis. If the correlation length between $0$ to $\lambda^*$ remains very large, one could argue that, in a pragmatic sense, there can still be a trajectory that flows from $0$ into the neighborhood of $\lambda^*$. However, this observation might raise some doubts about $\beta$ functions as the one shown in Figure~\ref{figbetafunc} and about the chances for success of high-order perturbation theory.

In short, perturbative renormalization relies on the scaling properties associated with the Gaussian fixed point $P_0$. Nonperturbative renormalization relies on the scaling properties associated with the critical point $\lambda^*$ of $\varphi^4$ theory.

If one is willing to work at a small reference length scale $\ell$ because this is sufficient to describe all phenomena of interest then one could try to formulate a model directly on that scale. Instead of producing the model $Q$ in Figure~\ref{figRGflow} by rescaling of the critical $\varphi^4$ theory one could try to introduce suitable interactions by hand to approximate $Q$. Such a procedure has the advantage that one can also consider non-renormalizable interactions and one avoids all divergencies, but the unlimited possibilities of introducing such interactions without any theoretical guidance is often considered as a severe disadvantage.

If there exists a physically distinguished length scale $\ell$, such as the Planck length $\ell_{\rm P}$ (see p.\,\pageref{Plancklength}) relevant to quantum gravity, renormalizability is not a natural requirement. There nevertheless seems to be a tendency to insist on renormalizability because it comes with the universality of fixed-point models and hence with high predictive power. If we simply model on the length scale $\ell_{\rm P}$, we have enormous freedom in selecting interactions and coupling constants and hence low predictive power. If the selection of a few-parameter theory does not result from renormalization, it should come from the elegance of a theory based on geometric and symmetry principles. For example, in view of the elegance of the Yang-Mills theory \cite{YangMills54} for strong interactions, it is almost a waste that the theory flows away to some fixed point under renormalization-group transformations. Yang-Mills theory is too beautiful for being only a minimal model among many other possible ones, but that is what it seems to be.

\subsection{Symmetries}\label{secsymmetries}
Symmetries play an important role in quantum field theory. In the beginning of Section \ref{secQFTinterp} we realized that, in particle physics, we need to respect the principles of special relativity; in other words, we need to respect Lorentz symmetry and must hence guarantee that our equations are covariant under Lorentz transformations. Gauge symmetry is the key to the formulation of the fundamental interactions, or collision rules, in Lagrangian field theory and leads to the famous standard model; in particular, gauge symmetry is known to be closely related to renormalizability. Without going into great detail, we here offer a few remarks on how we wish to handle symmetries.

\subsubsection{Lorentz symmetry}
The development of special relativity is intimately related to the classical theory of \index{Electrodynamics}electromagnetism. \index{Maxwell equations}Maxwell's equations for the electric field $\bm{E}$ and the magnetic field $\bm{B}$ generated by an electric charge density $\rho$ and current density $\bm{j}$ can be written as follows,
\begin{equation}\label{Max1eq}
  \bm{\nabla} \cdot \bm{E} =  \rho \,,
\end{equation}
\begin{equation}\label{Max2eq}
  \bm{\nabla} \cdot \bm{B} =  0 \,,
\end{equation}
\begin{equation}\label{Max3eq}
  \frac{\partial \bm{E}}{\partial t} = - \bm{j} + \bm{\nabla} \times \bm{B} \,,
\end{equation}
\begin{equation}\label{Max4eq}
  \frac{\partial \bm{B}}{\partial t} = - \bm{\nabla} \times \bm{E} \,,
\end{equation}
where, in addition to our standard convention (\ref{hbarcconv}), we have assumed $\epsilon_0=1$ for the electric constant or permittivity of free space. In these units, the elementary electric charge in three space dimensions is given by  $e_0 = \sqrt{4 \pi \alpha} \approx 0.30282212$, where $\alpha$ is the dimensionless fine-structure constant.

Writing \index{Maxwell equations}Maxwell's equations in the above form is very practical for applications. However, their Lorentz covariance is not at all manifest. We here follow exactly the same strategy: we simply write the basic equations in one particular inertial system, so that we have separated concepts of space and time. By relying so strongly on explicit time-evolution equations in Section \ref{sectiondynamics}, we clearly have a hard time to verify Lorentz covariance. Even worse, the assumptions of a finite volume and a dissipative mechanism motivated by our \emph{horror infinitatis} even violate the principles of special relativity.\footnote{The discussion of relativistic hydrodynamics in Section~5.2 of \cite{hcobet} shows that a covariant formulation of dissipative equations is possible; as our dissipative smearing mechanism relies on spatial diffusion, additional variables are presumably required for a covariant formulation.} We can only make sure to verify the Lorentz covariance of all our properly chosen final quantities in the limit of infinite volume and zero dissipation.

Of course, the final Lorentz covariance will not arise accidentally. For example, the energy-momentum relationship (\ref{relenergmomrel}) for a relativistic particle is a crucial ingredient to obtain Lorentz symmetry. Whenever we formulate the collision rules associated with fundamental interactions, we have to keep an eye on the principles of special relativity. As in the case of \index{Maxwell equations}Maxwell's equations (\ref{Max1eq})--(\ref{Max4eq}), the symmetry need not be manifest but, in the end, in the final predictions resulting from our mathematical image of fundamental particles and their interactions, it has to be there.

\subsubsection{Gauge symmetry}
The prototype of a gauge theory is given by \index{Maxwell equations}Maxwell's equations (\ref{Max1eq})--(\ref{Max4eq}). In addition to the evolutions equations (\ref{Max3eq}), (\ref{Max4eq}) for $\bm{E}$ and $\bm{B}$, they include the constraints (\ref{Max1eq}), (\ref{Max2eq}), which can be taken into account by writing the magnetic field
\begin{equation}\label{MaxpotBeq}
  \bm{B} = \bm{\nabla} \times \bm{A} \,,
\end{equation}
and the electric field
\begin{equation}\label{MaxpotEeq}
  \bm{E} = - \bm{\nabla} \phi - \frac{\partial \bm{A}}{\partial t} \,,
\end{equation}
in terms of the vector potential $\bm{A}$ and the scalar potential $\phi$. Note, however, that these potentials are not unique because any so-called gauge transformation
\begin{equation}\label{MaxgaugeAeq}
  \bm{A} = \bm{A} + \bm{\nabla} f \,,
\end{equation}
\begin{equation}\label{Maxgaugephieq}
  \phi = \phi - \frac{\partial f}{\partial t} \,,
\end{equation}
with an arbitrary function $f$ leaves the physical fields $\bm{E}$ and $\bm{B}$ unchanged. This freedom constitutes gauge symmetry. To get unique potentials $\bm{A}$ and $\phi$, one needs to impose further constraints, known as gauge conditions. In terms of the potentials, \index{Maxwell equations}Maxwell's equations are reduced to
\begin{equation}\label{Maxpotphieq}
  \left( \frac{\partial^2}{\partial t^2} - \bm{\nabla}^2 \right) \phi = \rho
  + \frac{\partial}{\partial t} \left( \frac{\partial \phi}{\partial t} + \bm{\nabla} \cdot \bm{A} \right) \,,
\end{equation}
and
\begin{equation}\label{MaxpotAeq}
  \left( \frac{\partial^2}{\partial t^2} - \bm{\nabla}^2 \right) \bm{A} = \bm{j}
  - \bm{\nabla} \left( \frac{\partial \phi}{\partial t} + \bm{\nabla} \cdot \bm{A} \right) \,.
\end{equation}
These equations become particularly simple in the Lorenz gauge\footnote{Named after the Danish physicist Ludvig Lorenz (1829--1891), not to be confused with the Dutch physicist Hendrik Lorentz (1853--1928) after whom the Lorentz symmetry of the previous subsection is named; actually, the Lorenz condition is Lorentz invariant.}
\begin{equation}\label{MaxLorentzgauge}
  \frac{\partial \phi}{\partial t} + \bm{\nabla} \cdot \bm{A} = 0 \,.
\end{equation}
In this gauge, (\ref{Maxpotphieq}) and (\ref{MaxpotAeq}) highlight the existence of electromagnetic waves in the absence of electric charges and currents as well as the possibility of a \index{Maxwell equations!Lorentz covariant formulation}Lorentz covariant formulation of \index{Maxwell equations}Maxwell's equations and the Lorenz gauge condition in terms of the four-vector fields $(\phi,\bm{A})$ and $(\rho,\bm{j})$.

The wave equations (\ref{Maxpotphieq}), (\ref{MaxpotAeq}) provide a good starting point for quantization. They suggest to introduce field quanta associated with each of the four components of the potentials $(\phi,\bm{A})$. These field quanta would correspond to four types of photons: longitudinal and temporal ones in addition to the usual transverse ones. The right-hand sides of (\ref{Maxpotphieq}), (\ref{MaxpotAeq}) provide the collision rules for photons and charged particles, respecting Lorentz symmetry. In these remarks, we have ignored the gauge condition (\ref{MaxLorentzgauge}). As a general strategy, we impose gauge conditions by specifying rules for identifying the physical states. The evolution itself takes place in an `unphysically large' Fock space.

In the context of electromagnetism, such a scenario has indeed been elaborated by Bleuler \cite{Bleuler50} and Gupta \cite{Gupta50} in 1950. The practical usefulness of their ideas within the present approach has been been shown in \cite{hco214}. The Bleuler-Gupta theory uses a modification of the standard scalar product, which can be introduced as a rule for the construction of bra- from ket-vectors. If a basis vector of the ket space contains $n$ temporal photons, the corresponding standard bra-vector is multiplied by a factor $(-1)^n$. The use of an indefinite scalar product may be alarming because it might endanger the probabilistic interpretation of the results as, for example, the norm of an odd-number temporal photon state is negative. This loss of unitarity is the price to pay for the \index{Minkowski metric}Minkowski metric to show up in a simple way, that is, for proper relativistic behavior. However, no interpretational problems arise for physically admissible states, where admissibility is defined in terms of a proper version of the covariant Lorenz gauge condition. In particular, the admissibility condition implies that physical correlations involve equal numbers of longitudinal and temporal photons having the same momentum (see Chapter 17 of \cite{Boyarkin1}), which gives us an idea of how interpretational problems are avoided. The physical states have been listed in an elementary way in Eq.~(2.16) of the pioneering work \cite{Bleuler50}. An elegant way of characterizing the physical states has been given in Section V.C.3 of \cite{Cohenetcpp}.

The generalization of the ideas of Gupta and Bleuler to obtain a manifestly covariant canonical quantization of the Yang-Mills gauge theories \cite{YangMills54} for weak and strong interactions has actually been developed in a series of papers by Kugo and Ojima \cite{KugoOjima78a,KugoOjima78b,KugoOjima79a,KugoOjima79b}. As a key ingredient to their approach, Kugo and Ojima use the conserved charges generating \index{BRST quantization}BRST transformations \cite{BecchiRouetStora76,Tyutin75} to characterize the physical states. In addition to \index{Ghost particle}ghost particles, three- and four-gluon collisions make the collision rules for quantum chromodynamics more complicated than for quantum electrodynamics. For electroweak interactions, an additional complication arises: the Higgs particle needs to be included into our Fock-space description. Although the details are considerably more complicated, the approach proposed in this book can be generalized to the Yang-Mills theories for strong and electroweak interactions and hence to all parts of the celebrated standard model of Lagrangian field theory.\footnote{A very competently written, illuminating history of the theory of strong interactions can be found in \cite{Caosr}.}

\index{BRST quantization}BRST quantization (see the original papers \cite{BecchiRouetStora76,Tyutin75} and the pedagogical BRST primer \cite{Nemeschanskyetal86}) is the overarching modern framework for the quantization of gauge field theories. The general idea is to quantize in an enlarged Hilbert space and to characterize the physically admissible states in terms of BRST charges, which are the operators that generate BRST transformations and commute with the Hamiltonian. In this approach, \index{BRST quantization}BRST symmetry may be considered as the fundamental principle of nature that replaces gauge symmetry \cite{Nemeschanskyetal86}.

\subsection{Expansions}
We first describe the general procedure for constructing perturbation expansions for the evolution super-operators associated with the fundamental quantum master equation. The Laplace transform of the evolution super-operator is the most natural object to be expanded. The expansions for super-operators can then be used in the expressions for the experimentally accessible correlation functions of Section~\ref{secLapltranscorrfunc}. We derive a magical identity that unifies perturbation expansions, truncation procedures, and numerical integration schemes. Finally, a simplification of the irreversible contribution to the quantum master equation in perturbation expansions is proposed.

\subsubsection{Perturbation theory}\label{secperturbtheory}
Our starting point for the construction of perturbation expansions is the linear zero-temperature quantum master equation (\ref{QMEthermolowTi}) which can be written as the quantum master equation for the noninteracting theory with an additional inhomogeneous collision term,
\begin{equation}\label{QMEthermolowTpsp}
    \frac{\dR\Delta\rho_t}{\dR t} = -\iR \Qcommu{H^{\rm free}}{\Delta\rho_t}
    + \sum_{\bm{k} \in K^d} \gamma_k
    \left( 2 a_{\bm{k}} \Delta\rho_t a^\dag_{\bm{k}}
    - \Qantico{a^\dag_{\bm{k}} a_{\bm{k}}}{\Delta\rho_t} \right)
    + {\cal L}^{\rm coll}(\Delta\rho_t) \,,
\end{equation}
where $\Delta \rho_t = \rho_t - \rho_{\rm eq}$ and the collision operator ${\cal L}^{\rm coll}$ is given by
\begin{equation}\label{QMEsplitcoll}
    {\cal L}^{\rm coll}(\rho) = -\iR \Qcommu{H^{\rm coll}}{\rho}
    - \sum_{\bm{k} \in K^d} \frac{\gamma_k}{\omega_k} \bigg(
    \Qcommux{a_{\bm{k}}}{\rho \Qcommu{a^\dag_{\bm{k}}}{H^{\rm coll}}}
    + \Qcommux{a^\dag_{\bm{k}}}{\Qcommu{a_{\bm{k}}}{H^{\rm coll}} \rho} \bigg) \,.
\end{equation}
The difference $\Delta \rho_t$ has trace zero and relaxes to zero (remember that the quantum master equation conserves the trace).

Following standard techniques for solving linear ordinary differential equations, the solution to the inhomogeneous linear quantum master equation (\ref{QMEthermolowTpsp}) can be written as
\begin{equation}\label{intmastereq}
    {\cal E}_t(\Delta\rho_0) = \Delta\rho_t = {\cal E}^{\rm free}_t(\Delta\rho_0)
    + \int_0^t {\cal E}^{\rm free}_{t-t'}\Big({\cal L}^{\rm coll}\big(\Delta\rho_{t'}\big)\Big) \, \dR t' \,,
\end{equation}
where ${\cal E}_t$ is the evolution operator for the full quantum master equation (\ref{QMEthermolowTpsp}), whereas ${\cal E}^{\rm free}_t$ is the evolution operator for the free quantum master equation (\ref{QMEthermolowTfree}). By iterative solution of the integral equation (\ref{intmastereq}), we obtain the expansion
\begin{eqnarray}
    {\cal E}_t(\rho) &=& {\cal E}^{\rm free}_t(\rho)
    + \int_0^t \dR t' \, {\cal E}^{\rm free}_{t-t'}\Big({\cal L}^{\rm coll}\big({\cal E}^{\rm free}_{t'}(\rho) \big) \Big)
    \nonumber \\
    &+& \int_0^t \dR t' \int_0^{t'} \dR t'' \, {\cal E}^{\rm free}_{t-t'}\Bigg( {\cal L}^{\rm coll}\bigg(
    {\cal E}^{\rm free}_{t'-t''}\Big({\cal L}^{\rm coll}\big({\cal E}^{\rm free}_{t''}(\rho)
    \big) \Big) \bigg) \Bigg) \qquad
    \nonumber \\
    &+& \ldots \,.
\label{Eperturbexpansion}
\end{eqnarray}

In view of the nested convolutions appearing in this expansion, it is useful to rewrite (\ref{Eperturbexpansion}) as an expansion for the Laplace-transformed super-operators ${\cal R}_s$ defined in (\ref{Rsupopdef}). The perturbation expansion (\ref{Eperturbexpansion}) then becomes much simpler,
\begin{eqnarray}
    {\cal R}_s(\rho) &=& {\cal R}^{\rm free}_s(\rho)
    + {\cal R}^{\rm free}_s \Big({\cal L}^{\rm coll} \big({\cal R}^{\rm free}_s(\rho) \big) \Big)
    \nonumber \\
    &+& {\cal R}^{\rm free}_s \Bigg( {\cal L}^{\rm coll}\bigg(
    {\cal R}^{\rm free}_s \Big({\cal L}^{\rm coll}\big({\cal R}^{\rm free}_s(\rho)
    \big) \Big) \bigg) \Bigg) + \ldots \,,
\label{EperturbexpansionR}
\end{eqnarray}
in terms of ${\cal L}^{\rm coll}$ and the free version of the Laplace-transformed  evolution operator (\ref{Rsupopdef}),
\begin{equation}\label{Rfreesupopdef}
    {\cal R}^{\rm free}_s(\rho) = \int_0^\infty {\cal E}^{\rm free}_t(\rho) \, \eR^{- s t} \, \dR t \,.
\end{equation}

\subsubsection{A magical identity}\label{secmagicid}
Writing the linear zero-temperature quantum master equation (\ref{QMEthermolowT}) or (\ref{QMEthermolowTi}) in the form $d\rho_t/dt={\cal L}\rho_t$, the corresponding time-evolution super-operator is given by ${\cal E}_t = \eR^{{\cal L} t}$ and its formal Laplace transform (\ref{Rsupopdef}) becomes
\begin{equation}\label{Rsupopformal}
    {\cal R}_s = (s-{\cal L})^{-1} \,.
\end{equation}
Of course, (\ref{Rfreesupopdef}) translates into the analogous identity for the free theory. After a few further formal rearrangements,
\begin{eqnarray}
    ({\cal R}_s)^{-1} &=& s - {\cal L}^{\rm free} - {\cal L}^{\rm coll}
    = s + r - {\cal L}^{\rm free} - ( r+{\cal L}^{\rm coll} ) \nonumber \\
    &=& ({\cal R}^{\rm free}_{s+r})^{-1} - ( r+{\cal L}^{\rm coll} ) \nonumber \\
    &=& [1 - ( r+{\cal L}^{\rm coll} ) {\cal R}^{\rm free}_{s+r}] \, ({\cal R}^{\rm free}_{s+r})^{-1} \,,
\label{magicder}
\end{eqnarray}
we can formulate the following identity,
\begin{eqnarray}
    {\cal R}_s &=& \lim_{q \rightarrow 1} {\cal R}^{\rm free}_{s+r} \sum_{n=0}^{\infty}
    \left[ q r \left( 1+\frac{{\cal L}^{\rm coll}}{r} \right) {\cal R}^{\rm free}_{s+r} \right]^n
    \nonumber \\
    &=& \lim_{q \rightarrow 1} \sum_{n=0}^{\infty}
    \left[ q r \, {\cal R}^{\rm free}_{s+r} \left( 1+\frac{{\cal L}^{\rm coll}}{r} \right) \right]^n
    {\cal R}^{\rm free}_{s+r} \,.
\label{magic}
\end{eqnarray}

Equation (\ref{magic}) has a number of interesting features. For $r=0$ and $q=1$, it simply reproduces the perturbation expansion (\ref{EperturbexpansionR}) in a somewhat simpler notation. For $r=0$ and $q<1$, only a limited number of terms contributes significantly to the infinite sum, whereas higher-order terms are suppressed exponentially. Such a truncation with $q<1$ is not attractive for analytical perturbative calculations because many higher-order terms making only small contributions need to be calculated. For numerical calculations, however, truncation with $q<1$ may be expected to possess much better convergence properties than the usual truncation after a fixed number of terms. Finally, for $r>0$, each factor $(1+{\cal L}^{\rm coll}/r)$ may be interpreted as part of a numerical integration scheme with a time step $1/r$. Actually, $r$ may be interpreted as the rate at which time steps of size $1/r$ are performed stochastically, so that the expansion (\ref{magic}) characterizes a numerical integration scheme (see (\ref{jumpHcoll}) for further details). Perturbation theory can hence be regarded as a stochastic numerical integration scheme in the limit of large time steps.

As (\ref{magic}) unifies perturbation theory, a geometric truncation procedure, and a numerical integration scheme at stochastically chosen times, we refer to it as a magical identity. In particular, we propose to use it as a starting point for numerical simulations (see Section \ref{seccomputersim} for further details).

\subsubsection{Simplified irreversible dynamics}\label{secSID}
The collision term ${\cal L}^{\rm coll}$ introduced in (\ref{QMEsplitcoll}) and appearing in the magical identity (\ref{magic}) is, at least for some purposes, unnecessarily complicated. For example, in a perturbation expansion, only a finite number of factors ${\cal L}^{\rm coll}$ occur. In the final limit $\gamma_k \rightarrow 0$, we hence expect that we can use the much simpler collision operator
\begin{equation}\label{QMEsplitcollSID}
    {\cal L}^{\rm coll}(\rho) = -\iR \Qcommu{H^{\rm coll}}{\rho} \,,
\end{equation}
consisting only of the reversible contribution from collisions. Among the irreversible contributions, only the one associated with the free theory is retained. We refer to the approximation (\ref{QMEsplitcollSID}) for the collision operator as \emph{simplified irreversible dynamics} (SID). As explained above, we expect SID to be exact for perturbation theory.

As SID fully accounts for the irreversible behavior of the free theory, it leads to a smoothed or regularized free time-evolution operator. This seems to be sufficient to provide a proper ultraviolet regularization for the interacting theory.

However, SID does not respect the full thermodynamic structure of our fundamental quantum master equation. In particular, the equilibrium density matrix gets slightly modified. This modification disappears only in the final limit of vanishing dissipation. We here adopt SID for many practical calculations; however, if a serious problem arises in any of the further developments, we can always return to the fully consistent thermodynamic quantum master equation with the full collision operator (\ref{QMEsplitcoll}).

\subsection{Unravelings}
We now describe the basic idea of \emph{unraveling} a quantum master equation (for more details and more rigor see, for example, Chapter~6 of \cite{BreuerPetru}). The basic idea is to obtain the time-dependent density matrix or statistical operator $\rho_t$ solving a master equation as a second moment or expectation,
\begin{equation}\label{unravel}
    \rho_t = E( \Dket{\psi_t} \Dbra{\psi_t} ) \,,
\end{equation}
where $\Dket{\psi_t}$ is a suitably defined stochastic process in the underlying Hilbert space. Such a process consists of random quantum jumps and, between jumps, a deterministic Schr\"odinger-type evolution, modified by a dissipative term.

The construction of an unraveling is not unique. For reasons of simplicity, we would like to construct an unraveling in which $\Dket{\psi_t}$ at any time $t$ is a complex multiple of one of the base vectors $(\ref{Fockstatecreate})$ of the Fock space. This idea would lead to a particularly natural and intuitive image of fundamental particles and their interactions. Remember that, according to Boltzmann's \index{Pluralism, scientific}scientific pluralism, for an image of nature there is no obligation to be unique.

\subsubsection{Free theory: one-process unraveling}\label{secunravelfreef}
To explain the basic idea, we first consider the zero-temperature master equation (\ref{QMEthermolowTfree}) for the noninteracting theory. We follow the construction of an unraveling given in Section 6.1 of \cite{BreuerPetru}. The continuous evolution equation for the state vector $\Dket{\psi_t}$ is taken to be of the deterministic form
\begin{equation}\label{Schroeirrfree}
    \frac{\dR}{\dR t} \Dket{\psi_t} = - \iR H^{\rm free} \Dket{\psi_t} -
    \sum_{\bm{k} \in K^d} \gamma_k \big( 1 - \Dket{\psi_t}\Dbra{\psi_t}\big)
    a^\dag_{\bm{k}} a_{\bm{k}} \Dket{\psi_t} \,,
\end{equation}
where the friction or damping term has been introduced to reproduce the last two terms in (\ref{QMEthermolowTfree}), but with an additional transverse projector $\big( 1 - \Dket{\psi_t}\Dbra{\psi_t}\big)$ to keep $\Dket{\psi_t}$ normalized. It turns out that the resulting nonlinear term (in $\Dket{\psi_t}$) is exactly what is required to compensate for the loss of existing states from jumps. These quantum jumps are of the form
\begin{equation}\label{jumpfree}
    \Dket{\psi_t} \rightarrow \frac{a_{\bm{k}} \Dket{\psi_t}}{\|a_{\bm{k}} \Dket{\psi_t}\|}
    \quad \mbox{with rate} \quad 2 \Dbra{\psi_t} a^\dag_{\bm{k}} a_{\bm{k}} \Dket{\psi_t} \gamma_k \,,
\end{equation}
where the normalization of the final state is no problem for all the transitions that can actually occur, that is, for those occurring with a positive rate and hence with $\|a_{\bm{k}} \Dket{\psi_t}\| > 0$.

Note that the jumps produce the following contribution to the time derivative of $\Dket{\psi_t}\Dbra{\psi_t}$ in terms of gain and loss terms,
\begin{eqnarray}
    2 \Dbra{\psi_t} a^\dag_{\bm{k}} a_{\bm{k}} \Dket{\psi_t} \gamma_k
    \left( \frac{a_{\bm{k}} \Dket{\psi_t}\Dbra{\psi_t} a^\dag_{\bm{k}}}{\|a_{\bm{k}} \Dket{\psi_t}\|^2}
    - \Dket{\psi_t}\Dbra{\psi_t} \right) = \hspace{4em} && \nonumber\\
    2 \gamma_k a_{\bm{k}} \Dket{\psi_t}\Dbra{\psi_t} a^\dag_{\bm{k}} \,\, - \,\,
    2 \gamma_k \Dket{\psi_t} \Dbra{\psi_t} a^\dag_{\bm{k}} a_{\bm{k}} \Dket{\psi_t} \Dbra{\psi_t} \,. &&
\label{unravelproof}
\end{eqnarray}
The last term cancels the contribution resulting from the nonlinear term in the modified Schr\"odinger equation (\ref{Schroeirrfree}) so that, after averaging, we indeed recover (\ref{QMEthermolowTfree}).

\begin{figure}[t]
\centerline{\includegraphics[height=8cm]{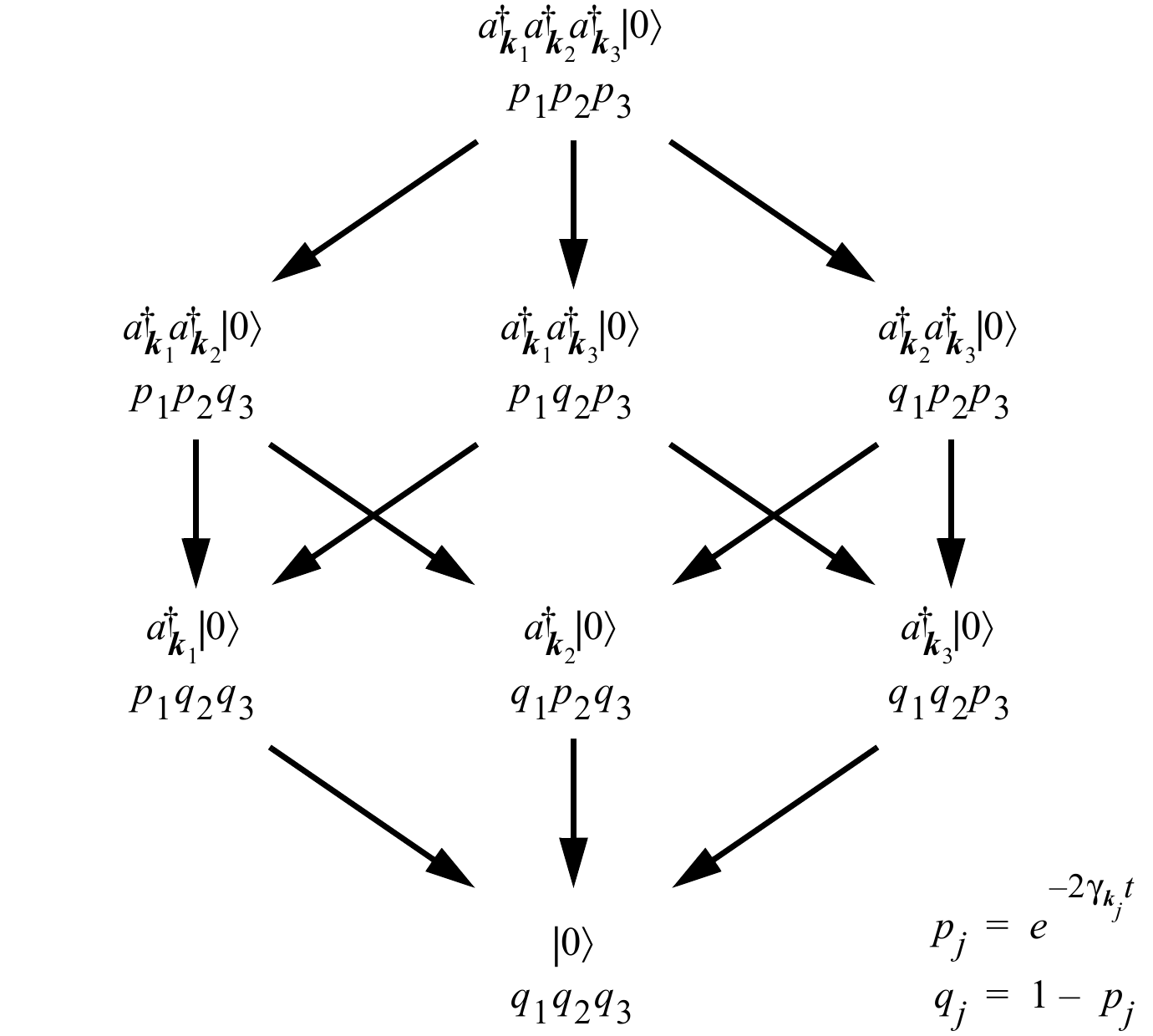}} \caption[ ]
{Decaying three-particle initial state with probabilities for finding the respective possible states after time $t$ (to avoid normalization issues, all three momenta are assumed to be different; see Figure~\ref{figfreevacdecayx} for degenerate cases).} \label{figfreevacdecay}
\end{figure}

\begin{figure}[t]
\centerline{\includegraphics[height=8cm]{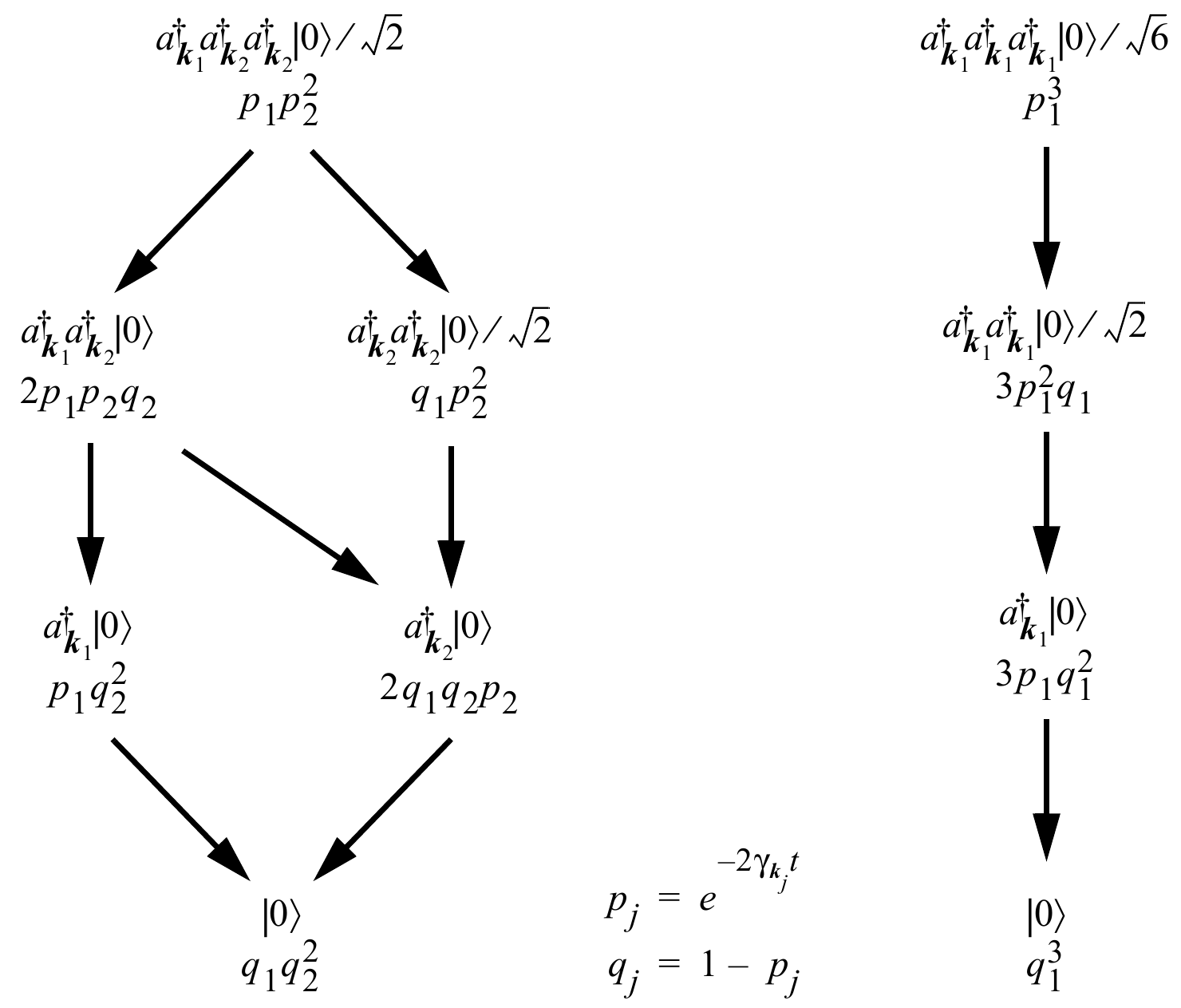}} \caption[ ]
{Decaying three-particle initial state with probabilities for finding the respective possible states after time $t$ for the degenerate cases $\bm{k}_2=\bm{k}_3$ and $\bm{k}_1=\bm{k}_2=\bm{k}_3$.} \label{figfreevacdecayx}
\end{figure}

If $\Dket{\psi_t}$ is a multiple of a base vector of the Fock space, it is an eigenstate of the operator $a^\dag_{\bm{k}} a_{\bm{k}}$ counting the particles with momentum $\bm{k}$ and (\ref{Schroeirrfree}) reduces to the free Schr\"odinger equation. Moreover, $\Dket{\psi_t}$ then is an eigenvector of the free Hamiltonian. The jump process consists of a limited number of relaxation modes, namely those, for which one of the particles present in $\Dket{\psi_t}$ can be eliminated. Note that, if we start with a Fock base vector, we always keep a multiple of a single normalized Fock base vector in our unraveling. In the course of time, particles are removed from the state $\Dket{\psi_t}$ until eventually the vacuum state $\Dket{0}$ is reached. At any finite time, one can calculate the probability for finding any state that can be reached by removing a number of particles from the initial Fock state (see Figures~\ref{figfreevacdecay} and \ref{figfreevacdecayx}).

\subsubsection{Free theory: two-process unraveling}\label{secunravelfreef2}
In the preceding section, we discussed a one-process unraveling for which any change of the state $\Dket{\psi_t}$ affects the bra- and ket-sides of the tensor $\Dket{\psi_t} \Dbra{\psi_t}$ in the same way. Such an unraveling can be used to evaluate correlation functions of the symmetric form (\ref{observableform2}) by performing jumps $\Dket{\psi_t} \rightarrow A_j \Dket{\psi_t}$ at times $t_j$; however, they cannot be used to evaluate correlation functions of the form (\ref{observableform3s}), where all the action is on the ket-side only. Such averages would need to be calculated by means of the polarization identity. As an alternative, two-process unravelings, which allow for different jumps on the bra- and ket-sides, have been introduced (this idea has been used in Section 6.1.4 of \cite{BreuerPetru} to evaluate multi-time correlation functions and in \cite{hco205,hco202} to simulate nonlinear master equations). Two-process unravelings are based on a more general second-moment representation of the density matrix than (\ref{unravel}), namely
\begin{equation}\label{unravel2}
    \rho_t = E( \Dket{\phi_t} \Dbra{\psi_t} ) \,,
\end{equation}
where $\Dket{\phi_t}$ and $\Dket{\psi_t}$ are two stochastic trajectories in Fock space simultaneously with (potentially) different jumps.

To obtain a two-process unraveling of the free theory, we consider the jumps
\begin{equation}\label{jump2pfree}
    \Dket{\phi_t} \rightarrow \frac{a_{\bm{k}} \Dket{\phi_t}\|\Dket{\phi_t}\|}{\|a_{\bm{k}} \Dket{\phi_t}\|} ,\,
    \Dket{\psi_t} \rightarrow \frac{a_{\bm{k}} \Dket{\psi_t}\|\Dket{\psi_t}\|}{\|a_{\bm{k}} \Dket{\psi_t}\|}
    \quad \mbox{with rate} \quad 2 i_{\bm{k}} (\Dket{\phi_t},\Dket{\psi_t}) \, \gamma_k \,,
\end{equation}
where
\begin{equation}\label{ikdef}
    i_{\bm{k}} (\Dket{\phi},\Dket{\psi}) =
    \frac{\|a_{\bm{k}} \Dket{\phi}\| \, \|a_{\bm{k}} \Dket{\psi}\|}{
    \|\Dket{\phi}\| \, \|\Dket{\psi}\|} \,,
\end{equation}
and the continuous evolution equations
\begin{equation}\label{Schroeirrfree1}
    \frac{\dR}{\dR t} \Dket{\phi_t} = - \iR H^{\rm free} \Dket{\phi_t} -
    \sum_{\bm{k}} \gamma_k \left[ a^\dag_{\bm{k}} a_{\bm{k}}
    - \iR_{\bm{k}} (\Dket{\phi_t},\Dket{\psi_t}) \right] \Dket{\phi_t} \,,
\end{equation}
and
\begin{equation}\label{Schroeirrfree2}
    \frac{\dR}{\dR t} \Dket{\psi_t} = - \iR H^{\rm free} \Dket{\psi_t} -
    \sum_{\bm{k}} \gamma_k \left[ a^\dag_{\bm{k}} a_{\bm{k}}
    - i_{\bm{k}} (\Dket{\phi_t},\Dket{\psi_t}) \right] \Dket{\psi_t} \,.
\end{equation}
The equivalence of this two-process unraveling with the zero-temperature master equation (\ref{QMEthermolowTfree}) can be checked by means of the arguments previously used for the one-process unraveling. Jumps can occur only if both $\Dket{\phi_t}$ and $\Dket{\psi_t}$ contain a free particle with the same momentum $\bm{k}$. In the jumps (\ref{jump2pfree}), the norms of $\Dket{\phi_t}$ and $\Dket{\psi_t}$ are conserved.

If the vectors $\Dket{\phi_t}$ and $\Dket{\psi_t}$ initially are equal unit vectors, we recover the one-process unraveling of Section~\ref{secunravelfreef}. More generally, for initial states involving only a single Fock base vector,
\begin{equation}\label{Schroeirrfree1ini}
    \Dket{\phi_0} = c(0) \, a^\dag_{\bm{k}_1} \ldots a^\dag_{\bm{k}_n} \Dket{0} \,,
\end{equation}
\begin{equation}\label{Schroeirrfree2ini}
    \Dket{\psi_0} = c'(0) \, a^\dag_{\bm{k}'_1} \ldots a^\dag_{\bm{k}'_{n'}} \Dket{0} \,,
\end{equation}
the time-dependent solutions of (\ref{Schroeirrfree1}), (\ref{Schroeirrfree2}) remain multiples of these base vectors and the time-dependent coefficients are
\begin{equation}\label{Schroeirrfree1ct}
    c(t) = \exp \left\{ \left( -\iR \sum\nolimits_{j=1}^n \omega_{k_j}
    - \sum\nolimits_{j=1}^n \gamma_{k_j} + \bar{\gamma} \right) t \right\} \, c(0) \,,
\end{equation}
\begin{equation}\label{Schroeirrfree2ct}
    c'(t) = \exp \left\{ \left( -\iR \sum\nolimits_{j=1}^{n'} \omega_{k'_j}
    - \sum\nolimits_{j=1}^{n'} \gamma_{k'_j} + \bar{\gamma} \right) t \right\} \, c'(0) \,.
\end{equation}
The contribution $\bar{\gamma}$, which arises from $i_{\bm{k}} (\Dket{\phi_t},\Dket{\psi_t})$ and is given by
\begin{eqnarray}
    \bar{\gamma} &=& \sum_{j=1}^n \sum_{j'=1}^{n'} \delta_{\bm{k}_j \bm{k}'_{j'}} \, \gamma_{k_j} \,
    \frac{\| a^\dag_{\bm{k}_1} \ldots a^\dag_{\bm{k}_{j-1}} a^\dag_{\bm{k}_{j+1}} \ldots a^\dag_{\bm{k}_n} \Dket{0}
    \|}{\| a^\dag_{\bm{k}_1} \ldots a^\dag_{\bm{k}_n} \Dket{0} \|}
    \nonumber \\ && \hspace{6.7em} \times \,
    \frac{\| a^\dag_{\bm{k}'_1} \ldots a^\dag_{\bm{k}'_{j'-1}} a^\dag_{\bm{k}'_{j'+1}} \ldots a^\dag_{\bm{k}'_{n'}}
    \Dket{0} \|}{\| a^\dag_{\bm{k}'_1} \ldots a^\dag_{\bm{k}'_{n'}} \Dket{0} \|} \,, \qquad
\label{gammabarexpr}
\end{eqnarray}
eliminates the exponential decay associated with those momenta appearing in both $\Dket{\phi_0}$ and $\Dket{\psi_0}$. For $\Dket{\phi_0}=\Dket{\psi_0}$, that is, for the one-process unraveling, all decay rates are eliminated. It is particularly easy to see that if all momenta in $\Dket{\phi_0}$ and all momenta in $\Dket{\psi_0}$ are different (but the momenta in $\Dket{\phi_0}$ are allowed to occur in $\Dket{\psi_0}$) because then all the norms in (\ref{gammabarexpr}) are equal to unity. For multiple occurrences of the same momentum in one of the states, the norms and the counting of matches become slightly more involved.

\subsubsection{Calculation of correlation functions}
\begin{figure}
\centerline{\includegraphics[width=8.5cm]{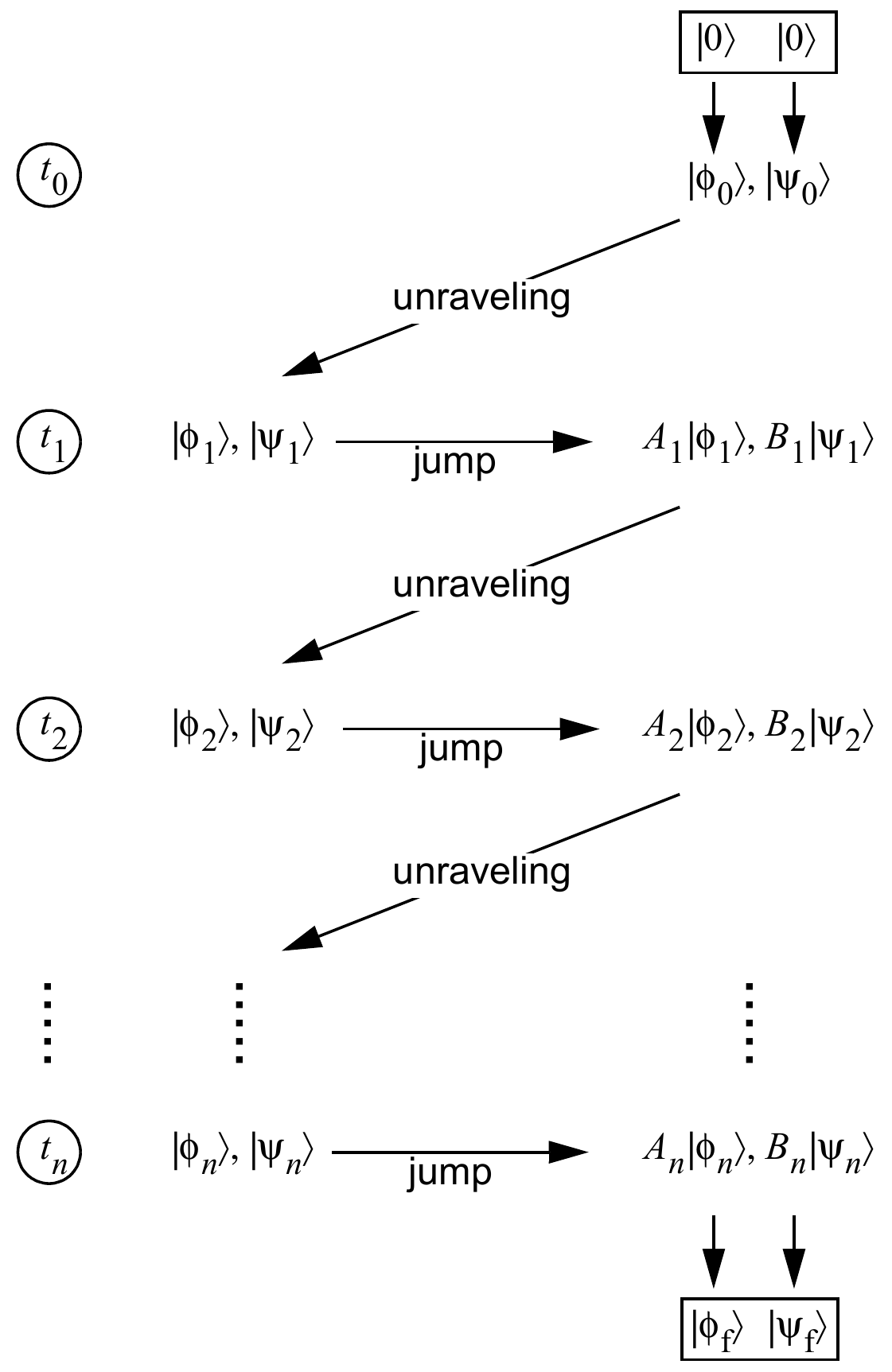}} \caption[ ]
{Procedure for calculating multi-time correlation functions from two-process unravelings.} \label{figunravelmulti}
\end{figure}

An important advantage of two-process unravelings is that the evolution operator ${\cal E}_t$ can naturally be extended from density matrices to dyadics of the form $\Dket{\phi} \Dbra{\psi}$ as initial conditions. This observation can be used to calculate the multi-time correlation functions defined in Section~\ref{secdefcorfcts} in a straightforward way. The procedure is illustrated in Figure~\ref{figunravelmulti}.

The first step from $t_0$ to $t_1$ can be seen as an `equilibration phase.' For the free theory, we simply stay in the ground state. The operators $A_j$ and $B_j$ are introduced through jumps of $\Dket{\phi_j}$ and $\Dket{\psi_j}$ at the times $t_j$ where, between these jumps, the two trajectories in Fock space are evolved according to the two-process unraveling. At the end, we read off the final states $\Dket{\phi_{\rm f}}$ and $\Dket{\psi_{\rm f}}$. We can then evaluate the multi-time correlation function (\ref{observableform3}) as
\begin{equation}\label{observableform8}
    {\rm tr} \left\{ A_n {\cal E}_{t_n-t_{n-1}}\Big( \ldots \, A_2 {\cal E}_{t_2-t_1} \big( A_1
    {\cal E}_{t_1-t_0}(\Dket{0} \Dbra{0}) B^\dag_1 \big)  B^\dag_2 \, \ldots \Big) B^\dag_n \right\} =  E\left[ \Dbraket{\psi_{\rm f}}{\phi_{\rm f}} \right] \,.
\end{equation}
Whenever $\|A_j\Dket{\phi_j}\|=0$ or $\|B_j\Dket{\psi_j}\|=0$ for some $j$, the corresponding trajectory does not contribute to the correlation function.

Note that in Figure~\ref{figunravelmulti} there is no final `equilibration phase' analogous to the initial one. This observation was our motivation for the passage from (\ref{observableform3}) to (\ref{observableform3l}). Based on (\ref{observableform3l}), we suggest to further evolve the states $\Dket{\phi_{\rm f}}$, $\Dket{\psi_{\rm f}}$ at time $t_n$ to $\Dket{\phi'_{\rm f}}$, $\Dket{\psi'_{\rm f}}$ at time $\tau/2$ by means of the unraveling. By fully mirroring the initial equilibration process starting at $t_0=-\tau/2$, we suggest to evaluate $\Dbraket{0}{\phi'_{\rm f}}\Dbraket{\psi'_{\rm f}}{0}$, which is the overlap of the `equilibrated pair of states' with the free vacuum. This overlap generally differs from $\Dbraket{0}{\phi_{\rm f}}\Dbraket{\psi_{\rm f}}{0}$ because the latter two states are freshly biased by the jump operators $A_j$, $B_j$. For an `equilibrated pair of states' we expect a unique overlap with the free vacuum, except for a linear dependence on the conserved initial scalar product $E[\Dbraket{\psi_{\rm f}}{\phi_{\rm f}}]$. We can hence rewrite (\ref{observableform8}) as
\begin{eqnarray}
    {\rm tr} \left\{ A_n {\cal E}_{t_n-t_{n-1}}\Big( \ldots \, A_2 {\cal E}_{t_2-t_1} \big( A_1
    {\cal E}_{t_1-t_0}(\Dket{0} \Dbra{0}) B^\dag_1 \big)  B^\dag_2 \, \ldots \Big) B^\dag_n \right\} = \qquad
    && \nonumber \\
    \frac{E\left[ \Dbraket{0}{\phi'_{\rm f}}\Dbraket{\psi'_{\rm f}}{0} \right]}{
    E\big[ \Dbraket{0}{\phi''_{\rm f}}\Dbraket{\psi''_{\rm f}}{0} \big]} \,, \qquad &&
\label{observableform9}
\end{eqnarray}
where the states $\Dket{\phi''_{\rm f}}$, $\Dket{\psi''_{\rm f}}$ result from the unraveling evolution of the pair $\Dket{0}$, $\Dket{0}$ from $-\tau/2$ to $\tau/2$. In the deterministic super-operator formulation, the equivalence of (\ref{observableform8}) and (\ref{observableform9}) is expressed by the equality of (\ref{observableform3}) and (\ref{observableform3l}).

\subsubsection{Interacting theory}\label{secunravelinteract}
With the help of the commutators (\ref{Hcollcommua}) and (\ref{Hcollcommuad}), the low-temperature quantum master equation (\ref{QMEthermolowTi}) for the $\varphi^4$ theory can be written as
\begin{eqnarray}
    \frac{\dR\Delta\rho_t}{\dR t} &=& -\iR \Qcommu{H}{\Delta\rho_t}
    + \sum_{\bm{k} \in K^d} \gamma_k \left[ {\cal X}_{\bm{k}}(a_{\bm{k}})
    + \frac{\lambda'}{\omega_k^2} \Big( {\cal X}_{\bm{k}}(a_{\bm{k}}) + {\cal X}_{\bm{k}}(a^\dag_{-\bm{k}}) \Big) \right]
    \nonumber \\
    &+& \frac{\lambda}{24 V} \sum_{\bm{k},\bm{k}_1,\bm{k}_2,\bm{k}_3 \in K^d}
    \frac{\gamma_k}{\omega_k}
    \frac{\delta_{\bm{k}_1+\bm{k}_2+\bm{k}_3,\bm{k}}}{\sqrt{\omega_k\omega_{k_1}\omega_{k_2}\omega_{k_3}}}
    \Big[ {\cal X}_{\bm{k}}(a_{\bm{k}_1} a_{\bm{k}_2} a_{\bm{k}_3}) \nonumber \\
    &+& 3 {\cal X}_{\bm{k}}(a^\dag_{-\bm{k}_1} a_{\bm{k}_2} a_{\bm{k}_3})
    + 3 {\cal X}_{\bm{k}}(a^\dag_{-\bm{k}_1} a^\dag_{-\bm{k}_2} a_{\bm{k}_3})
    + {\cal X}_{\bm{k}}(a^\dag_{-\bm{k}_1} a^\dag_{-\bm{k}_2} a^\dag_{-\bm{k}_3}) \Big] \,, \nonumber \\ &&
\label{QMEthermolowTexpl}
\end{eqnarray}
with the definition
\begin{equation}\label{Xkdef}
    {\cal X}_{\bm{k}}(A) = a_{\bm{k}} \Delta\rho_t A^\dag + A \Delta\rho_t a^\dag_{\bm{k}}
    - a^\dag_{\bm{k}} A \Delta\rho_t - \Delta\rho_t A^\dag a_{\bm{k}} \,.
\end{equation}
An alternative compact formulation of this quantum master equation is given by
\begin{equation}\label{QMEthermolowTx}
    \frac{\dR\Delta\rho_t}{\dR t} = -\iR \Qcommu{H}{\Delta\rho_t}
    + \sum_{\bm{k} l} \Gamma_{\bm{k} l} \Big( a_{\bm{k}} \Delta\rho_t A^\dag_l + A_l \Delta\rho_t a^\dag_{\bm{k}}
    - a^\dag_{\bm{k}} A_l \Delta\rho_t - \Delta\rho_t A^\dag_l a_{\bm{k}} \Big) \,,
\end{equation}
where the rates $\Gamma_{\bm{k} l}$ and the jump operators $A_l$, which are normal-ordered products of creation and annihilation operators, can be identified by comparison with (\ref{QMEthermolowTexpl}).

A great advantage of the two-process unraveling compared to the one-process unraveling is that it is applicable also to the difference $\Delta\rho_t = \rho_t - \rho_{\rm eq}$, which has a vanishing trace. Moreover, it can easily handle different jump operators $A_l \neq a_{\bm{k}}$ acting on the bra- and ket-sides of density matrices.

If we construct an unraveling for $\Delta\rho_t$ instead of $\rho_t$, the procedure for calculating correlation functions changes. We cannot allow for an equilibration phase because $\Delta\rho_t$ is expected to converge to zero. As noted before, $\rho_{\rm eq}$ is needed as an input for the linear quantum master equations at finite or zero temperature; only the full nonlinear quantum master equation can produce $\rho_{\rm eq}$ as an output. We hence need to find an independent method to simulate the equilibrium density matrix which, in the zero-temperature limit, is given by the ground state, $\rho_{\rm eq} = \Dket{\Omega} \Dbra{\Omega}/\Dbraket{\Omega}{\Omega}$.

We illustrate the modified calculation of correlation functions for the two-time correlation function $\hat{C}^{A_2 \, A_1}_{t_2 \, t_1}$ defined in (\ref{multitimecorpert}). This correlation function can be written as
\begin{equation}\label{twotimecorrDeltavers}
    \hat{C}^{A_2 \, A_1}_{t_2 \, t_1} = \bar{A}_1 \bar{A}_2 +
    {\rm tr} \left\{ A_2 {\cal E}_{t_2-t_1} [( A_1 - \bar{A}_1 ) \rho_{\rm eq} ] \right\} \,,
\end{equation}
where
\begin{equation}\label{Abaravdef}
    \bar{A} = {\rm tr} ( A \rho_{\rm eq} ) \,,
\end{equation}
and ${\cal E}_t$ is the evolution operator associated with the linear quantum master equation (\ref{QMEthermolowTx}) for $\Delta\rho_t$. Note that the evolution operator in (\ref{twotimecorrDeltavers}) indeed acts on a traceless operator, not on a density matrix. If we assume $\bar{A}_1=0$, the two-time correlation function takes the even simpler form
\begin{equation}\label{twotimecorrDeltavers0}
    \hat{C}^{A_2 \, A_1}_{t_2 \, t_1} =
    {\rm tr} \left[ A_2 {\cal E}_{t_2-t_1} ( A_1 \rho_{\rm eq}) \right] \,.
\end{equation}

We now construct a two-process unraveling associated with the master equation (\ref{QMEthermolowTx}) for the interacting theory. The jump rates should vanish if the jumps annihilate one of the current states, which suggests to make them proportional to the norms of the states resulting after a jump. The coupled quantum jumps are hence introduced by the following generalization of (\ref{jump2pfree}),
\begin{equation}\label{jumpinteract1}
    \left. \begin{array}{c}
       \Dket{\phi_t} \rightarrow
       \frac{\displaystyle A_l \Dket{\phi_t}\|\Dket{\phi_t}\|}{\displaystyle \|A_l \Dket{\phi_t}\|} \\
       \mbox{ } \\
       \Dket{\psi_t} \rightarrow
       \frac{\displaystyle a_{\bm{k}} \Dket{\psi_t}\|\Dket{\psi_t}\|}{\displaystyle \|a_{\bm{k}} \Dket{\psi_t}\|}
    \end{array} \right\}
    \quad \mbox{with rate} \quad \frac{\|A_l \Dket{\phi_t}\| \, \|a_{\bm{k}} \Dket{\psi_t}\|}{
    \|\Dket{\phi_t}\| \, \|\Dket{\psi_t}\|} \, \Gamma_{\bm{k} l} \,,
\end{equation}
and
\begin{equation}\label{jumpinteract2}
    \left. \begin{array}{c}
       \Dket{\phi_t} \rightarrow
       \frac{\displaystyle a_{\bm{k}} \Dket{\phi_t}\|\Dket{\phi_t}\|}{\displaystyle \|a_{\bm{k}} \Dket{\phi_t}\|} \\
       \mbox{ } \\
       \Dket{\psi_t} \rightarrow
       \frac{\displaystyle A_l \Dket{\psi_t}\|\Dket{\psi_t}\|}{\displaystyle \|A_l \Dket{\psi_t}\|}
    \end{array} \right\}
    \quad \mbox{with rate} \quad \frac{\|a_{\bm{k}} \Dket{\phi_t}\| \, \|A_l \Dket{\psi_t}\|}{
    \|\Dket{\phi_t}\| \, \|\Dket{\psi_t}\|} \, \Gamma_{\bm{k} l} \,.
\end{equation}
In order to reproduce the quantum master equation (\ref{QMEthermolowTx}), we choose the continuous evolution equations
\begin{equation}\label{Schroeirrinteract1}
    \frac{\dR}{\dR t} \Dket{\phi_t} = - \iR H \Dket{\phi_t} -
    \sum_{\bm{k} l} \Gamma_{\bm{k} l} \left( a^\dag_{\bm{k}} A_l
    - \frac{\|A_l \Dket{\phi_t}\| \, \|a_{\bm{k}} \Dket{\psi_t}\|}{
    \|\Dket{\phi_t}\| \, \|\Dket{\psi_t}\|} \right) \Dket{\phi_t} \,,
\end{equation}
and
\begin{equation}\label{Schroeirrinteract2}
    \frac{\dR}{\dR t} \Dket{\psi_t} = - \iR H \Dket{\psi_t} -
    \sum_{\bm{k} l} \Gamma_{\bm{k} l} \left( a^\dag_{\bm{k}} A_l
    - \frac{\|a_{\bm{k}} \Dket{\phi_t}\| \, \|A_l \Dket{\psi_t}\|}{
    \|\Dket{\phi_t}\| \, \|\Dket{\psi_t}\|} \right) \Dket{\psi_t} \,.
\end{equation}
Again, we need to solve two coupled nonlinear Schr\"odinger equations for the continuous evolution of $\Dket{\phi_t}$ and $\Dket{\psi_t}$. During this evolution, the norms of $\Dket{\phi_t}$ and $\Dket{\psi_t}$ change. However, when a jump of the type (\ref{jumpinteract1}) or (\ref{jumpinteract2}) occurs, the norms remain unchanged. A particularly nice feature of the unraveling (\ref{jumpinteract1})--(\ref{Schroeirrinteract2}) is that fixed multiples of $\Dket{\phi_t}$ and $\Dket{\psi_t}$ solve exactly the same evolution equations. In spite of the nonlinear character of the evolution equations, the linear nature of the underlying quantum master equation is partially retained.

Careful inspection of the jump rates in (\ref{jumpinteract1}) and (\ref{jumpinteract2}) reveals that only those rates $\Gamma_{\bm{k} l} \propto \gamma_k$ contribute for which $\Dket{\phi_t}$ or $\Dket{\psi_t}$ contains at least one particle with momentum $\bm{k}$. On the other hand, there can be a large number of possible values of $l$, in particular, when $A_l$ creates three particles. The jump rates for newly created particles with large momenta decay only weakly with $|\bm{k}|$, so that one might be concerned that the many possible jump channels associated with $A_l$ might endanger the limit of increasing the maximum momentum, $Z_L \rightarrow \infty$. However, one should realize that, according to the dissipative contribution for free particles illustrated in Figures~\ref{figfreevacdecay} and \ref{figfreevacdecayx}, high-momentum particles disappear very quickly.

\subsubsection{Various splittings and unravelings}
In Section~\ref{sectiondynamics} on dynamics, we introduced a quantum master equation consisting of reversible and irreversible contributions,
\begin{equation}\label{QMEsplit1}
    \frac{\dR\rho_t}{\dR t} = {\cal L}_{\rm rev}(\rho_t) + {\cal L}_{\rm irr}(\rho_t) \,.
\end{equation}
In Section~\ref{secperturbtheory} on perturbation theory, we made use of the alternative splitting
\begin{equation}\label{QMEsplit2}
    \frac{\dR\rho_t}{\dR t} = {\cal L}^{\rm free}(\rho_t) + {\cal L}^{\rm coll}(\rho_t) \,.
\end{equation}
For the low-temperature master equation (\ref{QMEthermolowT}), or for (\ref{QMEthermolowTi}) with $\Delta\rho_t$ instead of $\rho_t$, we can simultaneously perform both splittings and write
\begin{equation}\label{QMEsplit3}
    \frac{\dR\rho_t}{\dR t} = {\cal L}^{\rm free}_{\rm rev}(\rho_t) + {\cal L}^{\rm free}_{\rm irr}(\rho_t)
    + {\cal L}^{\rm coll}_{\rm rev}(\rho_t) + {\cal L}^{\rm coll}_{\rm irr}(\rho_t) \,,
\end{equation}
with
\begin{equation}\label{QMEsplit2a}
    {\cal L}^{\rm free}_{\rm rev}(\rho) = -\iR \Qcommu{H^{\rm free}}{\rho} \,,
\end{equation}
\begin{equation}\label{QMEsplit2b}
    {\cal L}^{\rm coll}_{\rm rev}(\rho) = -\iR \Qcommu{H^{\rm coll}}{\rho} \,,
\end{equation}
\begin{equation}\label{QMEsplit2c}
    {\cal L}^{\rm free}_{\rm irr}(\rho) =
    \sum_{\bm{k} \in K^d} \gamma_k
    \left( 2 a_{\bm{k}} \rho a^\dag_{\bm{k}} - a^\dag_{\bm{k}} a_{\bm{k}} \rho
    - \rho a^\dag_{\bm{k}} a_{\bm{k}} \right) \,,
\end{equation}
and
\begin{equation}\label{QMEsplit2d}
    {\cal L}^{\rm coll}_{\rm irr}(\rho) =
    - \sum_{\bm{k} \in K^d} \frac{\gamma_k}{\omega_k} \bigg(
    \Qcommux{a_{\bm{k}}}{\rho \Qcommu{a^\dag_{\bm{k}}}{H^{\rm coll}}}
    + \Qcommux{a^\dag_{\bm{k}}}{\Qcommu{a_{\bm{k}}}{H^{\rm coll}} \rho} \bigg) \,.
\end{equation}

Still another type of splitting is introduced by unravelings: continuous evolution versus jumps arise as a further distinguishing feature. Roughly speaking, reversible dynamics is continuous, whereas irreversible dynamics is associated with jumps. However, the modification of the Schr\"odinger equation through irreversible terms in (\ref{Schroeirrfree}) or (\ref{Schroeirrinteract1}), (\ref{Schroeirrinteract2}) shows that things are not so straightforward. One even has the freedom to treat continuous terms as jump terms. For example, in Hamiltonian dynamics, $H^{\rm free} + H^{\rm coll}$ usually acts continuously. If $H^{\rm coll}$ is small compared to $H^{\rm free}$, it might be useful to let $H^{\rm free}$ act continuously, but $H^{\rm coll}$ only in discrete jumps. More precisely, the stochastic jumps
\begin{equation}\label{jumpHcoll}
    \Dket{\psi_t} \rightarrow \Dket{\psi_t} - \frac{1}{r} \iR H^{\rm coll} \Dket{\psi_t}
    \quad \mbox{with rate } r \,,
\end{equation}
are equivalent to a continuously acting $H^{\rm coll}$ in the Schr\"odinger equation. In such jumps, one can actually choose one of the contributions to a Hamiltonian like (\ref{Hcolk}) in a probabilistic manner.

The possibilities in constructing unravelings are even richer. Instead of solving a continuous equation between jumps, we can try to give closed form expressions for the finite steps between jumps. For example, we can treat the collision effects ${\cal L}^{\rm coll}_{\rm rev}$ and, if we do not assume SID, also ${\cal L}^{\rm coll}_{\rm irr}$ through jumps and solve the free problem associated with ${\cal L}^{\rm free} = {\cal L}^{\rm free}_{\rm rev} + {\cal L}^{\rm free}_{\rm irr}$ between jumps in closed form. The discussion in Section~\ref{secunravelfreef} shows how this works (for more details see Section~\ref{secRfreeevol}). With all these possibilities of constructing unravelings one can, in particular, make sure that an unraveling consists only of complex factors times Fock base vectors. The choice of a particular unraveling may depend on practical requirements, like numerical efficiency in stochastic simulations, or on theoretical arguments, like particularly natural or elegant formulations leading to a convincing \index{Ontology}ontology. From a theoretical point of view, it is appealing to treat only the free Hamiltonian evolution as a continuous process and to realize all interactions among field quanta and with the heat bath through jumps. From a practical point of view, the ideas to be presented in Section~\ref{seccomputersim} should be useful for developing efficient computer simulations.

\subsubsection{Computer simulations}\label{seccomputersim}
Unravelings provide a convenient starting point for designing computer simulations. For calculating correlation functions in the presence of interactions, we need to simulate two coupled stochastic processes in Hilbert space. For practical purposes, it is important to avoid states that involve a large or increasing number of Fock base vectors. Ideally, the states visited during a simulation involve only one base vector, that is, the unraveling consists only of complex multiples of Fock base vectors. We have seen in Sections~\ref{secunravelfreef} and \ref{secunravelfreef2} that, for the one- and two-process unravelings of the free theory, stochastic processes restricted to multiples of base vectors arise naturally. If we treat the reversible effect of collisions according to the Schr\"odinger equation, a typical state acquires nonzero components for all base vectors. In order to avoid that, we need to treat collisions by jumps and to select one of the terms in the Hamiltonian stochastically. If we need to treat all collision effects by jumps, then it is tempting to treat the free theory in closed form, including the irreversible effects of the heat bath. In that situation, the simulation of Laplace-transformed correlation functions is particularly simple because the free irreversible dynamics happens according to a rate parameter.

For simplicity, let us consider the Laplace transform of the two-time correlation function (\ref{twotimecorrDeltavers0}),
\begin{equation}\label{twotimecorR}
    \tilde{C}^{A_2 A_1}_s =
    {\rm tr} \left[ A_2 {\cal R}_s ( A_1 \rho_{\rm eq} ) \right] \,.
\end{equation}
Equation (\ref{twotimecorR}) may be read as a recipe. We begin with $\Dket{\phi}$- and $\Dket{\psi}$-processes representing $\rho_{\rm eq}$. The operators $A_1$ and $A_2$ simply act on the $\Dket{\phi}$-process. In between, the operator ${\cal R}_s$ acts on both processes and we only need to describe how such an action can be simulated. We can do this according to the magical identity (\ref{magic}).

We choose $q$ slightly smaller than unity and select the value of $n$ with probability $(1-q) \, q^n$. A reasonable choice of $q$ in exploring a physical time scale $1/s$ with time step $1/r$ is $q=1-s/r$. We then have to apply alternating super-operators ${\cal R}^{\rm free}_{s+r}$ and $(1+{\cal L}^{\rm coll}/r)$, a total of $2n+1$ super-operators. The free evolution is usually well understood so that we can assume to have an analytical formula for applying ${\cal R}^{\rm free}_{s+r}$. For the SID approximation (\ref{QMEsplitcollSID}), the action of $(1+{\cal L}^{\rm coll}/r)$ on $\Dket{\phi_t}\Dbra{\psi_t}$ is given by
\begin{equation}\label{collactsim}
    \Dket{\phi_t}\Dbra{\psi_t} \,\, - \,\, \frac{\iR}{r} H^{\rm coll} \Dket{\phi_t}\Dbra{\psi_t}
    \,\, + \,\, \frac{\iR}{r} \Dket{\phi_t}\Dbra{\psi_t} H^{\rm coll} \,.
\end{equation}
We can choose one of the contributions to a Hamiltonian like (\ref{Hcolk}) acting on either $\Dket{\phi_t}$ or $\Dket{\psi_t}$ in a probabilistic manner.


\subsection{Summary}\label{secapproachsum}
We now have presented the basic mathematical and physical elements of our approach to quantum field theory, guided by our metaphysical postulates. As a variety of ideas and concepts have been introduced, it is worthwhile to summarize where we stand.

\subsubsection{Basic ingredients}
According to the approach proposed in this chapter, a quantum field theory is defined by the following ingredients:
\begin{enumerate}
  \item \emph{The Fock space associated with all the fundamental quantum particles or field quanta of the theory.} Each creation and annihilation operator is labeled by a countable set of momenta and, except for spin-zero particles, by a spin component. The allowed occupation numbers in the Fock base states make the distinction between bosons and fermions. Further labels come with the extra spaces associated with interactions, such as the `color' space for strong interactions.
  \item \emph{A quantum master equation for the time-evolution of a density matrix in Fock space.} This quantum master equation is given in terms of a Hamiltonian, coupling operators, a temperature, and a friction parameter. The free Hamiltonian is given by the relativistic single-particle energy-momentum relation, the interaction Hamiltonian can be written in terms of collision rules, typically involving three or four particles.
  \item \emph{A countable list of multi-time correlation functions to be predicted by the theory.} These correlation functions are chosen as the quantities of interest and should be related to experimentally accessible quantities.
\end{enumerate}

\begin{figure}
\centerline{\includegraphics[width=9cm]{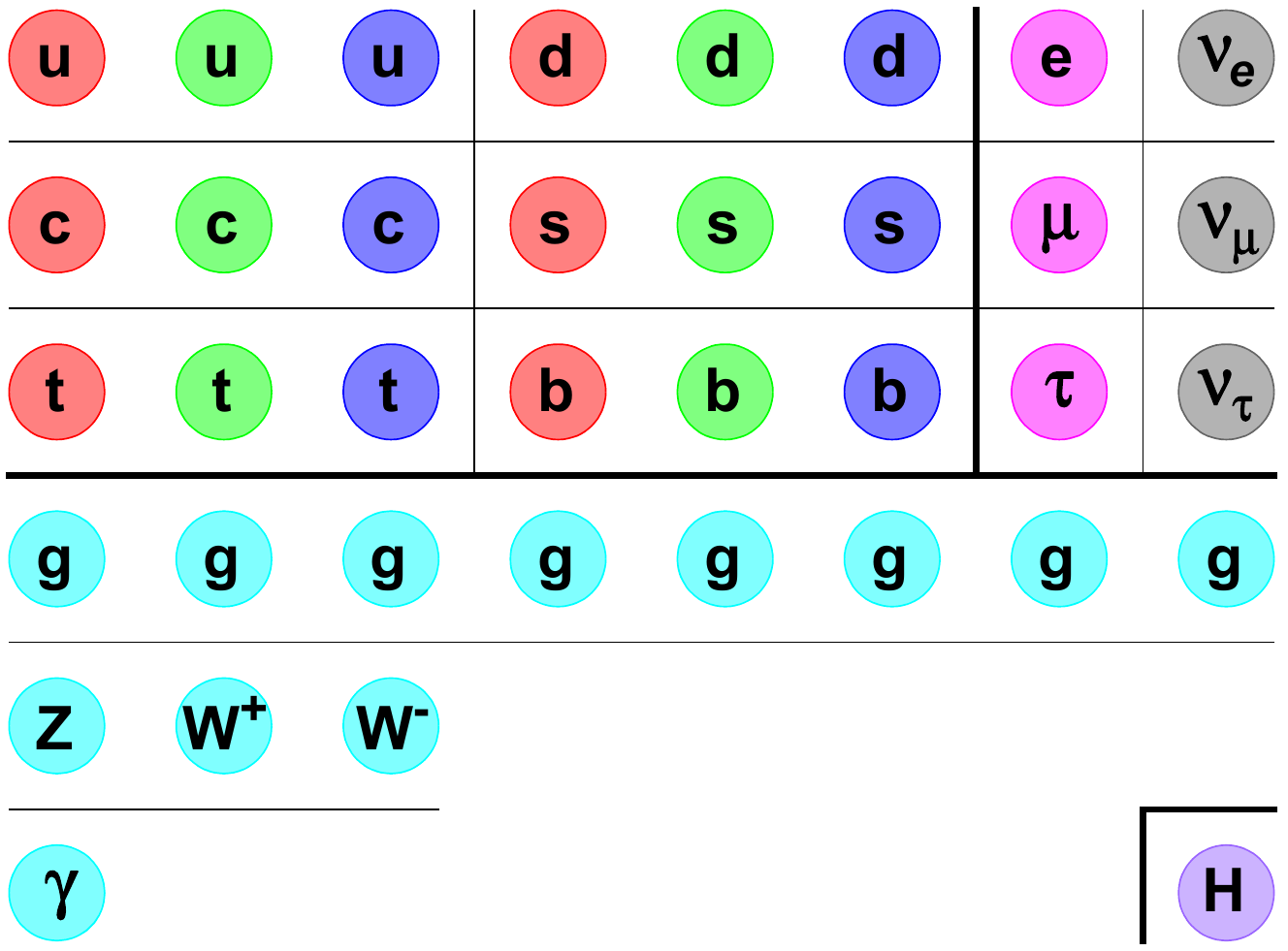}} \caption[ ]
{The zoo of fundamental particles in the standard model (see text for details).} \label{figparticlezoo}
\end{figure}

Let us now comment on these ingredients. The zoo of quantum particles to be included in the underlying Fock space is impressively large. The particle content of the standard model is summarized in Figure~\ref{figparticlezoo}. The upper half consists of fermions, the lower half of bosons. Each of the three fermion generations in the first three rows of Figure~\ref{figparticlezoo} consists of two quarks and two leptons. Each of the quarks comes in three `colors' (red, green, blue). The three quark generations are called up/down, charm/strange, and top/bottom; the three lepton generations are given by the electron, muon, and tau, each together with its partner neutrino. The neutrino masses are small, but nonzero. All the fermions in the upper half of Figure~\ref{figparticlezoo} are massive and possess spin $1/2$. In the lower half of Figure~\ref{figparticlezoo}, the gluons in eight different color combinations mediate strong interactions. The Z, $\rm W^+$, and $\rm W^-$ bosons mediate weak interactions, where $\rm W^+$ and $\rm W^-$ form a particle-antiparticle pair. The photon $\gamma$ mediates electromagnetic interactions. All these gauge bosons associated with fundamental interactions possess spin $1$. Photons and gluons are massless. Finally, the Higgs particle, which is required to give mass to the gauge bosons Z, $\rm W^+$, and $\rm W^-$, is a massive spin-zero boson. Figure~\ref{figparticlezoo} lists $37$ particles; not shown in the figure are the antiparticles of the three generations of fermions in the first three lines, so that the total number of fundamental particles according to the standard model is $37+24=61$. The list of fundamental particles should be understood as an \index{Ontological commitment}ontological commitment to be made in our image of nature. Organizing principles for such a large zoo of particles would clearly be desirable. In the toy example of $\varphi^4$ theory, we deal with just one kind of spin-zero bosons and we hence refer to it as scalar field theory.

The quantum master equation for the density matrix reflects the irreversible character of quantum field theories resulting from the elimination of fast small-scale degrees of freedom which, in our thermodynamic approach, are represented by a heat bath. Quantum field theories hence are of statistical nature and, consequently, they are used to evaluate multi-time correlation functions. The dissipation mechanism is formulated such that a nontrivial interplay between the free and interacting theories results, thus allowing us to group free particles into unresolvable clouds. Interacting particles have a very different status than free particles. The introduction of interacting particles depends on the dissipation mechanism, which naturally comes with renormalization and an underlying self-similarity. Stochastic unravelings offer an alternative representation of quantum field theories.

\subsubsection{Limits}
To keep our mathematical image of quantum field theory rigorous, we stay away from actual infinities in favor of potential infinities associated with limiting procedures (see p.\,\pageref{limitthermo} for an overview over the various limits). In order to obtain a countable set of momentum states, we consider a finite system volume that provides a low-energy (infrared) cutoff. The dissipative coupling to the heat bath provides a high-energy (ultraviolet) cutoff. A further ultraviolet cutoff that keeps the set of momentum states finite is required for the sake of mathematical rigor of intermediate calculations, but it is expected to be irrelevant in the final results because the dissipative coupling provides sufficient ultraviolet regularization, at least for dynamic properties. We hence have to perform \emph{two fundamental limits} in the end of all calculations: (i) The limit of infinite system volume $V$, in which the momentum states become dense in the continuum and sums become integrals. (ii) The limit of vanishing friction parameter $\gamma$. We call a quantum field theory for a countable set of quantities of interest $\cal C$ well defined if, for a properly chosen dependence of the model parameters on $V$ and $\gamma$, finite limits $V \rightarrow \infty$ and $\gamma \rightarrow 0$ exist for all quantities of interest in $\cal C$ and moreover possess all the required symmetries of the system. The remaining parameters in the limit are the physical parameters of the theory. Ideally, the order of the limits $V \rightarrow \infty$ and $\gamma \rightarrow 0$ would be irrelevant. In analytical calculations it is often convenient to perform the limit $V \rightarrow \infty$ first because integrals are easier to handle than sums. More subtle would be the case in which both limits have to be performed simultaneously. The limiting procedure needs to be analyzed thoroughly and is related to the usual renormalization program. The most reliable results are obtained if all calculations are performed at finite temperature and the zero-temperature limit is performed in the very end.

In principle, $V$ should be smaller than the volume of the universe and $\gamma^{1/3}$ should be larger than the Planck length. If a theory is formulated for these extreme values, all model parameters matter and all the required symmetries have to be established without any limiting procedures. These considerations might actually be relevant if gravity is to be included into quantum field theory. Even for effective quantum field theories, a relativistically covariant formulation of the dissipation mechanism would be appealing.

It should be noted that we make all efforts to keep the present approach mathematically rigorous although, in many aspects, it is close to Lagrangian field theory in the Hamiltonian formulation. We thus make a serious attempt to combine usefulness and rigor. There is no artificial elimination of divergencies. The entire procedure is motivated by a philosophical \emph{horror infinitatis}, which suggests to allow only for countable infinities and to rely on limiting procedures. Although these limiting procedures are related to the renormalization program, renormalization is not used as a trick to eliminate divergent integrals but rather as a systematic procedure to recognize well-defined limit theories and to refine perturbation expansions.

\subsubsection{Is there a measurement problem?}
We have seen in Section~\ref{secmeasureprob} that it is not so easy to explain what the measurement problem really is. Therefore, it is not so easy to say whether the dissipative approach to quantum field theory suffers from the measurement problem.

Dissipative quantum field theory, which is based on density matrices, suggests to consider correlation functions describing large ensembles of events. The most challenging form of the quantum measurement problem hence does not arise in this approach to fundamental particle physics. According to Margenau, the use of density matrices even has the potential to avoid problems with the description of the measurement process, which should not be discussed in terms of pure states or wave functions (see Sections 18.6 and 18.7 of \cite{Margenau}).

We actually do not expect any of the three of the statements formulated by Maudlin to specify the measurement problem (see p.\,\pageref{statementsmeasprob}) to be true: we use density matrices instead of wave functions, we rely on the potentially nonlinear thermodynamic quantum master equation, and all our quantities of interest are statistical in nature. There is no collapse of the wave function and no projection of the density matrix. In writing down the correlation functions (\ref{observableform2}), we introduce a very general form consistent with the quantum master equation, which can actually be interpreted in terms of imperfect measurements. Such imperfect measurements have also been considered by Kronz \cite{Kronz91} in his attempt to defend the projection postulate against various forms of criticism raised by Margenau.

The idea of unravelings allows us to argue why these correlation functions make perfect sense even in the absence of any observer performing measurements. In the correlation functions (\ref{observableform2}), we multiply a density matrix with $A_j$ from the left and with $A_j^\dag$ from the right. This corresponds exactly to the jumps of the type (\ref{jumpfree}) in a one-process unraveling. The jumps expressing irreversible behavior provide an intrinsic motivation to look at our correlation functions describing the correlation between such jumps. These operators are introduced by random fluctuations rather than by an experimenter.

The fact that the measurement problem does not seem to occur in our approach to quantum field theory, which nicely works without observers, by no means excludes or solves the measurement problem in other branches of quantum theory. For example, a much deeper understanding of the problem is certainly required in quantum information theory \cite{WhitakerA}. For that purpose, the knowledge interpretation, also known as \emph{epistemic interpretation}, is an attractive option. It is based on the idea that the wave function does not represent physical \index{Reality}reality, but our knowledge about physical reality. A detailed description, thorough interpretations, and thoughtful explanations of how various problems of quantum theory are resolved within the epistemic approach can be found in a fascinating book by Friederich \cite{Friederich}.

By using momentum eigenstates for the construction of the basic Fock space, our approach lacks any spatial resolution and hence does not directly suffer from issues with local reality and causality (as usually associated with the thought experiment of Einstein, Podolsky and Rosen \cite{EPR35} and the inequalities of Bell \cite{Bell64,Bell66}; see, for example, \cite{WhitakerA}). For free particles, the Fourier representation (\ref{phiexpression}) could be used to construct spatial dependencies, but this equation is not considered to be part of our mathematical image (see also the remarks in the last paragraph of Section \ref{secFourierfields}). In the presence of interactions, it seems to be impossible to achieve spatial resolution. This problem can be understood intuitively through the clouds of particles occurring in the interacting theory (see Section \ref{secpartfield} and also p.\,244 of \cite{Hegerfeldt98ip}).

\subsubsection{Remarks on historical roots}\label{sechistroots}
The intuitive foundations of quantum field theory have been laid in the years 1926 through 1932, right after the advent of quantum mechanics. Many key contributions have been made by Pascual Jordan (1902-1980) after his dissertation at the University of G\"ottingen (his adviser was Max Born). An enlightening, detailed discussion of the early papers on quantum field theory, which appeared mostly in the \emph{Zeitschrift f\"ur Physik} (in German), can be found in Chapters~1 and 2 of the textbook by Duncan \cite{Duncan}. Duncan emphasizes that Jordan's work ``pointed the way to a general procedure for extending the principles of quantum mechanics to field systems with infinitely many degrees of freedom'' (see p.\,27 of \cite{Duncan}) and ``Jordan was an early champion (\ldots) of the notion that wave-particle duality extended to a coherence in the mathematical formalisms used to describe radiation (specifically, the electromagnetic field) and matter'' (see p.\,40 of \cite{Duncan}).

Most famous is the so-called \emph{Dreim\"annerarbeit} (three-men work) published by Born, Heisenberg and Jordan in 1926 \cite{BornHeisJordan26}, which is generally considered as the first publication on quantum field theory. The last part of that work contains the quantization of the free electromagnetic field developed by Jordan (both Heisenberg and Born later expressed doubts about Jordan's calculations). The harmonic oscillators representing the modes of an electromagnetic field are treated in terms of position and momentum operators. The correct result for the energy fluctuations in blackbody radiation is obtained, where the proper calculation of interference effects takes some five pages. The wave-particle duality of photons is nicely cast into an elegant mathematical form.

An interaction between electromagnetic fields and electrons was introduced by Dirac \cite{Dirac27} in 1927, still within the non-relativistic quantum formalism. For that purpose, he uses creation and annihilation operators for the photons, but not for the electrons. In his approach, the number of photons is conserved because he assumes the Hamiltonian to be bilinear in the creation and annihilation operators; therefore, transitions between photon states of positive energy (physical) and states of zero energy (unphysical) are conveniently interpreted as creation and annihilation processes. As a highlight, Dirac's theory leads to Einstein's laws for the emission and absorption of radiation.

Also in 1927, Jordan and Klein \cite{JordanKlein27} introduced the `quantization of de Broglie waves', which is now known as \index{Second quantization}second quantization. Also the idea of normal ordering was introduced by these authors and, therefore, has frequently been referred to as the `Klein-Jordan trick'. The proper anticommutation relations for fermion creation and annihilation operators have been introduced by Jordan and Wigner \cite{JordanWigner28} in 1928. In 1929, the first relativistic formulation of electromagnetic fields interacting with electrons, based on \index{Second quantization}second quantization of the Dirac equation for the electron, was offered by Heisenberg and Pauli \cite{HeisenbgPauli29}; in this paper, also the formulation of Lagrangian quantum field theory was established.

A fully satisfactory treatment of electrons and positrons became possible with the introduction of Fock space \cite{Fock32} in 1932. By that time, a both intuitive and elegant formulation of quantum field theory was available. However, the proper handling of the many divergent expressions occurring in concrete calculations was not properly understood before the late 1940s.

\chapter{Scalar field theory}\label{chapphi4}
The development of the mathematical and physical elements in Section~\ref{secMPelements} has repeatedly been illustrated by the example of scalar field theory with quartic interactions or, shorter, $\varphi^4$ theory. Now it's high time to tell a more complete and coherent story of $\varphi^4$ theory. For our first discussion of a quantum field theory, we completely rely on perturbation theory. Whereas our approach is not particularly efficient for constructing perturbation expansions (unless supported by software for symbolic mathematical computation), such expansions provide the most intuitive and concrete results for illustration.

We first calculate a two-time, two-particle correlation function, known as the propagator, and discuss the limits of large system volume and small friction parameter to realize the connection to the renormalization program. We then calculate a two-time, four-particle correlation function containing the interaction vertex in order to find the critical coupling constant of $\varphi^4$ theory. Various terms appearing in perturbation expansions are illustrated by means of Feynman diagrams. All our calculations in this chapter are carried out in $d$ space dimensions. We show the equivalence of our results with those of Lagrangian field theory, which are obtained in $d+1$ dimensions.

The example of $\varphi^4$ theory is of conceptual importance because, in $d=1$ and $d=2$ space dimensions, the viability of axiomatic quantum field theory can be demonstrated by a constructive approach. Glimm and Jaffe \cite{GlimmJaffeII} have constructed the corresponding Heisenberg fields satisfying all axioms of quantum field theory in the limit of continuous space-time; in particular, these fields are associated with a nontrivial, interacting theory and exhibit the desired locality properties.

In the previous chapter, scalar $\varphi^4$ theory has frequently been used for illustration. It is based on only one kind of bosonic field quanta associated with the creation operators $a^\dag_{\bm{k}}$. The free Hamiltonian for massive quanta is given by (\ref{Hfree}) with the relativistic energy-momentum relation (\ref{relenergmomrel}), and the quartic interactions are described by the Hamiltonian (\ref{Hcolk}).

\section{Some basic equations}\label{secphi4basics}
We begin with a detailed discussion of the dissipative evolution of the free scalar field theory and introduce a convenient simplification. We then present the second-order perturbation expansion for a general Laplace-transformed two-time correlation function.

\subsection{Free dissipative evolution}\label{secRfreeevol}
For the explicit construction of the perturbation expansion (\ref{EperturbexpansionR}), we still need to find a closed-form expression for the super-operator ${\cal R}^{\rm free}_s$. The definition (\ref{Rfreesupopdef}) implies
\begin{equation}\label{Rfreesupopdefx}
    s \, {\cal R}^{\rm free}_s(\rho) = \int_0^\infty {\cal E}^{\rm free}_t(\rho)
    \left( - \frac{\dR}{\dR t}\right) \eR^{- s t} \, \dR t \,.
\end{equation}
Noting that $\rho_t={\cal E}^{\rm free}_t(\rho)$ by definition satisfies the free evolution equation (\ref{QMEthermolowTfree}) and defining the right-hand-side of that equation as ${\cal L}^{\rm free} \rho_t$, an integration by parts yields
\begin{equation}\label{Rsidentityfree}
    s \, {\cal R}^{\rm free}_s(\rho) - {\cal L}^{\rm free}\big({\cal R}^{\rm free}_s(\rho)\big) = \rho \,,
\end{equation}
which is nothing but the formal super-operator identity (\ref{Rsupopformal}), here written for the free theory,
\begin{equation}\label{Rsupopformalfree}
    {\cal R}^{\rm free}_s = (s-{\cal L}^{\rm free})^{-1} \,.
\end{equation}
Note that, with this identity, the domain of the super-operator ${\cal R}^{\rm free}_s$ can be extended from density matrices to arbitrary operators $X$.

As a next step, we split ${\cal L}^{\rm free}$, as occurring on the right-hand-side of (\ref{QMEthermolowTfree}), into ${\cal L}^{\rm free}_0$ and ${\cal L}^{\rm free}_-$ by defining
\begin{equation}\label{Lfreesplit1}
    {\cal L}^{\rm free}_0 X = -\iR \Qcommu{H^{\rm free}}{X} - \sum_{\bm{k} \in K^d} \gamma_k
    \left( a^\dag_{\bm{k}} a_{\bm{k}} X + X a^\dag_{\bm{k}} a_{\bm{k}} \right) \,,
\end{equation}
and
\begin{equation}\label{Lfreesplit2}
    {\cal L}^{\rm free}_- X = \sum_{\bm{k} \in K^d} 2 \gamma_k a_{\bm{k}} X a^\dag_{\bm{k}} \,.
\end{equation}
Equation (\ref{Rsupopformalfree}) can then be rewritten as
\begin{eqnarray}
    {\cal R}^{\rm free}_s &=& (s-{\cal L}^{\rm free}_0)^{-1}
    + (s-{\cal L}^{\rm free}_0)^{-1} {\cal L}^{\rm free}_- (s-{\cal L}^{\rm free}_0)^{-1}
    \nonumber \\
    &+& (s-{\cal L}^{\rm free}_0)^{-1} {\cal L}^{\rm free}_- (s-{\cal L}^{\rm free}_0)^{-1}
    {\cal L}^{\rm free}_- (s-{\cal L}^{\rm free}_0)^{-1} + \ldots \,.
\label{Rsupopformalfreeexp}
\end{eqnarray}

This latter identity provides a convenient starting point for evaluating the super-operator ${\cal R}^{\rm free}_s$. Our goal here is to evaluate ${\cal R}^{\rm free}_s(X)$ for operators of the general form
\begin{equation}\label{Xopform}
    X = a^\dag_{\bm{k}_1} \ldots a^\dag_{\bm{k}_n} \Dket{0} \Dbra{0}
    a_{\bm{k}'_{n'}} \ldots a_{\bm{k}'_1} \,,
\end{equation}
that is, for dyadic products of Fock base vectors. These operators $X$ form a basis of the space of linear operators on Fock space. The definition (\ref{Lfreesplit1}) is motivated by the fact that any $X$ of the form (\ref{Xopform}) is an eigenstate of the super-operator ${\cal L}^{\rm free}_0$,
\begin{equation}\label{LfreeXop1}
    {\cal L}^{\rm free}_0 X = \iR (\breve{\omega}_{k'_1}^* + \ldots + \breve{\omega}_{k'_{n'}}^*
    - \breve{\omega}_{k_1} - \ldots - \breve{\omega}_{k_n} ) X \,,
\end{equation}
so that $(s-{\cal L}^{\rm free}_0)^{-1}$ can be obtained in terms of eigenvalues. To obtain the compact formula (\ref{LfreeXop1}), we have used the definition
\begin{equation}\label{breveomdef}
    \breve{\omega}_k = \omega_k-\iR\gamma_k \,.
\end{equation}

From (\ref{Lfreesplit2}) we further obtain
\begin{equation}\label{LfreeXop2}
    {\cal L}^{\rm free}_- X =
    \sum_{j=1}^n \sum_{j'=1}^{n'} 2 \gamma_{k_j} \delta_{\bm{k}_j \bm{k}'_{j'}} X_{jj'} \,,\quad
\end{equation}
where $X_{jj'}$ has been introduced as a shorthand for the operator that arises from $X$ by removing the operators $a^\dag_{\bm{k}_j}$ and $a_{\bm{k}'_{j'}}$ from the sequences of creation and annihilation operators in (\ref{Xopform}), respectively. This observation reflects the jumps toward the vacuum state in Figures \ref{figfreevacdecay} and \ref{figfreevacdecayx} and motivates the subscript `$-$' in ${\cal L}^{\rm free}_-$ (${\cal L}^{\rm free}_0$ leaves the number of creation and annihilation operators unchanged). Note that a nonzero double-sum contribution in (\ref{LfreeXop2}) arises only if the same momentum vector occurs among both the creation and the annihilation operators. We now have provided a complete and convenient characterization of the free super-operator ${\cal R}^{\rm free}_s$.

As no matching pairs of momentum vectors can occur, we immediately find the simple special cases
\begin{equation}\label{Lfreeallcre}
    {\cal R}_s^{\rm free} \Big( a^\dag_{\bm{k}_1} \ldots a^\dag_{\bm{k}_n} \Dket{0} \Dbra{0} \Big) =
    \frac{1}{s+\iR(\breve{\omega}_{k_1}+\ldots+\breve{\omega}_{k_n})}
    a^\dag_{\bm{k}_1} \ldots a^\dag_{\bm{k}_n} \Dket{0} \Dbra{0} \,,
\end{equation}
\begin{equation}\label{Lfreeallann}
    {\cal R}_s^{\rm free} \Big( \Dket{0} \Dbra{0} a_{\bm{k}'_{n'}} \ldots a_{\bm{k}'_1} \Big) =
    \frac{1}{s-\iR(\breve{\omega}_{k'_1}^*+\ldots+\breve{\omega}_{k'_{n'}}^*)}
    \Dket{0} \Dbra{0} a_{\bm{k}'_{n'}} \ldots a_{\bm{k}'_1} \,,
\end{equation}
with the important particular examples of vacuum states,
\begin{equation}\label{Rsexample0}
    {\cal R}_s^{\rm free} \Big( \Dket{0} \Dbra{0} \Big) = \frac{1}{s} \Dket{0} \Dbra{0} \,,
\end{equation}
and single-particle states,
\begin{equation}\label{Rsexample1}
    {\cal R}_s^{\rm free} \Big( a^\dag_{\bm{k}} \Dket{0} \Dbra{0} \Big) =
    \frac{1}{s+\iR\breve{\omega}_k} a^\dag_{\bm{k}} \Dket{0} \Dbra{0} \,.
\end{equation}
A more general example is given by
\begin{eqnarray}
    {\cal R}_s^{\rm free} \Big( a^\dag_{\bm{k}} \Dket{0} \Dbra{0} a_{\bm{k}'_1} a_{\bm{k}'_2} \Big) &=&
    \frac{1}{s+\iR(\breve{\omega}_k-\breve{\omega}_{k'_1}^*-\breve{\omega}_{k'_2}^*)}
    \Bigg( a^\dag_{\bm{k}} \Dket{0} \Dbra{0} a_{\bm{k}'_1} a_{\bm{k}'_2} \nonumber \\
    && \hspace{-5em} + \, \frac{2 \gamma_k \delta_{\bm{k} \bm{k}'_2}}{s-\iR\breve{\omega}_{k'_1}^*}
    \Dket{0} \Dbra{0} a_{\bm{k}'_1}
    + \frac{2 \gamma_k \delta_{\bm{k} \bm{k}'_1}}{s-\iR\breve{\omega}_{k'_2}^*}
    \Dket{0} \Dbra{0} a_{\bm{k}'_2} \Bigg) \,.
\label{Rsexample2}
\end{eqnarray}

For any given $X$, the series (\ref{Rsupopformalfreeexp}) has only a finite number of terms, at most ${\rm min}(n,n')$. In the limit $\gamma_k \rightarrow 0$, the terms with fewer creation and annihilation operators should not matter because $\gamma_k$ appears as a factor in the numerator. For the purpose of developing perturbation theory, we hence make the simplifying assumption
\begin{equation}\label{Rfreesimplified}
    {\cal R}^{\rm free}_s(X) = (s-{\cal L}^{\rm free}_0)^{-1} X =
    \frac{X}{s+\iR (\breve{\omega}_{k_1} + \ldots + \breve{\omega}_{k_n}
    - \breve{\omega}_{k'_1}^* - \ldots - \breve{\omega}_{k'_{n'}}^* )} \,,
\end{equation}
where $X$ is assumed to be of the form (\ref{Xopform}). The regularizing effect of dissipation is still contained in the modified frequencies $\breve{\omega}_k = \omega_k-i\gamma_k$ and their complex conjugates in the denominator of (\ref{Rfreesimplified}). If such denominators appear outside sums over momentum vectors, in view of the final limit $\gamma_k \rightarrow 0$, we can safely replace the complex frequency $\breve{\omega}_k$ by the real frequency $\omega_k$. The approximation (\ref{Rfreesimplified}) is fully in line with the SID assumption proposed in Section~\ref{secSID}.

\subsection{Perturbation theory}
We would like to discuss $\varphi^4$ theory on the basis of the zero-temperature quantum master equation (\ref{QMEthermolowTi}). As this is a linear equation for the deviation from the equilibrium density matrix, it is convenient to consider the Laplace-transformed two-time correlation function (\ref{twotimecorR}) of the operators $A_1$ and $A_2$,
\begin{equation}\label{twotimecorRrep}
    \tilde{C}^{A_2 A_1}_s =
    {\rm tr} \left[ A_2 {\cal R}_s ( A_1 \rho_{\rm eq} ) \right] \,,
\end{equation}
where ${\rm tr} ( A_1 \rho_{\rm eq} ) = 0$. For the evolution super-operator ${\cal R}_s$ occurring in this correlation function, we have the perturbation expansion (\ref{EperturbexpansionR}) in the simplified composed super-operator notation
\begin{equation}\label{Reexpcompact}
    {\cal R}_s = {\cal R}^{\rm free}_s
    + {\cal R}^{\rm free}_s {\cal L}^{\rm coll} {\cal R}^{\rm free}_s
    + {\cal R}^{\rm free}_s {\cal L}^{\rm coll} {\cal R}^{\rm free}_s {\cal L}^{\rm coll} {\cal R}^{\rm free}_s
    + \ldots \,,
\end{equation}
where, within the SID assumption, ${\cal L}^{\rm coll}$ is given by (\ref{QMEsplitcollSID}) and ${\cal R}^{\rm free}_s$ by (\ref{Rfreesimplified}). The perturbation expansion for the zero-temperature equilibrium density matrix is obtained from (\ref{groundstatecond2sol}),
\begin{eqnarray}
    \rho_{\rm eq} &=& \Dket{0} \Dbra{0}
    - R^{\rm free}_0 H^{\rm coll} \Dket{0} \Dbra{0}
    - \Dket{0} \Dbra{0} H^{\rm coll} R^{\rm free}_0 \nonumber\\
    &+& \left(R^{\rm free}_0 H^{\rm coll}\right)^2 \Dket{0} \Dbra{0}
    + \Dket{0} \Dbra{0} \left(H^{\rm coll} R^{\rm free}_0\right)^2 \nonumber\\
    &+& R^{\rm free}_0 H^{\rm coll} \Dket{0} \Dbra{0} H^{\rm coll} R^{\rm free}_0 \nonumber\\
    &-& \Dket{0} \Dbra{0} \, \Dbra{0} H^{\rm coll} (R^{\rm free}_0)^2 H^{\rm coll} \Dket{0}
    + \ldots \,,
\label{rhoeqT0perturb}
\end{eqnarray}
where we have introduced the self-adjoint operator $R^{\rm free}_0 = (H^{\rm free})^{-1} P_0 = P_0 (H^{\rm free})^{-1} = P_0 (H^{\rm free})^{-1} P_0$ in terms of the free Hamiltonian and the projector suppressing the ground state. By putting these results together, we obtain the second-order perturbation expansion for the correlation function (\ref{twotimecorRrep}),
\begin{eqnarray}
    \tilde{C}^{A_2 A_1}_s &=& {\rm tr}
    \bigg\{ A_2 {\cal R}_s^{\rm free} \Big[ A_1 \Dket{0} \Dbra{0}
    + {\cal L}^{\rm coll} {\cal R}_s^{\rm free} \big( A_1 \Dket{0} \Dbra{0} \big)
    \nonumber \\
    &-& A_1 R^{\rm free}_0 H^{\rm coll} \Dket{0} \Dbra{0}
    - A_1 \Dket{0} \Dbra{0} H^{\rm coll} R^{\rm free}_0
    \nonumber \\
    &+& A_1 \left(R^{\rm free}_0 H^{\rm coll}\right)^2 \Dket{0} \Dbra{0}
    + A_1 \Dket{0} \Dbra{0} \left(H^{\rm coll} R^{\rm free}_0\right)^2
    \nonumber \\
    &+& A_1 R^{\rm free}_0 H^{\rm coll} \Dket{0} \Dbra{0} H^{\rm coll} R^{\rm free}_0
    - A_1 \Dket{0} \Dbra{0} \, \underline{\Dbra{0} H^{\rm coll} (R^{\rm free}_0)^2 H^{\rm coll} \Dket{0}}
    \nonumber \\
    &-& {\cal L}^{\rm coll} {\cal R}_s^{\rm free}
    \big(A_1 R^{\rm free}_0 H^{\rm coll} \Dket{0} \Dbra{0}
    + A_1 \Dket{0} \Dbra{0} H^{\rm coll} R^{\rm free}_0\big) \Big] \bigg\}
    \nonumber\\
    &+& \left({\cal L}^{\rm coll} {\cal R}_s^{\rm free}\right)^2 \big( A_1 \Dket{0} \Dbra{0} \big) \,.
\label{twotimecorperturb}
\end{eqnarray}
The underlined factor in this expansion can be evaluated after inserting (\ref{Hcolk}),
\begin{eqnarray}
    \Dbra{0} H^{\rm coll} (R^{\rm free}_0)^2 H^{\rm coll} \Dket{0} &=&
    \frac{{\lambda'}^2}{8} \sum_{\bm{k} \in K^d} \frac{1}{\omega_k^4}
    \nonumber\\
    && \hspace{-9em} + \,\, \frac{\lambda^2}{384} \, \frac{1}{V^2}
    \sum_{\bm{k}_1,\bm{k}_2,\bm{k}_3,\bm{k}_4 \in K^d}
    \frac{\delta_{\bm{k}_1+\bm{k}_2+\bm{k}_3+\bm{k}_4 , \bm{0}}}{
    \omega_{k_1}\omega_{k_2}\omega_{k_3}\omega_{k_4}}
    \frac{1}{(\omega_{k_1}+\omega_{k_2}+\omega_{k_3}+\omega_{k_4})^2} \,.
    \nonumber \\
    &&
\label{Hcoll2ndordfact}
\end{eqnarray}

\section{Propagator}\label{secphi4prop}
Let us now focus on the Laplace-transformed two-time correlation function (\ref{twotimecorRrep}) for the Fourier components of the field (\ref{phiFourier}), that is, for $A_1 = \varphi_{\bm{k}}$ and $A_2 = \varphi_{-\bm{k}}$,
\begin{equation}\label{2corrphi4defs}
    \tilde{\Delta}_{s \, \bm{k}} = \tilde{C}^{\varphi_{-\bm{k}} \, \varphi_{\bm{k}}}_{s}  \,,
\end{equation}
where, for symmetry reasons, we have ${\rm tr} ( \varphi_{\bm{k}} \rho_{\rm eq} ) = 0$. This correlation function is known as the propagator; it plays a fundamental role in quantum field theory. In principle, we should have introduced the physical propagator in terms of the operators $A_1 = \Phi_{\bm{k}}$ and $A_2 = \Phi_{-\bm{k}}$ defined in (\ref{phiFourierZ}) but, in view of the simple relationship
\begin{equation}\label{2corrphi4defsT}
    \tilde{C}^{\Phi_{-\bm{k}} \, \Phi_{\bm{k}}}_{s} =
    Z \, \tilde{C}^{\varphi_{-\bm{k}} \, \varphi_{\bm{k}}}_{s}  \,,
\end{equation}
it is more convenient to introduce a factor $Z$ translating between free particles and clouds at a later stage.

\subsection{Second-order perturbation expansion}\label{secphi4proppert2}
We can now evaluate the second-order perturbation expansion (\ref{twotimecorperturb}) for the propagator. The zeroth-order term, which is the very first term on the right-hand side of (\ref{twotimecorperturb}), can be evaluated directly by means of (\ref{Rsexample1}). All the occurring traces can be evaluated by repeated use of the commutation relations for creation and annihilation operators and the fact that the outcome of any annihilation operator acting on the free vacuum state vanishes. In the terms of first order in ${\cal L}^{\rm coll}$ or $H^{\rm coll}$, only the terms proportional to $\lambda'$ contribute. To this order we obtain
\begin{equation}\label{prop1storder}
    \tilde{\Delta}_{s \, \bm{k}} = \frac{1}{2 \omega_k} \, \frac{1}{s+\iR\breve{\omega}_k}
    \left\{ 1 - \frac{\iR \lambda'}{\omega_k} \left[ \frac{1}{s+\iR\breve{\omega}_k}
    + \frac{1}{\iR \omega_k}
    + \underline{\frac{(\breve{\omega}_k^*/\omega_k)-1}{s+\iR(\breve{\omega}_k-2\breve{\omega}_k^*)}}
    \right] \right\} \,.
\end{equation}

\begin{figure}[t]
\centerline{\includegraphics[width=10cm]{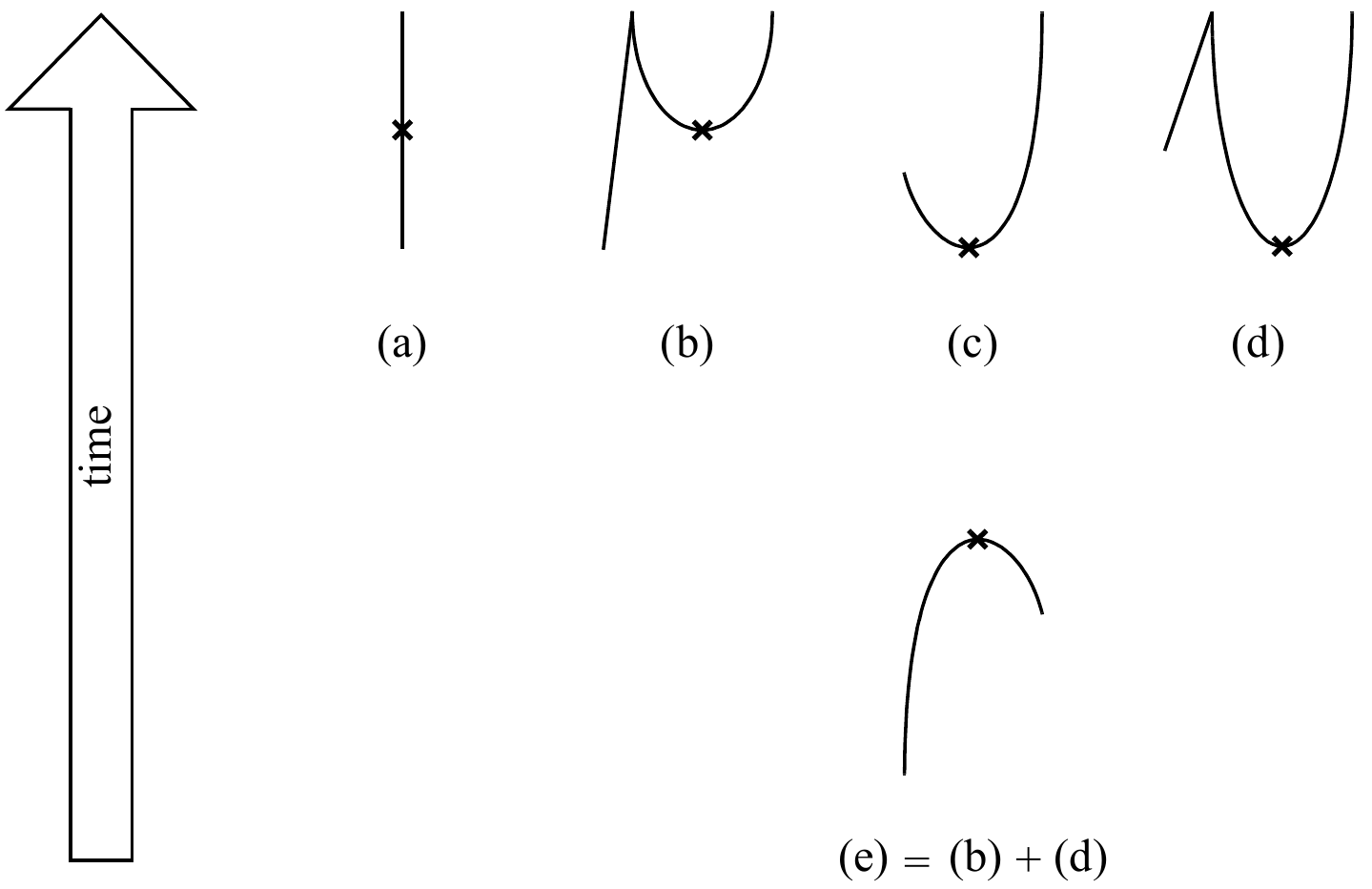}} \caption[ ]
{Feynman diagrams contributing to the propagator (\ref{prop1storder}) in first-order perturbation theory.} \label{figfeynmandia1}
\end{figure}

It is strongly recommended to check how the various terms in (\ref{prop1storder}) arise because the same kind of calculations occurs in perturbation theory over and over again. In constructing such a perturbation expansion, one can organize the various terms according to particular series of collisions with free evolution between collision events. It is helpful to represent each of these terms by a Feynman diagram. Evaluation of the first-order terms in the first two lines of (\ref{twotimecorperturb}) leads to the diagrams (a)--(d) in Figure~\ref{figfeynmandia1} (actually, both (a) and (b) result from the first-order term in the first line of (\ref{twotimecorperturb}), whereas (c) and (d) result from the two terms in the second line of (\ref{twotimecorperturb}), respectively). By definition, the event $A_1 = \varphi_{\bm{k}}$ always precedes the event $A_2 = \varphi_{-\bm{k}}$. In the diagram (a), a two-particle interaction occurs between $A_1$ and $A_2$ (a two-particle interaction is indicated by a cross in a Feynman diagram); in the diagram (c), an interaction occurs prior to $A_1$, an interaction required for reaching the equilibrium state. The structure of the general expression (\ref{twotimecorRrep}) for a two-time correlation function precludes an interaction after $A_2$, as $A_2$ is always the last event before performing the trace. Such an interaction after $A_2$, however, is mimicked by the diagrams (b) and (d) that contain cusps, suggesting reflections of parts of these diagrams. Indeed, if the contributions of the cusp diagrams (b) and (d) are added, the result can be represented by the Feynman diagram (e) in Figure~\ref{figfeynmandia1}. This combined diagram (e) leads to the same contribution as (c), except for a term that vanishes under the SID assumption introduced in Section~\ref{secSID} and (\ref{Rfreesimplified}) (which is the underlined term in (\ref{prop1storder})). The diagrams (a), (c), and (e) provide a nice representation of the first-order perturbation expansion for the propagator. Between the events $A_1$ and $A_2$, evolution is described by ${\cal R}_s^{\rm free}$, otherwise by $R^{\rm free}_0$.

\begin{figure}[t]
\centerline{\includegraphics[width=4cm]{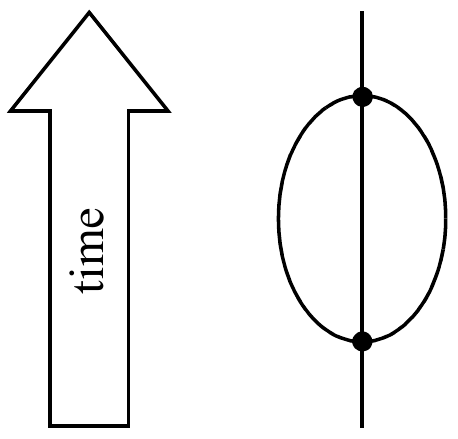}} \caption[ ]
{Topology of the connected Feynman diagrams contributing to the propagator in second-order perturbation theory.} \label{figfeynmandia2s}
\end{figure}

It turns out that Feynman diagrams are not particularly useful to guide or simplify perturbative calculations in the present setting based on a quantum master equation. The important role played by time and the temporal sequence of events makes the story of Feynman diagrams rather complicated. In the unified treatment of space and time in many standard textbooks on quantum field theory, much fewer diagrams are needed. Even better, factors are associated with lines rather than the evolution between two events involving several lines or segments of lines. The trick of reflecting parts of Feynman diagrams (see Figure~\ref{figfeynmandia1}) does no longer work in second-order perturbation theory. As the calculations follow simple systematic rules, it is much more convenient to base perturbation theory on symbolic computation. Learning how to write a proper symbolic code nowadays is probably more important than how to compute with Feynman diagrams. The distinction between connected and unconnected diagrams as well as the more detailed topological classification of diagrams, however, remain useful. For example, all connected second-order Feynman diagrams are topologically equivalent to the one shown in Figure~\ref{figfeynmandia2s}.

We are now ready to proceed to the second-order contributions to the propagator. We only need to consider the four-particle interactions proportional to $\lambda$ in the Hamiltonian (\ref{Hcolk}), because the terms proportional to $\lambda'$ and $\lambda''$ contribute only to higher-order terms. For $\lambda''$, this is obvious from the explicit formula (\ref{groundstatecondX}). For $\lambda'$, one should realize that there are no first-order contributions in $\lambda$ in (\ref{prop1storder}) and that the role of the $\lambda'$ term in (\ref{prop1storder}) is to compensate terms that would spoil convergence in the final limiting procedures; $\lambda'$ must hence be at least of order  $\lambda^2$. After a lengthy calculation and making the SID assumption introduced in Section~\ref{secSID} and (\ref{Rfreesimplified}), we find the following second-order contributions to the propagator,
\begin{eqnarray}
    && \hspace{-2em} - \frac{\lambda^2}{192 V^2 \omega_k^2} \sum_{\bm{k}_1,\bm{k}_2,\bm{k}_3 \in K^d}
    \frac{\delta_{\bm{k}_1+\bm{k}_2+\bm{k}_3, \bm{k}}}{\omega_{k_1}\omega_{k_2}\omega_{k_3}}
    \nonumber \\
    && \times \, \bigg\{ \frac{1}{\iR\bar{\omega}} \bigg[
    \frac{1}{\iR\omega_k} \left( \frac{1}{s+\iR\omega_k} + \frac{1}{s-\iR\omega_k} \right)
    \nonumber \\
    && \hspace{3em} + \, \frac{1}{\iR\bar{\omega}} \left( \frac{1}{s+\iR(\breve{\omega}-\breve{\omega}^*+\omega_k)}
    + \frac{1}{s+\iR(\breve{\omega}-\breve{\omega}^*-\omega_k)} \right) \bigg]
    \nonumber \\
    && + \, \frac{1}{\iR\bar{\omega}} \bigg[
    \frac{1}{s+\iR(\breve{\omega}_{k_1} + \breve{\omega}_{k_2} + \breve{\omega}_{k_3})}
    \left( \frac{1}{s+\iR\omega_k} - \frac{1}{s-\iR\omega_k} \right)
    \nonumber \\
    && \hspace{3em} + \, \frac{1}{s-\iR(\breve{\omega}_{k_1}^* + \breve{\omega}_{k_2}^* + \breve{\omega}_{k_3}^*)}
    \left( \frac{1}{s+\iR\omega_k} - \frac{1}{s-\iR\omega_k} \right) \bigg]
    \nonumber \\
    && + \, \frac{1}{s+\iR\omega_k} \bigg[
    \frac{1}{s+\iR(\breve{\omega}_{k_1} + \breve{\omega}_{k_2} + \breve{\omega}_{k_3})}
    \left( \frac{1}{s+\iR\omega_k} - \frac{1}{s-\iR\omega_k} \right)
    \nonumber \\
    && \hspace{3em} - \, \frac{1}{s-\iR(\breve{\omega}_{k_1}^* + \breve{\omega}_{k_2}^* + \breve{\omega}_{k_3}^*)}
    \left( \frac{1}{s+\iR\omega_k} - \frac{1}{s-\iR\omega_k} \right) \bigg] \bigg\} \,, \qquad
\label{prop2ndorderT2}
\end{eqnarray}
where
$\bar{\omega} = \omega_{k_1} + \omega_{k_2} + \omega_{k_3} + \omega_k$
and
$\breve{\omega} = \breve{\omega}_{k_1} + \breve{\omega}_{k_2} + \breve{\omega}_{k_3} + \breve{\omega}_k$. This second-order result consists of three blocks of four terms each. The first block results from the third and fourth lines of (\ref{twotimecorperturb}); the second and third blocks result from the fifth and sixth lines of (\ref{twotimecorperturb}), respectively. These blocks are sorted from more static to more dynamic, where the irreversible contribution to dynamics leads to increasing ultraviolet regularization. The purely static first contribution in (\ref{prop2ndorderT2}) actually diverges when the number of momentum states goes to infinity (that is, in the thermodynamic limit). The underlined contribution in the fourth line of (\ref{twotimecorperturb}) corresponds to an unconnected Feynman diagram; it is canceled by a contribution resulting from the other term in the same line. Also the terms in the fifth and sixth lines of (\ref{twotimecorperturb}) lead to some unconnected diagrams, but the corresponding contributions cancel within each term. In the end, unconnected diagrams do not contribute to the propagator $\tilde{\Delta}_{s \, \bm{k}}$.

If one expands the expression (\ref{prop2ndorderT2}) in $s$ then one realizes a simple structure of the expansion coefficients of the higher-order terms. This observation can be used to guide a rearrangement of the expression (\ref{prop2ndorderT2}), thus obtaining the explicit second-order perturbation expansion for $\tilde{\Delta}_{s \, \bm{k}}$ where, again, the SID assumption is made,
\begin{eqnarray}
    \tilde{\Delta}_{s \, \bm{k}} &=&
    \frac{1}{2 \omega_k} \, \frac{1}{s+\iR\omega_k}
    - \frac{\lambda'}{2\omega_k^3} \,
    \frac{s+2\iR\omega_k}{(s+\iR\omega_k)^2}
    + \frac{\lambda^2}{96 V^2 \omega_k^2}
    \sum_{\bm{k}_1,\bm{k}_2,\bm{k}_3 \in K^d}
    \frac{\delta_{\bm{k}_1+\bm{k}_2+\bm{k}_3, \bm{k}}}{\omega_{k_1}\omega_{k_2}\omega_{k_3}}
    \nonumber \\
    && \hspace{-3em} \times \,
    \bigg\{ - \frac{s}{\iR\bar{\omega}} \left[ \frac{1}{\iR\omega_k} \frac{1}{s^2+\omega_k^2}
    + \frac{1}{\iR\bar{\omega}} \frac{1}{[s+\iR(\breve{\omega}-\breve{\omega}^*)]^2+\omega_k^2} \right]
    \nonumber \\
    && \hspace{-3em} + \, \frac{2\iR\omega_k}{(s+\iR\omega_k)^2 (s-\iR\omega_k) \,
    [s+\iR(\breve{\omega}_{k_1} + \breve{\omega}_{k_2} + \breve{\omega}_{k_3})]}
    \bigg[ 1 + \frac{s}{\iR\bar{\omega}} \nonumber \\
    && \hspace{17em} + \, \frac{\iR(\omega_{k_1} + \omega_{k_2} + \omega_{k_3})}{s-
    \iR(\breve{\omega}_{k_1}^* + \breve{\omega}_{k_2}^* + \breve{\omega}_{k_3}^*)} \bigg]
    \nonumber \\
    && \hspace{-3em} - \, \frac{2\iR\omega_k \, \iR(\omega_{k_1} + \omega_{k_2} + \omega_{k_3})}{
    (s+\iR\omega_k)^2 (s-\iR\omega_k) \,
    [s+\iR(\breve{\omega}_{k_1} + \breve{\omega}_{k_2} + \breve{\omega}_{k_3})]
    [s-\iR(\breve{\omega}_{k_1}^* + \breve{\omega}_{k_2}^* + \breve{\omega}_{k_3}^*)]} \bigg\} \,.
    \nonumber \\ &&
\label{2corrphi4R012expl}
\end{eqnarray}
Note that the term in the last line cancels the last term in the preceding line. The term in the third line can be rewritten by means of the identity
\begin{equation}\label{2corrphisimplify}
    (s-\iR\omega_k) \left[ 1 + \frac{s}{\iR\bar{\omega}} \right] =
    s \, \frac{s+\iR(\omega_{k_1} + \omega_{k_2} + \omega_{k_3})}{\iR\bar{\omega}} - \iR \omega_k \,.
\end{equation}
We then recognize that, for $\gamma = 0$, a number of terms are antisymmetric in $s$. A more careful analysis shows that the symmetric part of these terms indeed vanishes in the limit of zero friction. As we eventually are interested only in the symmetric part, we from now on neglect all antisymmetric contributions to $\tilde{\Delta}_{s \, \bm{k}}$ and arrive at the compact expression
\begin{eqnarray}
    \tilde{\Delta}_{s \, \bm{k}} &=&
    \frac{-\iR}{2(s^2+\omega_k^2)}
    + \frac{\iR\lambda'}{(s^2+\omega_k^2)^2}
    \nonumber \\
    &-& \frac{\lambda^2}{48 V^2} \frac{\iR}{(s^2+\omega_k^2)^2}
    \sum_{\bm{k}_1,\bm{k}_2,\bm{k}_3 \in \bar{K}^d}
    \frac{\delta_{\bm{k}_1+\bm{k}_2+\bm{k}_3, \bm{k}}}{\omega_{k_1}\omega_{k_2}\omega_{k_3}} \,
    \frac{\breve{\omega}_{k_1} + \breve{\omega}_{k_2} + \breve{\omega}_{k_3}}{s^2 +
    (\breve{\omega}_{k_1} + \breve{\omega}_{k_2} + \breve{\omega}_{k_3})^2} \,.
    \nonumber \\ &&
\label{2corrphi4R012fin}
\end{eqnarray}

Equation (\ref{2corrphi4R012fin}) is the final result of all our efforts to construct a second-order perturbation expansion for the propagator. Note the remarkable simplicity of this result. In particular, we got rid of the divergent first term in (\ref{2corrphi4R012expl}) by ignoring antisymmetric contributions in $s$. The remaining sum in (\ref{2corrphi4R012fin}) is nicely convergent in the thermodynamic limit (note that we passed from $K^d$ to $\bar{K}^d$). Only the limits of infinite volume and vanishing friction remain to be performed.

Restricting $\tilde{\Delta}_{s \, \bm{k}}$ to its symmetric part is an example of making clever use of the freedom in choosing our quantities of interest. As a consequence of imposing an equilibrium initial condition, the Laplace transform is not a sufficiently dynamic quantity to be regularized by dissipation. The symmetric part, which is closely related to the Fourier transform, is much better regularized by dissipation and finite limits can be obtained.

Instead of inserting the expansion (\ref{Reexpcompact}) and truncating after a finite number of terms we could have inserted the magical identity (\ref{magic}). The high-order terms occurring for $q$ close to unity cannot be evaluated in closed form, but they can be handled by stochastic simulation techniques (the basic idea has been developed in \cite{hco214} and sketched in Section \ref{seccomputersim}). With the magic identity we expect better convergence behavior and we can pass from perturbation theory to a numerical integration method (see the discussion in Section \ref{secmagicid}).

\subsection{Going to the limits}
In the end of every calculation, we have to perform the \emph{two fundamental limits}: (i) The limit of infinite system volume $V$, in which the momentum states become continuous and sums become integrals. (ii) The limit of vanishing friction parameter $\gamma$. We now perform these limits for our fundamental result (\ref{2corrphi4R012fin}) for the propagator $\tilde{\Delta}_{s \, \bm{k}}$ of $\varphi^4$ theory in second-order perturbation theory.

The large-volume limit (i), which corresponds to a vanishing spacing of the momentum lattice, is easy to perform because we have massive particles so that no powers of $|\bm{k}_j|$ appear in the denominator and no divergencies for small wave vectors can arise. We can hence set $\gamma_0=0$ in (\ref{gammakconcr}) and perform that limit in the end. As already mentioned, we can also let the maximum momentum $Z_L K_L$ in (\ref{Kdlatticedef}) go to infinity because the sums over momenta in (\ref{2corrphi4R012fin}) converge for finite friction parameter $\gamma$. This can be seen as follows: Because of the Kronecker $\delta$, only two sums over momenta are unconstrained and we expect of the order of $Z_L^{2d}$ terms. In the denominator, we have three factors of order $Z_L K_L$ resulting from $\omega_{k_1}\omega_{k_2}\omega_{k_3}$. The frictional part of $\breve{\omega}_{k_1}+\breve{\omega}_{k_2}+\breve{\omega}_{k_3}$ grows as $(Z_L K_L)^4$ according to (\ref{gammakconcr}). The sum over momenta is hence expected to be perfectly convergent for $d<3.5$. We are actually interested only in $d\le 3$, so that we need more than six powers of large momenta in the denominator. Therefore, the $|\bm{k}|^4$-dependence of friction on momentum in (\ref{gammakconcr}) was exactly the right choice for providing regularization. A nonzero $\gamma$ is crucial for this argument.

In order to discuss the limit (ii) of vanishing friction parameter $\gamma$ for the sum in (\ref{2corrphi4R012fin}), we consider the expansion
\begin{equation}\label{phi4sexp}
    \frac{\breve{\omega}}{s^2 + \breve{\omega}^2} = \frac{1}{\breve{\omega}}
    \sum_{n=0}^{\infty} \left( - \frac{s^2}{\breve{\omega}^2} \right)^n \,,
\end{equation}
for $\breve{\omega} = \breve{\omega}_{k_1} + \breve{\omega}_{k_2} + \breve{\omega}_{k_3}$. For sufficiently large powers $n$ we observe convergence even for zero $\gamma$. The marginal case is $n=1$, which is expected to lead to logarithmic divergences for $d=3$. We hence need to worry about the first two terms only. We first determine the parameters $\lambda'$ and $Z$ introduced in (\ref{Hcolk}) and (\ref{phiFourierZ}), respectively, and then consider the convergence of the propagator $Z \tilde{\Delta}_{s \, \bm{k}}$ for vanishing friction parameter.

\subsubsection{Finding parameters}
Our quantities of interest are labeled by $s$ and $\bm{k}$. Whereas $\bm{k} \in K^d$ is a discrete set, we have introduced $s$ as a continuous variable and should hence choose a dense countable subset of $s$ values. For discussing the fundamental limits, it is natural to consider only a few quantities of interest because we only have the parameters $\lambda'$ and $Z$ to achieve convergence (two quantities are sufficient). Once these parameters are chosen, we can check convergence for a larger set of quantities.

Note that the parameters $\lambda'$ and $Z$ are not unique. Once we have found values for which the limits of certain quantities of interest exist, we can add any finite values to them without spoiling the existence of limits. We hence should choose particularly natural conditions or simple forms of $\lambda'$ and $Z$. Our particular choice affects the way we introduce the parameters of the limit model. An attractive alternative choice of $\lambda'$ and $Z$ will be discussed in Section~\ref{seconshellreno}.

As a first natural condition we impose the normalization requirement $Z \tilde{\Delta}_{0 \, \bm{0}} = \tilde{\Delta}_{0 \, \bm{0}}^{\rm free}$. Note that, as announced after (\ref{2corrphi4defsT}), we have reintroduced the factor $Z$ translating between free particles and clouds, the physical necessity of which we had discussed around (\ref{phiFourierZ}). From the general result (\ref{2corrphi4R012fin}) we get
\begin{equation}\label{Delta00}
    \iR Z \tilde{\Delta}_{0 \, \bm{0}} = \frac{Z}{2 m^2} - \frac{\lambda'}{m^4}
    + \frac{\lambda^2}{48 V^2 m^4} \sum_{\bm{k}_1,\bm{k}_2,\bm{k}_3 \in \bar{K}^d}
    \frac{\delta_{\bm{k}_1+\bm{k}_2+\bm{k}_3, \bm{0}}}{\omega_{k_1}\omega_{k_2}\omega_{k_3}}
    \, \frac{1}{\breve{\omega}_{k_1}+\breve{\omega}_{k_2}+\breve{\omega}_{k_3}} \,,
\end{equation}
where $Z$ in front of $\lambda'$ or $\lambda^2$ can be neglected because the difference from unity produces only higher-order terms, and hence we obtain our first condition for the parameters $\lambda'$ and $Z$,
\begin{equation}\label{Delta00c}
    \lambda' - \frac{1}{2} m^2 (Z-1) =
    \frac{\lambda^2}{48 V^2} \sum_{\bm{k}_1,\bm{k}_2,\bm{k}_3 \in \bar{K}^d}
    \frac{\delta_{\bm{k}_1+\bm{k}_2+\bm{k}_3, \bm{0}}}{\omega_{k_1}\omega_{k_2}\omega_{k_3}}
    \, \frac{1}{\breve{\omega}_{k_1}+\breve{\omega}_{k_2}+\breve{\omega}_{k_3}} \,.
\end{equation}

A second condition could be produced by considering another value of $s$ in addition to $s=0$ (still assuming $\bm{k}=\bm{0}$). However, the analysis is simpler and the result more elegant if we rather consider the term proportional to $s^2$ in an expansion of
\begin{equation}\label{ampupropphi4}
    (s^2+\breve{\omega}_k^2)^2 \, Z \tilde{\Delta}_{s \, \bm{k}} \,,
\end{equation}
where we have removed some factors occurring in the denominator of (\ref{2corrphi4R012fin}). Postulating that the $s^2$ term is not modified by the interaction (as we postulated before for the $s^0$ term), we find our second condition,
\begin{equation}\label{phi4Zchoice}
    Z = 1 + \frac{\lambda^2}{24 V^2} \sum_{\bm{k}_1,\bm{k}_2,\bm{k}_3 \in \bar{K}^d}
    \frac{\delta_{\bm{k}_1+\bm{k}_2+\bm{k}_3, \bm{0}}}{\omega_{k_1}\omega_{k_2}\omega_{k_3}}
    \frac{1}{(\breve{\omega}_{k_1}+\breve{\omega}_{k_2}+\breve{\omega}_{k_3})^3} \,.
\end{equation}
By inserting into (\ref{Delta00c}), we obtain an explicit expression for $\lambda'$,
\begin{equation}\label{phi4lpchoice}
    \lambda' = \frac{\lambda^2}{48 V^2} \sum_{\bm{k}_1,\bm{k}_2,\bm{k}_3 \in \bar{K}^d}
    \frac{\delta_{\bm{k}_1+\bm{k}_2+\bm{k}_3, \bm{0}}}{\omega_{k_1}\omega_{k_2}\omega_{k_3}}
    \left[\frac{1}{\breve{\omega}_{k_1}+\breve{\omega}_{k_2}+\breve{\omega}_{k_3}}
    + \frac{m^2}{(\breve{\omega}_{k_1}+\breve{\omega}_{k_2}+\breve{\omega}_{k_3})^3} \right] \,.
\end{equation}
Equations (\ref{phi4Zchoice}) and (\ref{phi4lpchoice}) give our choices for the parameters $Z$ and $\lambda'$ for which we can expect well-defined correlation functions in the limit of vanishing friction.

The sums in (\ref{phi4Zchoice}) and (\ref{phi4lpchoice}) depend on the particles mass $m$, the friction parameter $\gamma$, and the lattice spacing $K_L$ for the discrete momentum vectors, where we assume $Z_L \rightarrow \infty$. In the limit $K_L \rightarrow 0$, sums become integrals and dimensional arguments lead to
\begin{eqnarray}
  Z &=& 1 + (\lambda m^{d-3})^2 \, f_Z(\gamma^{1/3} m) \,,\\
  \lambda' m^{-2} &=& (\lambda m^{d-3})^2 \, f_{\lambda'}(\gamma^{1/3} m) \,.
\end{eqnarray}
We now have explicit results for the connection $Z$ between free particles and clouds and for the additional mass parameter $\lambda'/m$.

\subsubsection{General analysis}
Having determined the parameters $Z$ and $\lambda'$ in (\ref{phi4Zchoice}) and (\ref{phi4lpchoice}) from the limiting behavior of two quantities of interest, we should now check whether all quantities $Z \tilde{\Delta}_{s \, \bm{k}}$ have finite zero-friction limits for all $s$ and $\bm{k}$. By inserting $Z$ and $\lambda'$ into (\ref{2corrphi4R012fin}) we obtain
\begin{eqnarray}
    (s^2+\omega_k^2)^2 \, (Z \tilde{\Delta}_{s \, \bm{k}} -\tilde{\Delta}_{s \, \bm{k}}^{\rm free})&=&
    \frac{\lambda^2}{48 V^2} \sum_{\bm{k}_1,\bm{k}_2,\bm{k}_3 \in \bar{K}^d}
    \frac{\delta_{\bm{k}_1+\bm{k}_2+\bm{k}_3, \bm{0}}}{\omega_{k_1}\omega_{k_2}\omega_{k_3}}
    \nonumber \\
    &\times& \bigg[
    \frac{\iR}{\omega_{k_1}+\omega_{k_2}+\omega_{k_3}}
    - \frac{\iR(k^2+s^2)}{(\omega_{k_1}+\omega_{k_2}+\omega_{k_3})^3} \bigg]
    \nonumber \\
    && \hspace{-11em} - \, \frac{\lambda^2}{48 V^2}
    \sum_{\bm{k}_1,\bm{k}_2,\bm{k}_3 \in \bar{K}^d}
    \frac{\delta_{\bm{k}_1+\bm{k}_2+\bm{k}_3, \bm{k}}}{\omega_{k_1}\omega_{k_2}\omega_{k_3}}
    \, \frac{\iR(\omega_{k_1}+\omega_{k_2}+\omega_{k_3})}{s^2 +
    (\omega_{k_1}+\omega_{k_2}+\omega_{k_3})^2} \,. \qquad
\label{2corrphi4Rexplren}
\end{eqnarray}
Equation (\ref{2corrphi4Rexplren}) is our final result for the second-order contribution to the propagator\footnote{In Section \ref{secrelcovariance} we will see why, in view of the factor $(s^2+\omega_k^2)^2$, we can refer to the quantity defined in (\ref{2corrphi4Rexplren}) more appropriately as the second-order contribution to the `amputated propagator'.}\label{fnamputated} $Z \tilde{\Delta}_{s \, \bm{k}}$, which is proportional to $\lambda^2/V^2$ and has a further dependence on $s$, $\bm{k}$, $m$, and $K_L$ (remember that (\ref{KLchoiceL}) relates the spacing of the momentum lattice $K_L$ to the system size $L$ and hence to the volume $V=L^d$). The harmless limit $V \rightarrow \infty$, in which the dependence on $V$ and $K_L$ disappears, is performed by going from sums to integrals and from Kronecker $\delta$ symbols to Dirac $\delta$ functions,
\begin{eqnarray}
    (s^2+\omega_k^2)^2 \, (Z \tilde{\Delta}_{s \, \bm{k}} -\tilde{\Delta}_{s \, \bm{k}}^{\rm free})&=&
    \frac{\lambda^2}{48 (2\pi)^{2d}} \int \dR^dk_1 \dR^dk_2 \dR^dk_3
    \frac{\delta(\bm{k}_1+\bm{k}_2+\bm{k}_3)}{\omega_{k_1}\omega_{k_2}\omega_{k_3}}
    \nonumber \\
    &\times& \bigg[
    \frac{\iR}{\omega_{k_1}+\omega_{k_2}+\omega_{k_3}}
    - \frac{\iR(k^2+s^2)}{(\omega_{k_1}+\omega_{k_2}+\omega_{k_3})^3} \bigg]
    \nonumber \\
    && \hspace{-13em} - \, \frac{\lambda^2}{48 (2\pi)^{2d}} \int \dR^dk_1 \dR^dk_2 \dR^dk_3
    \frac{\delta(\bm{k}_1+\bm{k}_2+\bm{k}_3-\bm{k})}{\omega_{k_1}\omega_{k_2}\omega_{k_3}}
    \, \frac{\iR(\omega_{k_1}+\omega_{k_2}+\omega_{k_3})}{s^2 +
    (\omega_{k_1}+\omega_{k_2}+\omega_{k_3})^2} \,.
    \nonumber \\ & &
\label{2corrphi4Rexplrenc}
\end{eqnarray}

By using the expansion (\ref{phi4sexp}) for vanishing friction (that is, for $\breve{\omega}_{k_j}=\omega_{k_j}$) in (\ref{2corrphi4Rexplren}) or (\ref{2corrphi4Rexplrenc}), we obtain all terms in powers of $(\omega_{k_1}+\omega_{k_2}+\omega_{k_3})^{-1}$ and can thus identify all the potentially divergent terms. The terms proportional to $(\omega_{k_1}+\omega_{k_2}+\omega_{k_3})^{-1}$, which would lead to a power-law divergence for $d>2$, cancel. The terms proportional to $(\omega_{k_1}+\omega_{k_2}+\omega_{k_3})^{-3}$, which would lead to a logarithmic divergence for $d=3$, are more difficult to analyze. In order to avoid logarithmically divergent results, we need that the sums
\begin{equation}\label{verifyconv3}
    \sum_{\bm{k}_1,\bm{k}_2,\bm{k}_3 \in \bar{K}^3}
    \frac{1}{\omega_{k_1}\omega_{k_2}\omega_{k_3}}
    \left[ \frac{\delta_{\bm{k}_1+\bm{k}_2+\bm{k}_3, \bm{0}}-
    \delta_{\bm{k}_1+\bm{k}_2+\bm{k}_3, \bm{k}}}{\omega_{k_1}+\omega_{k_2}+\omega_{k_3}}
    - \frac{k^2\,\delta_{\bm{k}_1+\bm{k}_2+\bm{k}_3, \bm{0}}}{(\omega_{k_1}+\omega_{k_2}+\omega_{k_3})^3} \right]
\end{equation}
remain finite for an infinitely large lattice $\bar{K}^3$ of momentum vectors in $d=3$ dimensions. Numerical computations strongly suggest that this is indeed the case, but a more detailed analysis or analytical estimates are still required. It should also be clarified whether the sums remain finite only in the limit $V \rightarrow \infty$, which leads to the corresponding integral expression, or whether the fundamental limits $\gamma \rightarrow 0$ and $V \rightarrow \infty$ are interchangeable. All terms proportional to $(\omega_{k_1}+\omega_{k_2}+\omega_{k_3})^{-n}$ with $n > 3$ are nicely convergent for $d \le 3$.

\subsubsection{More than regularization}
The friction mechanism might be regarded as just another regularization procedure. For instance, one might say that $\ell = \gamma^{1/3}$ plays an analogous role to the small cutoff length $1/(Z_L K_L)$. However, there is more to dissipative regularization.

The thermodynamically consistent quantum master equation guarantees a controlled long-time behavior of the solutions. For any initial density matrix, convergence to the Gibbs equilibrium state can be demonstrated. This is a result of the robust thermodynamic structure of the evolution equations which is \emph{not} shared by the usual regularization methods. As an important consequence, correlation functions can be rigorously reformulated in terms of free vacuum expectation values (see Section \ref{secdefcorfcts}). In the zero-temperature limit, the proper interplay between ground state and dynamics is guaranteed.

Moreover, the thermodynamic regularization approach is special because it explicitly incorporates the eliminated small-scale  degrees of freedom as a heat bath. The thermodynamic regularization works in space and time. Ideally, one could formulate a dissipation mechanism that simultaneously respects the principles of nonequilibrium thermodynamics and relativistic covariance.

\subsection{Relativistic covariance}\label{secrelcovariance}
In Section \ref{secdefcorfcts}, we have realized that multi-time correlation functions are naturally defined for time-ordered sequences of operators. Only positive time differences occur and Laplace transforms hence turn out to be a natural tool (see Section~\ref{secLapltranscorrfunc}), in particular, for constructing perturbation theory (see Section~\ref{secperturbtheory}). We would now like to consider Fourier rather than Laplace transforms in the time domain because they go naturally with the Fourier transforms in the space domain. By considering functions of frequency $\omega$ and wave vector $\bm{k}$ we hope to recognize the relativistic covariance of scalar field theory.

Unfortunately, going from Laplace to Fourier transforms is calling for trouble. By passing from positive $s$ to imaginary $\pm i \omega$, that is, from exponential decay to oscillations, we might lose the convergence of integrals. One hence often adds an infinitesimally small positive quantity $\epsilon$ to $\pm i \omega$ to provide damping. Moreover, our experience with the definition of correlation functions and with the Feynman diagrams of perturbation theory (see, for example, Figure~\ref{figfeynmandia1}) teaches us that we must be careful with time-ordering when we wish to consider negative time differences.

\subsubsection{Definition of covariant propagator}
The Fourier transform is obtained by extending the Laplace transform (\ref{2corrphi4defs}) to the full domain of positive and negative time differences. In view of the structure of the time correlation functions (\ref{multitimecordef}), the extension to negative times $t$ is done most naturally by reversing the role of the two operators involved in the two-time correlation. We hence distinguish between positive and negative time differences and define
\begin{eqnarray}
    \iR \Delta_{\omega \, \bm{k}} &=& \int_{-\infty}^\infty
    \left[ \Theta(t) \, {\rm tr} \left\{ \varphi_{-\bm{k}} {\cal E}_t ( \varphi_{\bm{k}} \rho_{\rm eq} ) \right\}
    + \Theta(-t) \, {\rm tr} \left\{ \varphi_{\bm{k}} {\cal E}_{-t} ( \varphi_{-\bm{k}} \rho_{\rm eq} ) \right\}
    \right] \eR^{- \iR \omega t} \, \dR t
    \nonumber\\
    &=& \int_{-\infty}^\infty \left[ \Theta(t) \, \hat{C}^{-\bm{k} \bm{k}}_{t 0}
    + \Theta(-t) \, \hat{C}^{\bm{k} \, -\bm{k}}_{-t 0} \right] \eR^{- \iR \omega t} \, \dR t \,,
\label{2corrphi4defom}
\end{eqnarray}
where $\Theta$ is the Heaviside step function, that is, $\Theta(t)=0$ for $t<0$ and $\Theta(t)=1$ for $t \ge 0$. By comparison with (\ref{2corrphi4defs}), we find
\begin{equation}\label{2corrphi4Rfreesum}
    \iR\Delta_{\omega \, \bm{k}} = \tilde{\Delta}_{\iR\omega \, \bm{k}} + \tilde{\Delta}_{-\iR\omega \, -\bm{k}} \,.
\end{equation}
As our propagator (\ref{2corrphi4Rexplren}) has the symmetry property $\tilde{\Delta}_{s \, \bm{k}} = \tilde{\Delta}_{s \, -\bm{k}}$, (\ref{2corrphi4Rfreesum}) implies that we should symmetrize $\tilde{\Delta}_{s \, \bm{k}}$ also in $s$ for imaginary $s=\iR\omega$. Through the passage to Fourier transforms we automatically achieve a symmetrization of $\tilde{\Delta}_{s \, \bm{k}}$ for imaginary $s$, which seems to be safer than for real $s$ because that implies a combination of exponentially decreasing and increasing integrals. As we realized before, symmetrization leads to considerable simplification of the propagator and eliminates divergent terms.

Let us first consider the zeroth-order result for the free theory in (\ref{2corrphi4R012fin}), that is
\begin{equation}\label{2corrphi4Rfreesumxl}
    \iR\Delta_{\omega \, \bm{k}}^{\rm free} =
    \frac{\iR}{\omega^2-\omega_k^2} =
    \frac{\iR}{\omega^2-k^2-m^2} \,,
\end{equation}
which is known as the Feynman propagator. This propagator is relativistically invariant by depending only on the Lorentz scalar $\omega^2-k^2$. In the limits of vanishing friction and continuous and unbounded momentum vectors, the Lorentz invariance of the free theory thus becomes manifest. We now realize that the factor $1/\sqrt{2\omega_k}$, which was introduced in (\ref{phiexpression}) and leads to the prefactor $1/(2\omega_k)$ in (\ref{prop1storder}), is indeed crucial for Lorentz invariance. With the result (\ref{2corrphi4Rfreesumxl}), we furthermore realize that the two factors of $s^2 + \omega_k^2 = -\omega^2 + \omega_k^2 = -1/\Delta_{\omega \, \bm{k}}^{\rm free}$ introduced in (\ref{ampupropphi4}) and (\ref{2corrphi4Rexplren}) correspond to the amputation of the two free Feynman propagators associated with the dangling ends of the Feynamn diagrams in Figure~\ref{figfeynmandia2s} (see footnote on p.\,\pageref{fnamputated}).

Let us now include the next term in (\ref{2corrphi4R012fin}), which leads to the following accumulated result for the Feynman propagator,
\begin{equation}\label{2corrphi4Feynm1}
    \Delta_{\omega \, \bm{k}} = \frac{1}{\omega^2-\omega_k^2} +
    2 \lambda' \frac{1}{(\omega^2-\omega_k^2)^2} = \frac{1}{\omega^2-k^2-m^2}
    + 2 \lambda' \frac{1}{(\omega^2-k^2-m^2)^2} \,.
\end{equation}
Note that also the $\lambda'$ contribution is Lorentz invariant. We thus realize that the choice of $\lambda'$ has no influence on the relativistic invariance properties of the propagator, and neither has the choice of $Z$.

\subsubsection{Second-order propagator}
We finally focus on the invariance of the second-order term (\ref{2corrphi4Rexplrenc}) in the formulation with continuous momentum variables, including the contributions from $\lambda'$ and $Z$,
\begin{eqnarray}
    \Delta_{\omega \, \bm{k}}^{(2)} &=& \frac{1}{(\omega^2-\omega_k^2)^2} \,
    \frac{\lambda^2}{24 (2\pi)^{2d}} \int \frac{\dR^dk_1 \dR^dk_2 \dR^dk_3}{\omega_{k_1}\omega_{k_2}\omega_{k_3}}
    \nonumber \\
    &\times& \bigg\{ \delta(\bm{k}_1+\bm{k}_2+\bm{k}_3)
    \left[ \frac{1}{\omega_{k_1}+\omega_{k_2}+\omega_{k_3}}
    + \frac{\omega^2-k^2}{(\omega_{k_1}+\omega_{k_2}+\omega_{k_3})^3} \right]
    \nonumber \\
    &-& \delta(\bm{k}_1+\bm{k}_2+\bm{k}_3-\bm{k})
    \frac{\omega_{k_1}+\omega_{k_2}+\omega_{k_3}}{
    (\omega_{k_1}+\omega_{k_2}+\omega_{k_3})^2-\omega^2} \bigg\} \,.
\label{2propintexpr}
\end{eqnarray}
To be covariant, we would like this expression to depend on $\omega^2-k^2$ only. For the term in the last line, such a combined dependence $\omega$ and $\bm{k}$ is far from obvious. As $\bm{k}$ is involved in the $\delta$ function, we should pass from space to space-time, that is, from $d$ to $D=d+1$ dimensions. The $D$ components of a space-time vector $\underline{k}_j$ are written as $\underline{k}_j = (\kappa_j, \bm{k}_j)$, where $\kappa_j$ is real and $\bm{k}_j$ has $d$ components. Similarly, we use $\underline{k} = (\omega, \bm{k})$. The crucial identity for recognizing the covariance of $\Delta_{\omega \, \bm{k}}^{(2)}$ is
\begin{eqnarray}
    \int \frac{\dR^dk_1 \dR^dk_2 \dR^dk_3}{\omega_{k_1} \omega_{k_2} \omega_{k_3}} \,
    \delta(\bm{k}_1+\bm{k}_2+\bm{k}_3-\bm{k})
    \frac{\omega_{k_1}+\omega_{k_2}+\omega_{k_3}}{
    (\omega_{k_1}+\omega_{k_2}+\omega_{k_3}-\iR\epsilon)^2 - \omega^2} = \qquad
    && \nonumber\\
    \frac{1}{\pi^{2}} \int \dR^Dk_1 \dR^Dk_2 \dR^Dk_3 \frac{\delta(\underline{k}_1+\underline{k}_2+\underline{k}_3
    -\underline{k})}{(\kappa_1^2-\omega_{k_1}^2+\iR\epsilon)
    (\kappa_2^2-\omega_{k_2}^2+\iR\epsilon)(\kappa_3^2-\omega_{k_3}^2+\iR\epsilon)} \,.
    && \nonumber\\ &&
\label{phi4propcovarcheck}
\end{eqnarray}
This identity may be shown by replacing the factor $\delta(\kappa_1+\kappa_2+\kappa_3-\omega)$ contained in the $D$-dimensional $\delta$ function on the right-hand side of (\ref{phi4propcovarcheck}) by its one-dimensional Fourier representation and then performing the factorized integrations over $\kappa_1$, $\kappa_2$, and $\kappa_3$ by means of Cauchy's integral formula, where the infinitesimal quantity $\epsilon$ determines how integration paths have to be chosen around poles. Finally, one can perform the integration resulting from the Fourier representation of the $\delta$ function. It is this type of relations between $d$- and $D$-dimensional integrals derived by means of Cauchy's integral formula that is at the origin of the factor $1/\sqrt{2\omega_k}$ in (\ref{phiexpression}).

On the right-hand side of (\ref{phi4propcovarcheck}), we recognize that $\omega$ and $\bm{k}$ enter as a $D$-vector, together with the $D$-vectors $\underline{k}_j$ and their invariants. After integration, the result can only depend on the scalar invariant $\omega^2-k^2$. Note that the integrand on the right-hand side of (\ref{phi4propcovarcheck}) is the product of three free propagators. This observation reflects the fact that perturbation theory in $D$ dimensions is simpler than in $d$ dimensions, as has been pointed out in the context of Feynman diagrams in Section~\ref{secphi4proppert2}.

Once we know that (\ref{2propintexpr}) depends only on $\omega^2-k^2$, we can evaluate the functional dependence on that single variable for $\bm{k}=\bm{0}$ and we then obtain the crucial reformulation
\begin{eqnarray}
    \Delta_{\omega \, \bm{k}}^{(2)} &=& \frac{\lambda^2}{24 (2\pi)^{2d}} \,
    \frac{1}{(\omega^2-\omega_k^2)^2} \int \frac{\dR^dk_1 \dR^dk_2 \dR^dk_3}{\omega_{k_1}\omega_{k_2}\omega_{k_3}}
    \, \delta(\bm{k}_1+\bm{k}_2+\bm{k}_3) \nonumber \\
    &\times&
    \bigg[ \frac{1}{\omega_{k_1}+\omega_{k_2}+\omega_{k_3}}
    + \frac{\omega^2-k^2}{(\omega_{k_1}+\omega_{k_2}+\omega_{k_3})^3}
    \nonumber \\
    &-&
    \frac{\omega_{k_1}+\omega_{k_2}+\omega_{k_3}}{
    (\omega_{k_1}+\omega_{k_2}+\omega_{k_3})^2-(\omega^2-k^2)} \bigg] \,.
\label{2propintexpry}
\end{eqnarray}
This result can be further simplified to
\begin{eqnarray}
    \Delta_{\omega \, \bm{k}}^{(2)} &=& - \frac{\lambda^2}{24 (2\pi)^{2d}} \,
    \frac{(\omega^2-k^2)^2}{(\omega^2-\omega_k^2)^2}
    \int \frac{\dR^dk_1 \dR^dk_2 \dR^dk_3}{\omega_{k_1}\omega_{k_2}\omega_{k_3}}
    \, \delta(\bm{k}_1+\bm{k}_2+\bm{k}_3)
    \nonumber \\
    &\times& \frac{1}{(\omega_{k_1}+\omega_{k_2}+\omega_{k_3})^3}
    \, \frac{1}{(\omega_{k_1}+\omega_{k_2}+\omega_{k_3})^2-(\omega^2-k^2)} \,. \qquad \qquad
\label{2propintexprc}
\end{eqnarray}
This last equation produces a finite result. This observation proves that, at least in the continuum limit, the sums (\ref{verifyconv3}) indeed remain finite.

\subsubsection{Confrontation with the real world}
Equation (\ref{phi4propcovarcheck}) is very important not only for practical purposes like simplifying perturbation theory, but also for fundamental reasons. It allows us to verify the equivalence of our results with those of Lagrangian field theory. In one and two space dimensions, there moreover is the reassuring consistency with axiomatic quantum field theory.

As the standard model, which is formulated within Lagrangian field theory, describes fundamental particles and their electromagnetic, weak and strong interactions so impressively well, we here take agreement with this pragmatic approach as agreement with the real world. Of course, a direct confrontation with the real world becomes possible only when we apply our dissipative quantum field theory to electromagnetic, weak or strong interactions rather than the quartic interactions of scalar field theory.

\subsection{On shell renormalization}\label{seconshellreno}
The terms in the second line of (\ref{2propintexpry}) cancel the two lowest-order terms of an expansion of the term in the last line of (\ref{2propintexpry}) in $\omega^2-k^2$. Instead of expanding with respect to $\omega^2-k^2$, we could alternatively expand in terms of $\omega^2-k^2-m^2$ and again cancel the first two terms. This approach corresponds to what is known as `on shell renormalization' which indicates that one stays near the physical relation between the frequency $\omega$ and the wave vector $\bm{k}$ for a massive particle with mass $m$. Instead of (\ref{2propintexpry}), we then have
\begin{eqnarray}
    \Delta_{\omega \, \bm{k}}^{(2)} &=& \frac{\lambda^2}{24 (2\pi)^{2d}} \,
    \frac{1}{(\omega^2-\omega_k^2)^2} \int \frac{\dR^dk_1 \dR^dk_2 \dR^dk_3}{\omega_{k_1}\omega_{k_2}\omega_{k_3}}
    \delta(\bm{k}_1+\bm{k}_2+\bm{k}_3)
    \nonumber \\ &\times&
    \bigg\{ \frac{1}{(\omega_{k_1}+\omega_{k_2}+\omega_{k_3})^2-m^2}
    + \frac{\omega^2-\omega_k^2}{[(\omega_{k_1}+\omega_{k_2}+\omega_{k_3})^2-m^2]^2}
    \nonumber \\
    &-& \frac{1}{(\omega_{k_1}+\omega_{k_2}+\omega_{k_3})^2-(\omega^2-k^2)} \bigg\}
    \, (\omega_{k_1}+\omega_{k_2}+\omega_{k_3}) \,.
\label{2propintexprx}
\end{eqnarray}
This expression results from the following alternative choice of $Z$ and $\lambda'$,
\begin{equation}\label{phi4Zchoicex}
    Z = 1 + \frac{\lambda^2}{24 V^2} \sum_{\bm{k}_1,\bm{k}_2,\bm{k}_3 \in K^d}
    \frac{\delta_{\bm{k}_1+\bm{k}_2+\bm{k}_3, \bm{0}}}{\omega_{k_1}\omega_{k_2}\omega_{k_3}}
    \frac{\breve{\omega}_{k_1}+\breve{\omega}_{k_2}+\breve{\omega}_{k_3}}{
    [(\breve{\omega}_{k_1}+\breve{\omega}_{k_2}+\breve{\omega}_{k_3})^2-m^2]^2} \,,
\end{equation}
and,
\begin{equation}\label{phi4lpchoicex}
    \lambda' = \frac{\lambda^2}{48 V^2} \sum_{\bm{k}_1,\bm{k}_2,\bm{k}_3 \in K^d}
    \frac{\delta_{\bm{k}_1+\bm{k}_2+\bm{k}_3, \bm{0}}}{\omega_{k_1}\omega_{k_2}\omega_{k_3}}
    \frac{\breve{\omega}_{k_1}+\breve{\omega}_{k_2}+\breve{\omega}_{k_3}}{
    (\breve{\omega}_{k_1}+\breve{\omega}_{k_2}+\breve{\omega}_{k_3})^2-m^2} \,,
\end{equation}
instead of (\ref{phi4Zchoice}) and (\ref{phi4lpchoice}). The previous $Z$ is recovered for $m^2=0$, whereas the previous $\lambda'$ corresponds to the first-order expansion in $m^2$. The two possible choices for $Z$ and $\lambda'$ differ only by finite values and hence are equally good for guaranteeing the existence of finite limits.

\begin{figure}[t]
\centerline{\includegraphics[width=8cm]{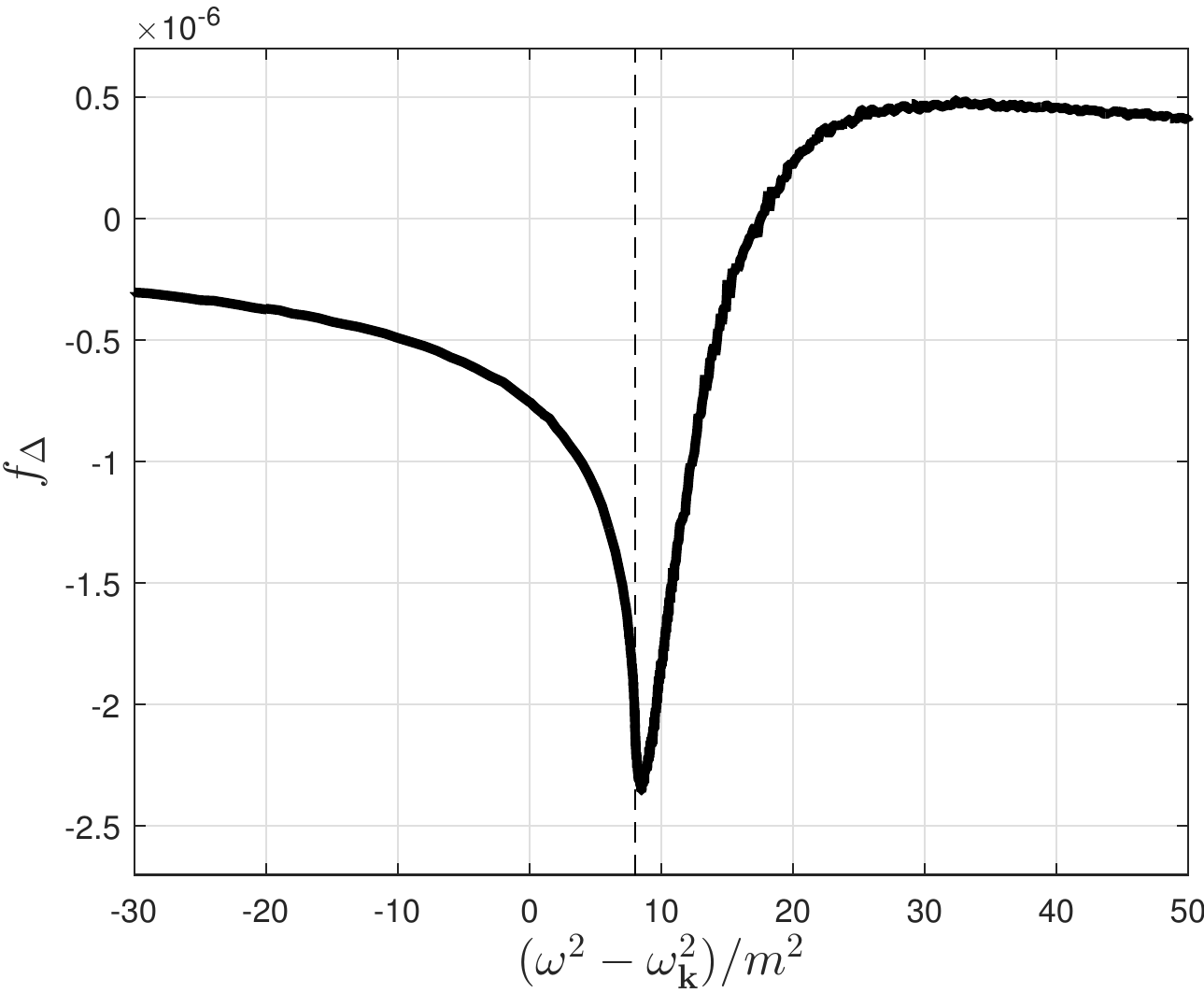}} \caption[ ]
{Scaling function $f_\Delta$ for the deviation from the free propagator introduced in (\ref{phi4propscaling}) for $d=2$. [Figure courtesy of M.\,Kr\"oger, ETH Z\"urich.]} \label{figppropscal2d}
\end{figure}

\begin{figure}[t]
\centerline{\includegraphics[width=8cm]{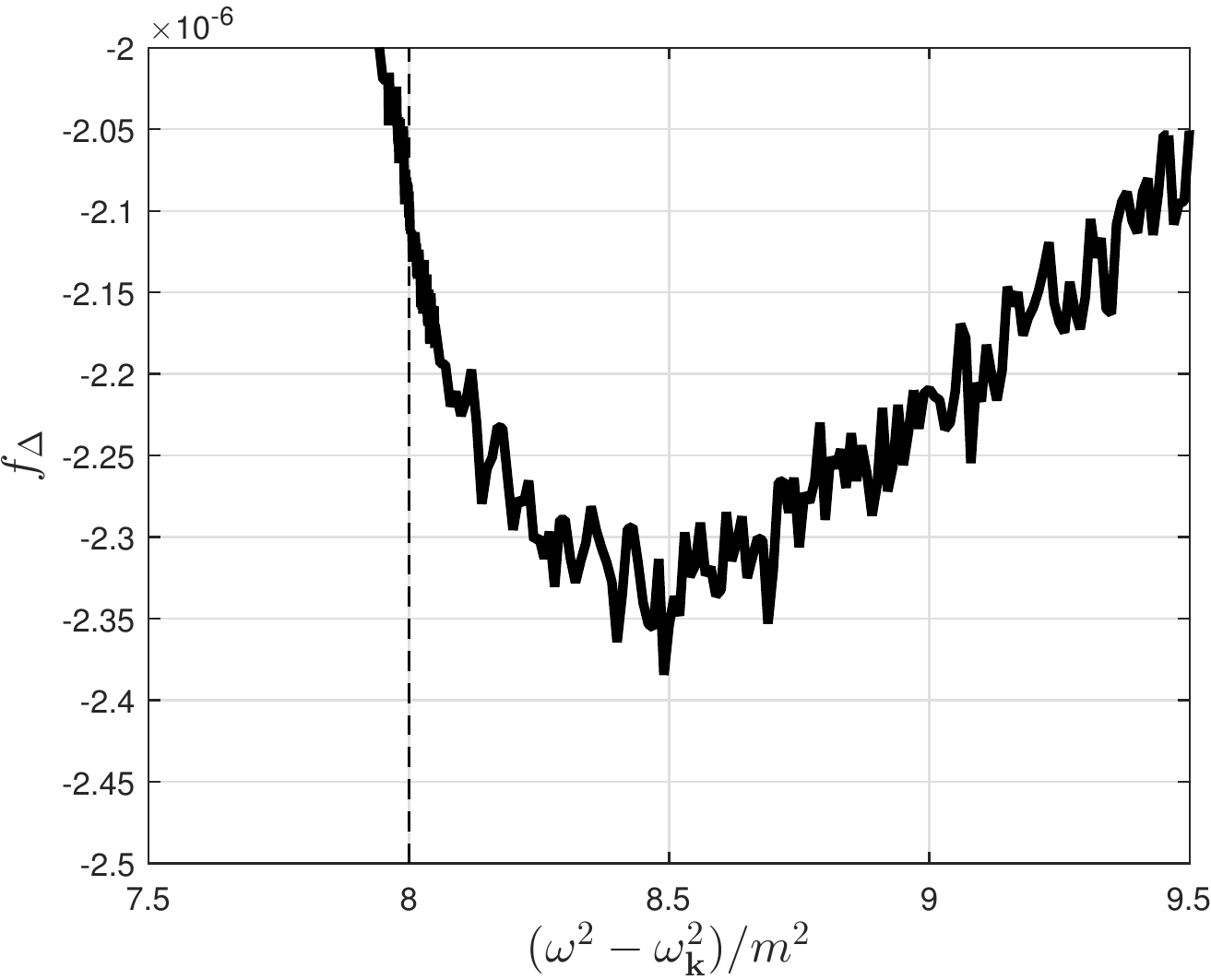}} \caption[ ]
{Closeup of the scaling function $f_\Delta$ shown in Figure~\ref{figppropscal2d} around its minimum. [Figure courtesy of M.\,Kr\"oger, ETH Z\"urich.]} \label{figppropscal2dcloseup}
\end{figure}

The result (\ref{2propintexprx}) can be rewritten in the compact form
\begin{eqnarray}
    \Delta_{\omega \, \bm{k}}^{(2)} &=& - \frac{\lambda^2}{24 (2\pi)^{2d}}
    \int \frac{\dR^dk_1 \dR^dk_2 \dR^dk_3}{\omega_{k_1}\omega_{k_2}\omega_{k_3}}
    \, \delta(\bm{k}_1+\bm{k}_2+\bm{k}_3)
    \nonumber \\
    &\times& \frac{\omega_{k_1}+\omega_{k_2}+\omega_{k_3}}{
    [(\omega_{k_1}+\omega_{k_2}+\omega_{k_3})^2-m^2]^2} \,
    \, \frac{1}{(\omega_{k_1}+\omega_{k_2}+\omega_{k_3})^2-(\omega^2-k^2)} \,.
    \nonumber \\ &&
\label{2propintexprxc}
\end{eqnarray}
It is a nice advantage of the on shell scheme that the prefactor diverging for $\omega^2 \approx \omega_k^2$ has disappeared in going from (\ref{2propintexprc}) to (\ref{2propintexprxc}). The total propagator is of the functional form
\begin{equation}\label{phi4propscaling}
    m^2 \Delta_{\omega \, \bm{k}} = \frac{m^2}{\omega^2-\omega_k^2} +
    (\lambda m^{d-3})^2 \, f_\Delta \left( \frac{\omega^2-\omega_k^2}{m^2} \right) \,.
\end{equation}
The function $f_\Delta$, which can be obtained from the integral expression in (\ref{2propintexprxc}), possesses regular behavior for small $(\omega^2-\omega_k^2)/m^2$, that is, for frequencies $\omega$ near the physical frequency $\omega_k$ for a particle with mass $m$ and wave vector $\bm{k}$. The scaling function $f_\Delta$ for $d=2$ is shown in Figure~\ref{figppropscal2d}. As the part between the minimum and the maximum of the curve is difficult to obtain by deterministic numerical methods for evaluating the integral in (\ref{2propintexprxc}) (as the integrand can diverge for $(\omega^2-\omega_k^2)/m^2 \ge 8$), the integral is obtained by the adaptive Monte Carlo methods implemented in \mica. A closeup of the region around the minimum is shown in Figure~\ref{figppropscal2dcloseup}. The error bars of the Monte Carlo calculation of the scaling function can be inferred from the fluctuations of the curves. Note that the pole in $\omega$ remains unchanged by the regular scaling function $f_\Delta$ so that we obtain a particularly convenient interpretation to the mass parameter.

\section{A four-particle correlation}
So far we have considered the second-order expansion of the propagator. We only needed to choose the parameters $Z$ and $\lambda'$ such that finite limits of this correlation function exist for infinite volume and vanishing dissipation. We now would like to benefit from the renormalization-group ideas developed in Section \ref{secrenormalization}, as they should come up naturally in perturbation theory. For that purpose, we need to construct an expansion for which several powers of the interaction parameter $\lambda$ occur. We here consider a simple quantity that contains terms of orders $\lambda$ and $\lambda^2$.

The propagator studied in Section \ref{secphi4prop} is a two-particle correlation function. We now calculate a simple four-particle correlation function where we do not attempt to achieve any spatial resolution. We consider only the Fourier modes $\varphi_{\bm{0}}$ with zero wave vector and we introduce the operator $Q$ proportional to the normal-ordered version of $\varphi_{\bm{0}}^2$,
\begin{equation}\label{4corrphi4Q}
    Q = a_{\bm{0}} a_{\bm{0}}
    + 2 a^\dag_{\bm{0}} a_{\bm{0}}
    + a^\dag_{\bm{0}} a^\dag_{\bm{0}} \,.
\end{equation}
As a special case of the two-time correlation function (\ref{twotimecorRrep}), we define
\begin{equation}\label{4corrphi4Lam}
    \Lambda_s = {\rm tr} \left\{ Q {\cal R}_s [ (Q-\bar{Q}) \rho_{\rm eq} ] \right\} \,,
\end{equation}
with $\bar{Q} = {\rm tr} ( Q \rho_{\rm eq} )$, which is the four-particle correlation to be investigated in the present section. We only consider contributions from connected Feynman diagrams in (\ref{4corrphi4Lam}). Therefore, we have $\Lambda_s^{\rm free} = 0$ rather than $\Lambda_s^{\rm free} = 2/(s+2im)$, where the latter expression corresponds to the creation and annihilation of two independent particles. The equilibrium average $\bar{Q}$ in (\ref{4corrphi4Lam}) could actually be omitted because it leads only to disconnected Feynman diagrams.

\subsection{Second-order perturbation expansion}
By means of (\ref{rhoeqT0perturb}) we obtain the following second-order result for the equilibrium average,
\begin{equation}\label{phisqeqav}
    \bar{Q} = - \frac{\lambda'}{m^2} +
    \frac{\lambda^2}{48 V^2 m^2} \sum_{\bm{k}_1,\bm{k}_2,\bm{k}_3 \in K^d}
    \frac{\delta_{\bm{k}_1+\bm{k}_2+\bm{k}_3, \bm{0}}}{\omega_{k_1}\omega_{k_2}\omega_{k_3}} \,
    \frac{\omega_{k_1}+\omega_{k_2}+\omega_{k_3}+2m}{
    (\omega_{k_1}+\omega_{k_2}+\omega_{k_3}+m)^2} \,.
\end{equation}
This result is given only for completeness because $\bar{Q}$ does not contribute to the connected Feynman diagrams in (\ref{4corrphi4Lam}). The sum in (\ref{phisqeqav}) diverges in the thermodynamic limit for three space dimensions. As a static quantity, it cannot be regularized by dissipative effects.

\begin{figure}[t]
\centerline{\includegraphics[width=4cm]{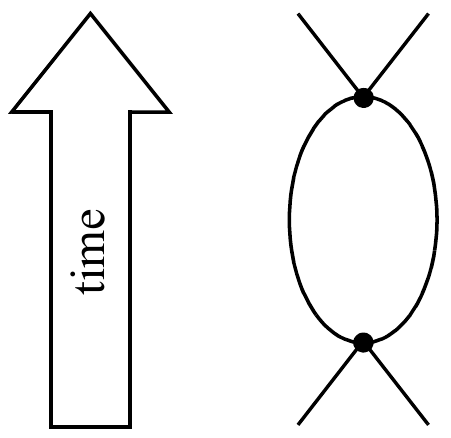}} \caption[ ]
{Topology of the connected Feynman diagrams contributing to the four-particle correlation $\Lambda_0$ (`effective interaction vertex') in second-order perturbation theory.} \label{figfeynmandia4s}
\end{figure}

The effort required for evaluating of the second-order perturbation theory for $\Lambda_0$ is comparable to what was required for calculating the propagator. On the one hand, there are a few more diagrams because we deal with more particles; on the other hand, the typical factors are simpler because we assume $\bm{k}=0$ and $s=0$. All connected Feynman diagrams contributing to $\Lambda_0$ in second-order perturbation theory are topologically equivalent to the diagram shown in Figure~\ref{figfeynmandia4s}. Our final result is
\begin{equation}\label{4corrphi4Lam1}
    \Lambda_0 = \frac{\iR}{8Vm^4} \bigg\{ \lambda + \frac{\lambda^2}{16V}
    \sum_{\bm{k} \in K^d} \frac{1}{\omega_k^2} \left[
    \frac{(3m+2\omega_k)m^2}{\omega_k(m+\omega_k)^3}
    - \frac{3}{\breve{\omega}_k}
    - \frac{3}{\breve{\omega}_k^*} \right] \bigg\} \,.
\end{equation}
We have written the terms that require regularization for $d=3$ in the simplest possible form; all other terms are evaluated in the limit of vanishing dissipation ($\breve{\omega}_k = \breve{\omega}_k^* = \omega_k$). Equation (\ref{4corrphi4Lam1}) suggests that we may interpret $8Vm^4 \, \Lambda_0 / \iR$ as an effective interaction vertex accounting for higher-order interaction effects. In the continuum limit ($K_L \rightarrow 0$), we find
\begin{equation}\label{4corrphi4Lam2}
    \Lambda_0 = \frac{\iR}{8Vm^4} \bigg\{ \lambda + \frac{\lambda^2}{16 (2\pi)^d}
    \int \frac{1}{\omega_k^2} \left[
    \frac{(3m+2\omega_k)m^2}{\omega_k(m+\omega_k)^3}
    - \frac{3}{\breve{\omega}_k}
    - \frac{3}{\breve{\omega}_k^*} \right] \dR^dk \bigg\} \,.
\end{equation}
Note that the quantity $\Lambda_0$ is imaginary.

\subsection{Critical coupling constant}
We next consider (\ref{4corrphi4Lam2}) in the limit of vanishing friction. For $d<3$, all the integrals are convergent. The leading-order corrections for $d$ close to $3$ result from the last two terms of the integrand,
\begin{eqnarray}
    \Lambda_0 &=& \frac{\iR}{8Vm^4} \bigg\{ \lambda + \frac{\lambda^2}{16 (2\pi)^d}
    \int \frac{1}{\omega_k^2} \left[ \frac{(3m+2\omega_k)m^2}{\omega_k(m+\omega_k)^3}
    - \frac{6}{\omega_k} \right] \dR^dk \nonumber \\
    &+& \frac{3}{8} \frac{\lambda^2}{(2\pi)^d} \ell^{3-d} \int \frac{q^3}{1+q^6} \dR^dq \bigg\}
\label{4corrphi4Lam3}
\end{eqnarray}
with $\ell=\gamma^{1/3}$, previously defined in (\ref{lengthdissipdef}). In order to find the correction term, we have used the identity
\begin{equation}\label{4corrphi4Lam2x}
    \int \frac{1}{\omega_k^2} \left( \frac{2}{\omega_k} - \frac{1}{\omega_k+\iR\gamma_k}
    - \frac{1}{\omega_k-\iR\gamma_k} \right) \dR^dk =
    2 \int \frac{1}{\omega_k^3} \, \frac{\gamma_k^2}{\omega_k^2+\gamma_k^2} \dR^dk \,,
\end{equation}
the approximation $\omega_k \approx k$ for the large values of $k$ dominating the integral, $\gamma_k = \gamma k^4$ (see (\ref{gammakconcr})), and the substitution $q = \ell k$.

\begin{figure}
\centerline{\includegraphics[width=8cm]{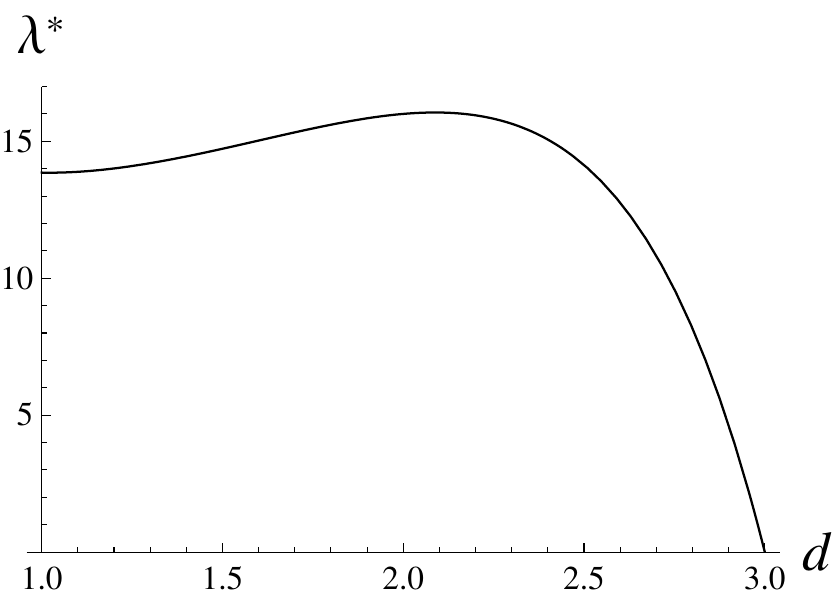}} \caption[ ]
{Fixed point value $\lambda^*$ of the dimensionless coupling constant resulting from second-order perturbation theory as a function of space dimensionality $d$.} \label{figphi4fixedpoint}
\end{figure}

By comparing (\ref{4corrphi4Lam3}) with the general structure of perturbation theory developed in (\ref{genpertexp3}), we find
\begin{equation}\label{alpharesult}
    \alpha = \epsilon = 3 - d ,
\end{equation}
and
\begin{equation}\label{lambdastar}
    \lambda^* = \frac{8}{3} (2\pi)^d \left[\int \frac{q^3}{1+q^6} \dR^dq\right]^{-1}
    = 2^{d+3} \pi^{\frac{d-2}{2}} \, \Gamma\left(\frac{d}{2}\right)
    \, \sin \left[ (3-d) \frac{\pi}{6} \right] \,.
\end{equation}
This result is shown in Figure \ref{figphi4fixedpoint}. For small $\epsilon$, that is, slightly below three space dimensions, we have the first order approximation
\begin{equation}\label{lambdastarex}
    \lambda^* \approx \frac{16}{3} \pi^2 \epsilon \,.
\end{equation}
Decreasing below $d=3$, the value of $\lambda^*$ quickly rises above $10$ and stabilizes around $15$. For $d=2$, we find the exact value $\lambda^*=16$. As such a value of $\lambda^*$ seems to be quite large one should notice that, in the perturbation expansions (\ref{2propintexpr}) and (\ref{4corrphi4Lam3}), $\lambda$ is typically divided by $(2\pi)^d$, where $\lambda^*/(2\pi)^d \approx 0.4$ for $d=2$.

It is quite remarkable that the detailed form of the integral in (\ref{lambdastar}) leads to a vanishing $\lambda^*$ in three space dimensions and to the well-known leading-order coefficient in the $\epsilon$ expansion of $\lambda^*$. Equations (\ref{alpharesult}) and (\ref{lambdastarex}) moreover lead to an explicit result for the $\beta$ function which agrees with the famous result for $\varphi^4$ theory as, for example, given in (11.17) of \cite{ZinnJustin} or in (18.5.7) of \cite{WeinbergQFT2}. Again, we can take agreement with the powerful Lagrangian approach to quantum field theory as a successful confrontation with the real world. The possibility to calculate the function $\lambda^*(d)$ in Figure \ref{figphi4fixedpoint} in such a straightforward way speaks for the power of the dissipative approach to quantum field theory.

The result (\ref{lambdastar}) deviates slightly from (108) of \cite{hco200}, except in the limit of small $\epsilon$. This is not surprising because the friction mechanism in \cite{hco200} is constructed in a different way. In particular, the definition of the length scale $\ell$ in (\ref{lengthdissipdef}) is completely different from the previous one, which depended not only on the friction parameter but also on the temperature. Only for small $\epsilon$ the precise definition of $\ell$ becomes irrelevant.

\section{Summary and discussion}\label{secphi4finalsummary}
In this chapter, we have calculated a two-particle and a four-particle correlation function for $\varphi^4$ theory within the framework of dissipative quantum field theory to second order in perturbation theory: the propagator $\Delta_{\omega \, \bm{k}}$ and the effective interaction vertex $\Lambda_0$. For properly chosen quantities of interest, finite results are obtained at any stage of the calculation and after performing all the proper limits. It is shown explicitly that the result for the propagator is relativistically covariant and in agreement with Lagrangian quantum field theory.

Of the limits listed and discussed on p.\,\pageref{listlimits}, we take the zero-temperature limit in the very beginning of the calculation by using the zero-temperature quantum master equation. Early limits come with a risk, but an early zero-temperature limit simplifies the calculations considerably and seems to be fine for perturbation theory. This still needs to be confirmed in higher-order calculations. If problems arise, the zero-temperature limit should be postponed to the very end of the calculations. However, the Feynman diagrams with the topologies shown in Figures \ref{figfeynmandia2s} and \ref{figfeynmandia4s}, which are associated with the second-order contributions to the propagator and the effective interaction vertex, are the only ones that lead to potentially diverging sums or integrals in the limit of vanishing dissipation. If a Feynman diagram does not contain subdiagrams of this type, no regularization is required to obtain a finite limit, as can be argued by power counting in the corresponding terms.

At zero temperature, $\varphi^4$ theory in $d$ space dimensions is characterized by the following parameters: the mass parameter $m$ characterizing the frequencies (\ref{relenergmomrel}) in the free Hamiltonian (\ref{Hfree}), the interaction parameters $\lambda$, $\lambda'$ and $\lambda''$ of the collisional Hamiltonian (\ref{Hcolk}), the factor $Z$ relating free particles to interacting particles (or clouds of free particles) introduced in (\ref{phiFourierZ}), the parameters $K_L$ (equivalent to the volume $V$) and $Z_L$ keeping momentum space discrete and finite [see (\ref{Kdlatticedef})], and the friction parameter $\gamma$ in (\ref{gammakconcr}). The number of parameters is reduced by the following considerations:
\begin{itemize}
  \item The parameter $\lambda''$ has previously been chosen to assign zero energy to the ground state of the interacting theory [see (\ref{groundstatecondX})].
  \item For finite $\gamma$ and `sufficiently dynamic' quantities of interest, the thermodynamic limit ($Z_L \rightarrow \infty$) can be performed at any stage of the calculation. `Sufficiently dynamic' means that ultraviolet regularization is provided by the friction mechanism. This meaning of `sufficiently dynamic' is illustrated by the fact that the Laplace transform in time is not sufficiently dynamic (because it depends on a static initial condition), whereas the Fourier transform is.
  \item The remaining fundamental limits of infinite volume and vanishing dissipation are associated with infrared and ultraviolet regularization, respectively. These limits seem to be interchangeable.
  \item The limit of infinite volume, or $K_L \rightarrow 0$, for the quantities of interest can be performed by passing from sums to integrals. For massive field quanta, no infrared problems arise so that this limit turns out to be harmless.
  \item For a finite limit of the propagator to exist for vanishing dissipation ($\gamma \rightarrow 0$), the parameters $Z$ and $\lambda'$ need to be chosen properly. Different choices of $Z$ and $\lambda'$ are possible [see (\ref{phi4Zchoice}), (\ref{phi4lpchoice}) or (\ref{phi4Zchoicex}), (\ref{phi4lpchoicex})]; these have an influence on the interpretation of the remaining parameters $m$ and $\lambda$ of the final quartic scalar field theory.
\end{itemize}

The dependence of some quantity of interest on $m$ and $\lambda$ can be further narrowed down by dimensional analysis. For example, the dimensionless propagator $m^2 \Delta_{\omega \, \bm{k}}$ can depend only on $\lambda m^{d-3}$ and the dimensionless arguments $\omega/m$ and $\bm{k}/m$ of the propagator. Relativistic covariance implies a combined dependence on $(\omega^2-\omega_k^2)/m^2$.

Renormalization-group theory suggests that even $\lambda$ should not be considered as a free parameter. In the nonperturbative approach illustrated in Figure~\ref{figRGflow}, there should be a well-defined model associated with a fixed point. From this perspective, $\varphi^4$ theory with a dimensionless coupling constant close to the critical value $\lambda^*$ only serves as a minimal model developing the universal features of scalar field theory on larger length and time scales. The only remaining parameter is a physical correlation length related to a physical mass (which we keep at $m$ by adjusting $\lambda'$). The critical value $\lambda^*$ can be estimated with the perturbative approach as described in Section~\ref{secrenobaseq}. Low-order perturbation expansions, such as (\ref{phi4propscaling}) or (\ref{4corrphi4Lam2}), should then be evaluated with the critical coupling constant directly at the physical length scale, that is, for $\lambda m^{d-3}=\lambda^*$. By analyzing correction terms in the perturbation expansion for the effective interaction vertex, we have actually found the lowest-order estimate for $\lambda^*$ as a function of space dimensionality $d$ (see Figure~\ref{figphi4fixedpoint}). For $\varphi^4$ theory it is actually known that an asymptotically safe theory associated with a nontrivial fixed point arises in one and two space dimensions \cite{GlimmJaffeII}. In three space dimensions, lowest-order perturbation theory gives $\lambda^*=0$ and hence suggests a noninteracting theory; however, higher-order terms might always change the situation.

\chapter{Quantum electrodynamics}
After developing the basic elements of a philosophically founded approach to quantum field theory and illustrating them for the toy model of scalar field theory, we are now ready to study one of the fundamental interactions of nature: electromagnetism. In doing so, we encounter two major challenges. First, we realize from (\ref{Maxpotphieq}) and (\ref{MaxpotAeq}) that we need to introduce four-vectors for the charge/current densities and the scalar/vector potentials of the electromagnetic field. We hence need to pay much closer attention to the behavior of fields under Lorentz transformations than for the scalar theory. Second, according to (\ref{MaxgaugeAeq}) and (\ref{Maxgaugephieq}), we need to deal with the evolution of gauge degrees of freedom that do not carry any direct physical information. In particular, we need to identify the physically relevant degrees of freedom. These two challenges result from the Lorentz and gauge symmetries of \index{Electrodynamics}electromagnetism.

For the discussion of the above issues in this chapter, we mostly restrict ourselves to reversible dynamics. A detailed formulation of irreversible dynamics will be included in the more complete version of this manuscript that will be published under the title ``A Philosophical Approach to Quantum Field Theory'' by Cambridge University Press in 2017. The more complete story will then be used to do some simple calculations for a number of specific problems (form factors and magnetic moments, electron-positron annihilation into two photons, hydrogen fine-structure, Lamb shift).

Throughout this chapter on electrodynamics, we extend our list (\ref{hbarcconv}) of natural units to $\hbar=c=\epsilon_0=1$, where $\hbar$ is the reduced Planck constant, $c$ is the speed of light, and $\epsilon_0$ is the electric constant or permittivity of free space. Only one further unit, naturally taken as mass or energy, remains to be specified. In $d$ space dimensions, $\hbar c \epsilon_0$ has the units of ${\rm C}^2 {\rm m}^{3-d}$ in terms of Coulombs and meters. For $d=3$, the electric charge in natural units becomes dimensionless and the elementary charge is given by $e_0 = \sqrt{4 \pi \alpha} \approx 0.30282212$, where $\alpha$ is the fine-structure constant. For $d=1$, the electric charge has units of energy.

\section{The Dirac equation}
As a classical starting point for the development of quantum electrodynamics, we so far have the Maxwell equations for electromagnetic fields (see Section~\ref{secsymmetries}). As a further input, we need a theory of electrons and positrons or, more generally, of fermions as the building blocks of matter. This is provided by the relativistic generalization of the Schr\"odinger equation, which is known as the Dirac equation. In this section, we first present and discuss Dirac's equation in three space dimensions. In the last subsection, we summarize the changes required in going from three dimensions to one space dimension. Our interest in one space dimension is motivated by the availability of nontrivial and useful exact results.

\subsection{Wave equations for free particles}
Wave equations for free particles are closely related to energy-momentum relations. For example, with the help of the replacements
\begin{equation}\label{Epreplaces}
    E_p \rightarrow \iR \frac{\partial}{\partial t} \,, \qquad
    \bm{p} \rightarrow \iR \bm{\nabla} \,,
\end{equation}
the nonrelativistic energy-momentum relation $E_p=\bm{p}^2/(2m)$ for a free particle of mass $m$ leads to the \emph{Schr\"odinger equation}
\begin{equation}\label{Schroedingereqwf}
    \iR \frac{\partial\psi}{\partial t} = - \frac{1}{2m} \bm{\nabla}^2 \psi \,,
\end{equation}
for the wave function $\psi = \psi_{\bm{x},t}$. In view of the quadratic form of the relativistic energy-momentum relation $E_p^2=m^2+\bm{p}^2$ for a particle of mass $m$, it is natural to expect that the relativistic generalization of the Schr\"odinger equation contains second-order derivatives with respect to both time and position. For the \emph{Klein-Gordon equation} for a classical scalar field $\varphi = \varphi_{\bm{x},t}$,
\begin{equation}\label{KleinGordoneq}
    \left( \frac{\partial^2}{\partial t^2} - \bm{\nabla}^2 + m^2 \right) \varphi = 0 \,,
\end{equation}
this indeed is the case.

It was one of Dirac's ingenious ideas to implement the relativistic energy-momentum relation in a first-order differential equation by introducing four matrices $\gamma^\mu$, $\mu=0,1,2,3$, satisfying the anticommutation relations
\begin{equation}\label{Diracgammaalgebra}
    \gamma^\mu \gamma^\nu + \gamma^\nu \gamma^\mu = - 2 \, \eta^{\mu\nu} \, 1 \,,
\end{equation}
where `$1$' stands for the unit matrix and $\eta^{\mu\nu}=\eta_{\mu\nu}$ represents the \index{Minkowski metric}Minkowski metric with signature $(-,+,+,+)$, that is, $\eta^{00}=\eta_{00}=-1$. These relations are chosen such that the operator identity $(\gamma^\mu \partial_\mu)^2 = - 1 \, \partial^\mu \partial_\mu$ holds. There are many ways of representing the anticommutation relations (\ref{Diracgammaalgebra}). However, any representation in 3+1 dimensions requires at least $4 \times 4$ matrices. We here make the particular choice
\begin{equation}\label{Diracmatrices0}
    \gamma^0 = \left( \begin{array}{rrrr}
      1 & 0 & 0 & 0 \\
      0 & 1 & 0 & 0 \\
      0 & 0 & -1 & 0 \\
      0 & 0 & 0 & -1 \\
    \end{array} \right) \,,
\end{equation}
and
\begin{equation}\label{Diracmatrices}
    \gamma^j = \left( \begin{array}{cc}
      0 & \sigma^j \\
      -\sigma^j & 0 \\
    \end{array} \right) \quad \mbox{for } j=1,2,3 \,,
\end{equation}
in terms of Pauli's famous $2 \times 2$ spin matrices,
\begin{equation}\label{Paulimatrices}
    \sigma^1 = \left( \begin{array}{rr}
      0 & 1 \\
      1 & 0 \\
    \end{array} \right) , \quad
    \sigma^2 = \left( \begin{array}{rr}
      0 & -\iR \\
      \iR & 0 \\
    \end{array} \right) , \quad
    \sigma^3 = \left( \begin{array}{rr}
      1 & 0 \\
      0 & -1 \\
    \end{array} \right) \,.
\end{equation}
Note the property ${\rm tr}(\gamma^\mu)=0$. For a representation in terms of $4 \times 4$ matrices, (\ref{Diracgammaalgebra}) implies
\begin{equation}\label{Diracgammaalgebratr}
    {\rm tr}(\gamma^\mu \gamma^\nu) = - 4 \, \eta^{\mu\nu} \,.
\end{equation}

Once one has chosen a representation of the anticommutation relations (\ref{Diracgammaalgebra}), the \emph{Dirac equation} can be written as
\begin{equation}\label{Diraceq}
    (\iR \gamma^\mu \partial_\mu - m ) \psi = 0 \,,
\end{equation}
where $\psi = \psi_{\bm{x},t}$ is a column vector of four complex-valued fields. By acting with $\iR \gamma^\mu \partial_\mu + m$ on (\ref{Diraceq}) one realizes that each component of $\psi$ satisfies the Klein-Gordon equation. Two components describe a spin $1/2$ fermion, the other two components its antiparticle.

\subsection{Relativistic covariance}
To establish the covariance of the Dirac equation (\ref{Diraceq}), we need to determine the transformation behavior of the four-component object $\psi$ under Lorentz transformations. Based on the anticommutation relations (\ref{Diracgammaalgebra}), one can show that the matrices
\begin{equation}\label{sigmagenerators}
    \sigma^{\mu\nu} = \frac{\iR}{4} ( \gamma^\mu \gamma^\nu - \gamma^\nu \gamma^\mu ) \,,
\end{equation}
provide a representation of the infinitesimal generators of the group of Lorentz transformations (see, for example, Section 2.2 of \cite{BjorkenDrellQM} or Section 3.2 of \cite{PeskinSchroeder}). It is hence interesting to consider the vector-space of four-component column vectors on which the matrices $\gamma^\mu$ and hence also $\sigma^{\mu\nu}$ are acting. The elements of this vector space are known as spinors to point out the fact that they come with a well-defined transformation behavior provided by the generators (\ref{sigmagenerators}) but that a rotation by $2\pi$ is represented by the negative of the unit matrix rather than the unit matrix expected for vectors.

For our choice of the matrices $\gamma^\mu$, the matrices $\sigma^{jk}$ for $j,k=1,2,3$ are given by
\begin{equation}\label{Paulimatrices4r}
    \sigma^{jk} = \frac{1}{2} \, \epsilon^{jkl} \left( \begin{array}{cc}
      \sigma^l & 0 \\
      0 & \sigma^l \\
    \end{array} \right) \,,
\end{equation}
along with $\sigma^{00} = 0$ and $\sigma^{0j} = - \sigma^{j0} = i \, \gamma^0 \gamma^j/2$, that is,
\begin{equation}\label{Paulimatrices4b}
    \sigma^{0j} = \frac{\iR}{2} \, \left( \begin{array}{cc}
      0 & \sigma^j \\
      \sigma^j & 0 \\
    \end{array} \right) \,.
\end{equation}
In (\ref{Paulimatrices4r}) we recognize two separate representations of rotations in the upper and lower halves of the four-component spinors, corresponding to two particles with spin $1/2$. We interpret these particles as electrons and positrons. The generators $\sigma^{0j}$ in (\ref{Paulimatrices4b}) mix space and time and hence represent the infinitesimal generators of Lorentz boosts in the $j$-direction.

The infinitesimal generator for the most general boost in spinor space,
\begin{equation}\label{boostspinor}
    \Theta_{\rm sp} = \iR \sum_{j=1}^3 \theta_j \sigma^{0j} =
    - \frac{1}{2} \left( \begin{array}{cccc}
      0 & 0 & \theta_3 & \theta_1-\iR\theta_2 \\
      0 & 0 & \theta_1+\iR\theta_2 & -\theta_3 \\
      \theta_3 & \theta_1-\iR\theta_2 & 0 & 0 \\
      \theta_1+\iR\theta_2 & -\theta_3 & 0 & 0 \\
    \end{array} \right) \,,
\end{equation}
where $\bm{\theta}=(\theta_1,\theta_2,\theta_3)$ is known as the rapidity vector, should be compared to the generator for boosts of four-vectors,
\begin{equation}\label{boostvector}
    \Theta_{\rm vec} = - \left( \begin{array}{cccc}
      0 & \theta_1 & \theta_2 & \theta_3 \\
      \theta_1 & 0 & 0 & 0 \\
      \theta_2 & 0 & 0 & 0 \\
      \theta_3 & 0 & 0 & 0 \\
    \end{array} \right) \,.
\end{equation}
Rather than elaborating the details of the algebraic arguments leading to the infinitesimal generators (\ref{boostspinor}) and (\ref{boostvector}), we here focus on studying their properties, relations, and consequences.

With the identities
\begin{equation}\label{boostvectorpowers}
    \Theta_{\rm sp}^2 = \frac{\bm{\theta}^2}{4} \, 1 \,, \qquad
    \Theta_{\rm vec}^3 = \bm{\theta}^2 \, \Theta_{\rm vec} \,,
\end{equation}
we find the representation of the most general finite boost acting on spinors,
\begin{equation}\label{boostspinorexp}
    \exp\{\Theta_{\rm sp}\} = \cosh \frac{\theta}{2}  \, 1
    + \frac{2 \sinh \theta/2}{\theta} \, \Theta_{\rm sp} \,,
\end{equation}
and on four-vectors,
\begin{equation}\label{boostvectorexp}
    \exp\{\Theta_{\rm vec}\} = 1 + \frac{\sinh \theta}{\theta} \, \Theta_{\rm vec}
    + \frac{\cosh \theta - 1}{\theta^2} \, \Theta_{\rm vec}^2 \,.
\end{equation}

A straightforward calculation gives the identity
\begin{equation}\label{boostrels}
    \exp\{\Theta_{\rm sp}\} \, \gamma^0 \, \exp\{\Theta_{\rm sp}\} = \gamma^0 \,,
\end{equation}
which actually is a direct consequence of the anticommutation relation $\gamma^0 \Theta_{\rm sp} + \Theta_{\rm sp} \gamma^0 = 0$, and
\begin{eqnarray}
    \exp\{\Theta_{\rm sp}\} \, \gamma^0 \gamma^\mu \exp\{\Theta_{\rm sp}\} &=&
    \gamma^0 \gamma^\mu + \frac{\sinh \theta}{\theta}
    (\Theta_{\rm sp} \gamma^0 \gamma^\mu + \gamma^0 \gamma^\mu \Theta_{\rm sp}) \nonumber\\
    &+& \frac{\cosh \theta - 1}{\theta^2}
    \left( 2 \Theta_{\rm sp} \gamma^0 \gamma^\mu \Theta_{\rm sp}
    + \frac{1}{2} \theta^2 \gamma^0 \gamma^\mu \right) \nonumber\\
    &=& {\left( \exp\{\Theta_{\rm vec}\} \right)^\mu}_\nu \gamma^0 \gamma^\nu \,.
\label{boostrelv}
\end{eqnarray}
These identities show how we can build Lorentz scalars and four-vectors from two spinors. Moreover, they are the key to establishing the relativistic covariance of the Dirac equation (\ref{Diraceq}). Finally, note that they can be combined into
\begin{equation}\label{boostrelc}
    \exp\{- \Theta_{\rm sp}\} \, \gamma^\mu \exp\{\Theta_{\rm sp}\} =
    {\left( \exp\{\Theta_{\rm vec}\} \right)^\mu}_\nu \gamma^\nu \,.
\end{equation}

By boosting the energy-momentum vector from the local rest frame to an arbitrary frame,
\begin{equation}\label{boostEpvector}
    \left( \begin{array}{c}
      E_p \\ \bm{p} \\
    \end{array} \right) =
    \exp\{\Theta_{\rm vec}\} \left( \begin{array}{c}
      m \\ 0 \\ 0 \\ 0 \\
    \end{array} \right) =
    \left( \begin{array}{c}
      m \cosh \theta \\ - m \sinh \theta \;\; \bm{\theta}/\theta \\
    \end{array} \right) \,,
\end{equation}
we can relate the rapidity and momentum vectors. With
\begin{equation}\label{coshident}
    \cosh \frac{\theta}{2} = \sqrt{\frac{1+\cosh \theta}{2}}  =
    \sqrt{\frac{E_p+m}{2m}} \,, \qquad
    \tanh \frac{\theta}{2} = \frac{\sinh \theta}{1+\cosh \theta} = \frac{p}{E_p+m} \,,
\end{equation}
we can rewrite (\ref{boostspinorexp}) as
\begin{equation}\label{boostspinorexpx}
    \exp\{\Theta_{\rm sp}\} = \sqrt{\frac{E_p+m}{2m}}
    \left( 1 + \frac{2p}{E_p+m} \frac{\Theta_{\rm sp}}{\theta} \right) \,,
\end{equation}
where, according to (\ref{boostEpvector}), $\bm{\theta}/\theta = -\bm{p}/p$.

\subsection{Planar wave solutions}\label{secDiracplanewave}
To obtain solutions of the Dirac equation (\ref{Diraceq}), we introduce the Fourier transform
\begin{equation}\label{spinorwvFou}
    \psi_{\bm{x}} = \frac{1}{\sqrt{V}} \sum_{\bm{p} \in K^3} \psi_{\bm{p}} \, \eR^{- \iR \bm{p} \cdot \bm{x}} \,,
\end{equation}
as in Section~\ref{secFourierfields}. We thus obtain
\begin{equation}\label{DiraceqFou}
    \iR \frac{\partial\psi_{\bm{p}}}{\partial t} = (m \gamma^0 - p_j \gamma^0 \gamma^j) \psi_{\bm{p}} \,.
\end{equation}
For every $\bm{p}$, (\ref{DiraceqFou}) is a set of coupled evolution equations for the four components of the spinor $\psi_{\bm{p}}$. For $\bm{p}=\bm{0}$, the components are uncoupled and we can introduce the four modes associated with the basis vectors
\begin{equation}\label{spinorsrest}
    \left( \begin{array}{c}
      1 \\
      0 \\
      0 \\
      0 \\
    \end{array} \right) \,, \quad
    \left( \begin{array}{c}
      0 \\
      1 \\
      0 \\
      0 \\
    \end{array} \right) \,, \quad
    \left( \begin{array}{c}
      0 \\
      0 \\
      1 \\
      0 \\
    \end{array} \right) \,, \quad
    \left( \begin{array}{c}
      0 \\
      0 \\
      0 \\
      1 \\
    \end{array} \right) \,,
\end{equation}

Starting from the four canonical base vectors (\ref{spinorsrest}) of the spinor space in the rest frame, we can now use (\ref{boostspinorexpx}) to generate the spinors for a particle with momentum $\bm{p}$. The result is given by
\begin{equation}\label{spinorsu1}
    u^{1/2}_{\bm{p}} = \sqrt{\frac{E_p+m}{2m}}
    \left( \begin{array}{c}
      1 \\
      0 \\
      \hat{p}_3 \\
      \hat{p}_1 + \iR \hat{p}_2 \\
    \end{array} \right) ,
\end{equation}
\begin{equation}\label{spinorsu2}
    u^{-1/2}_{\bm{p}} = \sqrt{\frac{E_p+m}{2m}}
    \left( \begin{array}{c}
      0 \\
      1 \\
      \hat{p}_1 - \iR \hat{p}_2 \\
      - \hat{p}_3 \\
     \end{array} \right) ,
\end{equation}
\begin{equation}\label{spinorsv1}
    v^{1/2}_{\bm{p}} = \sqrt{\frac{E_p+m}{2m}}
    \left( \begin{array}{c}
      \hat{p}_1 - \iR \hat{p}_2 \\
      - \hat{p}_3 \\
      0 \\
      1 \\
     \end{array} \right) ,
\end{equation}
and
\begin{equation}\label{spinorsv2}
    v^{-1/2}_{\bm{p}} = \sqrt{\frac{E_p+m}{2m}}
    \left( \begin{array}{c}
      \hat{p}_3 \\
      \hat{p}_1 + \iR \hat{p}_2 \\
      1 \\
      0 \\
    \end{array} \right) ,
\end{equation}
where we have introduced conveniently normalized momenta $\hat{p}_j = p_j/(E_p+m)$. The $u$-spinors describe electrons, the $v$-spinors describe positrons. In accordance with (\ref{Paulimatrices4r}), the superscripts $\pm 1/2$ correspond to spin values. For our sign convention in the Fourier transform (\ref{spinorwvFou}), one can verify that $u^{\pm 1/2}_{-\bm{p}}$ are eigenmodes of the right-hand side of (\ref{DiraceqFou}) with eigenvalues $+E_p$, whereas $v^{\pm 1/2}_{\bm{p}}$ are eigenmodes with eigenvalues $-E_p$. The existence of negative energy eigenvalues is considered as a problem of the Dirac equation and led to the interpretation of antiparticles as holes in a sea of particles (see, for example, Chapter~5 of \cite{BjorkenDrellQM}). However, the \index{Second quantization}second-quantization procedure with explicit particles and antiparticles, as developed in the following sections, actually works without any problems.

For future reference, we compile a number of useful properties of the spinors (\ref{spinorsu1})--(\ref{spinorsv2}). For $\bm{p}=\bm{0}$, the column vectors (\ref{spinorsu1})--(\ref{spinorsv2}) have been chosen to form an orthogonal basis of the four-dimensional vector space characterizing the spins of electrons and positrons. For arbitrary $\bm{p}$, we still have orthogonality and completeness relations. Actually, there are two versions, one for spinors with equal $\bm{p}$ and one for spinors with opposite $\bm{p}$. For example, one form of the orthogonality relations reads
\begin{equation}\label{upsorthog}
    \bar{u}^{\sigma}_{\bm{p}} \, u^{\sigma'}_{\bm{p}} =
    -\bar{v}^{\sigma}_{\bm{p}} \, v^{\sigma'}_{\bm{p}} =
    \delta_{\sigma\sigma'} , \quad
    \bar{u}^{\sigma}_{\bm{p}} \, v^{\sigma'}_{\bm{p}} =
    \bar{v}^{\sigma}_{\bm{p}} \, u^{\sigma'}_{\bm{p}} = 0 ,
\end{equation}
where we have introduced $\bar{u} = u^* \gamma^0$ and $\bar{v} = v^* \gamma^0$; an asterisk implies both complex conjugation and transposition of a vector or matrix and $\gamma^0$ is the $4\times 4$ matrix defined in (\ref{Diracmatrices0}).\footnote{The factor $\gamma^0$ in these definitions is motivated by the occurrence of $\gamma^0$ in (\ref{boostrelv}).} To establish the corresponding completeness relation, we consider the $4 \times 4$ matrices obtained as sums over tensor products of spinors and write them in a compact form in terms of the Pauli matrices and $2 \times 2$ unit matrices. We find
\begin{eqnarray}
    \Lambda_{\rm e}(\bm{p}) &=& \sum_\sigma u^{\sigma}_{\bm{p}} \bar{u}^{\sigma}_{\bm{p}} = \frac{1}{2m}
    \left( \begin{array}{cc}
      (E_p+m)\bm{1} & - p_j \sigma^j \\
      p_j \sigma^j & - (E_p-m)\bm{1} \\
    \end{array} \right) \nonumber \\
    &=& \frac{1}{2m} \, (E_p \gamma^0 - p_j \gamma^j + m \, \bm{1}) \,,
\label{projectorele}
\end{eqnarray}
and
\begin{eqnarray}
    \Lambda_{\rm p}(\bm{p}) &=& - \sum_\sigma v^{\sigma}_{\bm{p}} \bar{v}^{\sigma}_{\bm{p}} = \frac{1}{2m}
    \left( \begin{array}{cc}
      - (E_p-m)\bm{1} & p_j \sigma^j \\
      -p_j \sigma^j & (E_p+m)\bm{1} \\
    \end{array} \right) \nonumber \\
    &=& \frac{1}{2m} \, (- E_p \gamma^0 + p_j \gamma^j + m \, \bm{1}) \,.
\label{projectorpos}
\end{eqnarray}
For $\bm{p}=\bm{0}$, we realize that $\Lambda_{\rm e}(\bm{p})$ and $\Lambda_{\rm p}(\bm{p})$ may be regarded as projectors to the electron and positron degrees of freedom, respectively. In (\ref{upsorthog})--(\ref{projectorpos}), there is no integration over $\bm{p}$ and we indicate the summations over $\sigma$ explicitly. Equations (\ref{projectorele}) and (\ref{projectorpos}) imply the completeness relation
\begin{equation}\label{projectorsum}
    \Lambda_{\rm e}(\bm{p}) + \Lambda_{\rm p}(\bm{p}) = \bm{1} ,
\end{equation}
and the additional property
\begin{equation}\label{projectordif}
    \Lambda_{\rm e}(\bm{p}) - \Lambda_{\rm p}(-\bm{p}) = \frac{E_p}{m} \, \gamma^0 .
\end{equation}
We further find the identities
\begin{equation}\label{equalpspinsumu}
  \sum_\sigma \bar{u}^{\sigma}_{\bm{p}} \, \gamma^\mu \, u^{\sigma}_{\bm{p}}
  = {\rm tr}[\Lambda_{\rm e}(\bm{p}) \gamma^\mu] = 2 \frac{p^\mu}{m} \,,
\end{equation}
and
\begin{equation}\label{equalpspinsumv}
  \sum_\sigma \bar{v}^{\sigma}_{\bm{p}} \, \gamma^\mu \, v^{\sigma}_{\bm{p}}
  = - {\rm tr}[\Lambda_{\rm p}(\bm{p}) \gamma^\mu] = 2 \frac{p^\mu}{m} \,,
\end{equation}
where (\ref{Diracgammaalgebratr}) has been used for evaluating the traces.

As the spinors $v$ are obtained from the spinors $u$ by exchanging their upper and lower halves and flipping the spins, we immediately obtain the relations
\begin{equation}\label{uuvvident}
    \bar{u}^{\sigma}_{\bm{p}} \, \gamma^\mu \, u^{\sigma'}_{\bm{p}'} =
    \bar{v}^{-\sigma}_{\bm{p}} \, \gamma^\mu \, v^{-\sigma'}_{\bm{p}'} ,
\end{equation}
and similarly
\begin{equation}\label{uvvuident}
    \bar{u}^{\sigma}_{\bm{p}} \, \gamma^\mu \, v^{\sigma'}_{\bm{p}'} =
    \bar{v}^{-\sigma}_{\bm{p}} \, \gamma^\mu \, u^{-\sigma'}_{\bm{p}'} ,
\end{equation}
as well as
\begin{equation}\label{uvvuidentbare}
    \bar{u}^{\sigma}_{\bm{p}} \, v^{\sigma'}_{\bm{p}'} =
    - \bar{v}^{-\sigma}_{\bm{p}} \, u^{-\sigma'}_{\bm{p}'} .
\end{equation}
We further find the symmetry property
\begin{equation}\label{uuident}
    \bar{u}^{\sigma}_{\bm{p}} \, \gamma^\mu \, u^{\sigma'}_{\bm{p}'} =
    (\bar{u}^{\sigma'}_{\bm{p}'} \, \gamma^\mu \, u^{\sigma}_{\bm{p}})^* ,
\end{equation}
and, in view of Eq.~(\ref{uuvvident}), we similarly have
\begin{equation}\label{vvident}
    \bar{v}^{\sigma}_{\bm{p}} \, \gamma^\mu \, v^{\sigma'}_{\bm{p}'} =
    (\bar{v}^{\sigma'}_{\bm{p}'} \, \gamma^\mu \, v^{\sigma}_{\bm{p}})^* .
\end{equation}
These symmetry properties, as well as the further relation
\begin{equation}\label{uvident}
    \bar{u}^{\sigma}_{\bm{p}} \, \gamma^\mu \, v^{\sigma'}_{\bm{p}'} =
    (\bar{v}^{\sigma'}_{\bm{p}'} \, \gamma^\mu \, u^{\sigma}_{\bm{p}})^* ,
\end{equation}
are a direct consequence of the self-adjointness of the matrices $\gamma^0 \gamma^\mu$. These relationships can be checked explicitly by means of Tables \ref{tablecouplmatrix} and \ref{tablecouplmatriy}. The further identities
\begin{equation}\label{uupsumident}
    (p_j-p'_j) \, \bar{u}^{\sigma}_{\bm{p}} \, \gamma^j \, u^{\sigma'}_{\bm{p}'} =
    (E_p-E_{p'}) \, \bar{u}^{\sigma}_{\bm{p}} \, \gamma^0 \, u^{\sigma'}_{\bm{p}'} \,,
\end{equation}
and
\begin{equation}\label{uvpsumident}
    (p_j-p'_j) \, \bar{v}^{-\sigma}_{-\bm{p}} \, \gamma^j \, u^{\sigma'}_{\bm{p}'} =
    -(E_p+E_{p'}) \, \bar{v}^{-\sigma}_{-\bm{p}} \, \gamma^0 \, u^{\sigma'}_{\bm{p}'} \,,
\end{equation}
can also be checked by straightforward calculations.

\begin{table}
\caption[ ]{Components of $2m \, \bar{u}^{\sigma}_{\bm{p}} \, \gamma^\mu \, u^{\sigma'}_{\bm{p}'}/ \sqrt{(E_p+m)(E_{p'}+m)}$ in terms of $\hat{p}_j = p_j/(E_p+m)$, $\hat{p}_\pm = \hat{p}_1 \pm \iR\hat{p}_2$, $\hat{p}'_j = p'_j/(E_{p'}+m)$ and $\hat{p}'_\pm = \hat{p}'_1 \pm \iR\hat{p}'_2$; the pairs of spin values $\sigma,\sigma'$ are given in the first row, the values of the space-time index $\mu$ in the first column}
\begin{center}
\begin{tabular}{|c|c|c|c|c|}
\hline
& $\scriptstyle 1/2$, $\scriptstyle 1/2$ & $\scriptstyle -1/2$, $\scriptstyle -1/2$
& $\scriptstyle 1/2$, $\scriptstyle -1/2$ & $\scriptstyle -1/2$, $\scriptstyle 1/2$ \\
\hline
$\scriptstyle 0$ &
$\scriptstyle 1+\hat{p}_-\hat{p}'_++\hat{p}_3\hat{p}'_3$ &
$\scriptstyle 1+\hat{p}_+\hat{p}'_-+\hat{p}_3\hat{p}'_3$ &
$\scriptstyle \hat{p}_3\hat{p}'_--\hat{p}_-\hat{p}'_3$ &
$\scriptstyle -\hat{p}_3\hat{p}'_++\hat{p}_+\hat{p}'_3$ \\
$\scriptstyle 1$ &
$\scriptstyle \hat{p}_-+\hat{p}'_+$ &
$\scriptstyle \hat{p}_++\hat{p}'_-$ &
$\scriptstyle \hat{p}_3-\hat{p}'_3$ &
$\scriptstyle -\hat{p}_3+\hat{p}'_3$ \\
$\scriptstyle 2$ &
$\scriptstyle \iR\hat{p}_--\iR\hat{p}'_+$ &
$\scriptstyle -\iR\hat{p}_++\iR\hat{p}'_-$ &
$\scriptstyle -\iR\hat{p}_3+\iR\hat{p}'_3$ &
$\scriptstyle -\iR\hat{p}_3+\iR\hat{p}'_3$ \\
$\scriptstyle 3$ &
$\scriptstyle \hat{p}_3+\hat{p}'_3$ &
$\scriptstyle \hat{p}_3+\hat{p}'_3$ &
$\scriptstyle -\hat{p}_-+\hat{p}'_-$ &
$\scriptstyle \hat{p}_+-\hat{p}'_+$ \\
\hline
\end{tabular}
\end{center}
\label{tablecouplmatrix}
\end{table}

\begin{table}[t]
\caption[ ]{Components of $2m \, \bar{v}^{-\sigma}_{-\bm{p}} \, \gamma^\mu \, u^{\sigma'}_{\bm{p}'}/ \sqrt{(E_p+m)(E_{p'}+m)}$ in terms of $\hat{p}_j = p_j/(E_p+m)$, $\hat{p}_\pm = \hat{p}_1 \pm \iR\hat{p}_2$, $\hat{p}'_j = p'_j/(E_{p'}+m)$ and $\hat{p}'_\pm = \hat{p}'_1 \pm \iR\hat{p}'_2$; the pairs of spin values $\sigma,\sigma'$ are given in the first row, the values of the space-time index $\mu$ in the first column}
\begin{center}
\begin{tabular}{|c|c|c|c|c|}
\hline
& $\scriptstyle 1/2$, $\scriptstyle 1/2$ & $\scriptstyle -1/2$, $\scriptstyle -1/2$
& $\scriptstyle 1/2$, $\scriptstyle -1/2$ & $\scriptstyle -1/2$, $\scriptstyle 1/2$ \\
\hline
$\scriptstyle 0$ &
$\scriptstyle -\hat{p}_3+\hat{p}'_3$ &
$\scriptstyle \hat{p}_3-\hat{p}'_3$ &
$\scriptstyle -\hat{p}_-+\hat{p}'_-$ &
$\scriptstyle -\hat{p}_++\hat{p}'_+$ \\
$\scriptstyle 1$ &
$\scriptstyle -\hat{p}_3\hat{p}'_+-\hat{p}_-\hat{p}'_3$ &
$\scriptstyle \hat{p}_3\hat{p}'_-+\hat{p}_+\hat{p}'_3$ &
$\scriptstyle 1-\hat{p}_-\hat{p}'_-+\hat{p}_3\hat{p}'_3$ &
$\scriptstyle 1-\hat{p}_+\hat{p}'_++\hat{p}_3\hat{p}'_3$ \\
$\scriptstyle 2$ &
$\scriptstyle \iR\hat{p}_3\hat{p}'_+-\iR\hat{p}_-\hat{p}'_3$ &
$\scriptstyle \iR\hat{p}_3\hat{p}'_--\iR\hat{p}_+\hat{p}'_3$ &
$\scriptstyle -\iR(1+\hat{p}_-\hat{p}'_-+\hat{p}_3\hat{p}'_3)$ &
$\scriptstyle \iR(1+\hat{p}_+\hat{p}'_++\hat{p}_3\hat{p}'_3)$ \\
$\scriptstyle 3$ &
$\scriptstyle 1+\hat{p}_-\hat{p}'_+-\hat{p}_3\hat{p}'_3$ &
$\scriptstyle -1-\hat{p}_+\hat{p}'_-+\hat{p}_3\hat{p}'_3$ &
$\scriptstyle -\hat{p}_3\hat{p}'_--\hat{p}_-\hat{p}'_3$ &
$\scriptstyle -\hat{p}_3\hat{p}'_+-\hat{p}_+\hat{p}'_3$ \\
\hline
\end{tabular}
\end{center}
\label{tablecouplmatriy}
\end{table}

\subsection{Interaction with electromagnetic fields}\label{secpartEMfield}
So far, we have considered free particles. To incorporate the effect of an electromagnetic field into the Dirac equation, we need to know the interaction energy. In the rest frame, the energy of and electron is $-e_0 \phi = e_0 A_0$, and the energy for a positron is $e_0 \phi = - e_0 A_0$. Both cases can be combined and generalized in a Lorentz invariant form by considering $e_0 \gamma^\mu A_\mu$. One hence generalizes the Dirac equation (\ref{Diraceq}) in the presence of an electromagnetic field as follows,
\begin{equation}\label{DiraceqA}
    (\iR \gamma^\mu \partial_\mu - e_0 \gamma^\mu A_{\bm{x} \, \mu} - m ) \psi_{\bm{x}} = 0 \,.
\end{equation}
The formal replacement $\partial_\mu \rightarrow \partial_\mu + \iR e_0 A_\mu$ is known as minimal coupling (see, for example, (1.26) of \cite{BjorkenDrellQM}).

Equation (\ref{DiraceqA}) describes the effect of an electromagnetic field on charged fermions. Conversely, we would like to describe the effect of charged fermions on the electromagnetic field. According to (\ref{Maxpotphieq}), (\ref{MaxpotAeq}), this can be done by specifying the electric charge and current densities. The identity (\ref{boostrels}) implies that $\bar{\psi}_{\bm{x}} \psi_{\bm{x}}$ is a Lorentz scalar, whereas $\psi_{\bm{x}}^* \psi_{\bm{x}} = \bar{\psi}_{\bm{x}} \gamma^0 \psi_{\bm{x}}$ is the zero-component of the four-vector implied by the identity (\ref{boostrelv}) [note that $\Theta_{\rm sp}^* = \Theta_{\rm sp}$]. We identify this quantity as the electric current density four-vector,
\begin{equation}\label{currentclass}
    J_{\bm{x}}^\mu = -e_0 \bar{\psi}_{\bm{x}} \gamma^\mu \psi_{\bm{x}} \,.
\end{equation}
This completes our description of the mutual coupling between particles and electromagnetic fields.

\subsection{From 3+1 to 1+1 dimensions}\label{secfrom3to1}
Up to now we have considered the Dirac equation only in three space dimensions. For a more general development of quantum electrodynamics it is helpful to consider also the simpler case of one space dimension, for which even some closed-form results can be obtained.

\paragraph*{Matrices and wave equation.}
In one space dimension, the anticommutation relations (\ref{Diracgammaalgebra}) can be satisfied by $2 \times 2$ matrices. We use the following matrices,
\begin{equation}\label{Schwingermatrices}
    \eta = \left( \begin{array}{rr}
      -1 & 0 \\
      0 & 1 \\
    \end{array} \right) \,, \quad
    \gamma^0 = \left( \begin{array}{rr}
      1 &  0 \\
      0 & -1 \\
    \end{array} \right) \,, \quad
    \gamma^1 = \left( \begin{array}{rr}
      0 & 1 \\
      -1 & 0 \\
    \end{array} \right) \,.
\end{equation}
Spinors thus have two components so that electrons and positrons lose their spin index. With the matrices (\ref{Schwingermatrices}), the Dirac equation (\ref{DiraceqA}) for the two-component spinor $\psi_x$ can be written in the more explicit form
\begin{equation}\label{DiraceqAS}
    \left( \begin{array}{cc}
     \iR \frac{\partial}{\partial t} + e_0 A^0_x - m & \iR \frac{\partial}{\partial x} - e_0 A^1_x \\
     \iR \frac{\partial}{\partial x} - e_0 A^1_x & \iR \frac{\partial}{\partial t} + e_0 A^0_x + m \\
    \end{array} \right) \psi_x =
    \left( \begin{array}{c}
      0 \\
      0 \\
    \end{array} \right) \,.
\end{equation}
For $m=0$, this equation is symmetric under the exchange of the two components of the spinor $\psi_x$.

\paragraph*{Relativistic covariance and modes.}
We next need to establish the Lorentz covariance of the Dirac equation in one space dimension. Boosts with rapidity $\theta$ are represented by the following infinitesimal generators and their exponentials:
\begin{equation}\label{boostvectorS}
    \Theta_{\rm vec} = - \left( \begin{array}{cc}
      0 & \theta \\
      \theta & 0 \\
    \end{array} \right) \,, \qquad
    \exp\{\Theta_{\rm vec}\} = \left( \begin{array}{rr}
      \cosh \theta & -\sinh \theta \\
      -\sinh \theta & \cosh \theta \\
    \end{array} \right) \,,
\end{equation}
for the transformation of two-vectors, and
\begin{equation}\label{boostspinorS}
    \Theta_{\rm sp} = - \left( \begin{array}{cc}
      0 & \frac{\theta}{2} \\
      \frac{\theta}{2} & 0 \\
    \end{array} \right) \,, \qquad
    \renewcommand{\arraystretch}{1.25}
    \exp\{\Theta_{\rm sp}\} = \left( \begin{array}{rr}
      \cosh \frac{\theta}{2} & -\sinh \frac{\theta}{2} \\
      -\sinh \frac{\theta}{2} & \cosh \frac{\theta}{2} \\
    \end{array} \right) \,,
    \renewcommand{\arraystretch}{1}
\end{equation}
for the transformation of spinors. With these choices, we keep the identities (\ref{boostrels}) and (\ref{boostrelv}), which are the basis for introducing Lorentz scalars and vectors in terms of spinors and for showing the covariance of the Dirac equation. With the identification $p=-m\sinh\theta$ and $E_p=m\cosh\theta$, we obtain
\begin{equation}\label{boostspinorSexpx}
    \exp\{\Theta_{\rm sp}\} = \sqrt{\frac{E_p+m}{2m}} \left( \begin{array}{cc}
      1 & \frac{p}{E_p+m} \\
      \frac{p}{E_p+m} & 1 \\
    \end{array} \right) \,.
\end{equation}
The electron and positron spinors are hence given by
\begin{equation}\label{spinorsuv1d}
    u_p = \sqrt{\frac{E_p+m}{2m}} \left( \begin{array}{c}
      1 \\
      \frac{p}{E_p+m} \\
    \end{array} \right) \,, \qquad
    v_p = \sqrt{\frac{E_p+m}{2m}} \left( \begin{array}{c}
      \frac{p}{E_p+m} \\
      1 \\
    \end{array} \right) \,,
\end{equation}
respectively. The spinors $u_{-p}$ and $v_p$ provide the eigenmodes of the Fourier transformed Dirac equation,
\begin{equation}\label{DiraceqFouS}
    \iR \frac{\partial\psi_{\bm{p}}}{\partial t} = \left( \begin{array}{cc}
       m & -p \\
      -p & -m \\
    \end{array} \right) \psi_{\bm{p}} \,,
\end{equation}
where the eigenvalues are $+E_p$ and $-E_p$, respectively.

\paragraph*{Massless fermions.}
We have a special interest in quantum electrodynamics in one space dimension for massless fermions. For this special case, Schwinger has shown that analytical solutions can be found and it is hence referred to as the Schwinger model. We can no longer transform to a local rest frame of the fermions. The spinors (\ref{spinorsuv1d}) become degenerate and should be normalized differently. After multiplying them by $\sqrt{2m}$ we can perform the limit $m \rightarrow 0$. We need to distinguish between positive $p$ (right movers) and negative $p$ (left movers),
\begin{equation}\label{spinors1dR}
    u_p =  v_p = \sqrt{|p|} \left( \begin{array}{r}
     1 \\
     1 \\
    \end{array} \right) \qquad \mbox{for } p>0 \,,
\end{equation}
\begin{equation}\label{spinors1dL}
    u_p =  - v_p = \sqrt{|p|} \left( \begin{array}{r}
     1 \\
     -1 \\
    \end{array} \right) \qquad \mbox{for } p<0 \,.
\end{equation}
The much simpler counterpart of Table~\ref{tablecouplmatrix} for massless fermions in one space dimension is given in Table~\ref{tablecouplmatrix1d}. The identities $v_p = {\rm sgn}(p) \, u_p$ and $\bar{v}_p = {\rm sgn}(p) \, \bar{u}_p$ can be used to obtain related expressions and symmetry relations involving $v$ spinors.

\begin{table}
\caption[ ]{Components of $\bar{u}_p \, \gamma^\mu \, u_{p'} / (2 \sqrt{|p p'|})$ in terms of $\Theta_{p,p'} = 1$ for $p p' > 0$ and $\Theta_{p,p'} = 0$ for $p p' < 0$}
\begin{center}
\begin{tabular}{|c|c|}
\hline
$\mu$ & $\bar{u}_p \, \gamma^\mu \, u_{p'} / (2 \sqrt{|p p'|})$ \\
\hline
$0$ & $\Theta_{p,p'}$ \\
$1$ & ${\rm sgn}(p) \, \Theta_{p,p'}$ \\
\hline
\end{tabular}
\end{center}
\label{tablecouplmatrix1d}
\end{table}

\section{Elements of our mathematical image}
To develop a mathematical image of quantum electrodynamics, we follow the philosophically founded approach outlined in Chapter~\ref{chapQFTapproach}. The discussion of the Dirac equation in the previous section showed that, unlike for the scalar theory studied in Chapter~\ref{chapphi4}, it is not so straightforward to work in $d$ space dimensions for arbitrary $d$. Therefore, we here develop electromagnetism in $d=3$ dimensions and, in Section~\ref{secSchwinger}, specify the modifications required for $d=1$ which, for massless leptons, lead us to the Schwinger model.

\subsection{Fock space}\label{secQED3dFocksp}
In defining the Fock space, we need to select the particles of interest and their basic properties, that is, the labels by which the creation and annihilation operators are characterized. We here focus on electrons, positrons, and photons. Further particles, such as other lepton generations or quarks (see Figure~\ref{figparticlezoo}), can be introduced in an analogous manner. Some additional ghost particles will be required in Section~\ref{secBRSTfree} to handle gauge degrees of freedom.

In order to achieve a Lorentz covariant treatment of the four-vector potential, we use the four-photon quantization of the electromagnetic field. This elegant idea was originally developed in 1950 by Gupta \cite{Gupta50} for free electromagnetic fields and by Bleuler \cite{Bleuler50} in the presence of charged matter.

The four photon creation and annihilation operators, $a^{\alpha \, \dag}_{\bm{q}}$, $a^\alpha_{\bm{q}}$ for $\alpha=0,1,2,3$ and $\bm{q} \in K^3_\times$ satisfy the fundamental commutation relations for bosons,
\begin{equation}\label{acommutator}
    \Qcommu{a^\alpha_{\bm{q}}}{a^{\alpha' \, \dag}_{\bm{q}'}} = \delta_{\alpha\alpha'} \, \delta_{\bm{q}\bm{q}'} \,,
    \qquad \Qcommu{a^\alpha_{\bm{q}}}{a^{\alpha'}_{\bm{q}'}} =
    \Qcommu{a^{\alpha \, \dag}_{\bm{q}}}{a^{\alpha' \, \dag}_{\bm{q}'}} = 0 \,.
\end{equation}
We can hence construct the four-photon Fock space in the usual way (see Section~\ref{sectionFock}). Note that the origin is excluded from the lattice $K^3_\times$ of momentum vectors because photons are massless.

The Bleuler-Gupta approach uses a modification of the canonical inner product (\ref{caninnerproddef}). The goal of this modification is to introduce the minus sign associated with the temporal component of the \index{Minkowski metric}Minkowski metric. If the basis vector  $\Dket{n^\alpha_{\bm{q}}}$ contains a total of $N=\sum_{\bm{q} \in K^3_\times} n^0_{\bm{q}}$ temporal photons, one introduces a factor $(-1)^N$,
\begin{equation}\label{relinnerproddef}
    s^{\rm sign}(|{{n'}^{\alpha'}_{\bm{q}'}}\rangle,\Dket{n^\alpha_{\bm{q}}}) =
    (-1)^N \; s^{\rm can}(|{{n'}^{\alpha'}_{\bm{q}'}}\rangle,\Dket{n^\alpha_{\bm{q}}})
    = (-1)^N \prod_{\bm{q}, \alpha} \delta_{n^\alpha_{\bm{q}} {n'}^\alpha_{\bm{q}}} \,.
\end{equation}
Of course, the modification $s^{\rm sign}$ no longer is a proper inner product because it assigns the negative norm $-1$ to base vectors with an odd number of temporal photons. The use of an indefinite inner product may be alarming because it endangers the probabilistic interpretation of the results. However, no interpretation problems arise for properly restricted physically admissible states. The physical states have been identified in terms of a proper version of the covariant Lorentz gauge condition and listed in an elementary way in (2.16) of the pioneering work \cite{Bleuler50}. In particular, the admissibility condition implies that physical states involve equal average numbers of longitudinal and temporal photons having the same momentum (see Chapter 17 of \cite{Boyarkin1}), which gives us an idea of how interpretational problems are avoided. An elegant way of characterizing the physical states has been given in Section V.C.3 of \cite{Cohenetcpp} by transforming from longitudinal and temporal to left (gauge) and right (droite) photons,
\begin{equation}\label{gphotondef}
    a^{\rm g}_{\bm{q}} = \frac{1}{\sqrt{2}} ( a^0_{\bm{q}} + a^3_{\bm{q}} ) \,,
    \qquad
    a^{{\rm g} \, \dag}_{\bm{q}} = \frac{1}{\sqrt{2}} ( a^{0 \, \dag}_{\bm{q}} + a^{3 \, \dag}_{\bm{q}} ) \,,
\end{equation}
and
\begin{equation}\label{dphotondef}
    a^{\rm d}_{\bm{q}} = \frac{\iR}{\sqrt{2}} ( a^3_{\bm{q}} - a^0_{\bm{q}} )  \,,
    \qquad
    a^{{\rm d} \, \dag}_{\bm{q}} = \frac{\iR}{\sqrt{2}} ( a^{0 \, \dag}_{\bm{q}} - a^{3 \, \dag}_{\bm{q}} ) \,.
\end{equation}
For free fields, physically admissible states can contain an arbitrary number of g-photons, but no d-photons.

On the one hand, we had separately defined creation and annihilation operators and realized that they are adjoints with respect to the canonical inner product. On the other hand, in Section \ref{subsecobservables}, the physically relevant correlation functions have been expressed in terms of bra- and ket-vectors. We keep the superscript $\dag$ for the adjoint operator associated with the canonical inner product. For the adjoint operator associated with the signed inner product, we use the superscript $\ddag$. For example, the adjoint of the temporal photon annihilation operator $a^0_{\bm{q}}$ would be $a^{0 \, \ddag}_{\bm{q}} = -a^{0 \, \dag}_{\bm{q}}$, whereas the adjoints of spatial photon annihilation operators $a^j_{\bm{q}}$ would coincide for both inner products. Finally, we define the bra-vector $\Dbra{\phi}$ as the linear form $s^{\rm sign}(\Dket{\phi},\cdot)$.

We here do not attempt to derive or justify the rules for identifying the physically admissible states, nor do we show that we recover a well-defined inner product by restricting $s^{\rm sign}$ to physical states and introducing equivalence classes of states differing by states of vanishing signed norm. We rather derive the validity of the Bleuler-Gupta approach within the far more general framework of \index{BRST quantization}BRST quantization in a later section (see Section~\ref{secBRST4EM}).

We now turn to the leptons. The operators $b^{\sigma \, \dag}_{\bm{p}}$ and $b^\sigma_{\bm{p}}$ create and annihilate an electron of momentum $\bm{p}$ and spin $\sigma$. Similarly, the operators $d^{\sigma \, \dag}_{\bm{p}}$ and $d^\sigma_{\bm{p}}$ create and annihilate a positron of momentum $\bm{p}$ and spin $\sigma$. As we are now dealing with fermions instead of the bosonic photons, we need to specify the fundamental anticommutation relations for the creation and annihilation operators in accordance with Section~\ref{sectionFock},
\begin{equation}\label{banticommutator}
    \Qantico{b^\sigma_{\bm{p}}}{b^{\sigma' \, \dag}_{\bm{p}'}} =
    \delta_{\sigma\sigma'} \, \delta_{\bm{p}\bm{p}'} \,, \qquad
    \Qantico{b^\sigma_{\bm{p}}}{b^{\sigma'}_{\bm{p}'}} =
    \Qantico{b^{\sigma \, \dag}_{\bm{p}}}{b^{\sigma' \, \dag}_{\bm{p}'}} = 0 \,,
\end{equation}
\begin{equation}\label{danticommutator}
    \Qantico{d^\sigma_{\bm{p}}}{d^{\sigma' \, \dag}_{\bm{p}'}} =
    \delta_{\sigma\sigma'} \, \delta_{\bm{p}\bm{p}'} \,, \qquad
    \Qantico{d^\sigma_{\bm{p}}}{d^{\sigma'}_{\bm{p}'}} =
    \Qantico{d^{\sigma \, \dag}_{\bm{p}}}{d^{\sigma' \, \dag}_{\bm{p}'}} = 0 \,,
\end{equation}
and
\begin{equation}\label{bdanticommutator}
    \Qantico{b^\sigma_{\bm{p}}}{d^{\sigma'}_{\bm{p}'}} =
    \Qantico{b^\sigma_{\bm{p}}}{d^{\sigma' \, \dag}_{\bm{p}'}} =
    \Qantico{b^{\sigma \, \dag}_{\bm{p}}}{d^{\sigma'}_{\bm{p}'}} =
    \Qantico{b^{\sigma \, \dag}_{\bm{p}}}{d^{\sigma' \, \dag}_{\bm{p}'}} = 0 \,.
\end{equation}
In words, all fermion operators anticommute with each other except for the creation and annihilation operators of the same type with identical spin and momentum super- and subscripts, which satisfy canonical anticommutation relations. As we now deal simultaneously with both boson and fermion operators, we further need to specify the general rule that all boson operators commute with all fermion operators,
\begin{equation}\label{abcommutator}
    \Qcommu{a^\alpha_{\bm{q}}}{b^{\sigma'}_{\bm{p}'}} =
    \Qcommu{a^\alpha_{\bm{q}}}{b^{\sigma' \, \dag}_{\bm{p}'}} =
    \Qcommu{a^{\alpha \, \dag}_{\bm{q}}}{b^{\sigma'}_{\bm{p}'}} =
    \Qcommu{a^{\alpha \, \dag}_{\bm{q}}}{b^{\sigma' \, \dag}_{\bm{p}'}} = 0 \,,
\end{equation}
and
\begin{equation}\label{adcommutator}
    \Qcommu{a^\alpha_{\bm{q}}}{d^{\sigma'}_{\bm{p}'}} =
    \Qcommu{a^\alpha_{\bm{q}}}{d^{\sigma' \, \dag}_{\bm{p}'}} =
    \Qcommu{a^{\alpha \, \dag}_{\bm{q}}}{d^{\sigma'}_{\bm{p}'}} =
    \Qcommu{a^{\alpha \, \dag}_{\bm{q}}}{d^{\sigma' \, \dag}_{\bm{p}'}} = 0 \,.
\end{equation}
Although simple general rules fix the proper commutators or anticommutators for all creation and annihilation operators, in the present formulation of the Fock space, we have listed them all explicitly to illustrate these general rules and to guarantee clarity.

As before, the canonical inner product in the bigger Fock space of photons, electrons and positrons is modified by a factor $-1$ if a base vectors contains an odd number of temporal photons. This modification is referred to as the signed inner product and used to define bra-vectors.

\subsection{Fields}
For the purpose of heuristic derivations and of clarifying the connection to standard Lagrangian formulations of quantum electrodynamics, we here provide the expressions for the spatial fields associated with photons, electrons, and positrons. As in the case of scalar field theory, we do not consider these spatial fields as integral parts of our image of nature. However, they are useful in formulating the proper Hamiltonian describing interactions between photons and leptons.

In a cubic box of volume $V$, the position-dependent four-vector potential with components $A_{\bm{x} \, \mu}$ is introduced as a Fourier series in terms of polarization states and the corresponding creation and annihilation operators  [cf.\ (\ref{phiexpression})],
\begin{equation}\label{4vecpot}
    A_{\bm{x}} = \frac{1}{\sqrt{V}} \sum_{\bm{q} \in K^3_\times} A_{\bm{q}} \,
    \eR^{- \iR \bm{q} \cdot \bm{x}} \,,
\end{equation}
where the Fourier components are given by
\begin{equation}\label{4vecpotFou}
    A_{\bm{q}} = \frac{1}{\sqrt{2 q}}
    \left( \bar{n}^\alpha_{\bm{q}} a^{\alpha \, \dag}_{\bm{q}}
    + \varepsilon_{\alpha} \bar{n}^\alpha_{-\bm{q}} a^\alpha_{-\bm{q}} \right) \,.
\end{equation}
As photons are massless, we here use $q=|\bm{q}|$ instead of $\omega_q$. The temporal unit four-vector,
\begin{equation}\label{polarization0}
    \bar{n}^0_{\bm{q}} =
    \left( \begin{array}{c}
      1 \\
      0 \\
      0 \\
      0 \\
    \end{array} \right) ,
\end{equation}
is actually independent of $\bm{q}$. The three orthonormal spatial polarization vectors are chosen as
\begin{equation}\label{polarization1}
    \bar{n}^1_{\bm{q}} = \frac{1}{\sqrt{q_1^2+q_2^2}}
    \left( \begin{array}{c}
      0 \\
      q_2 \\
      - q_1 \\
      0 \\
    \end{array} \right) ,
\end{equation}
\begin{equation}\label{polarization2}
    \bar{n}^2_{\bm{q}} = \frac{1}{q \sqrt{q_1^2+q_2^2}}
    \left( \begin{array}{c}
      0 \\
      q_1 q_3 \\
      q_2 q_3\\
      - q_1^2 - q_2^2 \\
    \end{array} \right) ,
\end{equation}
and
\begin{equation}\label{polarization3}
    \bar{n}^3_{\bm{q}} = \frac{1}{q}
    \left( \begin{array}{c}
      0 \\
      q_1\\
      q_2\\
      q_3\\
    \end{array} \right) .
\end{equation}
The polarization vectors $\bar{n}^1_{\bm{q}}$ and $\bar{n}^2_{\bm{q}}$ correspond to transverse photons, $\bar{n}^3_{\bm{q}}$ corresponds to longitudinal photons. Note the symmetry property
\begin{equation}\label{polarsym}
    \bar{n}^\alpha_{-\bm{q}} = (-1)^\alpha \, \bar{n}^\alpha_{\bm{q}} \,.
\end{equation}
The sign $\varepsilon_{\alpha}$ in (\ref{4vecpotFou}) (defined as $+1$ for transverse and longitudinal photons, $-1$ for temporal photons) leads to a non-self-adjoint nature of the four-vector potential (\ref{4vecpot}). The spatial components are self-adjoint, whereas the temporal component is anti-self-adjoint. The latter statement can be expressed as $A_{\bm{x} \, 0}^\dag = - A_{\bm{x} \, 0} = A_{\bm{x}}^0$ where, as before, we assume a \index{Minkowski metric}Minkowski metric $\eta^{\mu\nu}=\eta_{\mu\nu}$ with signature $(-,+,+,+)$, that is, $\eta^{00}=\eta_{00}=-1$. It is hence important to note that (\ref{4vecpot}) and (\ref{polarization0}) specify $A_0$ with a lower index. The Lorentz covariant gauge condition (\ref{MaxLorentzgauge}) implies $A^0=\phi$, $A_0=-\phi$. For the signed inner product, all components of the four-vector potential are self-adjoint, $A_{\bm{x} \, \mu}^\ddag = A_{\bm{x} \, \mu}$.

To understand the normalization of the field (\ref{4vecpot}), we consider
\begin{equation}\label{Aexpressionnorm}
  \int_V \Dbra{N} : A^\ddag_{\bm{x} \, \mu} A^\mu_{\bm{x}} : \Dket{N} \dR^3x =
  \sum_{\bm{q} \in K^3_\times} \Dbra{N} : A^\ddag_{\bm{q} \, \mu} A^\mu_{\bm{q}} : \Dket{N} =
  \sum_{\bm{q} \in K^3_\times} \frac{1}{q} \Dbra{N} a^{\alpha \, \dag}_{\bm{q}} a^\alpha_{\bm{q}} \Dket{N} \,,
\end{equation}
where $\Dket{N}$ is a Fock space eigenvector with a total of $N$ particles. Of course, a nonrelativistic limit for interpreting the normalization of $A_{\bm{x}}$ in terms of particle numbers (see Section~\ref{secFourierfields}) is not possible for the massless photons.

The Fourier transform of the magnetic field, as obtained from the rotation of the spatial part of the vector potential (\ref{4vecpot}), is given by the spatial components of
\begin{equation}\label{magnfield}
    B_{\bm{q}} = \iR \sqrt{\frac{q}{2}} \,
    \Big( \bar{n}^1_{\bm{q}} a^{2 \, \dag}_{\bm{q}} - \bar{n}^1_{-\bm{q}} a^2_{-\bm{q}}
    - \bar{n}^2_{\bm{q}} a^{1 \, \dag}_{\bm{q}} + \bar{n}^2_{-\bm{q}} a^1_{-\bm{q}} \Big) \,.
\end{equation}
Note that the magnetic field involves only transverse photon operators. In view of the symmetries (\ref{polarsym}), the physically relevant Fourier components are associated with $a^{1 \, \dag}_{\bm{q}} - a^1_{-\bm{q}}$ and $a^{2 \, \dag}_{\bm{q}} + a^2_{-\bm{q}}$ which coincide with the transverse Fourier components of the vector potential in Eq.~(\ref{4vecpotFou}). For the Fourier transform of the free electric field we have
\begin{eqnarray}
    E_{\bm{q}} &=& \iR \sqrt{\frac{q}{2}} \,
    \Big[ - \bar{n}^1_{\bm{q}} a^{1 \, \dag}_{\bm{q}} + \bar{n}^1_{-\bm{q}} a^1_{-\bm{q}}
    - \bar{n}^2_{\bm{q}} a^{2 \, \dag}_{\bm{q}} + \bar{n}^2_{-\bm{q}} a^2_{-\bm{q}}
    \nonumber \\
    && \hspace{5em} - \bar{n}^3_{\bm{q}} \big( a^{3 \, \dag}_{\bm{q}} + a^{0 \, \dag}_{\bm{q}} \big)
    + \bar{n}^3_{-\bm{q}} \big( a^3_{-\bm{q}} - a^0_{-\bm{q}} \big) \Big] \,.
\label{elecfield}
\end{eqnarray}
The last term does not contribute for physically admissible states containing no d-photons. The divergence of the free electric field is given in terms of g-photons. Its transverse components are given in terms of $a^{1 \, \dag}_{\bm{q}} + a^1_{-\bm{q}}$ and $a^{2 \, \dag}_{\bm{q}} - a^2_{-\bm{q}}$.

In Section~\ref{secDiracplanewave}, we have compiled all the ingredients to define the four-component spinor field associated with annihilating an electron or creating a positron, each coming with the two values $\pm 1/2$ of the spin. It is given by the Fourier representation (see Eq.~(13.50) of \cite{BjorkenDrell})
\begin{equation}\label{spinorfieldFou}
    \psi_{\bm{x}} = \frac{1}{\sqrt{V}} \sum_{\bm{p} \in K^3} \psi_{\bm{p}} \, \eR^{- \iR \bm{p} \cdot \bm{x}} \,, \qquad
    \psi_{\bm{p}} = \sqrt{\frac{m}{E_p}}
    \left( v^{\sigma}_{\bm{p}}  d^{\sigma \, \dag}_{\bm{p}}
    + u^{-\sigma}_{-\bm{p}} b^{-\sigma}_{-\bm{p}} \right) \,.
\end{equation}
With the definitions $\bar{u} = u^* \gamma^0$ and $\bar{v} = v^* \gamma^0$, we obtain the following Fourier representation for $\bar{\psi}_{\bm{x}} = \psi_{\bm{x}}^\dag \gamma^0$ (note the operator generalization of the bar-operation),
\begin{equation}\label{spinorfieldbarFou}
    \bar{\psi}_{\bm{x}} = \frac{1}{\sqrt{V}} \sum_{\bm{p} \in K^3} \bar{\psi}_{-\bm{p}}
    \, \eR^{- \iR \bm{p} \cdot \bm{x}} \,, \qquad
    \bar{\psi}_{-\bm{p}} = \sqrt{\frac{m}{E_p}}
    \left( \bar{v}^{-\sigma}_{-\bm{p}}  d^{-\sigma}_{-\bm{p}}
    + \bar{u}^{\sigma}_{\bm{p}} b^{\sigma \, \dag}_{\bm{p}} \right) \,.
\end{equation}
Note that $\bar{\psi}_{-\bm{p}}$ rather than $\bar{\psi}_{\bm{p}}$ is taken as the Fourier transform of $\bar{\psi}_{\bm{x}}$; with this convention, $\bar{\psi}_{\bm{p}}$ is obtained from $\psi_{\bm{p}}$ by the previously defined bar-operation, $\bar{\psi}_{\bm{p}} = \psi_{\bm{p}}^\dag \gamma^0$.

The identity (\ref{projectordif}) implies the anticommutation relations
\begin{equation}\label{psianticommutator}
    \Qantico{\psi_{\bm{p}}}{\bar{\psi}_{\bm{p}'}} =
    \gamma^0 \, \delta_{\bm{p}\bm{p}'} \,, \qquad
    \Qantico{\psi_{\bm{p}}}{\psi_{\bm{p}'}} =
    \Qantico{\bar{\psi}_{\bm{p}}}{\bar{\psi}_{\bm{p}'}} = 0 \,,
\end{equation}
where the relation $\Qantico{\psi_{\bm{p}}}{\psi^\dag_{\bm{p}'}} = 1 \, \delta_{\bm{p}\bm{p}'}$ looks slightly more natural. The normalization of the fields is revealed by
\begin{eqnarray}
  \int_V \Dbra{N} : \bar{\psi}_{\bm{x}} \psi_{\bm{x}} : \Dket{N} \dR^3x &=&
  \sum_{\bm{p} \in K^3} \Dbra{N} : \bar{\psi}_{\bm{p}} \psi_{\bm{p}} : \Dket{N}
  \nonumber\\
  &=& \sum_{\bm{p} \in K^3} \frac{m}{E_p} \Dbra{N}
  \left( b^{\sigma \, \dag}_{\bm{p}} b^\sigma_{\bm{p}}
  + d^{\sigma \, \dag}_{\bm{p}} d^\sigma_{\bm{p}} \right) \Dket{N} \,, \qquad
\label{spinorfieldnorm}
\end{eqnarray}
where $\Dket{N}$ is a Fock space eigenvector with $N$ particles. For fermions, normal ordering comes with the corresponding sign changes. We can  interpret $: \bar{\psi}_{\bm{x}} \psi_{\bm{x}} :$ as the total particle number density, at least, if we ignore relativistic subtleties (see Section~\ref{secFourierfields}). A cleaner interpretation of the normalization is obtained from
\begin{eqnarray}
  -e_0 \int_V \Dbra{N} : \bar{\psi}_{\bm{x}} \gamma^0 \psi_{\bm{x}} : \Dket{N} \dR^3x &=&
  -e_0 \sum_{\bm{p} \in K^3} \Dbra{N} : \bar{\psi}_{\bm{p}} \gamma^0 \psi_{\bm{p}} : \Dket{N}
  \nonumber\\
  &=& e_0 \sum_{\bm{p} \in K^3} \Dbra{N} \left( d^{\sigma \, \dag}_{\bm{p}} d^\sigma_{\bm{p}}
  - b^{\sigma \, \dag}_{\bm{p}} b^\sigma_{\bm{p}} \right) \Dket{N} \,. \qquad\qquad
\label{spinorfieldnormq}
\end{eqnarray}
The charge density is obtained by counting the positrons and the electrons and taking the difference.

\subsection{Hamiltonian and current density}
The free Hamiltonian for our massless photons is given by
\begin{equation}\label{H0gendef}
    H_{\rm EM}^{\rm free} = \sum_{\bm{q} \in K^3_\times} q  \,
    a^{\alpha \, \dag}_{\bm{q}} a^\alpha_{\bm{q}} \,,
\end{equation}
where a summation over the possible polarization states $\alpha$ is implied by the same index occurring twice. All four photons are clearly treated on an equal footing. Their energy is given by the relativistic expression for massless particles. The free Hamiltonian (\ref{H0gendef}) is self-adjoint both for the canonical and for the signed inner products.

According to (13.59) of \cite{BjorkenDrell}, the Hamiltonian associated with the Dirac equation for the free electron/positron can be written as
\begin{equation}\label{H0Diracdef}
    H^{\rm free}_{\rm e/p} = \sum_{\bm{p} \in K^3} E_p \left( b^{\sigma \, \dag}_{\bm{p}} b^\sigma_{\bm{p}}
    + d^{\sigma \, \dag}_{\bm{p}} d^\sigma_{\bm{p}} \right) \,,
\end{equation}
where $E_p = \sqrt{\bm{p}^2 + m^2}$ gives the energy of a relativistic particle with mass $m$ and momentum $\bm{p}$, and a summation over the possible spin values $\sigma =\pm 1/2$ is implied by the same index occurring twice.

The interaction between charged leptons and photons is given by the Hamiltonian (see the discussion of energy in Section~\ref{secpartEMfield})
\begin{eqnarray}
    H^{\rm coll} &=& - \int J^\mu_{\bm{x}} A_{\bm{x} \, \mu} \, \dR^3x + e'' V
    = - \sum_{\bm{q} \in K^3_\times} J^\mu_{\bm{q}} A_{-\bm{q} \,\mu}  + e'' V\nonumber \\
    &=& - \sum_{\bm{q} \in K^3_\times} \frac{1}{\sqrt{2 q}} J^\mu_{\bm{q}}
    \left( \bar{n}^\alpha_{-\bm{q} \,\mu} a^{\alpha \, \dag}_{-\bm{q}}
    + \varepsilon_{\alpha} \bar{n}^\alpha_{\bm{q} \,\mu} a^\alpha_{\bm{q}} \right) + e'' V \,.
\label{Hintexpression}
\end{eqnarray}
As in (\ref{Hcolk}) for scalar field theory, we have added a constant background energy $e'' V$ where $e''$ is an energy per unit volume. We again choose the energy density such that the lowest energy eigenvalue associated with the ground state of the interacting theory is equal to zero. The electric current density four-vector $J^\mu_{\bm{x}}$ has been defined in (\ref{currentclass}) and its Fourier expansion is given by $J^\mu_{\bm{q}}$. As $J^\mu_{\bm{x}}$ and $A_{\bm{x} \, \mu}$ are four-vectors, $H^{\rm coll}$ is a Lorentz scalar. In the coordinate system in which the local current density vanishes, the local contribution to $H^{\rm coll}$ is the energy associated with the charge density in the electric field. Note that, in the canonical inner product, the Hamiltonian $H^{\rm coll}$ is not self-adjoint because $A_0$ is anti-self-adjoint. However, in the signed inner product, the Hamiltonian $H^{\rm coll}$ becomes self-adjoint. In terms of the Fourier transforms (\ref{spinorfieldFou}) and (\ref{spinorfieldbarFou}), we find
\begin{equation}\label{Jqexplicit0}
    J^\mu_{\bm{q}} = - \frac{e_0}{\sqrt{V}}\sum_{\bm{p}, \bm{p}' \in K^3}
    \delta_{\bm{p}-\bm{p}',\bm{q}} \,
    \bar{\psi}_{-\bm{p}} \, \gamma^\mu \, \psi_{-\bm{p}'} \,,
\end{equation}
or, by means of (\ref{equalpspinsumu}), in normal-ordered form,
\begin{eqnarray}
    J^\mu_{\bm{q}} &=& - \sum_{\bm{p}, \bm{p}' \in K^3} \delta_{\bm{p}-\bm{p}',\bm{q}}
    \frac{m e_0}{\sqrt{V E_p E_{p'}}} \Big(
    \bar{u}^{\sigma}_{\bm{p}} \, \gamma^\mu \, u^{\sigma'}_{\bm{p}'} \;
         b^{\sigma \, \dag}_{\bm{p}} b^{\sigma'}_{\bm{p}'}
    - \bar{u}^{\sigma'}_{\bm{p}'} \, \gamma^\mu \, u^{\sigma}_{\bm{p}} \;
         d^{-\sigma \, \dag}_{\bm{p}} d^{-\sigma'}_{\bm{p}'}
    \nonumber \\
    && \hspace{1em} + \bar{v}^{-\sigma}_{-\bm{p}} \, \gamma^\mu \, u^{\sigma'}_{\bm{p}'} \;
         d^{-\sigma}_{-\bm{p}} b^{\sigma'}_{\bm{p}'}
    + \bar{v}^{-\sigma'}_{-\bm{p}'} \, \gamma^\mu \, u^{\sigma}_{\bm{p}} \;
         b^{\sigma' \, \dag}_{-\bm{p}'} d^{-\sigma \, \dag}_{\bm{p}} \Big)
    - 2 \delta_{\bm{q},\bm{0}} \sqrt{V} {e'}^\mu \,,
    \nonumber \\ &&
\label{Jqexplicit}
\end{eqnarray}
with
\begin{equation}\label{backcurrentdens}
    {e'}^\mu = \frac{e_0}{V} \sum_{\bm{p} \in K^3} \frac{p^\mu}{E_p} \,.
\end{equation}
As the parameter (\ref{z12def}) arising from normal ordering in $\varphi^4$ theory, we treat the current density four-vector ${e'}^\mu$ as an independent model input. In the center-of-mass system, the spatial components of ${e'}^\mu$ vanish. Choosing ${e'}^\mu=0$ corresponds to the standard normal-ordering procedure suggested in Eq.~(13.61) of \cite{BjorkenDrell}.

The symmetry relations (\ref{uvvuident}), (\ref{uuident}) and (\ref{uvident}) imply the identity
\begin{equation}\label{Jqadjoint}
    \left( J^\mu_{\bm{q}} \right)^\dag = J^\mu_{-\bm{q}} \,,
\end{equation}
which reflects the self-adjointness of $J^\mu_{\bm{x}}$. These symmetry relations are also the reason why we can express all contributions to $J^\mu_{\bm{q}}$ in terms of only the two quantities $\bar{u}^{\sigma}_{\bm{p}} \, \gamma^\mu \, u^{\sigma'}_{\bm{p}'}$ and $\bar{v}^{-\sigma}_{-\bm{p}} \, \gamma^\mu \, u^{\sigma'}_{\bm{p}'}$. By means of (\ref{uupsumident}) and (\ref{uvpsumident}), we obtain the fundamental operator identity
\begin{equation}\label{localelchargecons}
     q_j J^j_{\bm{q}} = \Qcommu{H^{\rm free}_{\rm e/p}}{J^0_{\bm{q}}} \,,
\end{equation}
which expresses the conservation of electric charge in the form of a local continuity equation. By adding up the various contributions, we obtain the total Hamiltonian of quantum electrodynamics,
\begin{equation}\label{HQEDtotaldef}
    H = H_{\rm EM}^{\rm free} + H^{\rm free}_{\rm e/p} + H^{\rm coll} \,.
\end{equation}

Let us verify that our heuristically motivated Hamiltonian indeed leads to a quantized version of electrodynamics. By means of the total Hamiltonian (\ref{HQEDtotaldef}), the Fourier components (\ref{4vecpotFou}) of the four-vector potential, and the canonical commutation relations for the photon creation and annihilation operators, we obtain
\begin{equation}\label{quantumMaxwell}
    \Qcommu{H}{\Qcommu{H}{A_{\bm{q}\,\mu}}} = q^2 A_{\bm{q}\,\mu} - J_{\bm{q}\,\mu} \,.
\end{equation}
As the double commutator with the Hamiltonian can be interpreted as a double time derivative (with a minus sign), (\ref{quantumMaxwell}) may be regarded as the Fourier-transformed version of the Maxwell equations in the form (\ref{Maxpotphieq}), (\ref{MaxpotAeq}) on the operator level and hence as the verification of a successful implementation of quantum electrodynamics. In order to arrive at (\ref{quantumMaxwell}), we did not need to assume a specific form of the electric current four-vector $J_{\bm{q}}$; however, it is important to assume that $J_{\bm{q}}$ commutes with the photon creation and annihilation operators.

\subsection{Schwinger term}\label{secSchwingerterm}
This subsection is dedicated to a discussion of the commutators $\Qcommu{J^\mu_{\bm{q}}}{J^0_{\bm{q}'}}$, which play an important role in the further development. An argument offered by Schwinger \cite{Schwinger59} shows that $\Qcommu{J^\mu_{\bm{q}}}{J^0_{\bm{q}'}}$ cannot be equal to zero for all $\mu$. Otherwise, the charge conservation (\ref{localelchargecons}) and the property $H \Dket{\Omega} =0$ for the full vaccum state would imply
\begin{equation}\label{JJcommutation0impl}
    0 = q_j \Dbra{\Omega} \Qcommu{J^j_{\bm{q}}}{J^0_{\bm{q}'}} \Dket{\Omega} =
    \Dbra{\Omega} \Qcommu{\Qcommu{H}{J^0_{\bm{q}}}}{J^0_{\bm{q}'}} \Dket{\Omega} =
    - \Dbra{\Omega} J^0_{\bm{q}} H J^0_{\bm{q}'} + J^0_{\bm{q}'} H J^0_{\bm{q}} \Dket{\Omega} \,.
\end{equation}
Assuming $\bm{q}'=\bm{q}$, all $J^0_{\bm{q}} \Dket{\Omega}$ would be restricted to the zero-energy state(s) of $H$. In view of (\ref{groundstatecond2sol}) for the ground state and (\ref{Jqexplicit}) for the charge density, this appears to be impossible so that $\Qcommu{J^\mu_{\bm{q}}}{J^0_{\bm{q}'}}$ cannot be identically zero.

The nontrivial result for the commutator $\Qcommu{J^\mu_{\bm{q}}}{J^0_{\bm{q}'}}$ required by the above argument is given by what is known as the \index{Schwinger term}Schwinger term. It has been found to be a complex number, not an operator \cite{NishijimaSasaki75}. Construction of the Schwinger term \label{Schwingertermpage} requires subtle limiting procedures so that even the Jacobi identity for nested commutators needs to be verified explicitly \cite{Kubo94}.

From (\ref{Jqexplicit0}) and (\ref{psianticommutator}), we obtain the explicit formula
\begin{equation}\label{JJcommucareful}
  \Qcommu{J^\mu_{\bm{q}}}{J^0_{\bm{q}'}} = \frac{e_0^2}{V}
  \sum_{\bm{p}, \bm{p}', \bar{\bm{p}} \in K^3}
  ( \delta_{\bar{\bm{p}},\bm{p}+\bm{q}} - \delta_{\bar{\bm{p}},\bm{p}+\bm{q}'} ) \,
  \delta_{\bm{p}',\bm{p}+\bm{q}+\bm{q}'} \,
  \bar{\psi}_{\bm{p}} \gamma^\mu \psi_{\bm{p}'} \,.
\end{equation}
The summation over $\bar{\bm{p}}$ implies that the difference of two Kronecker symbols in parenthesis is nonzero only if either $\bm{p}+\bm{q} \in K^3$, $\bm{p}+\bm{q}' \notin K^3$ or $\bm{p}+\bm{q} \notin K^3$, $\bm{p}+\bm{q}' \in K^3$. This situation can only occur near the boundary of $K^3$. Had we extended the summation over $\bar{\bm{p}}$ to the unbounded set $\bar{K}^3$, such a situation could not occur and the result for the sum would turn out to be zero. As the sum is note absolutely convergent, however, this result depends on the particular way in which the infinite sum is carried out. The combination of unambiguous finite sums and well-defined limiting procedures is clearly preferable.

The right-hand side of (\ref{JJcommucareful}) contains operators at very high momenta (in a boundary layer of $K^3$). If the Schwinger term can only be a complex number, not an operator \cite{NishijimaSasaki75}, then we can obtain it as the vacuum expectation $\Dbra{0} \Qcommu{J^\mu_{\bm{q}}}{J^0_{\bm{q}'}} \Dket{0}$. In the spirit of our previous treatment of operator products, we can equivalently bring $\bar{\psi}_{\bm{p}} \gamma^\mu \psi_{\bm{p}'}$ into a normal-ordered form and keep only the constant parts arising in that procedure. Using (\ref{equalpspinsumv}), we thus arrive at
\begin{equation}\label{JJcommucarefulNO}
  \Qcommu{J^\mu_{\bm{q}}}{J^0_{\bm{q}'}} = 2 \frac{e_0^2}{V} \, \delta_{-\bm{q},\bm{q}'}
  \sum_{\bm{p}, \bar{\bm{p}} \in K^3}
  ( \delta_{\bar{\bm{p}},\bm{p}+\bm{q}} - \delta_{\bar{\bm{p}},\bm{p}-\bm{q}} ) \,
  \frac{p^\mu}{E_p} \,.
\end{equation}
The substitutions $\bm{p} \rightarrow - \bm{p}$, $\bar{\bm{p}} \rightarrow - \bar{\bm{p}}$ show that this commutator vanishes for $\mu = 0$,
\begin{equation}\label{JJcommucarefuls0}
  \Qcommu{J^0_{\bm{q}}}{J^0_{\bm{q}'}} = 0 \,.
\end{equation}The same substitution can be used to simplify the results for the spatial components $\mu = j$,
\begin{equation}\label{JJcommucarefulsj}
  \Qcommu{J^j_{\bm{q}}}{J^0_{\bm{q}'}} = 4 \frac{e_0^2}{V} \, \delta_{-\bm{q},\bm{q}'}
  \sum_{\bm{p}, \bar{\bm{p}} \in K^3}
  \delta_{\bar{\bm{p}},\bm{p}+\bm{q}} \, \frac{p^j}{E_p} \,.
\end{equation}
If $\bm{q}$ points in one of the coordinate directions, only a stripe along one of the six surfaces of the cube $K^3$ contributes to the sum. For large $Z_L$, we thus obtain
\begin{equation}\label{JJcommucarefulsjf}
  \Qcommu{J^j_{\bm{q}}}{J^0_{\bm{q}'}} = - 2 C_J (2Z_L+1)^2 \, \frac{e_0^2}{\pi} \,
  \frac{L}{V} \, q^j \delta_{-\bm{q},\bm{q}'} \,,
\end{equation}
where $C_J$ is the average projection of a radial unit vector onto the normal of one of the sides of a cube,
\begin{equation}\label{CJnprojectioncube}
  C_J = \frac{1}{4} \int_{-1}^1 \dR x_1 \int_{-1}^1 \dR x_2 \frac{1}{\sqrt{1+x_1^2+x_2^2}}
  = \frac{1}{4} \left[ \ln (97+56\sqrt{3}) - \frac{2\pi}{3} \right] \approx 0.79 \,.
\end{equation}

In three dimensions, the Schwinger term in (\ref{JJcommucarefulsjf}) diverges with increasing momentum cutoff and the constant $C_J$ depends on the fact that we have chosen the finite lattices in momentum space in the shape of cubes. Physical predictions should not depend on such singular values or arbitrary choices. Indeed, it has been found that the \index{Schwinger term}Schwinger term is of no physical relevance (see p.\,950 of \cite{Kinoshita}). From now on, we hence assume that the Schwinger term can be assumed to vanish for quantum electrodynamics in three space dimensions,
\begin{equation}\label{JJcommucarefulNO0}
  \Qcommu{J^\mu_{\bm{q}}}{J^0_{\bm{q}'}} = 0 \,.
\end{equation}
Most textbooks on quantum field theory don't even discuss the Schwinger term or the reasons why (\ref{JJcommucarefulNO0}) can be assumed. Note that the subtleties of the Schwinger term are much harder to recognize and resolve if one employs the setting of continuous fields from the very beginning. The discussion of this term will be continued in the context of the simplified Schwinger model in Section~\ref{secSchwingerterm1d}.

\subsection{BRST quantization}\label{secBRST4EM}
\index{BRST quantization}At this point, we would like to have a closer look at how the gauge transformations (\ref{MaxgaugeAeq}), (\ref{Maxgaugephieq}) and the gauge condition (\ref{MaxLorentzgauge}) of classical electrodynamics can be taken into account in the corresponding quantum theory. As the four-vector potential $A^\mu = (\phi, \bm{A})$ becomes operator-valued, it is natural to elevate also the function $f$, which characterizes the gauge transformation in (\ref{MaxgaugeAeq}), (\ref{Maxgaugephieq}), to the level of operators. By considering separate operator generalizations of $f$ for the creation and annihilation contributions $a^{\alpha \, \dag}_{\bm{q}}$ and $a^\alpha_{\bm{q}}$ to the four-vector potential (\ref{4vecpot}), (\ref{4vecpotFou}), namely $D^\dag_{\bm{q}}$ and $B_{\bm{q}}$, respectively, one arrives at the idea of \index{BRST quantization}BRST quantization (see the original papers \cite{BecchiRouetStora76,Tyutin75} and the pedagogical BRST primer \cite{Nemeschanskyetal86}). Among other illuminating insights, BRST quantization provides a nicely general and systematic justification of the Bleuler-Gupta approach sketched in Section~\ref{secQED3dFocksp}.

Before we elaborate the details, let us briefly reconsider the situation encountered in Section \ref{secsymmetries}. The `most intuitive' description of \index{Electrodynamics}electromagnetism is in terms of \emph{six} fields, because the components of the electric and magnetic fields ($\bm{E}$ and $\bm{B}$) may be considered as directly observable. The \index{Maxwell equations}Maxwell equations imply that these fields can be represented in terms of the potentials ($\phi$, $\bm{A}$). These \emph{four} fields provide the `most systematic' description because they actually form a four-vector under Lorentz transformations. But gauge symmetry suggests that not all of these four fields contain physically relevant information. In the end, only \emph{two} fields are required for the `most reduced' description of electromagnetic fields which, for free fields, correspond to the two transverse photon degrees of freedom. Now, shall we base our approach to \index{Electrodynamics}electromagnetism on two, four, or six fields? Shall we go for the `most reduced', the `most systematic', or the `most intuitive' description?

There seems to be good reason to go for the `most reduced' description because we should focus on the physical degrees of freedom. But can we be sure that we shall not discover further degrees of freedom as physical in the future? Couldn't there even be other forms of life to whom these degrees of freedom are more accessible? In this situation, our \index{Metaphysical postulates!first metaphysical postulate}first metaphysical postulate on p.\,\pageref{metaphys1} comes to our rescue: Let's go for the `most elegant' description.

Assuming that the `most elegant' description is provided by \index{BRST quantization}BRST quantization, this description is given in terms of six fields, just like the `most intuitive' description. The `most reduced' description based on transverse fields would be awkward, in particular, when one is interested in more complicated gauge theories than electromagnetism. The `most systematic' description based on the four-vector potential seems to be elegant but, unfortunately, only one of the constraints can be written in the elegant Lorentz invariant form of (\ref{MaxLorentzgauge}). In a sense, starting from the four-vector potential, we have two options for treating constraints: Lagrange's approach of the first and second kind obtained by including additional Lagrange multipliers or by parameterizing only the physical degrees of freedom, respectively. \index{BRST quantization}BRST quantization corresponds to the method of Lagrange multipliers (or Lagrange's equations of the first kind) \cite{Nemeschanskyetal86}. One introduces even more degrees of freedom which correspond to Lagrange multipliers, or constraint forces, and which eventually allow us to eliminate all unphysical degrees of freedom. Let us now elaborate the details of the procedure for achieving such an elimination.

\subsubsection{Free electromagnetic fields}\label{secBRSTfree}
We begin by introducing momentum-dependent creation and annihilation operators associated with gauge transformations by specifying the canonical anticommutation relations\index{BRST quantization}
\begin{equation}\label{BRSTanticomrelB}
    \Qantico{B_{\bm{q}}}{B^\dag_{\bm{q}'}} = \delta_{\bm{q}\bm{q}'} \,,
    \qquad \Qantico{B_{\bm{q}}}{B_{\bm{q}'}} =
    \Qantico{B^\dag_{\bm{q}}}{B^\dag_{\bm{q}'}} = 0 \,,
\end{equation}
\begin{equation}\label{BRSTanticomrelD}
    \Qantico{D_{\bm{q}}}{D^\dag_{\bm{q}'}} = \delta_{\bm{q}\bm{q}'} \,,
    \qquad \Qantico{D_{\bm{q}}}{D_{\bm{q}'}} =
    \Qantico{D^\dag_{\bm{q}}}{D^\dag_{\bm{q}'}} = 0 \,,
\end{equation}
and
\begin{equation}\label{BRSTanticomrelBD}
    \Qantico{B_{\bm{q}}}{D_{\bm{q}'}} =
    \Qantico{B_{\bm{q}}}{D^\dag_{\bm{q}'}} =
    \Qantico{B^\dag_{\bm{q}}}{D_{\bm{q}'}} =
    \Qantico{B^\dag_{\bm{q}}}{D^\dag_{\bm{q}'}} = 0 \,.
\end{equation}
The interplay with the creation and annihilation operators for photons is governed by the commutators
\begin{equation}\label{BRSTanticomrelaB}
    \Qcommu{a^\alpha_{\bm{q}}}{B_{\bm{q}'}} =
    \Qcommu{a^\alpha_{\bm{q}}}{B^\dag_{\bm{q}'}} =
    \Qcommu{a^{\alpha \, \dag}_{\bm{q}}}{B_{\bm{q}'}} =
    \Qcommu{a^{\alpha \, \dag}_{\bm{q}}}{B^\dag_{\bm{q}'}} = 0 \,,
\end{equation}
and
\begin{equation}\label{BRSTanticomrelaD}
    \Qcommu{a^\alpha_{\bm{q}}}{D_{\bm{q}'}} =
    \Qcommu{a^\alpha_{\bm{q}}}{D^\dag_{\bm{q}'}} =
    \Qcommu{a^{\alpha \, \dag}_{\bm{q}}}{D_{\bm{q}'}} =
    \Qcommu{a^{\alpha \, \dag}_{\bm{q}}}{D^\dag_{\bm{q}'}} = 0 \,.
\end{equation}

\index{BRST quantization}The anticommutation relations (\ref{BRSTanticomrelB})--(\ref{BRSTanticomrelBD}) imply fermionic behavior, whereas the absence of a spin or polarization index on the additional creation and annihilation operators implies spin zero. As these additional field quanta, which characterize the gauge transformations, violate the spin-statistics theorem (according to which, in three space dimensions, fermions possess half-integer spins and bosons possess integer spins), these quanta are usually referred to as \index{Ghost particle}`ghost particles'. As they are introduced in the context of the electromagnetic field, we can actually think of them as \index{Ghost photon}`ghost photons'. They should better not appear in physical correlation functions. The violation of the spin-statistics theorem is possible because we deal with a signed inner product in evaluating correlation functions.

\index{Ghost particle}The ghost particles are assumed to be massless so that the Hamiltonian (\ref{H0gendef}) of the free electromagnetic field in the enlarged description is generalized to
\begin{equation}\label{HEMBRST}
    \tilde{H}_{\rm EM}^{\rm free} = \sum_{\bm{q} \in K^3_\times} q
    \Big( a^{\alpha \, \dag}_{\bm{q}} a^\alpha_{\bm{q}}
    + B^\dag_{\bm{q}} B_{\bm{q}} + D^\dag_{\bm{q}} D_{\bm{q}} \big) \,,
\end{equation}
where, as always, we exclude zero momentum for massless particles. Also the signed inner product $s^{\rm sign}$ defined in (\ref{relinnerproddef}) needs to be extended properly. Compared to the canonical inner product, we introduce an additional factor of $-1$ for each \index{Ghost particle}ghost particle generated by some $B^\dag_{\bm{q}}$ in exactly the same way as we do for the temporal photons generated by some $a^{0 \, \dag}_{\bm{q}}$.

From a formal perspective, the occurrence of four photons and two ghost photons in (\ref{HEMBRST}) might suggest an underlying six-dimensional space-time. The ghost photons would mediate electromagnetic interactions in the extra dimensions. The signed inner product suggests that two dimensions are time-like (those associated with $a^0_{\bm{q}}$ and $B_{\bm{q}}$) and four dimensions are space-like. The two extra dimensions associated with the ghost photons are clearly distinguished by the anticommutation relations for the ghosts. It may hence be unsurprising that we perceive these two dimensions in a very different way, if at all. Motion and special relativity are still given in terms of four-vectors.

We can now write down the \index{BRST quantization}BRST transformation, which is the quantum counterpart of the gauge transformation (\ref{MaxgaugeAeq}), (\ref{Maxgaugephieq}), and we do that in terms of the BRST charge operator
\begin{equation}\label{BRSTcharge}
    Q = \sum_{\bm{q} \in K^3_\times} q \Big[ \big( a^{0 \, \dag}_{\bm{q}} + a^{3 \, \dag}_{\bm{q}} \big) B_{\bm{q}}
    + D^\dag_{\bm{q}} \big( a^0_{\bm{q}} - a^3_{\bm{q}} \big) \Big] \,,
\end{equation}
which, according to (\ref{gphotondef}) and (\ref{dphotondef}), creates g-photons and annihilates d-photons (in exchange with ghost photons). In terms of $Q$, the \index{BRST quantization}BRST transformation is obtained as
\begin{equation}\label{BRSTtransform}
    \renewcommand{\arraystretch}{1.75}
    \begin{array}{ll}
       \delta a^0_{\bm{q}} = \iR\Qcommu{Q}{a^0_{\bm{q}}} = - \iR q \, B_{\bm{q}} \,, \qquad &
          \delta a^{0 \, \dag}_{\bm{q}} = \iR\Qcommu{Q}{a^{0 \, \dag}_{\bm{q}}} = \iR q \, D^\dag_{\bm{q}} \,, \\
       \delta a^1_{\bm{q}} = \iR\Qcommu{Q}{a^1_{\bm{q}}} = 0 \,, \qquad &
          \delta a^{1 \, \dag}_{\bm{q}} = \iR\Qcommu{Q}{a^{1 \, \dag}_{\bm{q}}} = 0 \,, \\
       \delta a^2_{\bm{q}} = \iR\Qcommu{Q}{a^2_{\bm{q}}} = 0 \,, \qquad &
          \delta a^{2 \, \dag}_{\bm{q}} = \iR\Qcommu{Q}{a^{2 \, \dag}_{\bm{q}}} = 0 \,, \\
       \delta a^3_{\bm{q}} = \iR\Qcommu{Q}{a^3_{\bm{q}}} = -\iR q \, B_{\bm{q}} \,, \qquad &
          \delta a^{3 \, \dag}_{\bm{q}} = \iR\Qcommu{Q}{a^{3 \, \dag}_{\bm{q}}} = - \iR q \, D^\dag_{\bm{q}} \,, \\
       \delta B_{\bm{q}} = \iR\Qantico{Q}{B_{\bm{q}}} = 0 \,, \qquad &
          \delta B^\dag_{\bm{q}} = \iR\Qantico{Q}{B^\dag_{\bm{q}}} =
          \iR q \, \big( a^{0 \, \dag}_{\bm{q}} + a^{3 \, \dag}_{\bm{q}} \big) \,, \\
       \delta D_{\bm{q}} = \iR\Qantico{Q}{D_{\bm{q}}} = \iR q \big( a^0_{\bm{q}} - a^3_{\bm{q}} \big) \,, \qquad &
          \delta D^\dag_{\bm{q}} = \iR \Qantico{Q}{D^\dag_{\bm{q}}} = 0 \,.
    \end{array}
    \renewcommand{\arraystretch}{1}
\end{equation}
Note that the gauge transformation (\ref{MaxgaugeAeq}), (\ref{Maxgaugephieq}) affects the temporal and the longitudinal parts of the four-vector potential, which is mimicked by the first and fourth lines of the \index{BRST quantization}BRST transformation (\ref{BRSTtransform}) in an operator sense.\footnote{\label{fntimederop}The time derivative of an operator $A$ occurring in (\ref{Maxgaugephieq}) should be interpreted as $\partial A/\partial t = \iR \Qcommu{H}{A}$, which differs from the reversible evolution of a density matrix by the minus sign corresponding to the difference between the Heisenberg and Schr\"odinger pictures. The commutators $\iR \Qcommu{\tilde{H}_{\rm EM}^{\rm free}}{B_{\bm{q}}} = - \iR|\bm{q}| \, B_{\bm{q}}$ and $\iR \Qcommu{\tilde{H}_{\rm EM}^{\rm free}}{D^\dag_{\bm{q}}} = \iR|\bm{q}| \, D^\dag_{\bm{q}}$ show that the transformations in the first line of (\ref{BRSTtransform}) can be interpreted as the time derivatives of $B_{\bm{q}}$ and $D^\dag_{\bm{q}}$; the operators $-\iR|\bm{q}| \, B_{\bm{q}}$ and $- \iR|\bm{q}| \, D^\dag_{\bm{q}}$ in the fourth line of (\ref{BRSTtransform}) reflect the longitudinal nature of the gradient in (\ref{MaxgaugeAeq}) in Fourier space.} As announced, the operators $B_{\bm{q}}$ and $D^\dag_{\bm{q}}$ appear as the operator generalizations of gauge transformations. Their own transformation behaviour is governed by g- and d-photons.

The \index{BRST quantization}BRST transformation (\ref{BRSTtransform}) has a number of interesting properties. As the quantum generalization of the gauge transformation (\ref{MaxgaugeAeq}), (\ref{Maxgaugephieq}), we expect it to lead to a symmetry of the theory of electromagnetism that can be expressed as
\begin{equation}\label{BRSTconservation1}
    \Qcommu{\tilde{H}_{\rm EM}^{\rm free}}{Q} = 0 \,,
\end{equation}
which can indeed be verified by inserting (\ref{HEMBRST}), (\ref{BRSTcharge}) and using the various commutation and anticommutation relations for creation and annihilation operators. Also for the anti-BRST charge operator\index{BRST quantization}
\begin{equation}\label{antiBRSTcharge}
    Q^\ddag = \sum_{\bm{q} \in K^3_\times} q \Big[ B^\dag_{\bm{q}} \big( a^0_{\bm{q}} - a^3_{\bm{q}} \big) -
    \big( a^{0 \, \dag}_{\bm{q}} + a^{3 \, \dag}_{\bm{q}} \big) D_{\bm{q}} \Big] \,,
\end{equation}
which is the signed adjoint of $Q$ obtained by introducing minus signs in the canonical adjoints of  $a^0_{\bm{q}}$, $a^{0 \, \dag}_{\bm{q}}$ and $B_{\bm{q}}$, one finds a vanishing commutator with the Hamiltonian,
\begin{equation}\label{BRSTconservation2}
    \Qcommu{\tilde{H}_{\rm EM}^{\rm free}}{Q^\ddag} = 0 \,.
\end{equation}
Thanks to the minus signs associated with temporal photons in the inner product, as $Q$ also $Q^\ddag$ creates g-photons and annihilates d-photons. A further important property is that $\delta^2 \cdot = - \Qcommu{Q^2}{\cdot}$ vanishes for each of the basic creation and annihilation operators; in other words, $Q^2=0$, or the \index{BRST quantization}BRST transformation is nilpotent. Note that the anticommutation of \index{Ghost particle}ghost particle operators is crucial for establishing the nilpotency of $Q$ and $Q^\ddag$.

As $Q$ and $Q^\ddag$ are conserved by the free Hamiltonian $\tilde{H}_{\rm EM}^{\rm free}$, we can find the constrained dynamics simply by fixing the values of $Q$ and $Q^\ddag$. We define a state $\Dket{\psi}$ to be physical if
\begin{equation}\label{BRSTphysical}
    Q \Dket{\psi} = Q^\ddag \Dket{\psi} = 0 \,.
\end{equation}
We moreover define two physical states to be equivalent if the signed norm of their difference vanishes. For density matrices $\rho$, the conditions (\ref{BRSTphysical}) are replaced by\index{BRST quantization}
\begin{equation}\label{BRSTphysicalrho}
    \Qcommu{Q}{\rho} = \Qcommu{Q^\ddag}{\rho} = 0 \,.
\end{equation}
The condition (\ref{BRSTphysicalrho}) for the zero-temperature Gibbs state, together with the nilpotency of $Q$, implies (\ref{BRSTphysical}) for the ground state.

Note that the conditions (\ref{BRSTphysical}) are satisfied if $\Dket{\psi}$ does not contain any \index{Ghost particle}ghost particles or d-photons. In other words, $\Dket{\psi}$ contains only transverse and g-photons. A more detailed analysis shows that these conditions are not only sufficient, but also necessary. All physical states have a nonnegative signed norm, where the norm zero occurs whenever a state contains at least one g-photon. Any state $\Dket{\psi}$ containing a g-photon can be written as
\begin{equation}\label{psiQsomething}
    \Dket{\psi} = Q \Dket{\phi} = Q^\ddag \Dket{\bar{\phi}} \,,
\end{equation}
so that
\begin{equation}\label{nil4nothing}
    \Dbraket{\psi}{\psi} = \Dbra{\bar{\phi}} Q^2 \Dket{\phi} = 0 \,,
\end{equation}
as a consequence of the nilpotency of $Q$. For example, we have
\begin{equation}\label{psiQsomethingex}
    a^{{\rm g} \, \dag}_{\bm{q}} a^{{\rm g} \, \dag}_{\bm{q}'} \Dket{0} =
    Q \frac{1}{\sqrt{2}\,q} B^\dag_{\bm{q}} a^{{\rm g} \, \dag}_{\bm{q}'} \Dket{0} =
    Q^\ddag \frac{-1}{\sqrt{2}\,q} D^\dag_{\bm{q}} a^{{\rm g} \, \dag}_{\bm{q}'} \Dket{0} \,.
\end{equation}
Any state in the Fock space of transverse and g-photons can be written as $\Dket{\psi}_0+\Dket{\psi}_{1+}$, where $\Dket{\psi}_0$ is a linear combination of Fock base vectors containing only transverse photons and $\Dket{\psi}_{1+}$ is a linear combination of Fock base vectors containing at least one g-photon. Each vector $\Dket{\psi}_{1+}$ has a vanishing signed norm. Each vector $\Dket{\psi}_0$ defines a class of equivalent physical states. All the ingredients of the Bleuler-Gupta approach are thus elegantly reproduced within the framework of \index{BRST quantization}BRST quantization.

To prepare the \index{BRST quantization}BRST quantization in the presence of electric charges, we rewrite the BRST charges (\ref{BRSTcharge}) and (\ref{antiBRSTcharge}) as
\begin{equation}\label{BRSTchargegau}
    Q = \sum_{\bm{q} \in K^3_\times} \Big\{ \big( \Qcommu{\tilde{H}_{\rm EM}^{\rm free}}{a^{0 \, \dag}_{\bm{q}}}
    + q a^{3 \, \dag}_{\bm{q}} \big) B_{\bm{q}}
    - D^\dag_{\bm{q}} \big( \Qcommu{\tilde{H}_{\rm EM}^{\rm free}}{a^0_{\bm{q}}}
    + q a^3_{\bm{q}} \big) \Big\} \,,
\end{equation}
and
\begin{equation}\label{antiBRSTchargegau}
    Q^\ddag = - \sum_{\bm{q} \in K^3_\times} \Big\{ B^\dag_{\bm{q}} \big( \Qcommu{\tilde{H}_{\rm EM}^{\rm free}}{a^0_{\bm{q}}}
    + q a^3_{\bm{q}} \big)
    + \big( \Qcommu{\tilde{H}_{\rm EM}^{\rm free}}{a^{0 \, \dag}_{\bm{q}}}
    + q a^{3 \, \dag}_{\bm{q}} \big) D_{\bm{q}} \Big\} \,.
\end{equation}
In this version of the BRST charges, the gauge condition on the four-vector potential is more easily recognizable (see footnote on p.\,\pageref{fntimederop}).

In short, the general idea is to quantize in an enlarged Hilbert space and to characterize the physically admissible states in terms of BRST charges, which generate \index{BRST quantization}BRST transformations and commute with the Hamiltonian. BRST symmetry has been considered as a fundamental principle that replaces gauge symmetry \cite{Nemeschanskyetal86}. In the author's opinion, \index{BRST quantization}BRST symmetry simply provides a straightforward quantum implementation of gauge symmetry.

\subsubsection{Electromagnetic fields and charged matter}\label{secBRSTmatter}
The gauge transformation (\ref{MaxgaugeAeq}) involves a spatial derivative. We should hence expect that the implementation of gauge or \index{BRST quantization}BRST symmetry requires consideration of the high momenta associated with effects on small length scales. In general, we expect to find Lorentz and gauge symmetry only in the proper limits (vanishing friction, infinite and dense momentum lattice). However, for free electromagnetic fields, we have seen that one can actually implement BRST symmetry rigorously on a finite lattice of momenta. In the presence of charges, we need to consider the infinite lattice $\bar{K}^3_\times$ introduced on page~\pageref{Kbardefpage} to implement BRST symmetry of the reversible dynamics in a rigorous manner.

\index{BRST quantization}The BRST charges (\ref{BRSTchargegau}) and (\ref{antiBRSTchargegau}) can be generalized by adding $H^{\rm coll}$ to $\tilde{H}_{\rm EM}^{\rm free}$ (note that $H^{\rm free}_{\rm e/p}$ commutes with $a^0_{\bm{q}}$ and $a^{0 \, \dag}_{\bm{q}}$). By using (\ref{Hintexpression}), we find the commutators
\begin{equation}\label{Qwithmatterinputs}
    \Qcommu{H^{\rm coll}}{a^{0 \, \dag}_{\bm{q}}} = \frac{1}{\sqrt{2 q}} J^0_{\bm{q}} \,, \qquad
    \Qcommu{H^{\rm coll}}{a^0_{\bm{q}}} = \frac{1}{\sqrt{2 q}} J^0_{-\bm{q}} \,,
\end{equation}
and we hence obtain the following BRST charges in the presence of charged matter,
\begin{equation}\label{BRSTchargem}
    Q = \sum_{\bm{q} \in \bar{K}^3_\times} q
    \left[ \left( a^{0 \, \dag}_{\bm{q}} + a^{3 \, \dag}_{\bm{q}} \right) B_{\bm{q}}
    + D^\dag_{\bm{q}} \left( a^0_{\bm{q}} - a^3_{\bm{q}} \right) \right]
    + \sum_{\bm{q} \in \bar{K}^3} \frac{1}{\sqrt{2 q}} J^0_{\bm{q}} ( B_{\bm{q}}-D^\dag_{-\bm{q}} ) \,,
\end{equation}
and
\begin{equation}\label{antiBRSTchargem}
    Q^\ddag = \sum_{\bm{q} \in \bar{K}^3_\times} q \Big[ B^\dag_{\bm{q}} \big( a^0_{\bm{q}} - a^3_{\bm{q}} \big) -
    \big( a^{0 \, \dag}_{\bm{q}} + a^{3 \, \dag}_{\bm{q}} \big) D_{\bm{q}} \Big]
    - \sum_{\bm{q} \in \bar{K}^3_\times} \frac{1}{\sqrt{2 q}} J^0_{\bm{q}} ( B^\dag_{-\bm{q}}+D_{\bm{q}} ) \,.
\end{equation}
\index{BRST quantization}

We next need to supplement the fundamental (anti)commutation relations (\ref{BRSTanticomrelB})--(\ref{BRSTanticomrelaD}) by trivial anticommutation relations between \index{Ghost particle}ghost particles and fermions,
\begin{equation}\label{BRSTanticomrelbB}
    \Qantico{b^\sigma_{\bm{p}}}{B_{\bm{q}}} =
    \Qantico{b^\sigma_{\bm{p}}}{B^\dag_{\bm{q}}} =
    \Qantico{b^{\sigma \, \dag}_{\bm{p}}}{B_{\bm{q}}} =
    \Qantico{b^{\sigma \, \dag}_{\bm{p}}}{B^\dag_{\bm{q}}} = 0 \,,
\end{equation}
\begin{equation}\label{BRSTanticomrelbD}
    \Qantico{b^\sigma_{\bm{p}}}{D_{\bm{q}}} =
    \Qantico{b^\sigma_{\bm{p}}}{D^\dag_{\bm{q}}} =
    \Qantico{b^{\sigma \, \dag}_{\bm{p}}}{D_{\bm{q}}} =
    \Qantico{b^{\sigma \, \dag}_{\bm{p}}}{D^\dag_{\bm{q}}} = 0 \,,
\end{equation}
\begin{equation}\label{BRSTanticomreldB}
    \Qantico{d^\sigma_{\bm{p}}}{B_{\bm{q}}} =
    \Qantico{d^\sigma_{\bm{p}}}{B^\dag_{\bm{q}}} =
    \Qantico{d^{\sigma \, \dag}_{\bm{p}}}{B_{\bm{q}}} =
    \Qantico{d^{\sigma \, \dag}_{\bm{p}}}{B^\dag_{\bm{q}}} = 0 \,,
\end{equation}
\begin{equation}\label{BRSTanticomreldD}
    \Qantico{d^\sigma_{\bm{p}}}{D_{\bm{q}}} =
    \Qantico{d^\sigma_{\bm{p}}}{D^\dag_{\bm{q}}} =
    \Qantico{d^{\sigma \, \dag}_{\bm{p}}}{D_{\bm{q}}} =
    \Qantico{d^{\sigma \, \dag}_{\bm{p}}}{D^\dag_{\bm{q}}} = 0 \,.
\end{equation}
To find the generalization of the \index{BRST quantization}BRST transformations (\ref{BRSTtransform}) for the lepton operators, it is useful to have the commutators following from (\ref{Jqexplicit}) and (\ref{spinorfieldFou}), (\ref{spinorfieldbarFou}),
\begin{equation}\label{Jqcommutatorbc}
    \Qcommu{J^\mu_{\bm{q}}}{b^{\sigma \, \dag}_{\bm{p}}} = - \frac{e_0}{\sqrt{V}} \,
    \sqrt{\frac{m}{E_p}} \, \bar{\psi}_{-\bm{q}-\bm{p}} \, \gamma^\mu \, u^{\sigma}_{\bm{p}} \,,
\end{equation}
\begin{equation}\label{Jqcommutatordc}
    \Qcommu{J^\mu_{\bm{q}}}{d^{\sigma \, \dag}_{\bm{p}}} = \frac{e_0}{\sqrt{V}} \,
    \sqrt{\frac{m}{E_p}} \, \bar{v}^{\sigma}_{\bm{p}} \, \gamma^\mu \, \psi_{\bm{q}+\bm{p}} \,,
\end{equation}
\begin{equation}\label{Jqcommutatorba}
    \Qcommu{J^\mu_{\bm{q}}}{b^{\sigma}_{\bm{p}}} = \frac{e_0}{\sqrt{V}} \,
    \sqrt{\frac{m}{E_p}} \, \bar{u}^{\sigma}_{\bm{p}} \, \gamma^\mu \, \psi_{\bm{q}-\bm{p}} \,,
\end{equation}
\begin{equation}\label{Jqcommutatorda}
    \Qcommu{J^\mu_{\bm{q}}}{d^{\sigma}_{\bm{p}}} = - \frac{e_0}{\sqrt{V}} \,
    \sqrt{\frac{m}{E_p}} \, \bar{\psi}_{-\bm{q}+\bm{p}} \, \gamma^\mu \, v^{\sigma}_{\bm{p}} \,.
\end{equation}
For the finite lattice $K^3$, it is not clear that with $\bm{q}$ and $\bm{p}$ also $\pm\bm{q}\pm\bm{p}$ are contained in $K^3$. For the infinite lattice $\bar{K}^3$, this is guaranteed. For example, we then obtain
\begin{equation}\label{BRSTtransformpsi}
  \delta \psi^\dag_{\bm{p}} = \iR\Qantico{Q}{\psi^\dag_{\bm{p}}} =
  \iR \frac{e_0}{\sqrt{V}} \sum_{\bm{q} \in \bar{K}^3_\times} \frac{1}{\sqrt{2 q}}
  \sum_{\bm{p'} \in \bar{K}^3} \delta_{\bm{p}'+\bm{q},\bm{p}} \,
  \psi^\dag_{\bm{p}'} ( B_{\bm{q}}-D^\dag_{-\bm{q}} ) \,.
\end{equation}
Note that the ghost photon operators in the BRST transformation (\ref{BRSTtransformpsi}) are multiplied by $\psi^\dag_{\bm{p}'}$; this observation suggests that the transformation acts on a phase in an exponential. The convolution in momentum space implies that this phase shift is local in position space. The commutation relations (\ref{Jqcommutatorbc})--(\ref{Jqcommutatorda}) furthermore imply the vanishing commutator
\begin{equation}\label{Jqconservation}
    \bigg[ J^\mu_{\bm{q}} , \sum_{\bm{p} \in \bar{K}^3} \Big( d^{\sigma \, \dag}_{\bm{p}} d^{\sigma}_{\bm{p}}
    - b^{\sigma \, \dag}_{\bm{p}} b^{\sigma}_{\bm{p}} \Big) \bigg] = 0 \,,
\end{equation}
which expresses the conservation of the total charge of all electrons and positrons by the electric flux four-vector.

Nilpotency of the operators $Q$ and $Q^\ddag$ defined in (\ref{BRSTchargem}) and (\ref{antiBRSTchargem}) is an immediate consequence of the anticommutation rules assumed for \index{Ghost photon}ghost photons and of the commutation relation (\ref{JJcommucarefuls0}). The proper generalization of the symmetries (\ref{BRSTconservation1}) and (\ref{BRSTconservation2}) in the presence of charged matter is given by
\begin{equation}\label{BRSTconservation1q}
    \Qcommu{\tilde{H}_{\rm EM}^{\rm free}+H^{\rm free}_{\rm e/p}+H^{\rm coll}+e''V}{Q} = 0 \,,
\end{equation}
and
\begin{equation}\label{BRSTconservation2q}
    \Qcommu{\tilde{H}_{\rm EM}^{\rm free}+H^{\rm free}_{\rm e/p}+H^{\rm coll}+e''V}{Q^\ddag} = 0 \,.
\end{equation}
To verify (\ref{BRSTconservation1q}), one can make use of the three intermediate results
\begin{equation}\label{BRSTconservation1qa}
    \Qcommu{\tilde{H}_{\rm EM}^{\rm free}}{Q} = - \sum_{\bm{q} \in \bar{K}^3_\times} \sqrt{\frac{q}{2}}
    J^0_{\bm{q}} \, ( B_{\bm{q}}+D^\dag_{-\bm{q}} ) \,,
\end{equation}
\begin{equation}\label{BRSTconservation1qb}
    \Qcommu{H^{\rm free}_{\rm e/p}}{Q} = \sum_{\bm{q} \in \bar{K}^3_\times} \frac{1}{\sqrt{2 q}}
    \Qcommu{H^{\rm free}_{\rm e/p}}{J^0_{\bm{q}}} \, ( B_{\bm{q}}-D^\dag_{-\bm{q}} ) \,,
\end{equation}
\begin{equation}\label{BRSTconservation1qc}
    \Qcommu{H^{\rm coll}}{Q} = \sum_{\bm{q} \in \bar{K}^3_\times} \sqrt{\frac{q}{2}}
    J^0_{\bm{q}} \, ( B_{\bm{q}}+D^\dag_{-\bm{q}} )
    - \sum_{\bm{q} \in \bar{K}^3_\times} \frac{1}{\sqrt{2 q}}
    q_j J^j_{\bm{q}} \, ( B_{\bm{q}}-D^\dag_{-\bm{q}} ) \,,
\end{equation}
in combination with local electric charge conservation (\ref{localelchargecons}). We can then proceed as in the case of free electromagnetic fields.

We finally remark that $Q \Dket{0} \neq 0 \neq Q^\ddag \Dket{0}$, so that $\Dket{0}$, unlike the physical ground state $\Dket{\Omega}$ [see remark after (\ref{BRSTphysicalrho})], is not a physical state of the interacting theory. To modify it properly, we rewrite the \index{BRST quantization}BRST charges (\ref{BRSTchargem}) and (\ref{antiBRSTchargem}) as
\begin{equation}\label{BRSTchargemx}
    Q = \sqrt{2} \sum_{\bm{q} \in \bar{K}^3_\times} q
    \left[ \left( a^{{\rm g} \, \dag}_{\bm{q}} + \frac{1}{2q^{3/2}} J^0_{\bm{q}} \right) B_{\bm{q}}
    + \iR D^\dag_{\bm{q}} \left( a^{\rm d}_{\bm{q}} + \frac{\iR}{2q^{3/2}} J^0_{-\bm{q}} \right) \right] \,,
\end{equation}
and
\begin{equation}\label{antiBRSTchargemx}
    Q^\ddag = \sqrt{2} \sum_{\bm{q} \in \bar{K}^3_\times} q
    \left[ \iR B^\dag_{\bm{q}} \left( a^{\rm d}_{\bm{q}} + \frac{\iR}{2q^{3/2}} J^0_{-\bm{q}} \right)
    - \left( a^{{\rm g} \, \dag}_{\bm{q}} + \frac{1}{2q^{3/2}} J^0_{\bm{q}} \right) D_{\bm{q}} \right] \,,
\end{equation}
and introduce the operator
\begin{equation}\label{Exp4physical}
    E = \exp \left\{ - \sum_{\bm{q}} \frac{\iR}{2q^{3/2}} J^0_{-\bm{q}} a^{{\rm d} \, \dag}_{\bm{q}} \right\} \,.
\end{equation}
This operator has the useful property
\begin{equation}\label{Exp4physical1}
    \Qcommu{a^{\rm d}_{\bm{q}}}{E} = - \frac{\iR}{2q^{3/2}} J^0_{-\bm{q}} \, E \,,
\end{equation}
or
\begin{equation}\label{Exp4physical2}
    \left( a^{\rm d}_{\bm{q}} + \frac{\iR}{2q^{3/2}} J^0_{-\bm{q}} \right) E = E \, a^{\rm d}_{\bm{q}} \,,
\end{equation}
which allows us to show that $Q E \Dket{\psi} = Q^\ddag E \Dket{\psi} = 0$ for every state $\Dket{\psi}$ that does not contain any \index{Ghost photon}ghost or d-photons. In other words, the operator $E$ allows us to modify physical states of the free theory such that they become physical states of the interacting theory by introducing the d-photons required in the presence of electric charges.

\subsection{Quantum master equation}
The general ideas for introducing small-scale dissipation into quantum field theory have been discussed in Section~\ref{sectiondynamicsirr} and elaborated in Section~\ref{secRfreeevol}, both in the context of scalar field theory. We here address several complications that arise in the context of quantum electrodynamics.

The most severe problem is caused by the necessity to introduce a signed inner product. As the Hamiltonian is self-adjoint with respect to the signed inner product, but not with respect to the canonical inner product, we expect the same behavior for the equilibrium density matrix (\ref{rhoeqdef}), $\rho_{\rm eq}^\ddag = \rho_{\rm eq}$. The safest and most physical way to avoid negative probabilities is to insist on BRST invariance of the density matrix according to (\ref{BRSTphysicalrho}), not only in the limit of vanishing friction but even for finite friction.

To obtain the self-adjointness property $\rho_t^\ddag = \rho_t$, we rewrite the general quantum master equation (\ref{QMEthermogen}) as
\begin{eqnarray}
    \frac{\dR \rho_t}{\dR t} &=& -\iR \Qcommu{H}{\rho_t} \nonumber\\
    &-& \sum_\alpha \int\limits_0^1 f_\alpha(u) \bigg(
    \Qcommux{Q_\alpha}{\rho_t^{1-u} \Qcommu{Q^\ddag_\alpha}{\mu_t} \rho_t^u}
    + \Qcommux{Q^\ddag_\alpha}{\rho_t^u \Qcommu{Q_\alpha}{\mu_t} \rho_t^{1-u}}
    \bigg) \dR u \,. \nonumber\\ &&
\label{QMEthermogensig}
\end{eqnarray}
As before, $\mu_t = H+ \kB T \ln \rho_t$ is the free energy operator driving the irreversible dynamics. According to this quantum master equation, the property $\Qcommux{Q}{\rho_t} = 0$ is preserved in time if all the coupling operators $Q_\alpha$ commute with the BRST charge $Q$ and the anticharge $Q^\ddag$,
\begin{equation}\label{QMEgenBRST}
    \Qcommu{Q}{Q_\alpha} = \Qcommu{Q^\ddag}{Q_\alpha} = 0 \,.
\end{equation}
An overview of possible coupling operators $Q_\alpha$ is given in Table~\ref{tableinvarcoupop}.

\begin{table}
\caption[ ]{BRST invariant operators that can be used as coupling operators in a quantum master equation}
\renewcommand{\arraystretch}{1.5}
\begin{center}
\begin{tabular}{|c|c|c|}
\hline
photons & leptons & mixed \\
\hline
$a^1_{\bm{q}}$, $a^{1 \, \dag}_{\bm{q}}$ & $J^\mu_{\bm{q}}$ & $X_{\bm{q}}$,
    $X^\ddag_{\bm{q}} = \sqrt{2}q \, a^{{\rm g} \, \dag}_{\bm{q}} + \frac{1}{\sqrt{2q}} J^0_{\bm{q}}$ \\
$a^2_{\bm{q}}$, $a^{2 \, \dag}_{\bm{q}}$ & $\sum_{\bm{p} \in \bar{K}^3}
    \big( d^{\sigma \, \dag}_{\bm{p}} d^{\sigma}_{\bm{p}}
    - b^{\sigma \, \dag}_{\bm{p}} b^{\sigma}_{\bm{p}} \big)$ & \\
$a^{\rm d}_{\bm{q}}$, $a^{{\rm d} \, \ddag}_{\bm{q}} = \iR a^{{\rm g} \, \dag}_{\bm{q}}$ & & \\
\hline
\end{tabular}
\end{center}
\renewcommand{\arraystretch}{1}
\label{tableinvarcoupop}
\end{table}

Ideally, all particles should be subject to dissipation so that we obtain as much dynamical regularization as possible. But not all degrees of freedom really need to be regularized. The ghost photons are unproblematic because they evolve as free particles, and other degrees of freedom are suppressed by BRST invariance. Whereas the photonic degrees of freedom seem to be sufficiently regularized in the same way as in scalar field theory, the situation for the electrons and positrons. According to Table~\ref{tableinvarcoupop}, the coupling of photons to the heat bath can be achieved through photon creation and annihilation operators, and we can rely on the tried and tested assumption $f_\alpha(u) = \beta \gamma_q \eR^{-u \beta q}$. For leptons, BRST invariance forces us into an entirely different coupling mechanism. Note that this mechanism conserves the total electric charge, which is a desirable feature. The mixed operators $X_{\bm{q}}$, $X^\ddag_{\bm{q}}$ offer an interesting alternative option because they possess the properties $\Qcommu{H}{X_{\bm{q}}} = - q X_{\bm{q}}$ and $\Qcommu{H}{X^\ddag_{\bm{q}}} = q X^\ddag_{\bm{q}}$.

As an alternative to rigorous BRST invariance, we could obtain BRST invariance only in the linit of vanishing friction, but make sure that nevertheless no sign problems arise. As the signed inner product is canonical in the entire fermion sector of the Fock space, we could allow for dissipative violation of BRST invariance in the fermion sector. So, we could introduce the standard dissipation mechanism of scalar field theory for electrons and positrons. If we still want to insist on charge conservation, we could use the coupling operators $b^\sigma_{\bm{p}} d^{\sigma'}_{\bm{p}'}$ with the functions $f_\alpha(u) = \beta \gamma_{\bm{p},\bm{p}'}^{\sigma,\sigma'} \eR^{-u \beta (\omega_p + \omega_{p'})}$.

\section{Schwinger model}\label{secSchwinger}
In this section, we consider quantum electrodynamics with massless fermions in one space dimension. This model allows us to clarify some of the issues encountered also in three space dimensions through closed-form solutions, as was first realized by Schwinger \cite{Schwinger62}. Moreover, the Schwinger model has been used as a role model for explaining quark \index{Confinement}confinement by strong interactions \cite{KogutSusskind75a}.

For this toy version of quantum electrodynamics, we follow exactly the same sequence of steps as in the presentation of the elements of our mathematical image of nature in the preceding section. By doing so, we can not only make the presentation more compact and efficient, but we can also clarify some of the issues encountered in the previous section by focusing on similarities and differences.

For massless fermions, we exclude the momentum $q=0$ throughout this section, as we previously did for photons. Therefore, we can in particular divide by $q$. Every massless particle comes with a minimum energy of $K_L=2\pi/L$, where $L$ is the size (length) of the one-dimensional box [see (\ref{Kdlatticedef}) and (\ref{KLchoiceL})]. The limit $L \rightarrow \infty$ is taken in the end of all calculations. Note that we previously used the symbol $q$ for the length of the vector $\bm{q}$. In one space dimension, $q \in K^1_\times$ is used for the momentum variable and its absolute value is explicitly indicated as $|q|$.

It turns out that the Schwinger model does not require any regularization. Therefore, we can work directly in the limit of vanishing dissipation. In other words, it is sufficient to discuss only the reversible version of the Schwinger model.

\subsection{Fock space}
In defining the Fock space, we always need to select the particles of interest and their basic properties. For our toy model in one space dimension, we once more focus on electrons, positrons, and photons.

To achieve a Lorentz covariant treatment of the electromagnetic field, we adapt the ideas of Bleuler \cite{Bleuler50} and Gupta \cite{Gupta50} (see Section~\ref{secQED3dFocksp}) from three space dimensions to one. We consider temporal and longitudinal photons which, in $1+1$ dimensions, are actually the only ones. The only non-vanishing commutation relations for the corresponding photon creation and annihilation operators are given by
\begin{equation}\label{acommutator1d}
    \Qcommu{a^0_q}{a^{0 \, \dag}_{q'}} = \delta_{qq'} \,,
    \qquad
    \Qcommu{a^1_q}{a^{1 \, \dag}_{q'}} = \delta_{qq'} \,,
\end{equation}
for momentum labels $q \in K^1_\times$. We can then construct the two-photon Fock space in the usual way (see Section~\ref{sectionFock}). As in three space dimensions (see Section~\ref{secQED3dFocksp}), we use a modification of the canonical inner product (\ref{caninnerproddef}), where the goal of this modification is to introduce the minus sign associated with the temporal component of the \index{Minkowski metric}Minkowski metric.

In Section~\ref{secfrom3to1} we found that electrons and positrons in one space dimension do not possess spin. We hence introduce the operators $b^\dag_p$ and $b_p$ that create and annihilate an electron of momentum $p$ and similarly the operators $d^\dag_p$ and $d_p$ that create and annihilate a positron of momentum $p$. As we are dealing with fermions, we specify the fundamental anticommutation relations for the creation and annihilation operators in accordance with Section~\ref{sectionFock},
\begin{equation}\label{BRSTanticomrelbd1d}
    \Qantico{b_p}{b^\dag_{p'}} = \delta_{pp'} \,,
    \qquad
    \Qantico{d_p}{d^\dag_{p'}} = \delta_{pp'} \,.
\end{equation}
All other anticommutators among fermion operators vanish. All fermion operators commute with all photon operators. If we include the ghost photons required for BRST symmetry, the total list of creation operators for our Fock space becomes $a^{0 \, \dag}_q$, $a^{1 \, \dag}_q$, $B^\dag_q $, $D^\dag_q$, $b^\dag_q $, $d^\dag_q$, and we have the additional nontrivial anticommutation relations
\begin{equation}\label{BRSTanticomrelBDg1d}
    \Qantico{B_q}{B^\dag_{q'}} = \delta_{qq'} \,,
    \qquad
    \Qantico{D_q}{D^\dag_{q'}} = \delta_{qq'} \,.
\end{equation}
All ghost photon operators commute with all bosonic photon operators and anticommute with all lepton operators. Like $a^{0 \, \dag}_q$, also $B^\dag_q $ contributes to the signed inner product.

\subsection{Fields}
We briefly characterize the spatial fields that can be constructed from the photon and fermion creation and annihilation operators. As always, they are not considered to be part of our image of the real world, but they are useful in motivating the contribution to the Hamiltonian that is associated with electromagnetic interactions.

For quantum electrodynamics in one space dimension, the photon Fock space consists only of longitudinal and temporal photons. The two-component potential $A_{x\,\mu}$ is constructed as in (\ref{4vecpot}),
\begin{equation}\label{2vecpot}
    A_x = \frac{1}{\sqrt{L}} \sum_{q \in K^1_\times} A_q \, \eR^{- \iR q x} \,,
\end{equation}
where the Fourier components for massless photons are given by
\begin{equation}\label{2vecpotFou}
    A_q = \frac{1}{\sqrt{2 |q|}}
    \left( \bar{n}^0_q a^{0 \, \dag}_q - \bar{n}^0_{-q} a^0_{-q}
    + \bar{n}^1_q a^{1 \, \dag}_q + \bar{n}^1_{-q} a^1_{-q} \right) \,,
\end{equation}
and the temporal and longitudinal polarizations can be chosen as
\begin{equation}\label{polarization01}
    \bar{n}^0_q =
    \left( \begin{array}{c}
      1 \\
      0 \\
    \end{array} \right) \,,
    \qquad
    \bar{n}^1_q = {\rm sgn}(q)
    \left( \begin{array}{c}
      0 \\
      1 \\
    \end{array} \right) \,.
\end{equation}
Equation (\ref{2vecpotFou}) can be rewritten in the component form
\begin{equation}\label{vecpotFou2d}
    A_{q \, 0} = - A^0_q = \frac{1}{\sqrt{2|q|}} ( a^{0 \, \dag}_q - a^0_{-q} ) \,, \qquad
    A_{q \, 1} = A^1_q = \frac{{\rm sgn}(q)}{\sqrt{2|q|}} ( a^{1 \, \dag}_q - a^1_{-q} ) \,.
\end{equation}
The normalization of the field (\ref{vecpotFou2d}) is consistent with (\ref{Aexpressionnorm}),
\begin{equation}\label{Aexpressionnorm1d}
  \sum_{\bm{q} \in K^1_\times} \Dbra{N} : (A^\ddag_{q \, 0} A^0_q + A^\ddag_{q \, 1} A^1_q) : \Dket{N} =
  \sum_{\bm{q} \in K^1_\times} \frac{1}{|q|}
  \Dbra{N} ( a^{0 \, \dag}_q a^0_q + a^{1 \, \dag}_q a^1_q ) \Dket{N} \,,
\end{equation}
where $\Dket{N}$ is a Fock space eigenvector with a total of $N$ particles.

The Dirac field $\psi_x$ is built up from the normal modes of the Dirac equation for massless fermions in one space dimension,
\begin{equation}\label{spinorfieldFouS}
    \psi_x = \frac{1}{\sqrt{L}} \sum_{p \in K^1_\times} \psi_p \, \eR^{- \iR p x} \,, \qquad
    \psi_p = \frac{1}{\sqrt{2|p|}} \left( v_p d^{\dag}_p + u_{-p} b_{-p} \right) \,,
\end{equation}
with the spinors defined in (\ref{spinors1dR}) and (\ref{spinors1dL}). The operator $\psi_p$ creates a positron with momentum $p$ or annihilates an electron with momentum $-p$. This is in the spirit of adding antiparticles being equivalent to removing particles with opposite properties. We further have
\begin{equation}\label{spinorfieldbarFouS}
    \bar{\psi}_x = \frac{1}{\sqrt{L}} \sum_{p \in K^1_\times} \bar{\psi}_{-p} \, \eR^{- \iR p x} \,, \qquad
    \bar{\psi}_{-p} = \frac{1}{\sqrt{2|p|}} \left( \bar{v}_{-p} d_{-p} + \bar{u}_p b^{\dag}_p \right) \,,
\end{equation}
where the matrix $\gamma^0$ involved in the bar-operation is now given by (\ref{Schwingermatrices}). The counterpart of the anticommutation relations (\ref{psianticommutator}) is
\begin{equation}\label{psianticommutator1d}
    \Qantico{\psi_p}{\bar{\psi}_{p'}} =
    \gamma^0 \, \delta_{pp'} \,, \qquad
    \Qantico{\psi_p}{\psi_{p'}} = \Qantico{\bar{\psi}_p}{\bar{\psi}_{p'}} = 0 \,.
\end{equation}

Checking the normalization of the Dirac field (\ref{spinorfieldbarFouS}) in the usual way, we find
\begin{equation}\label{spinorfieldnorm1d}
  \int_V \Dbra{N} : \bar{\psi}_x \psi_x : \Dket{N} \dR x =
  \sum_{p \in K^1_\times} \Dbra{N} : \bar{\psi}_p \psi_p : \Dket{N} = 0 \,.
\end{equation}
For massless fermions in one dimension, our usual interpretation of the normalization of the fields does not work. The situation is different for the massless photons in (\ref{Aexpressionnorm1d}) because the construction of Lorentz vectors works entirely different in the two cases. Equation (\ref{spinorfieldnorm1d}) suggests that rest mass rather than particle number is used for the normalization of $\psi_x$. A finite result is obtained for
\begin{eqnarray}
  -e_0 \int_V \Dbra{N} : \bar{\psi}_x \gamma^0 \psi_x : \Dket{N} \dR x &=&
  -e_0 \sum_{p \in K^1_\times} \Dbra{N} : \bar{\psi}_p \gamma^0 \psi_p : \Dket{N}
  \nonumber\\
  &=& e_0 \sum_{\bm{p} \in K^1_\times} \Dbra{N} \left( d^\dag_p d_p
  - b^\dag_p b_p \right) \Dket{N} \,, \qquad\qquad
\label{spinorfieldnormq1d}
\end{eqnarray}
which shows that $\psi_x$ is properly normalized for the charge density.

\subsection{Hamiltonian and current density}
To define reversible dynamics in the Schwinger model, we use the Hamiltonian
\begin{equation}\label{HSchwinger}
    H = \sum_{q \in K^1_\times} |q| \Big( a^{0 \, \dag}_q a^0_q + a^{1 \, \dag}_q a^1_q
    + B^\dag_q B_q + D^\dag_q D_q \Big)
    + H^{\rm free}_{\rm e/p} + H^{\rm coll} \,,
\end{equation}
where the contribution $H^{\rm free}_{\rm e/p}$ describes free electrons and positrons,
\begin{equation}\label{H0Diracdef1d}
    H^{\rm free}_{\rm e/p} = \sum_{q \in K^1_\times} |q| \left( b^\dag_q b_q + d^\dag_q d_q \right) \,,
\end{equation}
and the interaction term $H^{\rm coll}$ is given by
\begin{equation}\label{HcollSchwinger}
    H^{\rm coll} = \sum_{q \in K^1_\times}
    : \left[ \left(J^0_q - \frac{1}{2} m_0^2 A^0_q\right) A^0_{-q}
    - \left(J^1_q - \frac{1}{2} m_1^2 A^1_q\right) A^1_{-q} \right] : + e'' V \,.
\end{equation}
For the time being, we can assume the parameters $m_0$ and $m_1$ to be zero. We will eventually realize why a modification of the current vector $(J^0_q, J^1_q)$ needs to be introduced. The extra terms possess the same structure as the mass terms proportional to $\lambda'$ in (\ref{Hcolk}), and they also serve the same purpose: they make sure that a consistent model can be formulated. As before in scalar field theory, the additional interaction parameters depend on the fundamental coupling strength; we will actually find $m_{0,1} \propto e_0^2$, which should be compared to $\lambda' \propto \lambda^2$ according to (\ref{phi4lpchoice}). For the Schwinger model, however, this result is nonperturbative. The occurrence of $m_0$ and $m_1$ should be seen in conjunction with the subtle Schwinger term (see Section~\ref{secSchwingerterm}) which plays a different role in one dimension than in three dimensions.

The generalization of (\ref{Jqexplicit0}) is straightforward,
\begin{equation}\label{Jqexplicit1d}
    J^\mu_{\bm{q}} = - \frac{e_0}{\sqrt{L}}\sum_{p, p' \in K^1_\times}
    \delta_{p-p',q} \, \bar{\psi}_{-p} \, \gamma^\mu \, \psi_{-p'} \,,
\end{equation}
from which we obtain the following normal-ordered Fourier components of the electric charge and current densities,
\begin{equation}\label{SchwingerJ0}
    J^0_q = - \frac{e_0}{\sqrt{L}} \sum_{p, p' \in K^1_\times} \hspace{-.7em} \delta_{p-p',q} \Big[
    \Theta_{p,p'} ( b^{\dag}_p b_{p'} - d^{\dag}_{-p'} d_{-p} )
    - {\rm sgn}(p) \Theta_{p,-p'} ( d_{-p} b_{p'} - b^{\dag}_p d^{\dag}_{-p'} ) \Big] ,
\end{equation}
and
\begin{equation}\label{SchwingerJ1}
    J^1_q = - \frac{e_0}{\sqrt{L}} \sum_{p, p' \in K^1_\times} \hspace{-.7em} \delta_{p-p',q} \Big[
    {\rm sgn}(p) \Theta_{p,p'} ( b^{\dag}_p b_{p'} + d^{\dag}_{-p'} d_{-p} )
    + \Theta_{p,-p'} ( d_{-p} b_{p'} + b^{\dag}_p d^{\dag}_{-p'} ) \Big] ,
\end{equation}
where it is convenient to introduce
\begin{equation}\label{Thetaequalsigns}
    \Theta_{p,p'} = \frac{1}{2} \, [ 1 + {\rm sgn}(p p') ] \,.
\end{equation}
Instead of (\ref{localelchargecons}), we now have the following more symmetric relations between $J^0_q$ and $J^1_q$,
\begin{equation}\label{SchwingerevolJ0vS}
     \Qcommu{H^{\rm free}_{\rm e/p}}{J^0_q} = q \, J^1_q  \,,
\end{equation}
and
\begin{equation}\label{SchwingerevolJ1vS}
     \Qcommu{H^{\rm free}_{\rm e/p}}{J^1_q} = q \, J^0_q \,.
\end{equation}

Finally, we can derive the modified Maxwell equations
\begin{equation}\label{modMaxSchw0}
    \Qcommu{H}{\Qcommu{H}{A^0_q}} = (q^2 + m_0^2 ) A^0_q - J^0_q \,,
\end{equation}
and
\begin{equation}\label{modMaxSchw1}
    \Qcommu{H}{\Qcommu{H}{A^1_q}} = (q^2 + m_1^2 ) A^1_q - J^1_q \,.
\end{equation}
These equations suggest that the temporal and longitudinal photons can gain the masses $m_0$ and $m_1$, respectively.

\subsection{Schwinger term}\label{secSchwingerterm1d}
Equation (\ref{JJcommucareful}) can easily be adapted to one space dimension. We then obtain
\begin{equation}\label{JJcommucarefulNO1d}
  \Qcommu{J^\mu_q}{J^0_{q'}} = \frac{e_0^2}{L} \, \delta_{-q,q'}
  \sum_{p, \bar{p} \in K^1_\times}
  ( \delta_{\bar{p},p+q} - \delta_{\bar{p},p-q} ) \,
  \frac{p^\mu}{|p|} \,,
\end{equation}
which differs from (\ref{JJcommucarefulNO}) by a factor of two. This difference is a consequence of the two-dimensional rather than four-dimensional representation of spinors or, in other words, of the absence of lepton spin in one space dimension. As in (\ref{JJcommucarefuls0}) we conclude that $\Qcommu{J^0_q}{J^0_{q'}}$ must be zero, and the same is true for $\Qcommu{J^1_q}{J^1_{q'}}$. In one space dimension, however, the counterpart of (\ref{JJcommucarefulsjf}) becomes much simpler,
\begin{equation}\label{SchwingerJJviaS}
    \Qcommu{J^1_q}{J^0_{q'}} =  - \frac{e_0^2}{\pi} \, q \, \delta_{-q q'} \,.
\end{equation}
There is no divergence for large momentum cutoffs, there is no dependence on the system size, and there is no ambiguous prefactor. We hence accept the Schwinger term (\ref{SchwingerJJviaS}) as a physical result for quantum electrodynamics in one space dimension.

\subsection{BRST quantization}\label{secBRSTSchw}
In Section~\ref{secBRSTfree}, we had motivated the construction of BRST charges by the form of gauge transformations. By closer inspection of (\ref{BRSTcharge}) and (\ref{antiBRSTcharge}) as well as (\ref{BRSTchargemx}) and (\ref{antiBRSTchargemx}), we note the following formal structure of BRST charges and anticharges in terms of suitable bosonic operators $X_q$,
\begin{equation}\label{BRSTSchwinger0}
    Q = \sum_{q \in K^1_\times} \big( X^\ddag_q B_q - D^\dag_q X_q \big) \,,
\end{equation}
and
\begin{equation}\label{antiBRSTSchwinger0}
    Q^\ddag = - \sum_{q \in K^1_\times} \big( B^\dag_q X_q + X^\ddag_q D_q \big) \,.
\end{equation}
If the operators $X_q$ commute among each other and with the ghost photon operators $B_q$, $B^\dag_q$, $D_q$, $D^\dag_q$, we find
\begin{equation}\label{Q2genstructure}
    Q^2 = \sum_{q,q' \in K^1_\times} B_{q'} D^\dag_q \, \Qcommu{X_q}{X^\ddag_{q'}} \,.
\end{equation}
Nilpotency of the BRST charge $Q$ therefore requires $\Qcommu{X_q}{X^\ddag_{q'}}=0$.

A sufficient condition for the operators $Q$ and $Q^\ddag$ to commute with the full Hamiltonian $H$ in (\ref{HSchwinger}) is given by
\begin{equation}\label{BRSTconserveq}
    \Qcommu{H}{X_q} = - |q| X_q \,.
\end{equation}
These conditions look like the characteristic of annihilation operators for massless particles. They should be associated with particles that cannot occur in physically admissible states. Indeed, for free fields, we found the annihilation operators $X_q = |q| \left( a^1_q - a^0_q \right)$, which lead to the rule that physical states for free fields cannot contain any d-photons. Nilpotency of $Q$ relies on $a^{0 \, \ddag}_q = - a^{0 \, \dag}_q$ and thus provides a deep reason for introducing signed inner products.

In the presence of charged leptons, the construction of $X_q$ must be reconsidered because the Schwinger term now leads to the nonvanishing commutators
\begin{equation}\label{HcollJ0commu}
    \Qcommu{H^{\rm coll}}{J^0_{-q}} = \frac{e_0^2}{\pi} \, q \, A^1_{-q} \,,
\end{equation}
and
\begin{equation}\label{HcollJ1commu}
    \Qcommu{H^{\rm coll}}{J^1_{-q}} = - \frac{e_0^2}{\pi} \, q \, A^0_{-q} \,.
\end{equation}
We try the general \emph{ansatz}
\begin{equation}\label{BRSTSchwingerXq}
    X_q = |q| \left( a^1_q - a^0_q \right) + \frac{1}{\sqrt{2 |q|}} \Big[ (1+\alpha) J^0_{-q}
    + \beta \, {\rm sgn}(q) J^1_{-q} \Big] \,,
\end{equation}
where the coefficients $\alpha$ and $\beta$ remain to be determined [$\alpha=\beta=0$ corresponds to (\ref{BRSTchargem})]. The ${\rm sgn}(q)$ has been introduced so that all these operators $X_q$ commute among each other and the BRST charge is nilpotent. By means of (\ref{SchwingerevolJ0vS}) and (\ref{SchwingerevolJ1vS}), we obtain
\begin{equation}\label{BRSTSchwingerXqcom}
    \Qcommu{H}{X_q} = \Qcommu{H^{\rm coll}}{X_q} -|q| \left\{ |q| \left( a^1_q - a^0_q \right)
    + \frac{1}{\sqrt{2 |q|}} \Big[ \beta J^0_{-q}
    + (1+\alpha) {\rm sgn}(q) J^1_{-q} \Big] \right\} \,.
\end{equation}
Using (\ref{HcollJ0commu}), (\ref{HcollJ1commu}), we realize that we can achieve
\begin{equation}\label{BRSTSchwingerXqcomx}
    \Qcommu{H^{\rm coll}}{X_q} = -|q| \frac{1}{\sqrt{2 |q|}}
    \Big[ J^0_{-q} - {\rm sgn}(q) J^1_{-q} \Big] \,,
\end{equation}
provided that we chose
\begin{equation}\label{SchwingerM0m1choices1}
    m_0^2 = \beta \, \frac{e_0^2}{\pi} \,, \qquad
    m_1^2 = (1+\alpha) \, \frac{e_0^2}{\pi} \,.
\end{equation}
The property (\ref{BRSTconserveq}) is finally obtained for
\begin{equation}\label{SchwingerM0m1choices2}
    \alpha = \beta \,.
\end{equation}
These considerations show that, as a consequence of the nontrivial Schwinger term, we cannot achieve BRST symmetry without giving mass at least to the longitudinal photons. The simplest way of achieving BRST symmetry is by choosing $\alpha = \beta = 0$. In other words, we choose the BRST charge as close as possible to the one for quantum electrodynamics in three space dimensions. Then, temporal photons remain massless, $m_0=0$, whereas longitudinal photons have the mass $m_1 = e_0/\sqrt{\pi}$. This photon mass is a beautiful exact result due to Schwinger. Other choices of the photon masses would be possible, but we must always have $m_1^2 - m_0^2 = e_0^2/\pi$; $m_0=0$ is the most natural choice, which we always use from now on.

It is now very easy to identify the BRST invariant operators among the non-ghost operators. For any bosonic operator that commutes with all ghost operators, we have
\begin{equation}\label{BRSTchargeAcommu}
    \Qcommu{Q}{A} = \sum_{q \in K^1_\times}
    \big( \Qcommu{X^\ddag_q}{A} B_q - D^\dag_q \Qcommu{X_q}{A} \big) \,.
\end{equation}
If $A$ commutes with $X_q$ and $X^\ddag_q$, we thus obtain BRST invariance (also the commutator with $Q^\ddag$ vanishes). For our standard choice $\alpha = \beta = 0$, we find the invariant operators
\begin{equation}\label{BRSTinvariantlist}
  a^1_q - a^0_q \,, \qquad a^{1 \, \dag}_q + a^{0 \, \dag}_q \,, \qquad
  J^0_q \,, \qquad J^1_q - \frac{e_0^2}{\pi} A^1_q \,.
\end{equation}
The modified electric charge flux is expected to appear in the local charge conservation law, which must be BRST invariant to be physical. The correlation functions of physical interest involve the charge density and the modified charge flux. Note that any Gibbs state, including the zero-temperature state, commutes with the BRST charges. This implies $Q \Dket{\Omega} = Q^\ddag \Dket{\Omega} = 0$, that is, BRST invariance of the ground state $\Dket{\Omega}$ of the interacting theory [see remark after (\ref{BRSTphysicalrho})].

\subsection{Exact solution}
Equations (\ref{SchwingerevolJ0vS}) and (\ref{HcollJ0commu}) imply
\begin{equation}\label{SchwingerevolJ0}
     \Qcommu{H}{J^0_q} = q \left( J^1_q - \frac{e_0^2}{\pi} A^1_q \right) \,.
\end{equation}
In the same way, (\ref{SchwingerevolJ1vS}) and (\ref{HcollJ1commu}) lead to
\begin{equation}\label{SchwingerevolJ1}
     \Qcommu{H}{J^1_q} = q \left( J^0_q + \frac{e_0^2}{\pi} A^0_q \right) \,.
\end{equation}
Equation (\ref{SchwingerevolJ0}) is a modified version of the charge conservation law in three space dimensions, with a mass-modified charge flux. The list (\ref{BRSTinvariantlist}) of BRST invariant quantities shows that this conservation law involves only invariant quantities and hence is physical. The formal similarity of (\ref{SchwingerevolJ1}) to (\ref{SchwingerevolJ0}) and the simplicity of (\ref{SchwingerevolJ1}) are interesting features of quantum electrodynamics in one space dimension. Let us consider the following additional commutators with the Hamiltonian, which can be considered as Heisenberg time derivatives (see footnote on p.\,\pageref{fntimederop}),
\begin{equation}\label{Schwingerphot0d}
     \Qcommu{H}{a^{0 \, \dag}_q} = |q| a^{0 \, \dag}_q + \frac{1}{\sqrt{2|q|}} J^0_q \,,
\end{equation}
\begin{equation}\label{Schwingerphot0}
     \Qcommu{H}{a^0_{-q}} = - |q| a^0_{-q} + \frac{1}{\sqrt{2|q|}} J^0_q \,,
\end{equation}
\begin{equation}\label{Schwingerphot1d}
     \Qcommu{H}{a^{1 \, \dag}_q} = |q| a^{1 \, \dag}_q - \frac{{\rm sgn}(q)}{\sqrt{2|q|}}
     \left( J^1_q - \frac{e_0^2}{\pi} A^1_q \right) \,,
\end{equation}
\begin{equation}\label{Schwingerphot1}
     \Qcommu{H}{a^1_{-q}} = - |q| a^1_{-q} - \frac{{\rm sgn}(q)}{\sqrt{2|q|}}
     \left( J^1_q - \frac{e_0^2}{\pi} A^1_q \right) \,.
\end{equation}

In view of (\ref{vecpotFou2d}), the six equations (\ref{SchwingerevolJ0})--(\ref{Schwingerphot1}) form a closed set of linear Heisenberg evolution equations for the six operators $J^0_q$, $a^{0 \, \dag}_q$, $a^0_{-q}$, $J^1_q$, $a^{1 \, \dag}_q$, $a^1_{-q}$ for every value of $q$. This observation shows that we are dealing with an exactly solvable model. It is the simplicity of (\ref{SchwingerevolJ1}) which makes the model solvable; this equation has no such simple counterpart in three space dimensions. The ghost photons evolve independently as massless free particles. In this section, we elaborate on various aspects of the exact solvability of the Schwinger model.

\subsubsection{From fermions to bosons}
Closer inspection of the current vector in (\ref{SchwingerJ0}) and (\ref{SchwingerJ1}) reveals, that the pair creation and annihilation operators
\begin{equation}\label{SchwingerPairopd}
    P^\dag_q = \frac{1}{\sqrt{L}} \left\{ \begin{array}{l} \displaystyle
      \sum_{0<p<q} b^{\dag}_p d^{\dag}_{q-p}
      \quad \mbox{for } q > 0 \,, \\ \\ \displaystyle
      \sum_{0<p<|q|} b^{\dag}_{-p} d^{\dag}_{-|q|+p}
      \quad \mbox{for } q < 0 \,, \\
    \end{array} \right.
\end{equation}
and
\begin{equation}\label{SchwingerPairop}
    P_q = \frac{1}{\sqrt{L}} \left\{ \begin{array}{l} \displaystyle
      \sum_{0<p<q} d_{q-p} b_p
      \quad \mbox{for } q > 0 \,, \\ \\ \displaystyle
      \sum_{0<p<|q|} d_{-|q|+p} b_{-p}
      \quad \mbox{for } q < 0 \,, \\
    \end{array} \right.
\end{equation}
play an important role in the Schwinger model. For positive $q$, the operator $P^\dag_q$ creates an electron-positron pair, where both particles move in the positive direction (right movers) and their momenta add up to $q$. For negative $q$, both particles move in the negative direction (left movers) with total momentum $q$. The convolution of momentum states corresponds to a local product in position space. The importance of these operators in finding exact solutions to fermion problems was recognized in \cite{MattisLieb65}.

The remaining terms in the current vector can be expressed in terms of the operators
\begin{equation}\label{SchwingerIntopd}
    I^\dag_q = \frac{1}{\sqrt{L}} \left\{ \begin{array}{l} \displaystyle
      \sum_{p>0} \left( b^{\dag}_{p+q} b_p - d^{\dag}_{p+q} d_p \right)
      \quad \mbox{for } q > 0 \,, \\ \\ \displaystyle
      \sum_{p>0} \left( d^{\dag}_{-p-|q|} d_{-p} - b^{\dag}_{-p-|q|} b_{-p} \right)
      \quad \mbox{for } q < 0 \,, \\
    \end{array} \right.
\end{equation}
and
\begin{equation}\label{SchwingerIntop}
    I_q = \frac{1}{\sqrt{L}} \left\{ \begin{array}{l} \displaystyle
      \sum_{p>0} \left( b^{\dag}_p b_{p+q} - d^{\dag}_p d_{p+q} \right)
      \quad \mbox{for } q > 0 \,, \\ \\ \displaystyle
      \sum_{p>0} \left( d^{\dag}_{-p} d_{-p-|q|} - b^{\dag}_{-p} b_{-p-|q|} \right)
      \quad \mbox{for } q < 0 \,. \\
    \end{array} \right.
\end{equation}
For positive $q$, the operator $I^\dag_q$ increases the momentum of every electron or positron moving in the positive direction by $q$ (where, on a finite lattice, the maximum momentum cannot be exceeded), and the  operator $I_q$ decreases the momentum of every electron or positron moving in the positive direction by $q$ (provided that the resulting momentum is still positive). For negative $q$, the corresponding operations are performed for left movers.

A nice feature of equations (\ref{SchwingerPairopd})--(\ref{SchwingerIntop}) is that right and left movers are treated in an entirely separated and symmetric way. If we further introduce
\begin{equation}\label{SchwingerSumops}
    S^\dag_q = P^\dag_q + I^\dag_q \,, \qquad
    S_q = P_q + I_q \,,
\end{equation}
we can rewrite the current vector (\ref{SchwingerJ0}), (\ref{SchwingerJ1}) in the form
\begin{equation}\label{J0fromSops}
    J^0_q = - e_0 \, {\rm sgn}(q) ( S^\dag_q - S_{-q} ) \,,
\end{equation}
and
\begin{equation}\label{J1fromSops}
    J^1_q = - e_0 ( S^\dag_q + S_{-q} ) \,,
\end{equation}
with the inverse transformations
\begin{equation}\label{SopfromJsd}
  S^\dag_q = - \frac{1}{2 e_0} [ {\rm sgn}(q) J^0_q + J^1_q ] \,,
\end{equation}
and
\begin{equation}\label{SopfromJs}
  S_q = - \frac{1}{2 e_0} [ {\rm sgn}(q) J^0_{-q} + J^1_{-q} ] \,.
\end{equation}
The commutation relations for the components of the current vector found in Section~\ref{secSchwingerterm1d} imply
\begin{equation}\label{Scommurels}
    \Qcommu{S_q}{S^\dag_{q'}} = \frac{|q|}{2\pi} \, \delta_{q q'} \,, \qquad
    \Qcommu{S_q}{S_{q'}} = \Qcommu{S^\dag_q}{S^\dag_{q'}} = 0 \,.
\end{equation}
After proper normalization, $\hat{S}^\dag_q = \sqrt{2\pi/|q|} \, S^\dag_q$ and $\hat{S}_q = \sqrt{2\pi/|q|} \, S_q$, we obtain the canonical commutation relations
\begin{equation}\label{Shatcommurels}
    \Qcommu{\hat{S}_q}{\hat{S}^\dag_{q'}} = \delta_{q q'} \,, \qquad
    \Qcommu{\hat{S}_q}{\hat{S}_{q'}} = \Qcommu{\hat{S}^\dag_q}{\hat{S}^\dag_{q'}} = 0 \,.
\end{equation}
Instead of (\ref{SchwingerevolJ0}) and (\ref{SchwingerevolJ1}) we can now solve the equations
\begin{equation}\label{HJcreancom1vS}
     \Qcommu{H}{\hat{S}^\dag_q} = |q| \hat{S}^\dag_q
     + \frac{e_0}{2\sqrt{\pi}} \, {\rm sgn}(q)
     \big( a^{1 \, \dag}_q - a^1_{-q} + a^{0 \, \dag}_q - a^0_{-q} \big) \,,
\end{equation}
\begin{equation}\label{HJcreancom2vS}
     \Qcommu{H}{\hat{S}_{-q}} = - |q| \hat{S}_{-q}
     - \frac{e_0}{2\sqrt{\pi}} \, {\rm sgn}(q)
     \big( a^{1 \, \dag}_q - a^1_{-q} - a^{0 \, \dag}_q + a^0_{-q} \big) \,,
\end{equation}
for the pair creation and annihilation operators $\hat{S}^\dag_q$, $\hat{S}_{-q}$ together with (\ref{Schwingerphot0d})--(\ref{Schwingerphot1}) for the photon creation and annihilation operators $a^{0 \, \dag}_q$, $a^0_{-q}$, $a^{1 \, \dag}_q$, $a^1_{-q}$.

The story behind the above definitions of new operators is an important one. Equation (\ref{Shatcommurels}) tells us that we have succeeded in identifying canonical boson creation and annihilation operators $\hat{S}^\dag_q$ and $\hat{S}_q$. The electric charge and current densities can be expressed as linear combinations of these creation and annihilation operators. The possibility of this construction rests on the Schwinger term.

What kind of particles does $\hat{S}^\dag_q$ create? The building block $P^\dag_q$ in (\ref{SchwingerSumops}) suggests that $\hat{S}^\dag_q$ creates pairs of electrons and positrons moving in the same direction. The extra terms associated with $I^\dag_q$ in (\ref{SchwingerSumops}) are required to obtain the proper commutation relations for creation and annihilation operators. These bound pairs are a characteristic feature of electrodynamics in one space dimension. Only for $d>1$ the force between electrons and positrons decays with distance according to a power law. This pairing mechanism for $d=1$ has been used as a role model for explaining quark \index{Confinement}confinement \cite{KogutSusskind75a}.

The derivation of the canonical commutation relations in (\ref{Shatcommurels}) relies on the Schwinger term, the derivation of which is a subtle matter. It is hence worthwhile to find the commutation relations (\ref{Shatcommurels}) more directly from the definitions (\ref{SchwingerSumops}). The various transformations correspond to a very different summation scheme. A summation based on the strict separation of left and right movers in the definitions (\ref{SchwingerPairopd})--(\ref{SchwingerIntop}) may hence lead to more robust results. We find
\begin{equation}\label{Scommurels1}
    \Qcommu{S_q}{S^\dag_{q'}} = \frac{|q|}{2\pi} \, \delta_{q q'} + \Theta(q,q') \frac{1}{L}
    ( b^{\dag}_{q'} b_q - d_q d^{\dag}_{q'} ) \,,
\end{equation}
\begin{equation}\label{Scommurels2}
    \Qcommu{S_q}{S_{q'}} = {\rm sgn}(q) \, \Theta(q,q') \frac{1}{L}
    ( d_{q'} b_q - d_q b_{q'} ) \,,
\end{equation}
and
\begin{equation}\label{Scommurels3}
    \Qcommu{S^\dag_q}{S^\dag_{q'}} = {\rm sgn}(q) \, \Theta(q,q') \frac{1}{L}
    ( b^{\dag}_{q'} d^{\dag}_q - b^{\dag}_q d^{\dag}_{q'} ) \,.
\end{equation}
in the formal limit $L \rightarrow \infty$, we recover the results (\ref{Scommurels}) that are consistent with the Schwinger term.

\subsubsection{An alternative formulation}
Our present Fock space rests on the creation operators $a^{0 \, \dag}_q$, $a^{1 \, \dag}_q$, $B^\dag_q $, $D^\dag_q$, $b^\dag_q $, $d^\dag_q$. We would now like to switch to an alternative Fock space for which the separate electron and positron operators are replaced by the pair creation operator: $a^{0 \, \dag}_q$, $a^{1 \, \dag}_q$, $B^\dag_q $, $D^\dag_q$, $\hat{S}^\dag_q$. This requires a new vacuum state $\Dket{0}'$ that is annihilated by all the corresponding annihilation operators (note that $\hat{S}_q$ does not annihilate $\Dket{0}$ which, however, is annihilated by $b_q $ and $d_q$). Again we introduce a signed inner product for which the temporal and ghost photons $a^{0 \, \dag}_q$ and $B^\dag_q $ come with minus signs. For the Schwinger model, we naturally encounter two different Fock spaces with two different vacuum states.

Reversible dynamics on the new Fock space can be introduced by the Hamiltonian
\begin{eqnarray}
  H &=& \sum_{q \in K^1_\times} |q| \Big( a^{0 \, \dag}_q a^0_q + a^{1 \, \dag}_q a^1_q
  + B^\dag_q B_q + D^\dag_q D_q + \hat{S}^\dag_q \hat{S}_q \Big) \nonumber\\
  &+& \frac{e_0}{2\sqrt{\pi}} \, {\rm sgn}(q)
  \Big[ \big(\hat{S}^\dag_{-q} + \hat{S}_q\big) \big(a^{1 \, \dag}_q - a^1_{-q}\big)
  - \big(\hat{S}^\dag_{-q} - \hat{S}_q\big) \big(a^{0 \, \dag}_q - a^0_{-q}\big) \Big]
  \nonumber\\
  &+& \frac{e_0^2}{4\pi|q|} \Big( 2 a^{1 \, \dag}_q a^1_q
  - a^{1 \, \dag}_q a^{1 \, \dag}_{-q} - a^1_q a^1_{-q} \Big) + e'' V \,.
\label{SchwingerHalter}
\end{eqnarray}
The free part in the first consists of the contributions for massless temporal, longitudinal and ghost photons as well as for lepton pairs. The second line contains the fundamental electromagnetic interactions creating and annihilating electron-positron pairs in combination with photons. The terms in the last line give electromagnetic mass to the longitudinal photons and adjust the ground-state energy to zero. As the Hamiltonian (\ref{SchwingerHalter}) is purely bilinear in boson creation and annihilation operators it might not immediately be recognized as a representation of quantum electrodynamics.

The Hamiltonian (\ref{SchwingerHalter}) reproduces all the commutators (\ref{Schwingerphot0d})--(\ref{Schwingerphot1}), (\ref{HJcreancom1vS}), and (\ref{HJcreancom2vS}) correctly, and it also contains the proper free evolution of the massless ghost photons required to handle the gauge invariance of the theory. After rewriting the operator $X_q$ in (\ref{BRSTSchwingerXq}) for our standard choice $\alpha=\beta=0$ in the form
\begin{equation}\label{BRSTSchwingerXqalt}
    X_q = |q| \left( a^1_q - a^0_q \right) + \frac{e_0}{2\sqrt{\pi}} \, {\rm sgn}(q)
  \big(\hat{S}^\dag_{-q} - \hat{S}_q\big) \,,
\end{equation}
we can verify that the Hamiltonian (\ref{SchwingerHalter}) satisfies (\ref{BRSTconserveq}), and (\ref{BRSTSchwinger0}), (\ref{antiBRSTSchwinger0}) can hence serve as BRST charges on the new Fock space. With the help of (\ref{BRSTchargeAcommu}), the gauge transformation behavior of $\hat{S}^\dag_q$ and $\hat{S}_q$,
\begin{equation}\label{BRSTtransformpair1}
  \delta \hat{S}^\dag_q = \iR\Qcommu{Q}{\hat{S}^\dag_q} =
  - \frac{\iR}{2} \frac{e_0}{\sqrt{\pi}} \, {\rm sgn}(q)
  \big( B_{-q} - D^\dag_q \big) \,,
\end{equation}
\begin{equation}\label{BRSTtransformpair2}
  \delta \hat{S}_q = \iR\Qcommu{Q}{\hat{S}_q} =
  \frac{\iR}{2} \frac{e_0}{\sqrt{\pi}} \, {\rm sgn}(q)
  \big( B_q - D^\dag_{-q} \big) \,,
\end{equation}
turns out to be much simpler than for the lepton creation and annihilation operators. Note the property $\delta\hat{S}^\dag_q-\delta \hat{S}_{-q}=0$, which corresponds to the BRST invariance of the electric charge density.

The alternative construction of the present subsection is not entirely equivalent to our original formulation because we now implement the canonical commutation relations (\ref{Scommurels}) rather than the commutation relations (\ref{Scommurels1})--(\ref{Scommurels3}), which become canonical only in the limit $L \rightarrow \infty$. This has the enormous advantage that the entire construction of the alternative Fock space becomes perfectly rigorous for finite $L$.

With the results of the preceding and the present subsections, we are able to find all correlations involving components of the electric current vector and photons from linear Heisenberg evolution equations. Instead of the electric current vector, we can alternatively use electron-positron pair operators. If we are interested in correlation functions involving electrons or positrons we need to resolve the bosonic electron-positron pairs. This task is considered after providing the explicit solution of the coupled system of Heisenberg evolution equations.

\subsubsection{Decoupled equations}
Equations (\ref{SchwingerevolJ0})--(\ref{Schwingerphot1}) form a set of six coupled linear equations. In the further development, we have produced alternative sets of coupled linear equations. For practical purposes, we would now like to decouple these equations (for such a decoupling in a more conventional setting see, for example, Section~II of \cite{Halpern76}). As a first step, we go to the double-commutator formulas (\ref{modMaxSchw0}), (\ref{modMaxSchw1}) for $m_0^2 = 0$, $m_1^2 = e_0^2/\pi$. Together with (\ref{SchwingerevolJ0}), (\ref{SchwingerevolJ1}), we have four coupled linear equations for $J^0_q$, $A^0_q$, $J^1_q$, and $A^1_q$. By taking the commutator of $H$ with (\ref{SchwingerevolJ0}), multiplying (\ref{SchwingerevolJ1}) by $q$ and adding the results we obtain
\begin{equation}\label{Schwcurrent0qua}
    \Box_q J^0_q = q \frac{e_0^2}{\pi} ( \Qcommu{H}{A^1_q} - q A^0_q ) \,,
\end{equation}
where, for an arbitrary operator $A$, we define
\begin{equation}\label{quabladef}
    \Box_q A = q^2 A - \Qcommu{H}{\Qcommu{H}{A}} \,.
\end{equation}
By taking the commutator of $H$ with (\ref{SchwingerevolJ1}) and multiplying (\ref{SchwingerevolJ0}) by $q$, we similarly obtain
\begin{equation}\label{Schwcurrent1qua}
    \Box_q J^1_q = q \frac{e_0^2}{\pi} ( q A^1_q - \Qcommu{H}{A^0_q} ) \,.
\end{equation}
By acting with $\Box_q + e_0^2/\pi$ on (\ref{Schwcurrent0qua}), using (\ref{modMaxSchw0}), (\ref{modMaxSchw1}) and, subsequently, (\ref{Schwcurrent0qua}), we obtain
\begin{equation}\label{Schwcurrent0qua2}
    \left( \Box_q + \frac{e_0^2}{\pi} \right) \Box_q J^0_q = 0 \,.
\end{equation}
We similarly find
\begin{equation}\label{Schwcurrent1qua2}
    \Box_q \Box_q J^1_q = 0 \,.
\end{equation}
Once the solutions $J^0_q$ and $J^1_q$ of these two Heisenberg evolution equations are obtained, we can solve (\ref{modMaxSchw0}), (\ref{modMaxSchw1}) for $A^0_q$, $A^1_q$, which can now be written in the more compact form
\begin{equation}\label{modMaxSchw0c}
    \Box_q A^0_q = J^0_q \,,
\end{equation}
and
\begin{equation}\label{modMaxSchw1c}
    \left( \Box_q + \frac{e_0^2}{\pi} \right) A^1_q = J^1_q \,.
\end{equation}
We thus have achieved the desired dacoupling of the linear system of equations in (\ref{Schwcurrent0qua2})--(\ref{modMaxSchw1c}). It is easy to recognize the eigenfrequencies $\pm |q|$ and $\pm \omega_q$ associated with the above equations. Of course, the initial and boundary conditions need to be chosen such that they are consistent with the original system of equations (\ref{SchwingerevolJ0})--(\ref{Schwingerphot1}).

Once we know the possible eigenfrequencies $\chi$, we can also identify the corresponding eigenoperators satisfying the condition $\Qcommu{H}{A} = \chi A$,
\begin{eqnarray}
    E_q^1 &=& |q| (a^{0 \, \dag}_q + a^{1 \, \dag}_q) + \frac{1}{\sqrt{2|q|}} J^0_q \,,
    \nonumber \\
    E_q^2 &=& |q|(\omega_q+|q|) a^{0 \, \dag}_q - |q|(\omega_q-|q|) a^0_{-q}
    \nonumber \\
    && + \omega_q(\omega_q+|q|) a^{1 \, \dag}_q - \omega_q(\omega_q-|q|) a^1_{-q}
    - \sqrt{2|q|} \, {\rm sgn}(q) J^1_q \,, \nonumber \\
    &&
\label{Schwingerevec1}
\end{eqnarray}
with the positive eigenvalues $\chi=|q|$ and $\chi=\omega_q$, respectively. The operator $E_q^1 = X_q^\ddag$ previously occurred in the construction of the BRST charges (see (\ref{BRSTSchwingerXq}) for $\alpha = \beta =0$). The operators $E_{-q}^{1 \, \ddag} = X_{-q}$ and $E_{-q}^{2 \, \ddag}$ are eigenoperators with the negative eigenvalues $\chi=-|q|$ and $\chi=-\omega_q$, respectively.

We are now in a position to evaluate correlation functions. By using (\ref{multitimecordef}) and (\ref{HsuperopE}) for purely reversible dynamics and assuming zero temperature, we find the two-time correlation function
\begin{equation}\label{Schwinger2cort}
  \hat{C}^{A_2 A_1}_{t_2 t_1} =
   {\rm tr} \left\{ A_2 {\cal E}_{t_2-t_1} ( A_1 \rho_{\rm eq} ) \right\} =
   \frac{\Dbra{\Omega} \eR^{\iR H(t_2-t_1)} A_2 \eR^{-\iR H(t_2-t_1)} A_1
   \Dket{\Omega}}{\Dbraket{\Omega}{\Omega}} \,,
\end{equation}
where the ground state $\Dket{\Omega}$ of the interacting theory is BRST invariant. In this formula we recognize the usefulness of introducing time-dependent Heisenberg operators in the reversible approach. If $A_1$ is an eigenoperator of $H$ with $\Qcommu{H}{A_1} = \chi A_1$, we obtain
\begin{equation}\label{Schwinger2corte}
  \hat{C}^{A_2 A_1}_{t_2 t_1} = \eR^{-\iR \chi (t_2-t_1)} \,
  \frac{\Dbra{\Omega} A_2  A_1 \Dket{\Omega}}{\Dbraket{\Omega}{\Omega}} \,,
\end{equation}
where we have used the fact that the ground state $\Dket{\Omega}$ is an eigenstate of $H$ with eigenvalue zero. For the Laplace-transformed correlation function (\ref{multitimecorR1}) we finally obtain
\begin{equation}\label{Schwinger2cors}
  \tilde{C}^{A_2 A_1}_s = \frac{1}{s+\iR\chi} \,
  \frac{\Dbra{\Omega} A_2  A_1 \Dket{\Omega}}{\Dbraket{\Omega}{\Omega}} \,.
\end{equation}
We already know this functional form from free propagators. This formula can easily be generalized to any linear combination of eigenoperators, in particular, to the Fourier components of the current and photon fields. As the eigenvalues $\chi=\pm|q|$ are doubly degenerate, also additional terms of the functional forms $(t_2-t_1)\eR^{-\iR \chi (t_2-t_1)}$ and $1/(s+\iR\chi)^2$ can appear in (\ref{Schwinger2corte}) and (\ref{Schwinger2cors}). More precisely, we find the following explicit expressions for the time-dependent Heisenberg operators,
\begin{eqnarray}
  a^{0 \, \dag}_q(t) &=& (C^{1\,\ddag}_q\iR|q|t+C^{1\,\ddag}_q-C^{3\,\ddag}_q) \, \eR^{\iR |q| t}
  - \frac{1}{2} C^1_{-q} \, \eR^{-\iR |q| t} \nonumber\\
  && - \, C^{2\,\ddag}_q|q| (\omega_q+|q|) \, \eR^{\iR \omega_q t}
  + C^2_{-q}|q| (\omega_q-|q|) \, \eR^{-\iR \omega_q t} \,, \nonumber\\
  a^0_{-q}(t) &=& \frac{1}{2} C^{1\,\ddag}_q \, \eR^{\iR |q| t}
  + (C^1_{-q}\iR|q|t-C^1_{-q}+C^3_{-q}) \, \eR^{-\iR |q| t} \nonumber\\
  && - \, C^{2\,\ddag}_q|q| (\omega_q-|q|) \, \eR^{\iR \omega_q t}
  + C^2_{-q}|q| (\omega_q+|q|) \, \eR^{-\iR \omega_q t} \,, \nonumber\\
  a^{1 \, \dag}_q(t) &=& (-C^{1\,\ddag}_q\iR|q|t+C^{3\,\ddag}_q) \, \eR^{\iR |q| t}
  - \frac{1}{2} C^1_{-q} \, \eR^{-\iR |q| t} \nonumber\\
  && + \, C^{2\,\ddag}_q\omega_q (\omega_q+|q|) \, \eR^{\iR \omega_q t}
  + C^2_{-q}\omega_q (\omega_q-|q|) \, \eR^{-\iR \omega_q t} \,, \nonumber\\
  a^1_{-q}(t) &=& -\frac{1}{2} C^{1\,\ddag}_q \, \eR^{\iR |q| t}
  + (C^1_{-q}\iR|q|t+C^3_{-q}) \, \eR^{-\iR |q| t} \nonumber\\
  && + \, C^{2\,\ddag}_q\omega_q (\omega_q-|q|) \, \eR^{\iR \omega_q t}
  + C^2_{-q}\omega_q (\omega_q+|q|) \, \eR^{-\iR \omega_q t} \,,
\label{SchwingerHeissol}
  \\
  \hat{S}^\dag_q(t) &=& m_1 {\rm sgn}(q) \bigg\{
  \left[-\frac{2|q|}{m_1^2}C^{1\,\ddag}_q+ \frac{1}{2|q|} \left(C^{1\,\ddag}_q\iR|q|t-\frac{1}{2}C^{1\,\ddag}_q-C^{3\,\ddag}_q\right)\right]
  \, \eR^{\iR |q| t} \hspace{3.5em} \nonumber\\
  && + \, \frac{1}{2|q|} \left(C^1_{-q}\iR|q|t+\frac{1}{2}C^1_{-q}+C^3_{-q}\right) \, \eR^{-\iR |q| t} \nonumber\\
  && + \, C^{2\,\ddag}_q|q| \, \eR^{\iR \omega_q t}
  + C^2_{-q}|q| \, \eR^{-\iR \omega_q t} \bigg\} \,, \nonumber\\
  \hat{S}_{-q}(t) &=& m_1 {\rm sgn}(q) \bigg\{
  \frac{1}{2|q|} \left(C^{1\,\ddag}_q\iR|q|t-\frac{1}{2}C^{1\,\ddag}_q-C^{3\,\ddag}_q\right)
  \, \eR^{\iR |q| t} \nonumber\\
  && + \, \left[\frac{2|q|}{m_1^2}C^1_{-q}+ \frac{1}{2|q|} \left(C^1_{-q}\iR|q|t+\frac{1}{2}C^1_{-q}+C^3_{-q}\right)\right] \, \eR^{-\iR |q| t} \nonumber\\
  && - \, C^{2\,\ddag}_q|q| \, \eR^{\iR \omega_q t}
  - C^2_{-q}|q| \, \eR^{-\iR \omega_q t} \bigg\} \,, \nonumber
\end{eqnarray}
with the following combinations of initial creation and annihilation operators,
\begin{equation}\label{initialcombs1}
  C^{1\,\ddag}_q = \frac{1}{2} \big(a^{0 \, \dag}_q + a^{1 \, \dag}_q\big)
  - \frac{m_1}{4q} \big( \hat{S}^\dag_q - \hat{S}_{-q} \big)
  = \frac{1}{2|q|} \, E_q^1 \,,
\end{equation}
\begin{eqnarray}
  C^{2\,\ddag}_q &=& \frac{1}{4} \bigg[ \frac{1}{\omega_q (\omega_q-|q|)} a^{0 \, \dag}_q
  - \frac{1}{\omega_q (\omega_q+|q|)} a^0_{-q}
  + \frac{1}{|q| (\omega_q-|q|)} a^{1 \, \dag}_q \nonumber\\
  && - \, \frac{1}{|q| (\omega_q+|q|)} a^1_{-q}
  + \frac{{\rm sgn}(q)}{m_1 \omega_q} \big( \hat{S}^\dag_q + \hat{S}_{-q} \big)\bigg]
  = \frac{1}{4|q|\omega_q m_1^2} \, E_q^2 \,,\qquad\qquad
  \label{initialcombs2}
\end{eqnarray}
and
\begin{eqnarray}
  C^{3\,\ddag}_q &=& -\frac{1}{2} a^{0 \, \dag}_q
  - \frac{q^2}{m_1^2} \big(a^{0 \, \dag}_q + a^{1 \, \dag}_q\big)
  + \frac{1}{4} \big(a^0_{-q} + a^1_{-q}\big) \qquad \nonumber\\
  &&- \, \frac{q}{2m_1} \big( \hat{S}^\dag_q + \hat{S}_{-q} \big)
  - \frac{m_1}{8q} \big( \hat{S}^\dag_q - \hat{S}_{-q} \big) \,.
\label{initialcombs3}
\end{eqnarray}
The equations in (\ref{SchwingerHeissol}) imply
$C^{1\,\ddag}_q(t) = C^{1\,\ddag}_q \, \eR^{\iR |q| t}$ and
$C^{2\,\ddag}_q(t) = C^{2\,\ddag}_q \, \eR^{\iR \omega_q t}$,
as we would expect for these eigenoperators. Moreover, we find
\begin{equation}\label{Schwingernonmode}
  C^{3\,\ddag}_q(t) = \big( C^{3\,\ddag}_q -
  C^{1\,\ddag}_q\iR|q|t \big) \, \eR^{\iR |q| t} \,.
\end{equation}
By taking the adjoint $\ddag$ and replacing $q$ by $-q$, we obtain the corresponding operators $C^j_{-q}$ associated with negative frequencies. The results of this subsection can be used to reduce any time-dependent correlation function involving boson creation and annihilation operators to equilibrium averages.

\subsubsection{From bosons to fermions}\label{secfrombostoferm}
We would now like to calculate also correlation functions involving single leptons. How can we (re)construct fermion operators from boson operators?

On the formal level, the answer to this question is based on the following operator identities (see, for example, (3.5) of \cite{Mandelstam75}; these formulas are related to the Baker-Campbell-Hausdorff formula for the product of exponentials of operators, matrices, or elements of a Lie algebra):
\begin{equation}\label{expnormorderedp}
    :\eR^A: \, :\eR^B: = \eR^{\Qcommu{A_{\rm a}}{B_{\rm c}}} :\eR^{A+B}: \,,
\end{equation}
\begin{equation}\label{expnormordered}
    :\eR^A: \, :\eR^B: = \eR^{\Qcommu{A}{B}} :\eR^B: \, :\eR^A: \,,
\end{equation}
where, as usual, the colons around an operator indicate normal ordering and the operators $A=A_{\rm c}+A_{\rm a}$, $B=B_{\rm c}+B_{\rm a}$ are assumed to be the sum of creation and annihilation operators. To obtain these identities, we must assume that the commutators occurring in these identities are complex numbers. In the following, we consider linear combinations of the creation and annihilation operators $a^{0 \, \dag}_q$, $a^0_{-q}$, $a^{1 \, \dag}_q$, $a^1_{-q}$, $\hat{S}^\dag_q$, and $\hat{S}_{-q}$, including their time-dependence. If we manage to make $\Qcommu{A}{B}$ an odd multiple of $i\pi$, then the normal-ordered exponentials of boson operators become fermion operators.

On the physical level, a comparison of the BRST transformations (\ref{BRSTtransformpsi}) and (\ref{BRSTtransformpair1}), (\ref{BRSTtransformpair2}) reveals that the spinor phases are affected in the same way as the boson operators associated with electron-positron pairs or with the current density.
The boson operators can hence be interpreted as the phases of the fermion operators and we naturally arrive at exponentials of boson operators in reconstructing spinors.

To elaborate the details, we introduce the following scalar field $\varphi$ associated with the bosonic pair creation and annihilation operators $\hat{S}^\dag_q$, $\hat{S}_q$,
\begin{equation}\label{phi4fermions}
  \varphi_q = \frac{\iR}{\sqrt{2|q|}} \big( \hat{S}^\dag_q - \hat{S}_{-q} \big) \,.
\end{equation}
The conjugate momenta of this field (see p.\,\pageref{genmomentadef}) are given by
\begin{equation}\label{phi4fermionsc}
  \pi_q = \iR \Qcommu{H}{\varphi_q} = - \sqrt{\frac{|q|}{2}} \left[ \hat{S}^\dag_q + \hat{S}_{-q}
  + \frac{e_0}{\sqrt{\pi}} \frac{1}{q} \big( a^{1 \, \dag}_q - a^1_{-q} \big) \right] \,.
\end{equation}
The pair $\varphi_q$, $\pi_q$ contains the same information as the electric charge balance equation (\ref{SchwingerevolJ0}). More concretely, it is interesting to note that the scalar field $\varphi$ serves as a potential for the BRST invariant electric current vector \cite{KogutSusskind75a,Halpern76},
\begin{equation}\label{scalpotentialJ0}
  J^0_q = \frac{e_0}{\sqrt{\pi}} \, \iR q \varphi_q \,,
\end{equation}
\begin{equation}\label{scalpotentialJ1}
  J^1_q - \frac{e_0^2}{\pi} A^1_q = \frac{e_0}{\sqrt{\pi}} \pi_q
  = \frac{e_0}{\sqrt{\pi}} \, \iR \Qcommu{H}{\varphi_q} \,.
\end{equation}
The charge conservation law (\ref{SchwingerevolJ0}) serves as an integrability condition which guarantees that $J^0_q$ is minus the space derivative and $J^1_q - (e_0^2/\pi) A^1_q$ is the time derivative of a potential in the Fourier representation. As $\pi_q$ is proportional to the BRST invariant combination $J^1_q - (e_0^2/\pi) A^1_q$, we also introduce the operator $\tilde{\pi}_q$ proportional to $J^1_q$,
\begin{equation}\label{phi4fermionsct}
  \tilde{\pi}_q = - \sqrt{\frac{|q|}{2}} \big( \hat{S}^\dag_q + \hat{S}_{-q} \big) \,.
\end{equation}

The further construction is based on the canonical commutation relation (\ref{posmomcom}) between field operators and conjugate momenta,
\begin{equation}\label{commuformatch1}
    \Qcommu{\varphi_q}{\pi_{-q'}} = \Qcommu{\varphi_q}{\tilde{\pi}_{-q'}} = \iR \delta_{q q'} \,,
\end{equation}
or
\begin{equation}\label{commuformatch2}
    \Qcommu{\varphi_x}{\pi_y} = \Qcommu{\varphi_x}{\tilde{\pi}_y} = \iR \delta(x-y) \,.
\end{equation}
The field operators at different positions commute, and so do the conjugate momenta. Equation (\ref{expnormordered}) then implies that all the operators
\begin{equation}\label{expanticomset}
    \phi^\pm_x = \, \, : \exp\left\{ -i \sqrt{n_{\rm odd} \pi}
    \left( \int_{-L/2}^x \tilde{\pi}_y \, dy \pm \varphi_x \right) \right\} :
\end{equation}
for different $x$ anticommute with each other, provided that $n_{\rm odd}$ is an odd positive integer. The proper BRST transformation behavior (\ref{BRSTtransformpsi}) for the phase of a spinor is obtained for $n_{\rm odd}=1$. An additional motivation for the \emph{ansatz} (\ref{expanticomset}) arises from the fact that these operators characterize pointlike (or kink) \index{Soliton}solitons \cite{Mandelstam75}. Their relevance is a consequence of the fact the bosonic phase-type operators allow for transitions between degenerate vacuum states separated by $2\pi$.

The canonical commutation relations between $\varphi_x$ and $\tilde{\pi}_y$ do not change in time, as can be verified by means of (\ref{SchwingerevolJ0}) and (\ref{SchwingerevolJ1}). When evaluated with the solution of the Heisenberg evolution equations given in the previous subsection, all the $\phi^\pm_x(t)$ for different $x$ are anticommuting operators at any given time $t$. We further obtain
\begin{equation}\label{expanticomsetDt}
    \frac{\partial \phi^\pm_x}{\partial t} = - \iR \sqrt{\pi} \, : \phi^\pm_x
    \left( \int_{-L/2}^x \frac{\partial \tilde{\pi}_y}{\partial t} \, dy
    \pm \pi_x \right) : \,,
\end{equation}
and
\begin{equation}\label{expanticomsetDx}
    \frac{\partial \phi^\pm_x}{\partial x} = - \iR \sqrt{\pi} \, : \phi^\pm_x
    \left( \tilde{\pi}_x
    \pm \frac{\partial \varphi_x}{\partial x} \right) : \,.
\end{equation}
By means of these results we can derive the identities
\begin{equation}\label{expanticomsetdeq1}
    \left[ \iR \frac{\partial}{\partial t} - \iR \frac{\partial}{\partial x}
    + e_0 ( A^0_x + A^1_x ) \right] \phi^+_x = 0 \,,
\end{equation}
and
\begin{equation}\label{expanticomsetdeq2}
    \left[ \iR \frac{\partial}{\partial t} + \iR \frac{\partial}{\partial x}
    + e_0 ( A^0_x - A^1_x ) \right] \phi^-_x = 0 \,.
\end{equation}
I deriving these identities we have neglected the operator $J^0_x + (e_0^2/\pi) A^0_x$ at the boundary $x=-L/2$. A more satisfactory discussion would require some regularization. The identities (\ref{expanticomsetdeq1}), (\ref{expanticomsetdeq2}) imply that
\begin{equation}\label{Schwingerspinoransatz0}
    \psi^{\rm R}_x = \left( \begin{array}{c}
     \phi^-_x \\
     \phi^-_x \\
    \end{array} \right) \,, \qquad
    \psi^{\rm L}_x = \left( \begin{array}{c}
     \phi^+_x \\
     - \phi^+_x \\
    \end{array} \right) \,,
\end{equation}
provide solutions of the Dirac equation (\ref{DiraceqAS}) for massless fermions. These solutions are associated with right and left movers, respectively ([see (\ref{spinors1dR}), (\ref{spinors1dL})]). We hence combine these two degrees of freedom in the form
\begin{equation}\label{Schwingerspinoransatz}
    \psi_x = C_\psi \left( \begin{array}{c}
     \phi^-_x + \phi^+_x \\
     \phi^-_x - \phi^+_x \\
    \end{array} \right) \,,
\end{equation}
where $C_\psi$ is an appropriate normalization constant. Note that this identification depends on our choice of the Dirac matrices for the Schwinger model. With this formula for the fermion spinor in terms of the bosonic pair operators, we are now able to calculate also correlations involving fermion operators. The explicit time-dependence of the Heisenberg version of the spinors (\ref{Schwingerspinoransatz}) with the exponential operators (\ref{expanticomset}) is obtained from the last two equations in (\ref{SchwingerHeissol}).

\chapter{Perspectives}
Let us summarize what we have achieved, which mathematical problems remain to be solved, and where we should go in the future to achieve a faithful image of nature for fundamental particles and all their interactions that fully reflects the state-of-the-art knowledge of quantum field theory. Sections \ref{secopenmathprob} and \ref{secfuturedirec} may be considered as a program for future work.

\section{The nature of quantum field theory}
Lagrangian quantum field theory, the branch of quantum field theory that has delivered spectacular predictions in fundamental particle physics, is plagued by severe divergencies and lacks mathematical rigor. In this book we have shown that, when guided by some pertinent philosophical considerations, we naturally arrive at a perfectly healthy, intuitive version of quantum field theory. All the crucial ingredients to quantum field can be established \emph{a priori}, not just as an \emph{a posteriori} reaction to the occurrence of divergencies or other problems associated with the representation of an infinite number of degrees of freedom. We briefly summarize the key elements of the theory, that is, of our faithful mathematical image of nature, and discuss some implications for the ontology of the most fundamental theory of matter.

\subsection{The image}
We choose to consider space and time as prerequisites for developing physical theories. Although we clearly need to respect the space-time transformation behavior of special relativity, we can rely on separate concepts of space and time whenever we choose a particular reference frame.

Philosophical doubts about our ability to handle uncountable sets of degrees of freedom lead us to consider finite volumes of increasing size instead of an infinite space. We thus obtain a countable set of momentum vectors providing infrared regularization. Philosophical considerations moreover suggest that irreversibility occurs naturally. We thus obtain a dynamic smoothing of fields providing ultraviolet regularization. The distinction between actual and potential infinities suggests the fundamental importance of limiting procedures, which should be postponed to the end of all calculations. A philosophical analysis of the quantum particle concept shows that unobservable free particles need to be grouped into observable clouds of free particles. Dissipative smearing sets the scale beyond which clouds cannot be resolved.

\emph{A priori} contemplations thus lead to all the concepts that are usually introduced only to address the pressing problem of divergencies. The entire story of Lagrangian quantum field theory arises in a perfectly natural way. Divergencies get remedied before they have a chance to occur. We thus obtain a sound mathematical framework that qualifies as a genuine image of nature for fundamental particles and their interactions.

On a more concrete level, we have shown that a quantum field theory is defined by a thermodynamically consistent quantum master equation for a density matrix evolving on a suitable Fock space that is largely defined by \emph{a priori} considerations. Alternatively, by unraveling of the quantum master equations, an interacting quantum field theory can be thought of as two coupled stochastic processes evolving in Fock space. These processes consist of a deterministic continuous Schr\"odinger-type evolution of state vectors interrupted by stochastic jumps. In this approach `quantum jumps' acquire a very specific, well-defined meaning and motivate correlation functions as natural quantities of interest even in the absence of observers. The advantages of thermodynamic regularization compared to other regularization mechanisms, such as a cutoff at high momenta, have been elaborated.

What do we gain? Most importantly, we gain mathematical rigor and conceptual clarity, with some distinct ontological implications to be considered in the subsequent section. All calculations of correlation functions as quantities of interest are based on robust mathematics. We completely avoid the problematic construction of an uncountably infinite set of field operators with prescribed commutation relations and the trouble resulting from inequivalent representations. Finally, the idea of unravelings opens the door for an entirely new simulation methodology, thus offering us an alternative to the simulation of lattice gauge theories.

\subsection{Implications for ontology}\label{secontology}
\index{Ontology}According to Margenau \cite{Margenau}, the multiply connected constructs of a theory possess physical \index{Reality}reality. From the position of \emph{constructive structural realism}, even unobservable entities of a theory expressing fundamental structural relations are considered to be real (for an inspiring discussion of structural realism see pp.\,5--8 of \cite{Caosr}). For example, Cao asks the question ``Are quarks real?'' (see p.\,238 of \cite{Caosr}). Although the quarks of quantum chromodynamics are \index{Confinement}confined and hence are not individually detectable in the laboratory, physicists are fully confident of the physical reality of quarks, based on the validity of quantum chromodynamics. On the other hand, Boltzmann didn't even insist on claiming reality for atoms (see Section \ref{secimagesBoltz}). During the 20th century, the metaphysical criteria for \index{Reality}reality have clearly been revised in a fundamental way.

Boltzmann's hesitation seems natural in view of his \index{Pluralism, scientific}scientific pluralism, or the \index{Underdetermination}underdetermination of theories by empirical evidences. How can the constructs of a theory be real if a multitude of alternative theories can exist, either in the course of time or even simultaneously? These constructs clearly would need to appear in all successful theories. But, using an example given by Cao, can the electron discovered by Sir Thomson in experiments with cathode rays in 1897 be identified with the electron of quantum electrodynamics? Cao justifies an affirmative answer. With increasing knowledge of reality it can even happen that the fundamental entities are down-graded to derived entities, but they continue to possess reality. These remarks can actually be condensed into an additional metaphysical postulate ruling the implications of an image of nature:\\

\index{Metaphysical postulates!fifth metaphysical postulate}
\noindent\framebox[\textwidth]{\parbox{\inboxwidth}{\emph{Fifth Metaphysical Postulate:}\label{metaphys5} The well-entrenched fundamental structural entities of a validated mathematical image of nature are real, even if they cannot be observed directly.}}\\[1.1ex]

Possessing a well-defined mathematical image of quantum field theory, we can now look into its ontological implications. So, what are the fundamental constructs of our approach to quantum field theory? For which entities can we claim reality?

The fundamental arena of our approach is a Fock space, which incorporates the idea of field quanta or quantum particles. By construction, these particles are independent. These countable quantum objects, which are assumed to possess momentum, spin if they are not scalar, and possibly further properties such as the isospin or color associated with weak and strong interactions, are the most fundamental entities of the theory. The choice of the list of fundamental particles occurs as an \index{Ontological commitment}ontological commitment of the theory. In our approach, these quantum particles are characterized by their well-defined momenta so that we lose spatial resolution.

In a next step, these quantum particles are promoted from `independent' to `free'. This step requires the definition of a straightforward free Hamiltonian based on the relativistic energy-momentum relation. A further contribution to the Hamiltonian introduces interactions in the form of collision rules. The corresponding collisions lead to the concept of clouds of free particles. The dissipative mechanism introduced into our image of nature then distinguishes between physically resolvable and unresolvable clouds of particles by setting a largest momentum or smallest length scale; the unresolvably small clouds may be considered as `interacting' particles. This concept depends on the presence of dissipation.

The unresolvable nature of interacting particles leads to the fact that free particles cannot be observed. In the spirit of constructive structural realism, we nevertheless feel that the free particles constitute the most fundamental \index{Reality}reality because they are the construct providing the basic arena for the entire theory. Moreover, the difference between free and interacting particles is experimentally unresolvable and can theoretically be described by a simple multiplicative factor. As a consequence of self-similarity on different length scales, this multiplicative factor can be determined along with the properties of physically resolvable clouds. In that sense, the theoretical relationship between unobservable free particles and observable interacting particles is well-understood within the theory so that the free particles can be observed indirectly. This argument strongly supports the idea that the free quanta can be considered as fundamental entities.

Also the fluctuating vacuum state of the interacting theory should be considered as an \index{Ontological commitment}ontological commitment (see p.\,206 of \cite{Caosr}). It is a consequence of the irreversible contribution to dynamics, which couples the free and interacting theories and makes the free vacuum state as unobservable as free particles.

Whereas free particles, collisions, and unresolvable clouds resulting from irreversible dynamics and leading to a fluctuating vacuum state of the interacting theory undoubtedly are the key constructs of our image of particle physics, the fundamental quantum master equation for an evolving density matrix is unattractive for developing a primitive
\index{Ontology!primitive ontology}
ontology of the ultimate physical entities. On the other hand, unravelings of master equations using Fock base vectors, continuous evolution and random jumps, are fundamentally about quantum particles and collision processes. For interactions happening through random jumps, a \emph{\index{Ontology!flash ontology}flash-type \index{Ontology!primitive ontology}primitive ontology of collisions between quantum particles} is most natural \cite{Allorietal08}. In our momentum representation, the locality of interactions is expressed as conservation of total momentum. Flash-type collision events are also what one directly observes in collider experiments. The tracks of the particles are in direct correspondence with their momenta and the emergence from a common vertex in a unpredictable position at a random time reflects the locality of the interaction expressed by momentum conservation. The need for two-process unravelings in which the dissipative interactions leading to clouds require simultaneous collisions paired in well-defined ways calls for further interpretational efforts.

There may be philosophical reservations against an \index{Ontology!particle ontology}ontology of free quantum particles because, in the presence of interactions, these quantum particles can be created or annihilated and, in that sense, cannot be regarded as a permanent or eternal substance. Cao hence suggests to interpret the quantum particles, or field quanta, as the manifestation of a field \index{Ontology!field ontology}ontology with the fields as the fundamental substance (see p.\,211 of \cite{Cao}). I find this rather a matter of wording than a satisfactory foundation. In my opinion, the complete absence of the field construct from the image of nature developed in this book certainly rules out a field \index{Ontology!field ontology}ontology.

If an eternal substance must be identified in a free quantum particle \index{Ontology!particle ontology}ontology, I suggest that it is the energy of these quantum particles that qualifies as a well-defined conserved substance, at least, within the short-time fluctuations allowed by Heisenberg's uncertainty principle (see footnote on p.\,\pageref{timeenergyuncert}).\footnote{Mass does not seem to be a good candidate for this lasting substance because massless particles, such as photons, would then not be taken into account.} Of course, this argument requires a finite energy density and thus depends on the regularization implied by the dissipative dynamics occurring according to our \index{Metaphysical postulates!fourth metaphysical postulate}fourth metaphysical postulate. In short, matter or substance is the energy content of the free quantum particles which, as a consequence of self-similarity, are indirectly observable as unresolvable clouds of free particles.

\section{Open mathematical problems}\label{secopenmathprob}
A fully convincing image of nature should be based on rigorous mathematics. For the approach developed in this book, there are a number of mathematical questions that deserve further consideration:
\begin{itemize}
  \item We have introduced the Fock space associated with the finite discrete set of momentum states from the $d$-dimensional lattice (\ref{Kdlatticedef}). Are there any useful alternative implementations of our regularization procedures? In particular, could an equivalent approach (or an equally rigorous approach) be developed for position states?
  \item In view of the discrepancy between the quantum master equations (\ref{QMEthermolowT}) and (\ref{QMEthermolowTi}): Can the formal zero-temperature limit (\ref{QMEthermolowTi}) of the thermodynamically consistent quantum master equation (\ref{QMEthermos}) be used throughout our developments or must the zero-temperature limit be postponed to the end of all calculations?
  \item Can the dissipation mechanism be implemented in such a way that the resulting quantum master equation possesses relativistic covariance? This would be nice to have for effective field theories and essential for making an attempt at quantum gravity, for which the dissipation mechanism might become a key part of the physics.
  \item The simplified irreversible dynamics (SID), which is obtained by replacing the collision operator (\ref{QMEsplitcoll}) by (\ref{QMEsplitcollSID}), seems to be a minor modification, but it spoils the exact thermodynamic consistency of the fundamental quantum master equation. Can SID be established as a valid approximation for all purposes of interest, or are there situations that require full thermodynamic consistency?
  \item Is there a simpler and hence more appealing two-process unraveling of the quantum master equation (\ref{QMEthermolowT})? Are there more robust unravelings for simulation purposes?
  \item Symbolic computation is the ideal tool for constructing perturbation expansions like (\ref{2corrphi4R012fin}) or (\ref{4corrphi4Lam1}). In particular, one could construct third- and higher-order perturbation expansions to discuss the resulting flow of the coupling constant as a function of the friction parameter.
  \item Only connected Feynman diagrams contribute to the second-order expression for the propagator of scalar field theory. Are also higher-order expansions for the propagator determined entirely by the connected Feynman diagrams introduced in Section \ref{secphi4proppert2}? For finite friction [if we do not make the approximations (\ref{QMEsplitcollSID}) and (\ref{Rfreesimplified})] or only in the limit of vanishing friction parameter?
  \item Can the infinite sums (\ref{verifyconv3}), which determine the behavior of the propagator for scalar field theory, be shown to be finite? Or are they finite only in the limit $V \rightarrow \infty$ in which the sums become integrals?
  \item Can the formulation (\ref{JJcommucarefulNO}) of the Schwinger term be obtained in a more rigorous way from (\ref{JJcommucareful})?
  \item Can the construction of fermion operators from boson operators in Section~\ref{secfrombostoferm} be made rigorous or is it limited to being merely a heuristic argument?
  \item Can the thermodynamic master equation be used to resolve the measurement process in a similar way as claimed by Petrosky and Prigogine \cite{PetroPrigo97} (see Section~\ref{sectionirrev})?
  \item Can we offer a mathematical formulation of a meaningful energy content of particle systems to provide the background for the \index{Ontology}ontology proposed at the end of Section~\ref{secontology}?
\end{itemize}

\section{Future developments}\label{secfuturedirec}
For the confrontation of our mathematical image with the real world, we should of course proceed from scalar field theory and quantum electrodynamics to the quantum field theory of all the fundamental interactions. Many challenging steps remain to be done:
\begin{itemize}
  \item A nicely general and systematic justification of the Bleuler-Gupta approach has been given in the context of BRST quantization (the acronym derives from the names of the authors of the original papers \cite{BecchiRouetStora76,Tyutin75}; for a nicely pedagogical BRST primer, see \cite{Nemeschanskyetal86}). The general idea is to quantize in an enlarged Hilbert space and to characterize the physically admissible states in terms of BRST charges, which generate \index{BRST quantization}BRST transformations and commute with the Hamiltonian. In this approach, BRST symmetry may be considered as a fundamental principle that replaces gauge symmetry \cite{Nemeschanskyetal86}. The BRST approach should be the starting point for generalizations to weak and strong interactions.
  \item The BRST approach to the canonical quantization of Yang-Mills theories has actually been developed in a series of papers by Kugo and Ojima \cite{KugoOjima78a,KugoOjima78b,KugoOjima79a,KugoOjima79b}. In addition to \index{Ghost particle}ghost particles, three- and four-gluon collisions make the kinetic theory for quantum chromodynamics more complicated than for quantum electrodynamics. For electroweak interactions, an additional complication arises: the Higgs particle needs to be included into the kinetic-theory description. Although the details become considerably more complicated, the ideas developed in this book can be generalized to describe the Yang-Mills theories for quantum chromodynamics and electroweak interactions and hence all parts of the standard model. The issue of the vacuum state in the presence of symmetry breaking deserves special attention.
  \item Electromagnetic, weak and strong interactions are represented by effective quantum field theories. These possess self-similarity which one can handle by choosing an ambiguous smallest length scale and relating different choices by renormalization. For gravity, their exists an unambiguous smallest length scale, the Planck scale, and presumably also a finite volume of the universe. Instead of the heavy restrictions on possible interactions resulting from the renormalization program allowing us to choose convenient minimal models within a large universality class, we are forced to choose a fundamental theory of gravity with restrictions coming only from mathematical elegance and simplicity (in accordance with our \index{Metaphysical postulates!first metaphysical postulate}first metaphysical postulate). As there is no need for limiting procedures associated with the smallest length scale and the finite volume, we have to formulate a theory with finite physical `cutoffs' but nevertheless possessing the appropriate physical symmetries. This is, in particular, a challenge for the formulation of the dissipative mechanism associated with the Planck scale. The physics of the Planck scale would then be a spatio-temporal smearing caused by dissipation in the most fundamental laws of nature.
  \item Lagrangian quantum field theory is focused on the investigation of scattering problems. Can dissipative quantum field theory also be employed to handle bound states and their statistical features? These are, for example, of crucial importance in the presence of \index{Confinement}confinement, which is a very important characteristic of strong interactions.
  \item It still needs to be demonstrated that the stochastic simulation techniques based on the kinetic-theory representation of quantum field theories \cite{hco214} are so powerful that they can be used to solve relevant problems, in particular, when regularization is provided by dissipation. Can they be used to solve problems inaccessible to simulations of lattice gauge theories?
\end{itemize}

\cleardoublepage

\chapter*{Author index}
\addcontentsline{toc}{chapter}{Author index}
\begin{multicols}{2}
\printauthorindex
\end{multicols}

\end{document}